\documentclass[12pt,english,american]{article}
\usepackage[T1]{fontenc}
\usepackage[latin9]{inputenc}
\usepackage{geometry}
\usepackage{cancel}
\usepackage{color}
\usepackage{babel}
\usepackage{amsthm}
\usepackage{amstext}
\usepackage{amssymb}
\usepackage{enumitem}
\usepackage{amsmath}
\usepackage[normalem]{ulem}
\usepackage{simplewick}
\usepackage[unicode=true,pdfusetitle,
 bookmarks=true,bookmarksnumbered=false,bookmarksopen=false,
 breaklinks=true,pdfborder={0 0 0},backref=false,colorlinks=true]
 {hyperref}
\hypersetup{
 linkcolor=blue,citecolor=blue, urlcolor=blue}

\makeatletter

\newcommand{\sipty}{{(\si)=\emptyset}}
\newcommand{\Omt}{\hat\Om} 
\renewcommand{\ln}{\log}
\newcommand{\cpsi}{\psi}

\usepackage{enumitem}		% customizable list environments
\theoremstyle{plain}
\newtheorem{thm}{\protect\theoremname}[section]
\theoremstyle{definition}

\theoremstyle{plain}
\newtheorem{prop}[thm]{\protect\propositionname}
\theoremstyle{remark}

\theoremstyle{plain}

\theoremstyle{plain}

\newcommand{\mrm}{\mathrm}

\usepackage{txfonts,refstyle,xcolor}

\newref{con}{name = Conjecture\ }
\newref{prop}{name = Proposition\ }
\newref{def}{name = Definition\ }
\newref{sec}{name = Section\ }
\newref{sub}{name = Section\ }
\newref{thm}{name = Theorem\ }
\newref{lem}{name = Lemma\ }
\newref{cor}{name = Corollary\ }
\newref{fig}{name = Figure\ }

\usepackage{bbm}
\newcommand{\charf}{1}  %{\mathbbm{1}}

\usepackage[all]{xy}

\newcommand{\xyR}[1]{%
     \makeatletter
     \xydef@\xymatrixrowsep@{#1}
     \makeatother
}

\newcommand{\xyC}[1]{%
     \makeatletter
     \xydef@\xymatrixcolsep@{#1}
     \makeatother
}

\newcommand{\bc}{ \color{black} }
\newcommand{\mc}{\color{black}}

\newcommand{\ncol}[1]{\color{normalcolor}}

\makeatother

\addto\captionsamerican{\renewcommand{\corollaryname}{Corollary}}
\addto\captionsamerican{\renewcommand{\definitionname}{Definition}}
\addto\captionsamerican{\renewcommand{\lemmaname}{Lemma}}
\addto\captionsamerican{\renewcommand{\propositionname}{Proposition}}
\addto\captionsamerican{\renewcommand{\remarkname}{Remark}}
\addto\captionsamerican{\renewcommand{\theoremname}{Theorem}}
\addto\captionsenglish{\renewcommand{\corollaryname}{Corollary}}
\addto\captionsenglish{\renewcommand{\definitionname}{Definition}}
\addto\captionsenglish{\renewcommand{\lemmaname}{Lemma}}
\addto\captionsenglish{\renewcommand{\propositionname}{Proposition}}
\addto\captionsenglish{\renewcommand{\remarkname}{Remark}}
\addto\captionsenglish{\renewcommand{\theoremname}{Theorem}}
\providecommand{\corollaryname}{Corollary}
\providecommand{\definitionname}{Definition}
\providecommand{\lemmaname}{Lemma}
\providecommand{\propositionname}{Proposition}
\providecommand{\remarkname}{Remark}
\providecommand{\theoremname}{Theorem}

\renewcommand{\1}{\!\!\!}
\newcommand{\kal}{\kappa^{(\la_0)}}
\newcommand{\tikal}{\ti{\kappa}^{(\la_0)}}
\newcommand{\3}{7}

\newcommand{\chit}{\ti\chi}
\newcommand{\at}{\mathrm{a}}

\newcommand{\emp}{\emptyset}
\newcommand{\sib}{(\si)}
\newcommand{\hv}{\hat{\mathrm{v}} }
\newcommand{\nv}{\mathrm{v}}

\newcommand{\hW}{\hat\W }

\newcommand{\Sin}{\mathrm{Si}}
\newcommand{\np}{\ov{n}}
\newcommand{\hveta}{\hat{\veta}}
\newcommand{\veta}{\varphi}
\newcommand{\uneps}{\underline{\eps}}
\newcommand{\one}{\mathbf{1}}
\newcommand{\app}{\tilde{\mathbf{1}}}
\newcommand{\ellp}{\ell'}
\newcommand{\sn}{t}
\newcommand{\tn}{s}
\renewcommand{\P}{p}

\renewcommand{\S}{\mathrm S}
\newcommand{\W}{\mathcal W}
\newcommand{\PSS}{S}
\newcommand{\ww}{\mathrm{w}}
\newcommand{\tiga}{\ga_0}

\newcommand{\ba}{\,|\,}
\newcommand{\nF}{V}

\newcommand{\s}{\mathrm{s}}
\newcommand{\Gn}{G_{1,n}}
\newcommand{\Hm}{G_{2,m}}
\newcommand{\tin}{\ti n}
\newcommand{\tim}{\ti m}
\newcommand{\ovGtin}{\ov G_{1,\ti m} }
\newcommand{\ovHtim}{\ov  G_{2,\ti n}}
\newcommand{\Gtin}{G_{2,\ti n}}
\newcommand{\Htim}{G_{1,\ti m}}
\newcommand{\mcD}{\mathcal D}
\newcommand{\B}{C_0^{2}}
\newcommand{\tic}{c}
\newcommand{\sym}{\mathrm{sym} }
\newcommand{\PS}{S}
\newcommand{\maxi}{\mathrm{max}}
\newcommand{\fib}{\mathrm{fi}}
\newcommand{\free}{\mathrm{fr}}
\newcommand{\I}{\mathrm I}
\newcommand{\ain}{\mathrm a}
\newcommand{\cin}{\mathrm c}
\newcommand{\hc}{\mathrm{h.c.}}
\newcommand{\mcA}{\mathcal A}
\newcommand{\mcB}{\mathcal B}
\newcommand{\mcC}{\mathcal C}
\newcommand{\el}{\mathrm e}

\newcommand{\kas}{\kappa_{*}}

\newcommand{\sig}{\rho}
\newcommand{\ci}{i_*'}

\newcommand{\pha}{\phantom}
\newcommand{\cvv}{\check{v} }
\newcommand{\cH}{\check H}
\newcommand{\unk}{\un k}
\newcommand{\unr}{\un r}
\newcommand{\hk}{\hat k }    %^{\ti m_1}
\newcommand{\hr}{\hat r } % ^{\ti n_1}
\newcommand{\chk}{\check k } % ^{\ti n_2}
\newcommand{\chr}{\check r } %  ^{\ti m_2}
\newcommand{\uchk}{\un{\check k}  }  %_{(\ti n_2)}
\newcommand{\uchr}{\un{\check r}  }  %_{(\ti m_2)}
\newcommand{\sip}{\si'}
\newcommand{\pho}{\mathrm{f}}
\newcommand{\alf}{\overline{\al}}
\newcommand{\nr}{\reta}
\newcommand{\nn}{\eta}
\newcommand{\neta}{\eta}
\newcommand{\reta}{\hat\eta}

\newcommand{\bb}{a}

\newcommand{\vv}{v}

\newcommand{\g}{\la }

\newcommand{\ti}{\tilde}

\newcommand{\un}{\underline}

\newcommand{\vac}{\Omega}
\newcommand{\Om}{\Omega}
\newcommand{\ga}{\gamma}

\newcommand{\ka}{\kappa}

\newcommand{\f}{\te{f}}
\newcommand{\be}{\beta}
\newcommand{\wh}{\widehat}
\newcommand{\pa}{\partial}

\newcommand{\ov}{\overline}
\newcommand{\vp}{\varphi}
\newcommand{\mfh}{\mathfrak{h}}

\newcommand{\eps}{\varepsilon}
\newcommand{\de}{\delta}
\newcommand{\De}{\Delta}
\newcommand{\e}{e}

\newcommand{\te}{\textrm}
\newcommand{\nin}{\noindent}
\newcommand{\si}{\sigma}
\newcommand{\ph}{\phantom}
\newcommand{\h}{\fr{1}{2}}
\newcommand{\nat}{\mathbb{N}}
\newcommand{\hil}{\mathcal{H}}
\newcommand{\om}{\omega}

\newcommand{\mco}{\mathcal{O}}

\newcommand{\supp}{\mathrm{supp}}
\newcommand{\fr}[2]{\frac{#1}{#2}}
\newcommand{\al}{\alpha}
\newcommand{\real}{\mathbb{R}}

\newcommand{\la}{\lambda}
\newcommand{\non}{\nonumber}
\newcommand{\Ga}{\Gamma}

\newcommand{\lan}{\langle}
\newcommand{\ran}{\rangle}

\def\proof{\noindent{\bf Proof. }}
\def\qed{$\Box$\medskip}
\newtheorem{theoreme}{Theorem } [section]
\newtheorem{proposition}[theoreme]{Proposition}
\newtheorem{lemma}[theoreme]{Lemma}
\newtheorem{definition}[theoreme]{Definition}
\newtheorem{corollary}[theoreme]{Corollary}
\newtheorem{remark}[theoreme]{Remark}
\newtheorem{example}[theoreme]{Example}
\newtheorem{criterion}[theoreme]{Criterion}

\newcommand{\beq}{\begin{equation}}
\newcommand{\eeq}{\end{equation}}
\newcommand{\beqa}{\begin{eqnarray}}
\newcommand{\eeqa}{\end{eqnarray}}
\newcommand{\ben}{\begin{arabicenumerate}}
\newcommand{\een}{\end{arabicenumerate}}
\newcommand{\bex}{\begin{example}}
\newcommand{\eex}{\end{example}}
\newcommand{\ber}{\begin{remark}}
\newcommand{\eer}{\end{remark}}
\newcommand{\bec}{\begin{corollary}}
\newcommand{\eec}{\end{corollary}}
\newcommand{\bep}{\begin{proposition}}
\newcommand{\eep}{\end{proposition}}
\newcommand{\becr}{\begin{criterion}}
\newcommand{\eecr}{\end{criterion}}

\newcommand{\cO}{O}
\newcommand{\cB}{\mathcal{B}}

\newcommand{\nee}{e}

\def\bel{\begin{lemma}}
\def\eel{\end{lemma}}
\def\bet{\begin{theoreme}}
\def\eet{\end{theoreme}}
\def\bed{\begin{definition}}
\def\eed{\end{definition}}

\def\hal{\hat\la}

\begin{document}
\title{Coulomb scattering in the massless Nelson model IV. Atom-electron scattering.  } 

\author{
{\bf Wojciech Dybalski}\\
Faculty of Mathematics and Computer Science \\  Adam Mickiewicz University in Pozna\'n\\
ul. Uniwersytetu Pozna\'nskiego 4, 61-614 Pozna\'n, Poland.\\
E-mail: {\tt wojciech.dybalski@amu.edu.pl}
\and
{\bf Alessandro Pizzo}\\
Dipartimento di Matematica, Universit\`a di Roma ``Tor
Vergata''\\ 
Via della Ricerca Scientifica, 1 - I--00133 Roma, Italy.\\
E-mail: {\tt pizzo@axp.mat.uniroma2.it}}

\date{}

\maketitle

%\maketitle

\begin{abstract}
We consider the massless Nelson model with two types of massive particles which we call atoms and electrons.
The atoms interact with photons via an infrared regular form-factor and thus they are Wigner-type particles
with sharp mass-shells.  The electrons have an infrared  singular form-factor and thus they are infraparticles
accompanied by soft-photon clouds correlated with their velocities. In the  weak coupling regime we construct scattering 
states of one atom and one electron, and demonstrate their asymptotic clustering into individual particles. The proof relies
on the Cook's argument, clustering estimates{\color{black},} and the non-stationary phase method. The latter technique requires sharp
estimates on derivatives of the ground state wave functions of the fiber Hamiltonians of the model, which were proven in the earlier papers of
this series.  Although we rely on earlier {\color{black}studies of} the atom-atom and electron-photon scattering in  the Nelson model,  the paper 
is written in a self-contained manner. A perspective on the open problem of the electron-electron scattering in this model is also given.

\end{abstract}

%TO DO \\
%\begin{enumerate}
%\item $\ka^{\la_0} \to \ka^{(\la_0)}$, Last lemma, symmetrization on page 53.
%\end{enumerate}

\section{Introduction}
\setcounter{equation}{0}

%{\mc kolor}

The infrared problem is a family of long-standing difficulties related to large scales and low energies, which affect    spectral and scattering theory
of  models of Quantum Field Theory (QFT). 
One aspect of this problem is to provide a consistent mathematical description 
of scattering processes involving 
massless particles (`photons'){\color{black},} massive electrically charged particles (`electrons'){\color{black},} and
massive electrically neutral particles (`atoms'). On the side of Functional Analysis, these questions concern a construction
of wave operators for some sophisticated self-adjoint operators. 
Scattering of atoms and photons is relatively
well understood both in models of non-relativistic Quantum Electrodynamics (QED) \cite{FGS01,  DP12.0} and in the relativistic framework of algebraic
QFT \cite{Bu77, Dy05, Du17,  Du15, He14}. Also collisions of \emph{one} electron and photons found consistent descriptions in both frameworks,
in spite of a complicated structure of electrically charged particles (`infraparticle problem') \cite{Pi05, CFP07, AD15}.
The fact that the electron always moves slower than the photons is the key decoupling
mechanism in this situation. This mechanism is absent in the next more difficult case,
which is the electron-atom scattering. To our knowledge, such collision processes 
are out of control  in any rigorous setting of QFT. The main result of this paper is
a construction of atom-electron scattering states in the massless Nelson model, whose infrared behaviour is similar to QED.
Our analysis provides a solid basis for a future construction of two-electron scattering states,
which is the ultimate goal of this series of papers.

Let us now describe in informal terms the setting and results of the present paper. The Hilbert space of our model
is $\hil= \Ga(\mfh_{\at})\otimes \Ga(\mfh_{\el})\otimes\Ga(\mfh_{\pho})$ where the respective factors are the Fock spaces
of atom, electron, and photon degrees of freedom over the single-particle spaces $\mfh_{\at}, \mfh_{\el}, \mfh_{\pho}$. The
Hamiltonian of the Nelson model describing {\color{black}the interaction} of these particles is a self-adjoint operator on $\hil$, formally
given by
\beqa
H\1&:=&\1\int d^3p\, \Omt(p)\nn_1^*(p)\nn_1(p)+\int d^3p\, \Omt(p)\nn_2^*(p)\nn_2(p)+\int d^3k\, \om(k) a^*(k)a(k)\label{free-evolution-intro}\\
& &+ \int d^3p d^3k\,\bigg(\vv_1(k)\,\nn_1^*(p+k)\bb(k)\nn_1(p)+   \vv_2(k)\,\nn_2^*(p+k)\bb(k)\nn_2(p)  +  \hc\bigg).
\label{full-auxiliary-model-intro}
\eeqa
Here $\eta^{(*)}_1(p), \eta^{(*)}_2(p), a^{(*)}(k)$ are the improper creation/annihilation operators of atoms, electrons, and photons;  $\Omt(p):=p^2/2$, and  $\om(k):=|k|$ are the `bare' dispersion relations of the massive and massless particles, respectively. 
We remark that the  terms in (\ref{free-evolution-intro}) describe  the free evolution of the three species of particles,
while (\ref{full-auxiliary-model-intro})  are the interaction terms. The distinction
between the two massive particle species is encoded in the form-factors $\vv_1, \vv_2$. They have the form
\beqa
\vv_1(k):=\la\fr{ \chi_{[0,\ka)}(k) |k|^{\alf}   }{\sqrt{2|k|}}, \quad  \vv_2(k):=\la\fr{ \chi_{[0,\ka)}(k) }{\sqrt{2|k|}}, \label{form-factors-intro}
\eeqa
where  $\la$ is the coupling constant, $\chi_{[a,b)}$ is an approximate characteristic function of the region $a\leq |k|\leq b$, the parameter $\ka$ is a fixed ultraviolet
cut-off and the parameter $\alf>0$ regularizes the interaction between atoms and photons. By alluding to the underlying classical system,
 the presence of the factor $|k|^{\alf}$ in the form-factor of the atom can be physically interpreted as zero  charge of this particle, while its
absence in the form-factor of the electron indicates that this particle is charged.

We start our discussion of scattering theory with identifying \emph{physical} single-particle states of the massive particles, which 
may not be elements of $\mfh_{\at}, \mfh_{\el}$ in the presence of interaction. The fact that  
 the Hamiltonian $H$ conserves the numbers of  massive particles simplifies this analysis. By restricting
$H$ to the respective subspaces we obtain Hamiltonians $H^{(n_1,n_2)}$ describing $n_1$ atoms and $n_2$ electrons interacting 
with photons. Alluding to {\color{black}the} translation invariance of the model, we can write the direct integral decompositions  
\beqa
H^{(1,0)}\simeq \int^{\oplus}d^3p\, H_{1,p}, \quad  H^{(0,1)}\simeq \int^{\oplus}d^3p\, H_{2,p},
\eeqa 
of the single-atom and single-electron Hamiltonians into the `fiber Hamiltonians' $H_{1,p}$, $H_{2,p}$ at fixed
momentum $p$. 
The decisive difference between the atoms and the electrons, which can be traced back to the
distinct form factors in (\ref{form-factors-intro}), is that $H_{1,p}$ has a ground state vector $\psi_p$, while
$H_{2,p}$ does not, in the relevant region of momenta $p$. In other words, denoting by $E_{i,p}$, $i=1,2$,
the lower boundary of the spectrum of $H_{i,p}$, it turns out  that $E_{1,p}$ is an eigenvalue while $E_{2,p}$ is not \cite{KM12,HH08,Da18}. 
The absence of the ground states of $H_{2,p}$
is a manifestation of the \emph{infraparticle problem} for the electron in the Nelson model.

Physical states of our model describing one atom in empty space are readily obtained
\beqa
\psi_{1,h}\simeq \int^{\oplus} d^3p\, h(p)\psi_{1,p}, \quad h\in L^2(\real^3, d^3p).
\eeqa
For the construction of scattering states it is crucial to provide \emph{renormalised creation operators}
which create such physical single-atom states from the vacuum vector $\Om\in \hil$.  They are given by 
\beqa
\nr^*_{1}(h):=\sum_{m=0}^{\infty}\fr{1}{\sqrt{m!}}\int d^3p\,d^{3m}k\,h(p) f_{1,p}^{m}(k_1,\ldots,k_m)\bb^*(k_1)\ldots \bb^*(k_m)
\nn^*(p- \unk), \label{renormalized-new-intro}
\eeqa     
where $\un{k}=k_1+\cdots+k_m$ and  $\{ f_{1,p}^{m} \}_{m\in \nat_0}$ are the $m$-photon wave functions of the vectors $\psi_{1,p}$. 
We refer here to `virtual' photons which together with the `bare' atom state $\eta_1^*(h)\Om$ form the physical single-atom state.
The complicated form of (\ref{renormalized-new-intro}) is dictated {\color{black}by} the requirement that 
\beqa
 \psi_{1,h}=\nr^*_{1}(h)\Om, \label{single-atom-intro}
\eeqa
and it is easy to see that 
$e^{iHt} \nr^*_{1}(h)\Om=\nr^*_1(e^{iE_1t}h)$, where $ (e^{iE_1t}h)(p)=
e^{iE_{1,p}t} h(p)$.
Thus natural candidates for scattering states of two atoms have the form
\beqa
\Psi^+_{\mrm{at},h,h'}=\lim_{t\to\infty} e^{iHt} \nr^*_{1}(e^{-iE_{1}t }h)  \nr^*_{1}( e^{-iE_{1}t }h') \Om, \label{two-atom-scattering}
\eeqa
where the interacting evolution $e^{iHt}$ should be compensated by the free
evolution $e^{-iE_{1}t }$ at asymptotic times, similarly as in  the quantum-mechanical scattering. Up to certain technical modifications (which disappear in the limit $t\to\infty$) we proved the existence of the limit in (\ref{two-atom-scattering}) in \cite{DP12.0} as a test case for the present investigation. In the absence of infrared problems similar results had been obtained in \cite{Al73, Al72, Fr73.1}.

The single-particle problem for the electron is considerably more difficult, since the ground states of the fiber Hamiltonians
$H_{2,p}$ are not available. The construction of single-electron states in models of non-relativistic QED,  
developed by  Fr\"ohlich,  Pizzo{\color{black},} and Chen   in \cite{Fr73, Fr74.1, Pi03, Pi05, CFP07, CFP09}, proceeds as follows:
First, one introduces  an infrared cut-off $\si>0$  by replacing the form factor $\vv_2$ in (\ref{form-factors-intro}) with $\vv_2^{\si}:=\la\fr{ \chi_{[\si,\ka)}(k) }{\sqrt{2|k|}}$ and denotes the resulting fiber Hamiltonians  $H_{2,\si}$. These
Hamiltonians have ground states $\psi_{2,p,\si}$ given by families of wave functions $\{ f^{m}_{2,p,\si} \}_{m\in \nat_0}$ and
we can define their renormalized creation operators $\nr^*_{2,\si}(h)$ by replacing $f_{1,p}^{m} $ with $ f^{m}_{2,p,\si}$
in (\ref{renormalized-new-intro}). However, the vectors $\nr^*_{2,\si}(h)\Om$ tend weakly to zero as $\si\to 0$, thus we
cannot use the analogue of formula (\ref{single-atom-intro}) to construct single-electron states. The infrared cut-off $\si$
can be removed only after dressing the renormalized single-electron states  $\nr^*_{2,\si}(h)\Om$ with 
soft-photon clouds. As these clouds are correlated with the electron velocity, we first need to divide the momentum spectrum
of the electron into cubes $\{\Ga_j^{(t)}\}_{j\in\nat}$, whose volumes tend to zero as $t\to \infty$. Denoting by $\nv_j$ the velocity in the center of each cube,  the soft-photon cloud is given by
\beqa
\W_{\si}(\nv_j,t):=\exp\bigg\{-\int d^3k \, \vv^{\si}_2(k) \fr{a(k)e^{i|k|t}-a^*(k)e^{-i|k|t}}{|k|(1-e_k\cdot \nv_j ) }   \bigg\},
\eeqa
where $e_k:=k/|k|$. Now physical single-electron states have the form
\beqa
\psi^+_{2,h}:=\lim_{t\to \infty} e^{iHt}\sum_{j}  \W_{\si}(\nv_j,t) \nr^*_{2,\si_t}(  e^{-iE_{2,\si_t}t } e^{i\ga_{\si_t}(\nv_j,t)}h_j^{(t)}  )\Om,
\label{single-electron-state-intro}
\eeqa
where $h_j^{(t)}$ is the restriction of $h$ to the $j$-th cell of the partition at time $t$,  the phase $\ga_{\si_t}(\nv_j,t)$ is needed
for technical reasons  which will be explained below,
 and the time-dependent cut-off $\si_t$ tends to zero as $t\to \infty$ at a suffciently fast rate. 
The existence of the limit in (\ref{single-electron-state-intro})  was  shown in \cite{Pi05}. We  remark that 
the general form of the approximating sequence in (\ref{single-electron-state-intro}) is consistent with the Dollard
formalism of long-range scattering  \cite{Dy17}. 

With the single-particle problem settled, we can move on to the main result of the present paper, which is
the construction of atom-electron scattering states. Given formulas  (\ref{single-atom-intro}),  
(\ref{two-atom-scattering}), (\ref{single-electron-state-intro}) it is not difficult to guess that they have the form
\beqa
\Psi^+_{h_1,h_2}:=\lim_{t\to\infty } e^{iHt}\sum_{j}  \W_{\si_t}(\nv_j,t)\nr^*_{1,\si_t}(e^{-iE_{1,\si_t}t }h{\color{black}_1}) \nr^*_{2,\si_t}(  e^{-iE_{2,\si_t}t } e^{i\ga_{\si_t}(\nv_j,t)}h_{{\color{black}2,}j}^{(t)}  )\Om, \label{atom-electron-intro}
\eeqa
where {\color{black}$h_{2,j}^{(t)}$ is the restriction of $h_2$ to the $j$-th cell of the partition at time $t$}, and where we introduced the time-dependent cut-off also for the atom for technical reasons. Apart from the existence of the
limit in (\ref{atom-electron-intro}) we also show the clustering property
\beqa
\lan \Psi^+_{h_1,h_2},  \Psi^+_{h_1', h_2'}\ran= \lan \psi_{1,h_1}, \psi_{1,h_1'}\ran \, \lan \psi_{2,h_2}^+, \psi_{2,h_2'}^+\ran
\label{intro-clustering}
\eeqa
which  says  that the states (\ref{atom-electron-intro})  asymptotically decouple into the constituent particles.
The  existence of the limit in (\ref{atom-electron-intro}) is shown by a combination 
of methods from the work \cite{Pi05} on single-electron states and the paper \cite{DP12.0} on scattering of two atoms.
Also the estimates on the derivatives of the wave functions $f^{m}_{1/2, p,\si}$, established in the companion 
papers \cite{DP12, DP16}{\color{black},} play a crucial role.  These sophisticated estimates, obtained by
iterative analytic perturbation theory combined with the theory of non-commutative recurrence relations, enter whenever
we refer to the  non-stationary phase method in the discussion below.

To explain the main steps of the proof, let us introduce the auxiliary
approximating vectors
\begin{align}
\Psi_{\si,j}(\tn,\sn)&:=e^{iH\sn}\W_{\si}(\nv_j,\sn) \reta^*_{1, \si}(h_{1,\sn})\reta^*_{2,\si}(h_{2,j,\sn}^{(\tn)} e^{i\ga_{\si}(\nv_j,\sn)})\Om, \label{vec-s-t}\\
\Phi_{\si,j}(s,t)&:=e^{iHt} \reta^*_{1, \si}(h_{1,t})\reta^*_{2,\si}(h_{2,j,t}^{(s)} e^{i\ga_{\si}(\nv_j,t)})\Om,
\end{align}
which are in obvious relation to (\ref{atom-electron-intro}) given that  {\color{black}$h_{1,t}(p):=e^{-i E_{1,p,\si} t} h_1(p)$ and $h^{(s)}_{2,j,t}(p):=e^{-i E_{2,p,\si} t} h^{(s)}_{2,j}(p)$}. Anticipating the use
of the Cook's method, we determine the time derivative of $\Psi_{\si,j}(\tn,\sn)$ and estimate its dependence on the parameters $\si,t$. (We leave aside the dependence on the partition via $j, s$ for the purpose of this introductory discussion).
 This derivative can be expressed as follows:
\beqa
\pa_{t}\Psi_{\si,j}(\tn,\sn) = \pa_t(e^{iH\sn}\W_{\si}(\nv_j,\sn) e^{-iH\sn})  \Phi_{\si,j}(s,t)+ 
(e^{iH\sn}\W_{\si}(\nv_j,\sn)e^{-iH\sn}) \pa_t \Phi_{\si,j}(s,t). \label{intro-first-computation}
\eeqa
As for the second term on the r.h.s. of (\ref{intro-first-computation}), we obtain 
 \begin{align}
\pa_{\sn}\Phi_{\si,j}(s,t) &=
e^{iHt}i[[H_{\I}^{\ain}, \reta^*_{1, \si}(h_{1,\sn})], \reta^*_{2,\si}(h_{2,j,\sn}^{(\tn)} e^{i\ga_{\si}(\nv_j,\sn)})]\Om
\label{zero-double-comm-intro}\\
&\ph{44} +e^{iH \sn}\reta^*_{1, \si}(h_{1,\sn})\reta^*_{2,\si}(i (\pa_t\ga_{\si}(\nv_j,\sn))  h_{2,j,\sn}^{(\tn)} e^{i\ga_{\si}(\nv_j,\sn)})\Om \label{zero-phase-term-intro}\\
&\ph{44} +e^{iHt}\bigg(i \reta^*_{1, \si}(h_{1,\sn})\cH_{\I}^{\cin}\reta^*_{2,\si}(h_{2,j,\sn}^{(\tn)} e^{i\ga_{\si}(\nv_j,\sn)})\Om+\{1 \leftrightarrow 2\}\bigg). \label{mismatch-intro}
\end{align}
In (\ref{zero-double-comm-intro}), which involves the  photon-annihilator part $H_{\I}^{\ain}$ of the interaction Hamiltonian (\ref{full-auxiliary-model-intro}), we recognize the double commutator expression familiar from our work on scattering of two atoms
\cite{DP12.0}. By adapting the non-stationary phase analysis from this reference we obtain 
$(\ref{zero-double-comm-intro})=O(\si^{-\de_{\la_0}} t^{-1-\rho} )$ for some fixed  $\rho>0$ and a parameter $\de_{\la_0}>0$ which can be made arbitrarily small at the cost of reducing the maximal value $\la_0$ of the coupling constant.
The term in (\ref{mismatch-intro}) results from a mismatch between the forward time evolution $e^{itH}$ without the infrared cut-off and the backward time-evolution $e^{-i E_{i, \si }t }$ with an infrared cut-off. It  
 involves the creation part of the interaction Hamiltonian for $|k|\leq \si$ and can be estimated by $O(\si^{1-\de_{\la_0}})$. Since
ultimately we will set $\si=\si_t\sim t^{-\ga}$ for some $\ga>0$ sufficiently large, this term does not cause any problems.
However,  the term in (\ref{zero-phase-term-intro}), involving the time-derivative of the phase, is not meant to be estimated.
Like in the single-electron case studied in \cite{Pi05}, its role is to cancel a slowly decaying term which originates from the
first term on the r.h.s. of (\ref{intro-first-computation}). In fact, by a non-trivial generalization of the analysis from \cite{Pi05}   we obtain
\beqa 
& & \pa_t(e^{iH\sn}\W_{\si}(\nv_j,\sn) e^{-iH\sn})  \Phi_{\si,j}(s,t)\non\\
& &\ph{4444444444} =e^{iH\sn}\reta^*_{1, \si}(h_{1,\sn})\reta^*_{2,\si}(-i (\pa_t\ga_{\si})(\nv_j,\sn)  h_{2,j,\sn}^{(\tn)} e^{i\ga_{\si}(\nv_j,\sn)})\Om+O(\si^{-\de_{\la_0}} t^{-1-\rho}) \label{intro-phase-cancellation}
\eeqa
and note the cancellation between the first term on the r.h.s. of (\ref{intro-phase-cancellation}) and (\ref{zero-phase-term-intro}).
Thus altogether we get
\beqa
\pa_{\sn}{\color{black}\Psi_{\si,j}(s,t)}=O\bigg( \fr{1}{\si^{\de_{\la_0}}} \fr{1}{ t^{1+\rho}}+  \si^{1-\de_{\la_0}}  \bigg)   \label{Cook-estimate-intro}
\eeqa
and all the error terms above  are understood in the Hilbert space norm.

Another important ingredient in the proof  of existence of the atom-electron scattering states~(\ref{atom-electron-intro})
are clustering estimates. By extending the non-stationary phase arguments from \cite{DP12.0} so as to incorporate the
soft-photon clouds and phases, we are able to show for $\sip\geq \si>0$ 
\beqa
\lan  \W_{\sip}(\nv_l,t) \reta^*_{1, \sip}(h'_{1,t})\reta^*_{2,\sip}(h_{2,l,t}^{'(t)} e^{i\ga_{\sip}(\nv_l,t)})\Om, \W_{\si}(\hv_j,t) 
\reta^*_{1, \si}(h_{1,t})\reta^*_{2,\si}(h_{2,j,t}^{(t)} e^{i\ga_{\si}(\hv_j,t)})\Om\ran \ph{4444444444}\non\\
= \lan \reta^*_{1, \sip}(h_{1,t}')\Om, \reta^*_{1, \si}(h_{1,t})\Om\ran 
 \lan \W_{\sip}(\nv_l,t) \reta^*_{2,\sip}(h_{2,l,t}^{'(s)} e^{i\ga_{\si}(\nv_l,t)})\Om, \W_{\si}(\hv_j,t) \reta^*_{2,\si}(h_{2,j,t}^{(s)} e^{i\ga_{\si}(\hv_j,t)})\Om\ran \label{physical-clustering}  \quad \\
\ph{444444444444444444444444444444444444444}+ O\bigg(\fr{1}{\si^{\de_{\la_0}}}\bigg(\fr{1}{t \si^{\eps}}+\fr{1}{t^{1-\eps}} +(\sip)^{\rho} \bigg) \bigg),\quad\quad\quad
\label{off-diagonal-corr-main-x-intro}
\eeqa
where  $\eps>0$ can be chosen arbitrarily small  {\bc at the cost of reducing also $\rho>0$ and adjusting the time-scale of the partition, see Subsection~\ref{single-electron-subsection} below.}  Apart from giving the clustering property (\ref{intro-clustering}), this estimate
is a crucial ingredient of the proof of convergence of (\ref{atom-electron-intro}), as  we will see below.  As we know 
from \cite{Pi05}, for $\sip=\si$ further clustering is possible, replacing (\ref{physical-clustering}) with
\beqa
\lan \reta^*_{1, \sip}(h_{1,t}')\Om, \reta^*_{1, \si}(h_{1,t})\Om\ran 
\lan  \W_{\sip}(\nv_l,t)\Om,   \W_{\si}(\hv_j,t) {\color{black}\Omega}\ran \lan \reta^*_{2,\sip}(h_{2,l,t}^{'(t)} e^{i\ga_{\si}(\nv_l,\infty)})\Om, \reta^*_{2,\si}(h_{2,j,t}^{(t)} e^{i\ga_{\si}(\hv_j,\infty)})\Om\ran. \quad \label{further-clustering}
\eeqa
This observation from \cite{Pi05}, combining the Cook's method with the fact that single-electron states (at fixed infrared cut-off)
are vacua of the asymptotic photon fields \cite{FGS04}, is recalled in  Subsection~\ref{clustering-subsection} below. 
We remark that such `clustering out' of soft-photon clouds does not seem possible for $\sip\neq \si$. 

With the ingredients (\ref{Cook-estimate-intro})--(\ref{further-clustering}) we can proceed to
the actual proof of the existence  of  the atom-electron scattering states  (\ref{atom-electron-intro}). We follow the general
strategy from \cite{Pi05}: Denoting by $\Psi_{h_1,h_2,t}:=\sum_{j} \Psi_{\si_t,j}(t,t)$ the approximating sequence on the
r.h.s. of (\ref{atom-electron-intro}), we aim at the estimate
\beqa
\|\Psi_{h_1,h_2,t_2}-\Psi_{h_1,h_2,t_1}\|\leq  c_{\la_0} \sum_{i=1}^{M}\fr{t_2^{\eps_i}}{t_1^{\eta_i}}    \label{telescopic-intro}
\eeqa
for $t_2\geq t_1$ and $\eta_i>2\eps_i$. This bound gives convergence via {\color{black}a} telescopic argument (cf. proof of Theorem~\ref{main-general-result} below). Estimate (\ref{telescopic-intro}) is proven in three steps:
\begin{enumerate}
\item[(a)] \emph{Change of the partition.} Relies on the variant (\ref{further-clustering}) of the clustering estimate for equal infrared cut-offs.
\item[(b)] \emph{Cook's argument.} Relies on estimate (\ref{Cook-estimate-intro}).
\item[(c)] \emph{Shift of the infrared cut-off.}  Clustering estimate (\ref{physical-clustering})--(\ref{off-diagonal-corr-main-x-intro}) for
different infrared cut-offs reduces the problem to the shift of the cut-off in the single-atom and  single-electron case. The latter problem was solved in \cite{Pi05} and we recall the argument in Appendix~\ref{Cut-off-variation}.
\end{enumerate}

A natural future direction of our project is to construct scattering states of two electrons in the Nelson model. 
The candidate scattering states have the form
\beqa
& &\Psi^+_{\mrm{el}, {h',h}}=\lim_{t\to\infty } e^{iHt}\sum_{j,j'}  \W_{\si_t}(\nv_j,t)  \W_{\si_t}(\nv_{j'},t)  \times \non\\
& &\ph{444444444444444444}\times e^{i\theta_{j,j'}}\nr^*_{2,\si_t}(e^{-iE_{2,\si_t}t } e^{i\ga_{\si_t}(\nv_{j'},t)} {\color{black}h'}_{j'}^{(t)}) \nr^*_{2,\si_t}(  e^{-iE_{2,\si_t}t } e^{i\ga_{\si_t}(\nv_j,t)}h_j^{(t)}  )\Om, \label{electron-electron-intro}
\eeqa
where $\theta_{j,j'}$ denotes a suitable two-particle Coulomb phase, which remains to be specified. Since our main  estimates
 (\ref{Cook-estimate-intro})--(\ref{further-clustering}) rely in a crucial way on the assumption
that $\alf>0$ in the atom form-factor in (\ref{form-factors-intro}), we are not yet able to control the limit in (\ref{electron-electron-intro}). 
However, preliminary computations suggest that a proper choice of the Coulomb phase may provide a generalization of the 
Cook's method estimate (\ref{Cook-estimate-intro}) to the two-electron case. Furthermore, by  applying the 
method mentioned below (\ref{further-clustering}), a counterpart of the clustering estimate 
(\ref{physical-clustering})--(\ref{further-clustering}) should follow, \emph{but only for $\sip=\si$}. With these ingredients we could accomplish  steps 
(a) and (b) of the proof of convergence above in the two-electron case. However, in step (c) we would not be able to reduce the problem  to the single electron case, due to the absence of the clustering estimate for two electrons for $\sip>\si$. This latter
problem is currently the main technical obstruction to the proof of existence of two-electron scattering states. A possible
strategy to circumvent this difficulty is to replace the approximating sequences in (\ref{electron-electron-intro}) and (\ref{single-electron-state-intro}) by formulas from \cite{Dy17}, suggested by the Dollard formalism. The advantage of the latter is that they are expressed in terms of infrared-finite quantities and thus a priori do not require a time-dependent infrared cut-off.

Our paper is organized as follows: In Section~\ref{preliminaries-and-results} we define the model, discuss more precisely the single-atom and single-electron states{\color{black},} and state our main result which is the existence of atom-electron scattering states.
In Section~\ref{preparations} we derive the Cook's method estimate (\ref{Cook-estimate-intro}) and the clustering estimates
(\ref{physical-clustering})--(\ref{further-clustering}). Section~\ref{convergence-atom-electron} is devoted to the proof of
estimate (\ref{telescopic-intro}), following the steps (a),(b),(c), mentioned above. 
The more technical part of the discussion is postponed to the appendices. In Appendix~\ref{spectral-theory-app} we recall
our spectral results from the companion papers  \cite{DP12, DP16}. In Appendix~\ref{domain-questions} we show
the self-adjointness of $H$ and some relevant domain problems. Appendix~\ref{Fock-space-combinatorics} is concerned
with Fock space combinatorics, especially the problem of keeping track of `contractions' of pairs of creation/annihilation operators.
These methods are applied in Appendix~\ref{Vacuum-exp-ren-creation} to compute expectation values of the renormalised
creation operators of electrons and photons, resulting  from scalar products of scattering states approximants. The time-dependence of these expressions is analysed in Appendix~\ref{non-stationary-app} using the method of non-stationary phase.
These last four appendices recall and generalize the relevant material from \cite{DP12.0}. 
Appendices~\ref{analysis-of-phase-I}, \ref{analysis-of-phase-II}  are devoted to the problem of cancellation of the phase and constitute a substantial generalisation of the corresponding problem from \cite{Pi05}. 
In Appendix~\ref{asymptotic-annihilation}  we give a new proof of the known fact that the massive single-particle states are vacua of the asymptotic photon fields. This is used in the clustering problem, as mentioned below (\ref{further-clustering}).  
Finally Appendix~\ref{Cut-off-variation}  gives a simplified version of an argument from \cite{Pi05} concerning the shift of
the infrared cut-off.

\vspace{1cm}

\noindent{\bf Acknowledgment:}  The authors are  grateful to J. Fr\"ohlich for the unpublished notes that have
inspired this series of  papers. W.D. was supported {\color{black}by} the DFG within the grants DY107/2-1, DY107/2-2.

\section{Preliminaries and results} \label{preliminaries-and-results}
\setcounter{equation}{0}

\subsection{The model}

We consider an interacting system of two types of  massive spinless bosons, which we  call  `atoms'  and `electrons',  
and massless spinless bosons, which we  call `photons'.   Let $\mfh_{\at},  \mfh_{\el},  \mfh_{\pho}$
be the respective single-particle spaces (all naturally isomorphic to $L^2(\real^3, d^3p)$) and 
$\Ga(\mfh_{\at}),  \Ga(\mfh_{\el}),  \Ga(\mfh_{\pho})$ the corresponding symmetric Fock spaces.
The (improper) creation/annihilation operators on $\Ga(\mfh_{\at}), \Ga(\mfh_{\el}), \Ga(\mfh_{\pho})$
will be denoted by $\nn_1^{(*)}(p), \nn_2^{(*)}(p), \bb^{(*)}(k)$, respectively. They satisfy the canonical commutation relations:
\beqa
& &[\nn_1(p),\nn_1^*(p')]=\de(p-p'), \ \  [\nn_2(p),\nn_2^*(p')]=\de(p-p'), \ \ [\bb(k),\bb^*(k')]=\de(k-k'),
\eeqa
with all other commutators being zero.

The free Hamiltonians of the atoms, electrons{\color{black},} and photons are given by
\beqa
H_{\at}:=\int d^3p\, \Omt(p)\nn_1^*(p)\nn_1(p), \ \ H_{\el}:=\int d^3p\, \Omt(p)\nn_2^*(p)\nn_2(p),\ \ H_{\pho}:=\int d^3k\, \om(k) a^*(k)a(k),
\eeqa
where $\Omt(p)=\fr{p^2}{2}$ and $\om(k)=|k|$. We recall that these operators 
are essentially self-adjoint on  $\mcC_{\at}$, $\mcC_{\el}$, $\mcC_{\pho}$, respectively, where $\mcC_{\at/\el/\pho}\subset \Ga(\mfh_{\at/\el/\pho})$
are  dense subspaces consisting of finite linear combinations of symmetrized tensor products of elements of $C_0^{\infty}(\real^3)$.

The physical Hilbert space of our system is $\hil:=\Ga(\mfh_{\at})\otimes \Ga(\mfh_{\el})\otimes  \Ga(\mfh_{\pho})${\color{black},}
and we will follow the standard convention to denote operators of the form $A\otimes 1\otimes 1$, $1\otimes B\otimes 1$,
$1\otimes 1\otimes C$,  by $A$, $B$, $C$, respectively.
The Hamiltonian describing the free evolution of the composite system of electrons and photons is given by
\beqa
H_{\free}:=H_{\at}+H_{\el}+H_{\pho}{\color{black},}
\eeqa
and it is essentially self-adjoint on  
\begin{equation}\label{def-C}
{\color{black} \mcC:=\mcC_{\at}\otimes \mcC_{\el}\otimes \mcC_{\pho}. }
\end{equation}

Now we introduce the interaction  between  atoms{\color{black},}
electrons{\color{black},} and photons. Let $\g \in \real$ s.t. $0<|\g|< 1$ be the coupling constant,  $\ka=1$ be the  ultraviolet cut-off\footnote{We set $\ka=1$ to simplify
the proofs of Theorem~\ref{preliminaries-on-spectrum} and Theorem~\ref{main-theorem-spectral}, given in the companion paper \cite{DP12}. 
In the present paper we will write $\ka$ explicitly.}{\color{black},}
 and let $1/2\geq \alf> 0$ be a parameter which controls the infrared regularity of atoms. 
Given these parameters, we define the form-factors
\beqa
\vv_1(k):=\la \fr{\chi_{\kappa}(k)|k|^{\alf} }{(2|k|)^{1/2}}, \quad 
\vv_2(k):=\la \fr{\chi_{\kappa}(k) }{(2|k|)^{1/2}}. \label{form-factor-definitions}
\eeqa
Here
$\chi_{\ka}\in C_0^{\infty}(\real^3)$ is rotationally invariant, non-increasing in the radial direction, supported in $\mcB_\ka$ and equal to one on $\mcB_{(1-\eps_0)\ka}$, for some fixed $0<\eps_0<1$. (We denote by $\mcB_r$ the open ball of radius $r$ centered at zero).
The interaction Hamiltonian, defined  as a symmetric  operator on $\mcC$, is given by the following formula 
\beqa
H_{\I}:=\sum_{i\in{1,2}}\int d^3p d^3k\,\vv_i(k)\big(\nn_i^*(p+k)\bb(k)\nn_i(p)+\hc\big). \label{interaction-Hamiltonian}
\eeqa
For future reference we denote by $H_{\I}^\ain$ (resp. $H_{\I}^\cin$) the terms involving $\bb(k)$ (resp.  $\bb^*(k)$) on the r.h.s. of (\ref{interaction-Hamiltonian}). In view of the presence of the factor $|k|^{\alf}$ in (\ref{form-factor-definitions}), we will say that the
interaction of atoms and photons is \emph{infrared regular}. In contrast, the interaction of electrons and photons is \emph{infrared singular}.

Proceeding analogously as in \cite{Fr73,Fr74.1}, the full Hamiltonian $H:=H_{\free}+H_{\I}$ 
can be defined as a self-adjoint {\color{black}operator} on a dense domain in $\hil$.  We outline briefly this construction: First,
we note that both $H_{\free}$ and $H_{\I}$ preserve the number of atoms and electrons. Let us therefore define
$\hil^{(n_1,n_2)}:=\Ga^{(n_1)}(\mfh_{\at})\otimes \Ga^{(n_2)}(\mfh_{\el}) \otimes \Ga(\mfh_{\pho})$, where $\Ga^{(n)}(\mfh_{\at / \el})$
is the $n$-particle subspace of $\Ga( \mfh_{\at /\el})${\color{black},} and let $H_{\free}^{(n_1,n_2)}$ and $H_{\I}^{(n_1,n_2)}$ be the restrictions
of the respective operators to  $\hil^{(n_1,n_2)}$, defined on $\mcC^{(n_1,n_2)}:=\mcC\cap \hil^{(n_1,n_2)}$. As shown in Lemma~\ref{self-adjointness},
using the Kato-Rellich theorem, each $H^{(n_1,n_2)}=H_{\free}^{(n_1,n_2)}+H_{\I}^{(n_1,n_2)}$ can be defined as a  
self-adjoint operator on the domain of $H^{(n_1,n_2)}_{\free}$, which is bounded from below and essentially self-adjoint on $\mcC^{(n_1,n_2)}$. Then we can define
\beqa
H:=\bigoplus_{(n_1,n_2)\in \nat_0^{\times 2}} H^{(n_1,n_2)}
\eeqa 
as an operator on $\mcC$.  Since $H^{(n_1,n_2)}\pm i$ have  dense ranges  on $\mcC^{(n_1,n_2)}$, the operators $H\pm i$ have  dense ranges on $\mcC$, 
thus $H$ is essentially self-adjoint on this domain.

On $\mcC$ we have the following formula for $H$
\def\Omz{\Om_0}
\def\nnz{\nn_0}
\def\vvz{\vv_0}
\beqa
H\1&=&\1\int d^3k\, \om(k) a^*(k)a(k)  \non\\
& &+\sum_{i\in\{1,2\}}\bigg(\int d^3p\, \Omt(p)\nn_i^*(p)\nn_i(p)+ (\int d^3p d^3k\,\vv_i(k)\nn_i^*(p+k)\bb(k)\nn_i(p)+\hc)\bigg).
\label{full-auxiliary-model}
\eeqa
It  reduces to a more familiar expression on $\mcC^{(n_1,n_2)}$
\beqa \label{ham-12}
H^{(n_1,n_2)}\1&=&\1\sum_{j=1}^{n_1}\fr{(i\nabla_{x_{1,j}})^2}{2}+\sum_{j=1}^{n_2}\fr{(i\nabla_{x_{2,j}})^2}{2}+\int d^3k\, \om(k) a^*(k)a(k)\non\\
& &+\sum_{j=1}^{n_1}\int d^3k\, \vv_1(k)\, ( e^{ikx_{1,j}}a(k)+e^{-ikx_{1,j}}a^*(k) )\non\\
& &+\sum_{j=1}^{n_2}\int d^3k\, \vv_2(k)\, ( e^{ikx_{2,j}}a(k)+e^{-ikx_{2,j}}a^*(k) ),
\label{explicit-Hamiltonian}
\eeqa
where $x_{1,j}$ is the position operator of the $j$-th atom and $x_{2,j}$ is the position operator of the $j$-th electron. 

As auxiliary objects we will also need the Hamiltonian $H_{\si}$ with an infrared cut-off $0<\si\leq \ka$. It is constructed
starting from the form-factors
\beqa
\vv_1^{\si}(k):=\la \fr{\chi_{[\si,\kappa)}(k)|k|^{\alf} }{(2|k|)^{1/2}}, \quad 
\vv_2^{\si}(k):=\la \fr{\chi_{[\si,\kappa)}(k) }{(2|k|)^{1/2}}, \label{form-factor-definitions-x}
\eeqa
and repeating the steps above. Here $\chi_{[\si,\ka)}(k)=\one_{\mcB'_{\si} }(k)\chi_{\ka}(k)$, where $\mcB'_{\si}$
is the complement of the ball of radius $\si$ and $\one_{\De}$ is the characteristic function of the set $\De$.
For future convenience we also introduce a function $\chit_{[\si,\ka)}$ on $\real$ s.t. $\chi_{[\si,\ka)}(k)=\chit_{[\si,\ka)}(|k|)$.

Finally, we introduce the atom, electron and photon momentum operators
\beqa
P_{\at}^\ell:=\int d^3p\, p^\ell\nn_1^*(p)\nn_1(p), \quad   P_{\el}^\ell:=\int d^3p\, p^{\ell}\nn_2^*(p)\nn_2(p),\quad P_{\pho}^\ell:=\int d^3k\, k^\ell a^*(k)a(k),
\eeqa
for $\ell\in\{1,2,3\}$, which are essentially self-adjoint  on $\mcC$. We  recall that $H$ is translationally invariant, that is {\color{black}$H$} commutes with the
total momentum operators $P^{\ell}$, given by
\beqa
P^\ell:=P_{\at}^\ell+P_{\el}^\ell+P_{\pho}^\ell,\quad  \ell\in\{1,2,3\},
\eeqa
which are  essentially self-adjoint on $\mcC$ as well.

%%%%%%%%%%%%%%%%%%%%%%%%%%%%%%%%%%%%%%%%%
\subsection{Fiber Hamiltonians and renormalized creation operators}
%%%%%%%%%%%%%%%%%%%%%%%%%%%%%%%%%%%%%%%%%

Due to  translation invariance, $H^{(1,0)}$ and $H^{(0,1)}$ can be decomposed into fiber Hamiltonians the usual way:
\beqa
H^{(1,0)}=\Pi^*\int^{\oplus}d^3p\, H^{(1)}_{1,p}\,\Pi, \quad  H^{(0,1)}=\Pi^*\int^{\oplus}d^3p\, H^{(1)}_{2,p}\,\Pi,
\eeqa 
where $\Pi:=Fe^{iP_{\pho}x}$ and $F$ is the Fourier transform in the atom/electron variables, and
\beqa
H_{1/2,p}^{(1)}:=\Omt(p-P_{\f})+H_{\pho}+\int d^3k\, \vv_{1/2}(k)\, ( b(k)+b^*(k) ).
\eeqa
As auxiliary quantities we also introduce fiber Hamiltonians with a fixed infrared cut-off $\si$: 
\beqa \label{ham-12-sigma}
 H_{1/2,p,\si} :=  H^{(1)}_{1/2,p,\si}:=\Omt(p-P_{\f})+H_{\pho}+\int d^3k\, \vv^{\si}_{1/2}(k)\, ( b(k)+b^*(k) ).
\eeqa
Their normalized ground state vectors, denoted  $\psi_{1/2,p,\si}$, correspond to the eigenvalues
$E_{1/2,p,\si}$ and have phases fixed in Definition 5.3 of \cite{DP12}.  Namely, the phases of the vectors $\psi_{i,P,\si}$, $i=1,2$, are determined by the definitions
\beqa \label{def-checkpsi}
{\color{black}\psi_{i,p,\sigma}:= W^*_{i, p,\sigma}\check{\phi}_{i,p,\sigma}}, \quad
 \check{\phi}_{i,p,\sigma}:=\frac{\oint_{\gamma}\frac{d{z}}{H_{i, p,\sigma}^{{ \mrm{w} }}-{z}}\Omega}{\|\oint_{\gamma}\frac{dz}{H_{i, p,\sigma}^{\mrm{w}}- z}\Omega\|}, \quad
 W_{i,p,\si}:=\e^{-\int d^3k \, \vv^{\si}_i(k) \fr{b(k)-b^*(k)}{|k|(1-e_k\cdot \nabla E_{i,p,\si} ) }  },
% \label{phi-def}
\eeqa
where $H_{i,p,\sigma}^{{\mrm{w}}}:=W_{i,p,\sigma}H_{i,p,\sigma}W_{i,p,\sigma}^{*}$, $\Omega$ is the vacuum vector in the fiber Fock space 
$\Ga(\mfh_{\fib})$, and $\gamma$ is a circle  enclosing no other point of the spectrum of 
$H_{i, p,\sigma}^{{\mrm{w}}}$ apart {\color{black}from} $E_{i,p,\sigma}$.  

%{\color{red} COMMENT: I wonder whether these phases should be included in the present  paper.}.  
For $p\in S$,  where $S:=\{\, p\in \real^3\,|\, |p|<1/3\,\}$,  $\la\in (0,\la_0]${\color{black},} and $\si\in (0, \kappa_{\la_0}]$, where $\la_0$, $\ka_{\la_0}$ {\color{black}are} sufficiently small,  
the eigenvalues $E_{1/2,p,\si}$ and wave functions $\{ f_{1/2,p,\si}^n\}$ of $\psi_{1/2,p,\si}$ satisfy the regularity properties 
proven in \cite{DP12} and listed in Appendix~\ref{spectral-theory-app}.

Now we define the renormalized creation operators of the massive particles
\beqa
\nr^*_{1/2,\si}(h):=\sum_{m=0}^{\infty}\fr{1}{\sqrt{m!}}\int d^3p\,d^{3m}k\,h(p) f_{1/2,p,\si}^{m}(k_1,\ldots,k_m)\bb^*(k_1)\ldots \bb^*(k_m)
\nn_{1/2}^*(p- \unk)\,; \label{renormalized-new}
\eeqa     
this lengthy expression can be condensed as follows
\beq
\nr_{1/2,\si}^*(h):=\sum_{m=0}^{\infty}\fr{1}{\sqrt{m!}}\int d^3p\,d^{3m}k\,h(p) f^{m}_{1/2,p,\si}(k)\bb^*(k)^m \, \nn_{1/2}^*(p- \unk),
\label{renormalized}
\eeq 
using the short-hand notation, which will appear frequently below:
\beqa
& &f^m_{1/2,p,\si}(k):=f^m_{1/2,p,\si}(k_1,\ldots, k_m),\\
& &\bb^*(k)^m:=\bb^*(k_1)\ldots\bb^*(k_m), \\
& &\un k:=k_1+\cdots +k_m.
\eeqa
It is shown in Lemma~\ref{ren-creation-operator} that $\nr_{1/2,\si}^*(h)$ and $\nr_{1/2,\si}^*(h_1)\nr_{1/2,\si}^*(h_2)$, for ${\color{black}h}, h_1, h_2\in \B(\PS)$,
 are well defined operators  on $\mcC$. Since $\nr_{1/2,\si}(h):=(\nr_{1/2,\si}^*(h))^*$ are obviously well defined on $\mcC$, we also obtain that
$\nr_{1/2,\si}^*(h)$ are closable.

%%%%%%%%%%%%%%%%%%%%%%%%%%%%%%%%%%%%%
\subsection{Single-atom states}
%%%%%%%%%%%%%%%%%%%%%%%%%%%%%%%%%%%%%

Since atoms are non-relativistic counterparts of Wigner particles, {\color{black}the} construction of the single-atom states is quite simple. For any $h\in L^2(\real^3, d^3p)$, $\supp\, h\subset S$, we define \beqa
\psi_{1,h,\si}=\Pi^*\int^{\oplus}d^3p\,  h(p) \psi_{1,p,\si}. \label{single-particle-trivial-one}
\eeqa
 The space of all such single-atom states will be denoted $\hil_{1,\si}$. We also note a simple relation 
\beqa
\psi_{1,h,\si}=\nr^*_{1,\si}(h)\Om.  \label{single-particle-trivial-two}
\eeqa
With the help of Theorem~\ref{preliminaries-on-spectrum} it is easy to show that $\psi_{1,h}:=\lim_{\si\to 0} \psi_{1,h,\si}$ exists.
The resulting subspace of single-atom states without the infrared cut-off  will be denoted by $\hil_{1}\subset \hil^{ (1,0)}$ 
%{\color{red}COMMENT: Where is $ \hil^{(1)}$ defined? It seems on p. 8 between {\bc (2.6)} and {\bc (2.7)} but it is not clear.}. 
By Lemma~\ref{fiber-ground-states-lemma}
below, for any two single-atom states  $\psi_{1,h}, \psi_{1,h'}$ as above 
\beqa
\lan \psi_{1,h}, \psi_{1,h'}\ran=\lan h, h'\ran,
\eeqa
where the scalar product on the l.h.s. above is in $\hil^{(1,0)}$ and {\color{black} the one} on the r.h.s. in $L^2(\real^3)$. 

%%%%%%%%%%%%%%%%%%%%%%%%%%%%%%%%%%%%%
\subsection{Single-electron states} \label{single-electron-subsection}
%%%%%%%%%%%%%%%%%%%%%%%%%%%%%%%%%%%%%
\newcommand{\hatk}{e_k}

Since electrons are infraparticles, the construction of single-electron states is more difficult. We outline it here following \cite{Pi05}.
We denote by $\psi_{2,p,\si}$ the normalized  ground states of the Hamiltonians $H^{(1)}_{2,p,\si}$, {\color{black}whose phases are fixed in (\ref{def-checkpsi})}. 
{\color{black}It is well known that the Hamiltonians $H_{2,p}^{(1)}$, $p\in S$, do not
have ground states and that $\mathrm{w-}\lim_{\si\to 0}\psi_{2,p,\si}=0$. }
However, substitutes for these ground states can be constructed as follows: We introduce the Weyl operator for $p\in S$ and
$0<\si\leq \ka$
\beqa
W_{p,\si}=
\e^{-\int d^3k \, \vv_2^{\si}(k)  \fr{(b(k)-b^*(k))}{|k|(1-\hatk\cdot\nabla E_{2,p,\si} )}     }, \label{dressing-trafo-intro}
\eeqa 
where $\hatk:=k/|k|$.
Now we {\color{black}recall} the transformed Hamiltonian
\beqa
H_{2,\P,\si}^{(1),\ww}=W_{p,\si}  H_{2,\P,\si}^{(1)} 
W_{p,\si}^*.
\eeqa
We will denote by $\phi_{2,p,\si}:=W_{p,\si}\psi_{2,p,\si}$ the normalized eigenvectors of the transformed Hamiltonians. In this case we have
the existence of the limit $\phi_{2,p}:=\lim_{\si\to 0} \phi_{2,p,\si}$.

\vspace{0.2cm}

\nin\bf Cell partition: \rm Let us consider a region in momentum space, for convenience a cube $V$ of volume equal to one, centered at zero. 
We now construct for $ 1\leq t$ a cell partition  of  $V$, according to the following recipe:
At time $1 \leq t$ the linear dimension of each cell is $1/2^{\np}$, 
where $\np\in\nat$ is s.t.
\def\peps{\ov{\eps}}
\beq
(2^{\np})^{1/\peps}\leq t< (2^{\np+1})^{1/\peps} \label{cell-partition-def}
\eeq 
for a small exponent $0<\peps<1/15$ fixed a posteriori. (The upper bound $1/15$ is convenient in Corollary~\ref{off-diagonal-corr-new-corr}).  Thus there are $N(t):=2^{3\np}\leq t^{3\peps}$ cells.
Each such cell is denoted $\Ga^{(t)}_j$ and the collection of all cells $\Ga^{(t)}$. 
We will also need an approximate characteristic function $\app_{\Ga_j^{(t)}}$ 
supported inside of $\Ga_j^{(t)}$, which tends to the sharp characteristic function of $\Ga^{(t)}_j$ as $t\to\infty$. 
We construct this function as follows:
%%%%%%%%%%%%%%%%%%%%%%%%%%%%%%%%%%%%%%%%%%%%%
\bed\label{partition-def} Let $\veta\in C^{\infty}(\real)$ be equal to zero on $(-\infty, 0)$, 
equal to one on $(1,\infty)$ and monotonously increasing. Let us fix $a>0$, $0<\uneps<1${\color{black},}
and define a function on $\real^3$:
\beqa
\hveta_{a,\uneps}(p)=\prod_{i=1}^3\veta\bigg(\fr{(p^i+a)}{a\uneps}\bigg)
\veta\bigg(-\fr{(p^i-a)}{a\uneps}\bigg).
\eeqa
This function is equal to one on the cube $[-a(1-\uneps), a(1-\uneps)]^{\times 3}$ and vanishes 
outside of $[-a,a]^{\times 3}$.
Now let $p_{j(\np)}$ be the position of the center of the $j$-th cell in the $\np$-th partition. 
We set
\beqa \label{def-tilde-identity}
\app_{\Ga_{j(\np(t))}^{(t)}}(p){\color{black}:}=\hveta_{2^{-(\np(t)+1)},\uneps(t)}(p-p_{j(\np(t))}),
\eeqa
where the dependence $t\mapsto \np(t)$ is restricted by (\ref{cell-partition-def}) and $t\mapsto \uneps(t)=t^{-\3\peps}$, 
where $\peps$ appeared in~(\ref{cell-partition-def}).
\eed
%%%%%%%%%%%%%%%%%
\newcommand{\Omm}{\mu}
%%%%%%%%%%%%%%%%%%%%%%%%%%%%%%%%%%%%%%%%%%%%%%%%%%%%%%%%
\nin We recall that such smooth partition was not needed in \cite{Pi05} in the context of Compton scattering. We need it
here, because Coulomb scattering requires much more detailed information about the localization of particles in space. 
%%%%%%%%%%%%%%%%%%%%%%%%%%%%%%%%%%%%%%%%%%%%%%%%
\vspace{0.2cm}

\nin\bf Phases: \rm     Let us define the one-particle phase similarly as in \cite{Pi05}:
\beqa
\ga_{\si}(\nv_j,t)(p)\!:=\!\left\{ \begin{array}{ll} -\int_1^t\bigg\{\int_{\si}^{\si_{\tau}^\S} d|k| \int d\Omm(e_k) \,\vv_2^{\si}(k)^2(2|k|)
\bigg(\fr{\cos(k\cdot\nabla E_{2,p,\si}\tau-|k|\tau)}{1-\hatk\cdot \nv_j }\bigg)  \bigg\} d\tau, & \textrm{ for }  t \leq \big(\fr{\kal}{\si}\big)^{\fr{1}{\al}},  
\label{slow-cutoff}\\ 
-\int_1^{{ \big(\fr{\kal}{\si}\big)^{ \fr{1}{\al} } } }\bigg\{ \int_{\si}^{\si_{\tau}^\S} d|k| \int d\Omm(e_k) \,  \vv_2^{\si}(k)^2(2|k|)
\bigg(\fr{\cos(k\cdot\nabla E_{2,p,\si}\tau-|k|\tau)}{1-\hatk\cdot \nv_j }\bigg)  \bigg\} d\tau, & \textrm{ for }  t> (\fr{\kal}{\si})^{\fr{1}{\al}},
\end{array}\right.
\eeqa
where $\nee_k:=(\sin \theta \cos \vp, \sin \theta \sin \vp, \cos\theta)$ is the normal vector to the unit sphere,  $d\Omm:=\sin\,\theta d\theta d\phi$ is the measure on the unit sphere,  $\tau\mapsto \si^{\S}_{\tau}=\kal\tau^{-\al}$, $1/2<\al<1$ and
$\kal$ is specified in the `Standing assumptions and conventions' below.
We also write $\dot{\ga}_{\si}(\nv_j,t)(p):=\pa_t\ga_{\si}(\nv_j,t)(p)$.
%%%%%%%%%%%%%%%%%%%%%%%%%%%%%%%%%%%%%%%%%%%%%%%
\vspace{0.2cm}

\nin\bf Photon clouds: \rm  The photon cloud associated with the cube $\Ga_j^{(t)}$ 
has the form:
\beq
\W_{\si}(\nv_j,t):=\exp\bigg\{-\int d^3k \, \vv^{\si}_2(k) \fr{a(k)e^{i|k|t}-a^*(k)e^{-i|k|t}}{|k|(1-\hatk\cdot \nv_j ) }   \bigg\},
\eeq
where $\nv_j:=\nabla E_{2,p_j,\si}$ is the velocity in the center of the cube $\Ga_j^{(t)}$. 

\vspace{0.2cm}

\nin\bf Single-electron  states: \rm  Let us set 
$h_{j}^{(s)}(p):=\app_{\Ga^{(s)}_j}(p)h(p)$, 
and define:
\beqa
& &\psi_{2,\si,j}(s,t):=e^{iHt}\W_{\si}(\nv_j,t)\reta^*_{\si}(e^{-iE_{2,p,\si}t}  e^{i\ga_{\si}(\nv_j,t)}  h_{j}^{(s)})\Om,\\
& &\psi_{2,h,t}:=\sum_{j\in \Ga^{(t)}} \psi_{{\color{black}2,}\si_t,j}(t,t),
\eeqa
for a  fast dependence of $t\mapsto \si_t$, i.e., $\si_t=\kal t^{-\ga}$, $4<\ga\leq \ga_0$. 
%%%%%%%%%%%%%%%%%%%%%%%%%%%%%%%%%%
\bet\label{one-infraparticle-theorem} \cite{Pi05}  Let $h\in C_0^2(S)$ be such that $\nabla E_{2,p}\neq 0$ for $p\in \supp\, h$. 
Then, under our standing assumptions listed below, the following limits exist
\beqa
\psi_{2,h}^+:=\lim_{t\to\infty}\psi_{2,h,t}
\eeqa
and are called  single-electron states. Furthermore, for any two states $\psi_{2,h}^+$\,\,, ${\color{black}\psi_{2,h'}^+}$ as above
\beqa
\lan \psi_{2,h}^+\,\,, \psi^+_{2,h'}\ran=\lan h, h'\ran,
\eeqa
where the scalar product on the l.h.s. above is in $\hil^{(1,0)}$ and {\color{black}the one} on the r.h.s. in $L^2(\real^3)$. 
\eet
%%%%%%%%%%%%%%%%%%%%%%%%%%%%
\begin{remark} From rotation invariance and strict convexity of $p\mapsto\nabla E_{2,p}$ one concludes that $ \nabla E_{2,p}\neq 0$
is equivalent to $p\neq 0$ for $p\in S$. (Cf. Theorem~\ref{preliminaries-on-spectrum}).
\end{remark}
Strictly speaking the proof of Theorem~\ref{one-infraparticle-theorem} in \cite{Pi05} was obtained for a sharp partition in momentum space. 
However, it will be clear from the proof of our main result (Theorem~\ref{main-general-result} below) that this result remains valid also
in the present case. (Note that the single-electron case is considered in Section~\ref{preparations} in parallel with the electron-atom case).

%%%%%%%%%%%%%%%%%%%%%%%%%%%%%%%%%%%%%%%%%%%%
\subsection{Results: Atom-electron scattering states}\label{Results-subsection}
%%%%%%%%%%%%%%%%%%%%%%%%%%%%%%%%%%%%%%%%%%%%

Given $h\in C_0^2(S)$, we define the corresponding velocity supports
\beqa
V_i(h):=  \{ \nabla E_{i,p} \,|\, p\in \supp\, h \}, \label{velocity-support}
\eeqa 
where $i=1$ pertains to the atom and $i=2$ to the electron. Now let $h_1, h_2\in C_0^2(S)$ 
be s.t. $V_1(h_1) \cap V_2(h_2)=\emptyset$ and $V_1(h_1), V_2(h_2)$ do not contain zero.
For each cube $j$ we write 
\beq
\Psi_{\si,j}(s,t):=e^{iHt}\W_{\si}(\nv_j,t)\reta^*_{1, \si}(h_{1,t})\reta^*_{2,\si}(h_{2,j,t}^{(s)} e^{i\ga_{\si}(\nv_j,t)})\Om, \label{approximating-vectors}
\eeq
where 
\begin{equation}
{\color{black}h_{1,t}(p):=\e^{-i E_{1,p,\si}t}h_1(p)\,\,,\,\,h_{2,j,t}^{(s)}:=\e^{-i E_{2,p,\si}t}h_{2,j}^{(s)}(p).} \label{def-h1t-h2jt}
\end{equation}
Now we set  $\si_t:=\kal/ t^{\gamma}$, $4<\ga\leq \ga_0$ (the fast cut-off), and define
\beqa
\Psi_{h_1,h_2,t}:=\sum_{j\in \Ga^{(t)}} \Psi_{\si_t,j}(t,t).
\eeqa
Our main result is  the following: 
%%%%%%%%%%%%%%%%%%%%%%%%%%%%%%%%%%
\bet\label{main-general-result}  Let $h_1, h_2$ be such that $V_1(h_1)$, $V_2(h_2)$ are disjoint and do not contain zero.
Then,  under our standing assumptions listed below,  the following limit exists
\beqa
\Psi^+_{h_1,h_2}=\lim_{t\to\infty}\Psi_{h_1,h_2,t} \label{atom-electron-scattering}
\eeqa
and is called the atom-electron scattering state. Furthermore, for any two states $\Psi^+_{h_1,h_2}$, $\Psi^+_{h_1', h_2'}$ 
as above, we have
\beqa
\lan \Psi^+_{h_1,h_2},  \Psi^+_{h_1', h_2'}\ran= \lan \psi_{1,h_1}, \psi_{1,h_1'}\ran \, \lan \psi_{2,h_2}^+, \psi_{2,h_2'}^+\ran, \label{main-thm-clustering}
\eeqa
where $ \psi_{1,h_1}, \psi_{1,h_1'}$ and $\psi_{2,h_2}^+,  \psi_{2,h_2'}^+$ are the constituent single-atom and single-electron states of the states 
$ \Psi^+_{h_1,h_2},  \Psi^+_{h_1', h_2'}$, respectively.
\eet
%%%%%%%%%%%%%%%%%%%%%%%%%%%%%%%%%%%%%
\proof In Theorem~\ref{main-technical-result} below we show that for some finite $M\in \nat$ and $\eta_i>2\eps_i\geq 0$
\beqa\label{telescop}
\|\Psi_{h_1,h_2,t_2}-\Psi_{h_1,h_2,t_1}\|\leq  c_{\la_0} \sum_{i=1}^{M}\fr{t_2^{\eps_i}}{t_1^{\eta_i}}.   
\eeqa
From this bound we obtain convergence via the telescopic argument: We set $\Psi(t):=\Psi_{h_1,h_2,t}$ and proceed as in the proof of Theorem~3.1 of \cite{Pi05}:
Suppose $t_1^n\leq t_2<t_1^{n+1}$. Then we can write
\beqa
\|\Psi(t_2)-\Psi(t_1)\| \1&\leq&\1 \bigg(\sum_{k=1}^{n-1}\|\Psi(t_1^{k+1})-\Psi(t_1^k)\|\bigg)+\|\Psi(t_2)- \Psi(t_1^n)\|
\leq  c_{\la_0}\sum_{i=1}^M\sum_{k=1}^{n} \fr{t_1^{(k+1)\eps_i}}{t_1^{k\eta_i}} \non\\
\1&\leq&\1 c_{\la_0} \sum_{i=1}^M\fr{t_1^{\eps_i}}{t_1^{\eta_i-\eps_i} }\sum_{k'=0}^{\infty} 
\fr{1}{t_1^{k'(\eta_i - \eps_i) }}
\leq c_{\la_0}\sum_{i=1}^{M}\fr{1}{t_1^{\eta_i-2\eps_i}} \fr{1}{1-(1/t_1)^{\eta_i-\eps_i} }.   
\eeqa
Since the last expression tends to zero as $t_1\to \infty$, we obtain  convergence of $t\mapsto \Psi(t)$ as $t\to\infty$. 

The clustering estimate~(\ref{main-thm-clustering}) follows from the existence of the limit in (\ref{atom-electron-scattering}) and from Theorem~\ref{off-diagonal-corr-new}.
In this latter theorem one sets $s=t$ and $\si=\sip=\si_t$. \qed
%%%%%%%%%%%%%%%%%%%%%%%%%%%%%%%%%%%%%%%%%%%%%%
%\vspace{1.0cm}
\newpage

\noindent \bf Standing assumptions and conventions\rm: 
\begin{enumerate}

\item We will only consider outgoing asymptotic configurations ($t\to\infty$), since the incoming case ($t\to-\infty$) is analogous. The corresponding asymptotic quantities will be denoted by an upper index $+$, for example $\Psi^+_{h_1,h_2}$ in (\ref{atom-electron-scattering}). We restrict the time parameter to $t\geq 1$.

\item The  parameters $\P_{\maxi}=1/3$ and  $1/2\geq \alf>0$ {\color{black}(introduced in (\ref{form-factor-definitions}))} are kept fixed in the remaining part of the paper.

\item  As specified in Proposition~\ref{preliminaries-on-spectrum} and Theorem~\ref{main-theorem-spectral}, 
the maximal coupling constant $\g_0>0$ corresponds to the values of $\P_{\maxi}$ and $1/2\geq \alf>0$ fixed above.  The value of  $\g_0$ will be reduced in the course of our discussion without explicit notice, but it
will remain non-zero.

\item We will set $\al=(1+16\peps)^{-1}$ {\color{black}(introduced in (\ref{slow-cutoff}))} and $\peps>0$ {\color{black}(introduced in (\ref{cell-partition-def}))} sufficiently
small, as in Corollary~\ref{Cook-corollary}. Specifically,
we require that $7\peps\leq 3$ for the proof of Proposition~\ref{cook-arg-prop} and $7\peps\leq \alf$ for the proof of Proposition~\ref{change-of-cut-off}.
In Appendix~\ref{Cut-off-variation} the value of $\peps$ is again reduced.

\item { We always assume that the coupling constant satisfies $|\la|\in (0, \la_0]$, where $\la_0$ is s.t. Theorems~\ref{preliminaries-on-spectrum} 
and \ref{main-theorem-spectral} hold.  We recall that the maximal value of the {\color{black}(infrared)} cut-off admitted by these theorems is $0< \ka_{\la_0}\leq \ka$.
We first reduce this maximal value to $\tikal:=\min\{ \ka_{\la_0}, (1-\eps_0)\ka \}$ 
to ensure that the infrared cut-off does not interfere with the
smooth ultraviolet cut-off. Next, we fix the functions $h_1, h_2$, appearing in Theorem~\ref{main-general-result}, whose velocity supports $V_i(h_i)$, $i=1,2$,
are disjoint. Now we choose $0< \kal \leq  \tikal$ 
 s.t. for $0<\si\leq  \kal $ the approximate velocity supports 
$V_{i, \si}(h_{i}):=\{ \nabla E_{i,p,\si} \,|\, p\in \supp\, h_{i} \}$, $i=1,2$, are also disjoint and the same is true for $h_1',h_2'$ also appearing in Theorem~\ref{main-general-result}. Such a choice of $\kal$ is possible due to  relation~(\ref{additional-spectral-relation}). The infrared cut-offs will always be restricted to 
$ \kal \geq\sip\geq \si>0$.}

\item We denote by $\ga\in (4,\ga_0]$ the parameter which controls the time dependence of the (fast) infrared cut-off, i.e., $\si_t={\kal}/t^{\ga}$.
The parameter $\ga_0$ is kept fixed until  the proof of the first estimate  in (\ref{cut-off-shifts}), given in Appendix~\ref{Cut-off-variation}, where $\ga$
is chosen sufficiently large {(and $t>1$)} {\color{black} that implies $\ga_0$  sufficiently large}. The parameter $\ga_0$ appears also in the definition of the
slow infrared cut-off   $\si_{\s,t}:={\kal}(\si_t/{\kal})^{1/(8\ga_0)}$ in Subsection~\ref{non-stationary-subsection}. 
Another slow cut-off
$ \si^{\S}_{t}=\fr{\kal}{t^{\al}}$, $1/2<\al<1$ was defined below (\ref{slow-cutoff}) and controls the phase.  {We note that for $t\geq 1$ we have
$0\leq \si_t\leq   \si^{\S}_{t} \leq      \si_{\s, t}\leq \kal$}. 

\item $\ti\g_0\mapsto \de_{\ti\g_0}$, $\ti\g_0\mapsto \de'_{\ti\g_0}$, will denote positive functions of $\ti\g_0\in (0,\la_0]$, which may differ from line to line,
 and have the property
\beqa
\lim_{\ti\g_0\to 0} \de_{\ti\g_0}=0,  \quad \lim_{\ti\g_0\to 0} \de'_{\ti\g_0}=0.
\eeqa
We will assume that $0<\de_{\g_0}< 1/(8\ga_0)$ (cf. the proof of Lemma~\ref{second-contribution-decay-new-old}). {Furthermore, in our estimates $\de_{\la_0}$ will always be chosen
sufficiently small without further notice (at the cost of reducing $\la_0$)}.

\item We will denote by $c,c',c''$ numerical constants which may depend on $\PS$,  $\eps_0$, $\ka$, $\alf$,  $\tiga$, $\al$ and functions $h_1,h_2$ 
but not on $\si$, $t$, $\la$, $\la_0$ or the electron and photon momenta.  (Independence of $\la_0$ is used  in the proofs of Lemmas~\ref{summation}, \ref{summation-ell}. If a constant depends on one of the latter parameters, this will be indicated
by a subscript, e.g.  $c_{\la_0}$). The values of the constants may change from line to line.

\item We will denote by $(p,q)\mapsto D(p,q)$, $(p,q)\mapsto D'(p,q)$ smooth, compactly supported functions on $\real^3\times\real^3$,
which may depend of $\PS$, $\eps_0$,  $\ka$, $\alf$, $\tiga$, $\al$ and functions $h_1,h_2$, but not on $\si$, $t$, $\la$, $\g_0$   or the  photon momenta.
These functions may change from line to line.

\item We will denote by $k=(k_1,\ldots, k_m)\in \real^{3m}$ a collection of photon variables. {For any such $k_i\in \real^3$ we denote by $(k_i^1, k_i^2, k_i^3)$ the components.}
A lower or upper index $m$ of a function indicates that it is a symmetric  function of $(k_1,\ldots, k_m)$.  
For example:
\beqa
f^m(k):=f^{m}(k_1,\ldots, k_m).
\eeqa  
Similarly, we set $a^*(k)^m:=a^*(k_1)\ldots a^*(k_m)$. We note that the order in which the components of $k$ are
listed is irrelevant, since they enter always into symmetric expressions.

\item We separate the atom, electron{\color{black},} and photon variables $p_1,p_2\in\real^3$, $k\in \real^{3m}$ by semicolons.
For example:
\beqa
G_m(p_1;p_2; k):=G_m(p_1;p_2; k_1,\ldots, k_m). \label{notation-G-m}
\eeqa

\item Two collections consisting of (1) atom and photon variables: $p_1\in \real^3, k\in \real^{3m}$, (2) electron and
photon variables: $p_2\in \real^3, r\in \real^{3m}$ are separated by a bar. For example  
\beqa
F_{m,n}(p_1; k \ba p_2; r):=F_{m,n}(p_1; k_1,\ldots, k_m \ba p_2; r_1,\ldots, r_n).
\eeqa

\item Given $k=(k_1,\ldots, k_m)$ we write $\un k:= k_1+\cdots +k_m$.

\item We denote by $e$ a generic vector on a unit sphere in $\real^3$. For $k\in\real^3$ we write $e_k:=k/|k|$.

\item {If $X$ is an element of a Banach space  and $\|X\|\leq cY$ for some $Y\in \real_+$ then we write $X=O(Y)$.  }

\end{enumerate}

{\color{black}For the reader's convenience we list the main symbols used in the next sections with a brief description  (within parentheses) of their role:
\begin{itemize}
\item $\lambda$ (coupling constant)
\item  $\lambda_0$ (upper bound of the coupling constant)
\item $\peps$ (parameter regulating the cell-partition)
\item $\uneps $ (parameter regulating the support of the approximated characteristic function of each cell)
\item $\eps_0$ (parameter entering  the support of the approximated characteristic function of a ball in $\mathbb{R}^3$)
\item $\ka$ (ultraviolet cut-off)
\item $\sigma$ (infrared cut-off)
\item $\si_t$ (time-dependent fast infrared cut-off)
\item $\kal$ (parameter entering the fast infrared cut-off)
\item $\gamma$ (exponent of the time variable in  the fast infrared cut-off)
\item $\ga_0$ (maximal value of $\gamma$; enters the definition of the slow cut-off $\si_{\s,t}$)
\item $\alf$  (parameter regulating the atom form factor) 
\item $\al$ (exponent related to the slow infrared cut-off)
\end{itemize}}

%%%%%%%%%%%%%%%%%%%%%%%%%
\section{Preparations}\label{preparations}
%%%%%%%%%%%%%%%%%%%%%%%%%
\setcounter{equation}{0}

\subsection{A  domain}
%%%%%%%%%%%%%%%%%%%%%%%%%%%%%%%%%%%%%%%%%%%%%%%%
In Theorem~\ref{non-diagonal-theorem} below we derive bounds for $\pa_t \Psi_{\si,j}(s,t)$, where $\Psi_{\si,j}(s,t)$ appeared in (\ref{approximating-vectors}) above.
For the purpose  of this derivation we introduce a suitable domain: First, we fix $l_1,l_2\in\nat_0$, $r_1,r_2>0$ and consider vectors of the form
\beqa
\Psi_{l_1,l_2}^{r_1,r_2}{\color{black}:}=\sum_{m=0}^{\infty}\fr{1}{\sqrt{m!}}\int d^{3l_1}p_1 d^{3l_2}p_2 d^{3m}k\, F_{l_1,l_2,m}(p_1{\color{black};}p_2;k)\nn_1^*(p_1)^{l_1}\nn_2^*(p_2)^{l_2}\bb^*(k)^m\vac, \label{psi-l}
\eeqa
where $F_{l_1,l_2,m}\in L^2_{\sym}(\mcB_{r_1}^{\times {(l_1+l_2)} }  \times\mcB_{r_2}^{\times m}, d^{3l_1}p_1 d^{3l_2}p_2d^{3m}k)$, i.e., 
$F_{l_1,l_2,m}$ are square-integrable functions, symmetric (independently) in their atom, electron{\color{black},} and photon variables.  {\color{black}The support
in each atom/electron (resp. photon) variable is a ball of radius $r_1$ (resp. $r_2$). }
Moreover, the norms of $F_{l_1,l_2,m}$ satisfy the bound
\beqa
\|F_{l_1,l_2,m}\|_2\leq \fr{c^m}{\sqrt{m!}} \label{factorial-bound}
\eeqa
for some $c\geq 0$, independent of $m$, which guarantees that the vector (\ref{psi-l}) is well defined. Now we set
\beqa
\mcD:=\mathrm{Span}\{\,  \Psi_{l_1,l_2}^{r_1,r_2} \,|\, l_1,l_2\in\nat_0, r_1,r_2>0\,\}, \label{invariant-domain}
\eeqa
where $\mathrm{Span}$ means finite linear combinations. This domain is dense and it contains $\mcC$ {\color{black}(see (\ref{def-C}))}. 
In Lemma~\ref{ren-creation-operator} we show  that $H, H_{\at}, H_{\el}, H_{\pho}, H_{\I}^{\ain/\cin}$ and $\nr_{1/2,\si}^*(h)$, $h\in \B(S)$, 
are well-defined  on $\mcD$ and leave this domain invariant. All equalities of operators in this paper are understood to hold on $\mcD$,
unless stated otherwise. 
%%%%%%%%%%%%%%%%%%%%%%%%%%%%%%%%%%%%%%%%%%%%%%%%%%%%%%%%%%%%%%%%%%%%
\subsection{Time derivatives}\label{sect-time-der}
%%%%%%%%%%%%%%%%%%%%%%%%%%%%%%%%%%%%%%%%%%%%%%%%%%%%%%%%%%%%%%%%%%%% 
In this subsection we will consider time-derivatives of various time-dependent families of  vectors, which will be needed for application of the Cook's method in Section~\ref{convergence-atom-electron}. {\color{black}To be more specific, we are interested in the derivative w.r.t. time $t$ of the expression
\begin{equation}
e^{iH \sn}\W_{\si}(\hv_j,\sn) \reta^*_{1, \si}(h_{1,\sn})\reta^*_{2,\si}(h_{2,j,\sn}^{(\tn)} e^{i\ga_{\si}(\hv_j,\sn)})\Om
\end{equation}
that, for convenience, we re-express as
\begin{equation}\label{broken-exp}
(e^{iH \sn}\W_{\si}(\hv_j,\sn) e^{-iH \sn})\,(e^{iH \sn}
\reta^*_{1, \si}(h_{1,\sn})\reta^*_{2,\si}(h_{2,j,\sn}^{(\tn)} e^{i\ga_{\si}(\hv_j,\sn)})\,\Om)\,.
\end{equation}
The time derivative of  (\ref{broken-exp}) is finally estimated in Corollary \ref{Cook-corollary} resulting from Theorem \ref{non-diagonal-theorem}. In order to study the derivative of (\ref{broken-exp}),  firstly, in Lemma \ref{clouds-cook-proposition} and Proposition \ref{phases-main-proposition},  we compute and analyse the derivative of the first factor (from the left) in (\ref{broken-exp}). Next, in Lemma  \ref{Cook-method-proposition}  we study the derivative of the second factor. Finally, in Theorem   \ref{non-diagonal-theorem} we combine the foregoing ingredients.}

We start by recalling the following simple lemma  \cite[Lemma 2.3]{DP12.0}. 
%%%%%%%%%%%%%%%%%%%%%%%%%%%%%%%%%%%%%%%%%%%%%%%%%%%%%%%%%%%%%%%%%%%%%
\bel\label{fiber-ground-states-lemma}  Let $h\in \B(\PSS)$  and $\psi_{i,h,\si}:=\nr_{i,\si}^*(h)\vac$, $i=1,2$. 
Then 
\beqa
\psi_{i,h,\si}=\Pi^*\int^{\oplus}d^3\P\, h(\P) \psi_{i,\P,\si}.
\eeqa
 Consequently, $H^{(1,0)}_{\si} \psi_{1,h,\si}= \psi_{1, E_{1,\si}h,\si}$,   $H^{(0,1)}_{\si} \psi_{2,h,\si}= \psi_{2, E_{2,\si}h,\si}$,  where $(E_{i,\si}h)(p):=E_{i,p,\si}h(p)$ {\color{black}and $H^{(n_1,n_2)}_{\si} $ is defined   as in  (\ref{ham-12}) by introducing the infrared cut-off $\sigma$ in the form factors $ \vv_1(k)$, $ \vv_2(k)$}.
\eel
%%%%%%%%%%%%%%%%%%%%%%%%%%%%%%%%%%%%%%%%%%%%%%%%%%%%%%%%%%%%%%%%%%%%%%%%

\nin To simplify some statements and proofs in the remaining part of this subsection we introduce:

\vspace{0.3cm}

\nin\bf Notation: \rm We will use the index $\sib\in \{\si, \emp\}$ to distinguish quantities with and without the infrared cut-off. For $\sib=\emp$ the index can
simply be omitted. For example $H_{\sib}\in \{H_{\si}, H\}$ or  $\vv_i^{\sib}\in \{ \vv_i^{\si}, \vv_i\}$.  

\vspace{0.3cm}
 
%%%%%%%%%%%%%%%%%%%%%%%%%%%%%%%%%%%%%%%%%%%%%%%%%%%%
\bel\label{clouds-cook-proposition} The identity below holds true on $ e^{iH_{\sib}\sn}\mcD$ 
\beqa
{\color{black}\pa_{\sn}(e^{iH_{\sib}\sn}\W_{\si}(\hv_j,\sn) e^{-iH_{\sib}\sn})=e^{iH_{\sib}\sn}\W_{\si}(\hv_j,\sn)\sum_{i\in \{1,2\}}A_{i,{\color{black}\sigma}}e^{-iH_{\sib}\sn}}, \label{derivative-first-bracket-x}
\eeqa
where %{\color{yellow}COMMENT: I would change the symbol $F_{\si,\sn}(k)$ to $F_{j,\si,\sn}(k)$ for consistency}
\beqa
A_{i, {\color{black}\sigma}}:=\int d^3p d^3k\,\vv_i^{\si}(k) \nn_i^*(p+k)  F_{j,\si,\sn}(k) \nn_i(p),\ \ 
 F_{j,\si,\sn}(k):=\fr{ \vv_2^{\si}(k)}{|k|} \bigg(  \fr{e^{-i|k|\sn} }{(1-\hatk\cdot \hv_j ) }+\fr{e^{i|k|\sn} }{(1+\hatk\cdot \hv_j ) } \bigg).
\eeqa
\eel
%%%%%%%%%%%%%%%%%%%%%%%%%%%%%%%%%%%%%%%%%%%%%%%%%%%%
\proof We compute
\beqa
\pa_{\sn}(e^{iH_{\sib} \sn}\W_{\si}(\hv_j,\sn) e^{-iH_{\sib} \sn})\1 &=& \1 e^{iH_{\sib} \sn}( i[H_{\sib}, \W_{\si}(\hv_j,\sn)]+\pa_{\sn} \W_{\si}(\hv_j,\sn) ) e^{-iH_{\sib} \sn}\non\\
\1 &=&\1  e^{iH_{\sib} \sn} i[H_{\I,\sib}, \W_{\si}(\hv_j,\sn)] e^{-iH_{\sib} \sn},
\eeqa
where we have used 
\beqa
i[\int d^3k\, \om(k) a^*(k)a(k), \W_{\si}(\hv_j,\sn)]=-\pa_{\sn} \W_{\si}(\hv_j,\sn).
\eeqa
Now we recall that
\begin{align}
\W_{\si}(\hv_j,\sn)&=\exp\bigg\{-\int d^3k \, \vv_2^{\si}(k) \fr{a(k)e^{i|k|\sn}-a^*(k)e^{-i|k|\sn}}{|k|(1-\hatk\cdot \hv_j ) }   \bigg\},  \\
H_{\I,\sib }&= \sum_{i\in \{1,2\}}\int d^3p d^3k\,\vv_i^{\sib}(k) \nn_i^*(p+k)  \big(\bb(k)+\bb^*(-k)  \big)\nn_i(p),
\label{recall-definitions}
\end{align}
where $\vv_i^{\sib}\in \{ \vv_i^{\si},  \vv_i\}$. Making use of the formulas, where $a(g):=\int d^3k \, a(k) \ov{g}(k)$, $a^*(g):=a(g)^*$, 
%{\color{red}COMMENT: I have the impression that $a(g)$ has never been defined.}
\beqa
& &[a(g), e^{a^*(f)-a(f)} ]=e^{a^*(f)-a(f)}\lan g,f\ran, \\
& &[a^*(g), e^{a^*(f)-a(f)} ]=e^{a^*(f)-a(f)}\lan f,g\ran, 
\label{Weyl-standard}
\eeqa
we get
\begin{align}
i[H_{\I,\sib}, \W_{\si}(\hv_j,\sn)]&=\W_{\si}(\hv_j,\sn)\sum_{i\in \{1,2\}}\int d^3p d^3k\,\vv_i^{\sib}(k) \nn_i^*(p+k) F_{\si,\sn}(k) \nn_i(p) \label{cloud-commutator}\\
&=\W_{\si}(\hv_j,\sn)\sum_{i\in \{1,2\}}A_{{\mc i}, {\color{black}\sigma}}, 
\end{align}
where 
\beqa
F_{j,\si,\sn}(k):=
\fr{ \vv_2^{\si}(k)}{|k|} \bigg(  \fr{e^{-i|k|\sn} }{(1-\hatk\cdot \hv_j ) }+\fr{e^{i|k|\sn} }{(1+\hatk\cdot \hv_j ) } \bigg). 
\eeqa
We note that  $\vv_i^{\sib}(k)\vv_2^{\si}(k)=\vv_i^{\si}(k)\vv_2^{\si}(k)$, thus
we can replace $\vv_i^{\sib}$ with $\vv_i^{\si}$ on the r.h.s. of (\ref{cloud-commutator}). This concludes the proof. \qed
%%%%%%%%%%%%%%%%%%%%%%%%%%%%%%%%%%%%%%%%%%%%%%%%%%%%%%%%%%%%%%%%%%%%%%%%%%%%%%%%%
%%%%%%%%%%%%%%%%%%%%%%%%%%%%%%%%%%%%%%%%%%%%%%%%%%%%%

{\color{black} Next Proposition \ref{phases-main-proposition} (whose
proof is postponed to the appendices)  is concerned with the action of the operator $$ \sum_{i\in \{1,2\}} A_{i,\sigma}   e^{-iH_{\sib}\sn}$$ on the second factor of (\ref{broken-exp}). In this proposition, we set a positive exponent $\delta$ depending on $\alpha$ (the parameter driving the slow cut-off)  and entering the control of Cook's argument.  Eventually,  in Corollary \ref{Cook-corollary}, we will set $\al=(1+16\peps)^{-1}$ and pick $\de=8\peps$, with $\peps>0$ sufficiently
small, where $\peps$ drives the rate of convergence to zero of  the cell size. Hence we will tune the slow infrared-cut off and the time dependence of the cell partition so to provide the convergence (as $t\to \infty$) of the vectors $\psi^{\sib}_{2,\si,j}(\tn,\sn)$ and $\Psi^{\sib}_{\si,j}(\tn,\sn)$,  implied by (\ref{simplified-estimate-zero}) and (\ref{simplified-estimate}). }
%%%%%%%%%%%%%%%%%%%%%%%%%%%%%%%%%%%%%%%%%%%%%%%%%%
\bep\label{phases-main-proposition} Let $A_{{1},{\color{black}\sigma}}, A_{{2},{\color{black}\sigma}}$ be as in Lemma~\ref{clouds-cook-proposition}. Then, for any  $0<\de<(\al^{-1}-1)$, $1\leq s\leq t$,
\begin{align}
 [A_{{\mc 1},{\color{black}\sigma}}, \reta^*_{1, \si}(h_{1,\sn})]\reta^*_{2,\si}(h_{2,j,\sn}^{(\tn)} e^{i\ga_{\si}(\hv_j,\sn)})\Om&= R^1_j(\si, t, s),\label{phases-estimate-x}\\
\reta^*_{1, \si}(h_{1,\sn})[A_{{\mc 2},  {\color{black}\sigma}}, \reta^*_{2,\si}(h_{2,j,\sn}^{(\tn)} e^{i\ga_{\si}(\hv_j,\sn)})]\Om&=\reta^*_{1, \si}(h_{1,\sn})\reta^*_{2,\si}(-i \dot\ga_{\si}(\hv_j,\sn)  h_{2,j,\sn}^{(\tn)} e^{i\ga_{\si}(\hv_j,\sn)})\Om+R^2_j(\si, t, s), \,\,\,\,\, \label{two-Regular-result-x}\\
\,[A_{ {\mc 2}, {\color{black}\sigma}}, \reta^*_{2,\si}(h_{2,j,\sn}^{(\tn)} e^{i\ga_{\si}(\hv_j,\sn)})]\Om&=\reta^*_{2,\si}(-i \dot\ga_{\si}(\hv_j,\sn)  h_{2,j,\sn}^{(\tn)} e^{i\ga_{\si}(\hv_j,\sn)})\Om+R^2_j(\si, t, s), \label{two-singular-result-x}
\end{align}
where 
\begin{align}
\|R^1_j(\si, t, s)\| &\leq \fr{c}{\si^{\de_{\la_0}}}\bigg((\si^{\S}_{t})^{1+\alf}+t(\si^{\S}_t)^{3}+  \fr{(\si^{\S}_t)^{\de}}{t}+\fr{1}{(\si^{\S}_t)^{1+\de} t^2} \bigg),\\
\|R^2_j(\si, t, s)\|&\leq \fr{c_{t}^2}{\si^{\de_{\la_0}}}\bigg( t(\si^{\S}_t)^{3}+  \fr{(\si^{\S}_t)^{\de}}{t}+\fr{1}{(\si^{\S}_t)^{1+\de} t^2}  \bigg),\label{two-phases-estimate-x}
\end{align}
$c_{t}:=c(\uneps(t) t^{-\peps})^{-1}$  and the slow cut-off $\si^{\S}_t=\kal t^{-\al}$ appeared in (\ref{slow-cutoff}).
\eep
%%%%%%%%%%%%%%%%%%%%%%%%%%%%%%%%%%%%%%%%%%%%%%%%%%%%
\proof For estimates~(\ref{two-Regular-result-x}), (\ref{two-singular-result-x}) see 
Proposition~\ref{two-Regular-phases-proposition}.  For the estimate of~(\ref{phases-estimate-x}) see
Proposition~\ref{Regular-phases-proposition}. \qed
%%%%%%%%%%%%%%%%%%%%%%%%%%%%%%%%%%%%%%%%%%%%%%%%%%%%%%
\bel\label{Cook-method-proposition}  Let us set
\begin{align}
\phi^{\sib}_{2,\si,j}(\tn,\sn)&:=e^{iH_{\sib}\sn}\reta^*_{2,\si}(h_{2,j,\sn}^{(\tn)} e^{i\ga_{\si}(\nv_j,\sn)})\Om, \\
\Phi^{\sib}_{\si,j}(s,t)&:=e^{iH_{\sib}t} \reta^*_{1, \si}(h_{1,t})\reta^*_{2,\si}(h_{2,j,t}^{(s)} e^{i\ga_{\si}(\nv_j,t)})\Om.
\end{align}
The following equalities  hold true:
\begin{align}
\pa_t\phi^{\sib}_{2,\si,j}(\tn,\sn)&=e^{iH_{\sib} \sn}\reta^*_{2,\si}(i \dot\ga_{\si}(\nv_j,\sn)  h_{2,j,\sn}^{(\tn)} e^{i\ga_{\si}(\nv_j,\sn)})\Om
+e^{iH_{\sib}t}i \cH_{\I,\sib}^{\cin}\reta^*_{2,\si}(h_{2,j,\sn}^{(\tn)} e^{i\ga_{\si}(\nv_j,\sn)})\Om,\\
\pa_{\sn}\Phi^{\sib}_{\si,j}(s,t) &=
e^{iH_{\sib}t}i[[H_{\I,\sib}^{\ain}, \reta^*_{1, \si}(h_{1,\sn})], \reta^*_{2,\si}(h_{2,j,\sn}^{(\tn)} e^{i\ga_{\si}(\nv_j,\sn)})]\Om\label{zero-double-comm}\\
&\ph{44} +e^{iH_{\sib} \sn}\reta^*_{1, \si}(h_{1,\sn})\reta^*_{2,\si}(i \dot\ga_{\si}(\nv_j,\sn)  h_{2,j,\sn}^{(\tn)} e^{i\ga_{\si}(\nv_j,\sn)})\Om \label{zero-phase-term}\\
&\ph{44} +e^{iH_{\sib}t}\bigg(i \reta^*_{1, \si}(h_{1,\sn})\cH_{\I,\sib}^{\cin}\reta^*_{2,\si}(h_{2,j,\sn}^{(\tn)} e^{i\ga_{\si}(\nv_j,\sn)})\Om+\{1 \leftrightarrow 2\}\bigg).
\label{zero-double-comm-rest-one}
\end{align}
The operators $H_{\I,\sib}^{\ain}$ and $\cH^{\cin}_{\I,\sib}$ are defined on $\mcC$ by
\begin{align}
H_{\I,\sib}^{\ain}&:=\sum_{i\in\{1,2\}}\int d^3p d^3k\,\vv_i^{\sib}(k)\nn_i^*(p+k)\bb(k)\nn_i(p),\label{annihilation-hamiltonian-cutoff} \\
\cH_{\I,\sib}^{\cin}&:=\sum_{i\in\{1,2\}}\int d^3p d^3k\,\cvv^{\sib}_i(k)\nn_i^*(p-k)\bb^*(k)\nn_i(p), \label{check-hamiltonian}
\end{align}
where $\cvv_1^{\sib}(k)\in {\color{black}\{ 0, \g\fr{\mathbf{1}_{\mcB_\si}(k) |k|^{\alf} }{(2|k|)^{1/2}}\}}$,  
$\cvv_2^{\sib}(k)\in {\color{black} \{0, \g\fr{\mathbf{1}_{\mcB_\si}(k)}{(2|k|)^{1/2}}\}}$, $\vv_i^{\sib}\in \{ \vv_i^{\si}, \vv_i\}$.
(In particular  $\cH_{\I,\si}^{\cin}=0$). 
\eel
%%%%%%%%%%%%%%%%%%%%%%%%%%%%%%%%%%%%%%%%%%%%%%%%%%%%%%%%
\proof We consider only the case $\sib=\emp$, since $\sib=\si$ is analogous and simpler. Also, we derive only the formula for 
$\pa_{\sn}\Phi^{\sib}_{\si,j}(s,t)$, as the formula for $\pa_t\phi^{\sib}_{2,\si,j}(\tn,\sn)$ can be obtained by similar and simpler steps.

We write $h_{2,t}^{\ga}:=h_{2,j,t}^{(\tn)} e^{i\ga_{\si}(\nv_j,\sn)}$ and note that it depends on $t$ both via the free evolution and $\ga$.
We compute
\begin{align}
\pa_{\sn}\Phi_{\si,j}(s,t) &=e^{iHt}iH\nr_{1,\si}^*(h_{1,t})\nr_{2,\si}^*(h_{2,t}^{\ga})\vac+e^{iHt}\nr_{1,\si}^*(\pa_t h_{1,t})\nr_{2,\si}^*(h_{2,t}^{\ga})\vac
+e^{iHt}\nr_{1,\si}^*(h_{1,t})\nr_{2,\si}^*(\pa_t (h_{2,t}^{\ga}))\vac\non\\
&=e^{iHt}iH\nr_{1,\si}^*(h_{1,t})\nr_{2,\si}^*(h_{2,t}^{\ga} )\vac-e^{iHt}\nr_{2,\si}^*(h_{2,t})\nr_{1,\si}^*(i (E_{1,\si}h_1)_{t})\vac\non\\
&\ph{44}-e^{iHt}\nr_{1,\si}^*(h_{1,t})\nr_{2,\si}^*(i(E_{2,\si} h_2)^{\ga}_{t})\vac  
+e^{iH \sn}\reta^*_{1, \si}(h_{1,\sn})\reta^*_{2,\si}(i \dot\ga_{\si}(\nv_j,\sn)  h_{2,\sn}^{\ga} )\Om, \label{first-derivative-formula}
\end{align}
where $(E_{i,\si}h_i)(p):=E_{i,p,\si}h_i(p)$, { and  $(E_{2,\si} h_2)^{\ga}_{t}:= e^{-i E_{2,\si}t } E_{2,\si} h_2 e^{i\ga_{\si}(\nv_j,\sn)}$}, $i=1,2$.
The first term on the r.h.s. above is well defined by Lemma~\ref{ren-creation-operator}. 
The equality 
\beqa
(\pa_t\nr_{1,\si}^*( h_{1,t}))\nr_{2,\si}^*(h_{2,t}^{\ga})\vac=\nr_{1,\si}^*(\pa_t h_{1,t})\nr_{2,\si}^*(h_{2,t}^{\ga})\vac
\eeqa
can easily be justified with the help of Lemmas~\ref{norms-of-scattering-states} and \ref{summation}.
Now we note the following identity, which is meaningful due to Lemma~\ref{ren-creation-operator}:
\begin{align}
iH\nr_{1,\si}^*(h_{1,t})\nr_{2,\si}^*(h_{2,t}^{\ga})\vac&=i[[H, \nr_{1,\si}^*(h_{1,t})],\nr_{2,\si}^*(h_{2,t}^{\ga} )]\vac
+\nr_{1,\si}^*(h_{1,t})i H  \nr_{2,\si}^*(h_{2,t}^{\ga})\vac\non\\
&\ph{44}+\nr_{2,\si}^*(h_{2,t}^{\ga})i H \nr_{1,\si}^*(h_{1,t})\vac, \label{H-double-commutator}
\end{align}
where we made use of the fact that $H\Om=0$.
As for the first term on the r.h.s. of (\ref{H-double-commutator}), we note that $[H_{\free}, \nr_{1,\si}^*(h_{1,t})]$ and $[H_{\I}^{\cin}, \nr_{1,\si}^*(h_{1,t})]$
are sums of products of creation operators and therefore commute with $\nr_{2,\si}^*(h_{2,t}^{\ga})$. Thus we get
\beqa
i[[H, \nr_{1,\si}^*(h_{1,t})],\nr_{2,\si}^*(h_{2,t}^{\ga})]\vac=i[[H_{\I,{\color{black}(\sigma)=\emptyset}}^{\ain} , \nr_{1,\si}^*(h_{1,t})],\nr_{2,\si}^*(h_{2,t}^{\ga})]\vac,
\eeqa
where $H_{\I, {\color{black}(\sigma)}}^{\ain}$ is given by (\ref{annihilation-hamiltonian-cutoff}).

As for the second  term on the r.h.s. of (\ref{H-double-commutator}), we obtain 
\begin{align}
\nr_{1,\si}^*(h_{1,t})i H  \nr_{2,\si}^*(h^{\ga}_{2,t})\vac&=\nr_{1,\si}^*(h_{1,t})i (H-H_{\si}) \nr_{2,\si}^*(h^{\ga}_{2,t})\vac+\nr_{1,\si}^*(h_{1,t})
\nr_{2,\si}^*(i(E_{2,\si}h_2^{\ga})_{t})\vac\non\\
&=\nr_{1,\si}^*(h_{1,t})i \cH_{\I,\sipty}^{\cin} \nr_{2,\si}^*(h^{\ga}_{2,t})\vac
+i\nr_{1,\si}^*(h_{1,t})\nr_{2,\si}^*( (E_{2,\si}h_2^{\ga})_{t})\vac, \label{Lemma-enters}
\end{align}
{where we wrote $\sipty$ (rather than dropping $\si$) in order to keep $\si$ in the notation}.
Here in the first step we applied Lemma~\ref{fiber-ground-states-lemma} and in the last step we made use of the fact that
the operator
\beqa
\cH^{\ain}_{\I,\sipty}:=\sum_{i\in\{1,2\}}\int d^3p d^3k\,\cvv^{\sipty}_i(k)\nn_i^*(p+k)\bb(k)\nn_i(p)
\eeqa
annihilates $\nr_{2,\si}^*(h_{2,t})\vac$ due to the fact that $\cvv^{\sipty}_i$ are supported below the infrared cut-off. 
As the last term on the r.h.s. of (\ref{H-double-commutator}) can be treated analogously, this concludes the proof. \qed
%%%%%%%%%%%%%%%%%%%%%%%%%%%%%%%%%%%%%%%%%%%%%%%%%%%%%%%%%%%%%%%%%%%%%%%%%%%%%%%%%

%%%%%%%%%%%%%%%%%%%%%%%%%%%%%%%%%%%%%%%%%%%%%%%%%%%
\bet\label{non-diagonal-theorem}  We define for $1\leq s\leq t$
\begin{align}
\psi^{\sib}_{2,\si,j}(\tn,\sn)&:=e^{iH_{\sib}\sn}\W_{\si}(\hv_j,\sn)\reta^*_{2,\si}(h_{2,j,\sn}^{(\tn)} e^{i\ga_{\si}(\hv_j,\sn)})\Om,  \label{single-particle-vector-for-I}\\
\Psi^{\sib}_{\si,j}(\tn,\sn)&:=e^{iH_{\sib}\sn}\W_{\si}(\hv_j,\sn) \reta^*_{1, \si}(h_{1,\sn})\reta^*_{2,\si}(h_{2,j,\sn}^{(\tn)} e^{i\ga_{\si}(\hv_j,\sn)})\Om.
\end{align}
Then {\color{black}for $0<\de<(\al^{-1}-1)$}
\begin{align}
\|\pa_t \psi^{\sib}_{2,\si,j}(\tn,\sn)\|&\leq  \fr{c_{t}^2}{\si^{\de_{\la_0}}}\bigg(t(\si^{\S}_{t})^{3}+  \fr{(\si^{\S}_{t})^{\de}}{t}+\fr{1}{(\si^{\S}_{t})^{1+\de} {t}^2}  \bigg)\\
& \ph{444}+ c\si^{1-\de_{\g_0}}, \label{H-check-contribution-one}\\
\|\pa_t \Psi^{\sib}_{\si,j}(\tn,\sn)\|&\leq \fr{c_{t}^2}{\si^{\de_{\la_0}}}\bigg((\si^{\S}_{t})^{1+\alf}+t(\si^{\S}_{t})^{3}+  \fr{(\si^{\S}_{t})^{\de}}{t}+\fr{1}{(\si^{\S}_{t})^{1+\de} {t}^2}  \bigg) \label{non-diagonal-main-lemma-two-x} \\ 
& \ph{444}+\fr{c_{1,t} }{\si^{\de_{\g_0}}  }  \fr{1}{(\kal)^2} \bigg(    \fr{\si_{t}^{\alf/(8\tiga)  } }{t}+\fr{1}{t^{2}\si_{t}^{1/(4\tiga)}}   \bigg)
\label{non-diagonal-main-lemma-two-new-new} \\
& \ph{444}+ c\si^{1-\de_{\g_0}}, \label{H-check-contribution}
\end{align}
where (\ref{H-check-contribution-one}), (\ref{H-check-contribution})  can be omitted in the case $\sib=\si$. We set above 
$c_{1,t}:=c( (\uneps(t) t^{-\peps})^{-2}+t^{1-\al})$ and $c_{t}:=c(\uneps(t) t^{-\peps})^{-1}$ {\color{black}where $ \uneps(t)=t^{-\3\peps}$}.
\eet
%%%%%%%%%%%%%%%%%%%%%%%%%%%%%%%%%%%%%%%%%%%%%%%%%%%%
\proof We discuss only $\Psi^{\sib}_{\si,j}(\tn,\sn)$, as the analysis of  $\psi^{\sib}_{2,\si,j}(\tn,\sn)$ is analogous and simpler. We write
\beqa
\Psi^{\sib}_{\si,j}(\tn,\sn)=(e^{iH_{\sib}\sn}\W_{\si}(\hv_j,\sn) e^{-iH_{\sib}\sn})( e^{iH_{\sib}\sn}\reta^*_{1, \si}(h_{1,\sn})\reta^*_{2,\si}(h_{2,j,\sn}^{(\tn)} e^{i\ga_{\si}(\hv_j,\sn)})\Om)
\label{approx-vector}  
\eeqa
and compute the derivative w.r.t. $\sn$. We will consider separately two terms
\beqa
\Psi^{\sib}_{\si,j}(\tn,\sn)^{(1)}=\pa_{\sn}(e^{iH_{\sib}\sn}\W_{\si}(\hv_j,\sn) e^{-iH_{\sib}\sn})( e^{iH_{\sib} \sn}\reta^*_{1, \si}(h_{1,\sn})\reta^*_{2,\si}(h_{2,j,\sn}^{(\tn)} e^{i\ga_{\si}(\hv_j,\sn)})\Om),\label{derivative-first-bracket}\\
\Psi^{\sib}_{\si,j}(\tn,\sn)^{(2)}= 
(e^{iH_{\sib}\sn}\W_{\si}(\hv_j,\sn) e^{-iH_{\sib} \sn}) \pa_{\sn} ( e^{iH_{\sib} \sn}\reta^*_{1, \si}(h_{1,\sn})\reta^*_{2,\si}(h_{2,j,\sn}^{(\tn)} e^{i\ga_{\si}(\hv_j,\sn)})\Om). \label{derivative-second-bracket}
\eeqa
We start with the analysis of (\ref{derivative-first-bracket}). 
Making use of Lemma~\ref{clouds-cook-proposition} we get
\beqa
\Psi^{\sib}_{\si,j}(\tn,\sn)^{(1)}\1&=&\1 e^{iH_{\sib}\sn}  \W_{\si}(\hv_j,\sn)(A_{ {1}, {\color{black}\sigma}}+
A_{ {2}, {\color{black}\sigma}})\reta^*_{1, \si}(h_{1,\sn})\reta^*_{2,\si}(h_{2,j,\sn}^{(\tn)} e^{i\ga_{\si}(\hv_j,\sn)})\Om\non\\
\1&=&\1 e^{iH_{\sib}\sn}  \W_{\si}(\hv_j,\sn) [A_{{1},  {\color{black}\sigma}}, \reta^*_{1, \si}(h_{1,\sn})]\reta^*_{2,\si}(h_{2,j,\sn}^{(\tn)} e^{i\ga_{\si}(\hv_j,\sn)})\Om\label{A-1-term}\\
& &+ e^{iH_{\sib}\sn}  \W_{\si}(\hv_j,\sn) \reta^*_{1, \si}(h_{1,\sn})[A_{{2},  {\color{black}\sigma}},   \reta^*_{2,\si}(h_{2,j,\sn}^{(\tn)} e^{i\ga_{\si}(\hv_j,\sn)})]\Om. \label{A-2-term}
\eeqa
As for (\ref{A-1-term}), (\ref{A-2-term}), Proposition~\ref{phases-main-proposition}  gives 
\begin{align}
& \| [A_{{\mc 1},  {\color{black}\sigma} }, \reta^*_{1, \si}(h_{1,\sn})]\reta^*_{2,\si}(h_{2,j,\sn}^{(\tn)} e^{i\ga_{\si}(\hv_j,\sn)})\Om\|\leq  
\fr{c}{\si^{\de_{\la_0}}}\bigg((\si^{\S}_{t})^{1+\alf}+t(\si^{\S}_t)^{3}+  \fr{(\si^{\S}_t)^{\de}}{t}+\fr{1}{(\si^{\S}_t)^{1+\de} t^2}  \bigg),
\label{first-comm-estimate}\\
& \reta^*_{1, \si}(h_{1,\sn})[A_{ {\mc 2},  {\color{black}\sigma}}, \reta^*_{2,\si}(h_{2,j,\sn}^{(\tn)} e^{i\ga_{\si}(\hv_j,\sn)})]\Om
=\reta^*_{1, \si}(h_{1,\sn})\reta^*_{2,\si}(-i \dot\ga_{\si}(\hv_j,\sn)  h_{2,j,\sn}^{(\tn)} e^{i\ga_{\si}(\hv_j,\sn)})\Om+R^2_j(\si, t, s), \label{three-two-Regular-result}
\end{align}
where 
\beqa
\|R^2_j(\si, t, s)\|\leq \fr{c_{t}^2}{\si^{\de_{\la_0}}}\bigg( t(\si^{\S}_t)^{3}+  \fr{(\si^{\S}_t)^{\de}}{t}+\fr{1}{(\si^{\S}_t)^{1+\de} t^2}  \bigg). 
\label{three-two-phases-estimate}
\eeqa
Relations~(\ref{three-two-phases-estimate}) and (\ref{first-comm-estimate}) give the contribution~(\ref{non-diagonal-main-lemma-two-x}) 
to the bound in the statement of the theorem.

Let us now have a look at (\ref{derivative-second-bracket}): We have by Lemma~\ref{Cook-method-proposition}  
\beqa
\pa_{\sn} ( e^{iH_{\sib} \sn}\reta^*_{1, \si}(h_{1,\sn})\reta^*_{2,\si}(h_{2,j,\sn}^{(\tn)} e^{i\ga_{\si}(\hv_j,\sn)})\Om)=
e^{iH_{\sib}t}i[[H_{\I,\sib}^{\ain}, \reta^*_{1, \si}(h_{1,\sn})], \reta^*_{2,\si}(h_{2,j,\sn}^{(\tn)} e^{i\ga_{\si}(\hv_j,\sn)})]\Om & &\label{double-comm}\\
+e^{iH_{\sib} \sn}\reta^*_{1, \si}(h_{1,\sn})\reta^*_{2,\si}(i \dot\ga_{\si}(\hv_j,\sn)  h_{2,j,\sn}^{(\tn)} e^{i\ga_{\si}(\hv_j,\sn)})\Om & & \label{phase-term}\\
+e^{iH_{\sib}t}\bigg(i \reta^*_{1, \si}(h_{1,\sn})\cH_{\I,\sib}^{\cin}\reta^*_{2,\si}(h_{2,j,\sn}^{(\tn)} e^{i\ga_{\si}(\hv_j,\sn)})\Om+\{1 \leftrightarrow 2\}\bigg).
\label{low-mode-term}& &
\eeqa
The first term on the r.h.s. of (\ref{three-two-Regular-result}), substituted in (\ref{A-2-term}) cancels (\ref{phase-term}) (cancellation of the phase). 
Concerning~(\ref{double-comm}), we obtain from  Proposition~\ref{double-commutator-proposition} that
\beqa
\|[ [H_{\I,\sib}^{\ain}, \nr_{1,\si}^*(h_{1,t})  ], \nr_{2,\si}^*( h_{2,j,\sn}^{(\tn)} e^{i\ga_{\si}(\hv_j,\sn)}   )]\vac\|\leq  
\fr{c_{1,t} }{\si^{\de_{\g_0}}  } \fr{1}{(\kal)^2}\bigg(    \fr{\si_t^{\alf/(8\tiga)  } }{t}+\fr{1}{t^2\si_t^{1/(4\tiga)}}   \bigg).
\eeqa
This gives rise to
contribution~(\ref{non-diagonal-main-lemma-two-new-new}) in the statement of the theorem. 

As for (\ref{low-mode-term}), which is non-zero only in the case $\sib=\emp$, we obtain immediately from Proposition~\ref{check-contribution} that it gives rise to contribution~(\ref{H-check-contribution}) in the statement of the theorem. This concludes the proof. \qed
%%%%%%%%%%%%%%%%%%%%%%%%%%%%%%
\bec\label{Cook-corollary} For $\al=(1+16\peps)^{-1}$, $\de=8\peps${\color{black},} and $\peps>0$ sufficiently
small, we have
\begin{align}
\|\pa_t \psi^{\sib}_{2,\si,j}(\tn,\sn)\|&\leq \fr{c_{\la_0}}{\si^{\de_{\la_0}}}\fr{1}{t^{1+7\peps}}+c\si^{1-\de_{\g_0}}, \label{simplified-estimate-zero}\\
\|\pa_t \Psi^{\sib}_{\si,j}(\tn,\sn)\|&\leq \fr{c_{\la_0}}{\si^{\de_{\la_0}}}\fr{1}{t^{1+7\peps}}+c\si^{1-\de_{\g_0}}, \label{simplified-estimate}
\end{align}
where the terms $c\si^{1-\de_{\g_0}}$  can be omitted for $\sib=\si$ and ${\color{black}c_{\la_0}}$ means that the constant may depend on $\la_0$.
\eec
%%%%%%%%%%%%%%%%%%%%%%%%%%%%%%%
%%%%%%%%%%%%%%%%%%%%%%%%%%%%%%%
\proof We consider only (\ref{simplified-estimate}) as the proof of (\ref{simplified-estimate-zero}) is similar and simpler. It suffices to choose  {\color{black} the} parameters, 
within the specified restrictions, so that (\ref{non-diagonal-main-lemma-two-x})--(\ref{H-check-contribution}) give the required bound. 
Since we set $\uneps(t)={t}^{-\3\peps}$, we have $\fr{1}{\uneps(t) t^{-\peps} }\leq t^{8\peps}$ and therefore $c_t^2:={\color{black}[c(\uneps(t) t^{-\peps})^{-1}]^2}\leq {\color{black}c^2} t^{16\peps}$. 
To ensure that the four contributions to (\ref{non-diagonal-main-lemma-two-x}) can be bounded by the first term on the r.h.s. of (\ref{simplified-estimate}), 
we recall that $\si^{\mathrm{S}}_t=\kal t^{-\al}$ and   demand
\beqa
16\peps+7\peps\1&<&\1 \al(1+\alf)-1, \quad \al(1+\alf)>1,  \label{first-condition-new}\\
16\peps+7\peps \1&<&\1 3\al-2,\quad\quad\quad\quad 3\al-1>1,\\
16\peps+ 7\peps \1&<&\1 \de\al,\\
16\peps+7\peps \1&<&\1 1-(1+\de)\al ,\quad\quad  (1+\de)\al<1. \label{last-condition-new}
\eeqa
We note that $3\al-1>1$ always holds if $\al(1+\alf)>1$ is true, (since $1/2\geq\alf>0$). 
Setting  $0<\de<\alf/4$,
we demand that the following condition holds:
\beq
(1+\alf)^{-1}<\al<(1+\de)^{-1}, \label{restriction-on-al}
\eeq
 which ensures $\al(1+\alf)>1$  and $(1+\de)\al<1$.  To be specific, we set
\beq
\al=(1+2\de)^{-1}, \label{def-al-new}
\eeq
which also ensures $1/2<\al<1$.
Given this, we can choose $\peps$ sufficiently small (depending {\color{black}on} $\alf$) so that  conditions (\ref{first-condition-new})-(\ref{last-condition-new})  
are met. 

Let us now consider (\ref{non-diagonal-main-lemma-two-new-new}). By (\ref{def-al-new}), we have $(1-\al)\leq 2\de$, thus
for $\de= 8\peps$  we get
\beqa
c_{1,t}\leq c t^{16\peps}.
\eeqa
Consequently, to ensure the required bound on (\ref{non-diagonal-main-lemma-two-new-new}), we impose the conditions
\beqa
16\peps+7\peps\1&<&\1 \fr{\alf \ga}{8\ga_0}, \label{first-condition-peps} \\
16\peps+7\peps\1&<&\1 1- \fr{\ga}{4\ga_0}, \label{second-condition-peps}
\eeqa
which hold for $\peps$ sufficiently small. \qed

%%%%%%%%%%%%%%%%%%%%%%%%%%%%%%%%

%%%%%%%%%%%%%%%%%%%%%%%%%%%%%%%%%%%%%%
\subsection{Clustering estimates} \label{clustering-subsection}
%%%%%%%%%%%%%%%%%%%%%%%%%%%%%%%%%%%%%%

In this section we discuss clustering of scalar products of approximating vectors into scalar products of their basic building blocks 
(atom, bare electron{\color{black},} and the photon cloud).  We start with the following theorem, which can be considered our main technical result. It
ensures clustering into physical particles (atom and the electron dressed with the cloud). 
%%%%%%%%%%%%%%%%%%%%%%%%%%%%%%%%%%%%%%%%%%%
\bet \label{off-diagonal-corr-new} For $\kal\geq \sip\geq \si>0$, there holds the estimate
\beqa
& &\lan  \W_{\sip}(\nv_l,s) \reta^*_{1, \sip}(h'_{1,s})\reta^*_{2,\sip}(h_{2,l,s}^{'(s)} e^{i\ga_{\sip}(\nv_l,s)})\Om, \W_{\si}(\hv_j,s) 
\reta^*_{1, \si}(h_{1,s})\reta^*_{2,\si}(h_{2,j,s}^{(s)} e^{i\ga_{\si}(\hv_j,s)})\Om\ran\non\\
& &= \lan \reta^*_{1, \sip}(h_{1,s}')\Om, \reta^*_{1, \si}(h_{1,s})\Om\ran \times \non\\
& &\ph{44444}\times\lan \W_{\sip}(\nv_l,s) \reta^*_{2,\sip}(h_{2,l,s}^{'(\tn)} e^{i\ga_{\si}(\nv_l,s)})\Om, \W_{\si}(\hv_j,s) \reta^*_{2,\si}(h_{2,j,s}^{(\tn)} e^{i\ga_{\si}(\hv_j,s)})\Om\ran
+ R^{W}_{l,j}(\si',\si,s),\quad\quad\quad
\label{off-diagonal-corr-main-x}
\eeqa
where
\beqa
|R^{W}_{l,j}(\si',\si,s)|\leq \fr{c}{\si^{\de_{\la_0}}}\fr{1}{\kal} \bigg\{ 
  \fr{1}{  s\si^{ 1/(8\tiga) } } +  \fr{1}{s^{1-8\peps}}   +(\sip)^{\alf/(8\ga_0)} \bigg\}.
\eeqa
\eet
%%%%%%%%%%%%%%%%%%%%%%%%%%%%%%%%%%%%%%%%%%%%%%%%%%%%%%%%%%%%%%%%%%%%%%%%%
\proof {\color{black} Follows from (\ref{formula-strong-clustering-proposition}) in Proposition~\ref{strong-clustering-proposition}, assuming $\un\eps(s):=s^{-7\bar\eps}$ and $s\leq t$.} \qed
%%%%%%%%%%%%%%%%%%%%%%%%%%%%%%%%%%%%%%%%%%%%%%%%%%%%%%%%%%%%%%%%%%%%%%%%%

 In the remaining part of this section we will also cluster out the photon clouds. We will consider only the case $\sip=\si$
which suffices for our proposes. In this case it is possible to use the {\color{black}Cook} method to accelerate clustering  as  noted in \cite{Pi05}.
For the reader's convenience, we recall this argument here.

Let us first  introduce the clouds depending on a parameter $\hal$:
\beqa
\W_{\si}^{\hal}(\hv_j,t):=\exp\bigg\{-\hal\int d^3k \, \vv^{\si}_2(k) \fr{a(k)e^{i|k|t}-a^*(k)e^{-i|k|t}}{|k|(1-\hatk\cdot \hv_j ) }   \bigg\}.
\eeqa
We also set for $\kal \geq \si>0$
\beqa
\hW_{\si,l,j}^{\hal}(\sn):=\W^{\hal}_{\si}(\nv_l,t)^*\W^{\hal}_{\si}(\hv_j,t). \label{two-Weyl-operators}
\eeqa
(We  shall suppress the dependence of the l.h.s. on $\nv_l, \hv_j$  in our notation, as it will be clear from the context). 
 We can write 
\begin{align}
\hW_{\si,l,j}^{\hal}(\sn)&=\exp\big\{-\hal \big(a(h_{l,j,t}^{\si})-a^*(h_{l,j,t}^{\si}) \big)    \big\},  \\
h_{l,j,t}^{\si}(k)&:=\fr{\vv_2^{\si}(k) f_{\hv_j,\nv_l}(e_k)}{|k|}\e^{-i|k|t},\label{h-photon-def}\\
f_{\hv_j,\nv_l}(e_k)&:= 
\fr{ \hatk\cdot(\hv_j-\nv_l)}{(1-\hatk\cdot \hv_j) (1-\hatk\cdot \nv_l)}. 
\end{align}
With these definitions, we  also have
\begin{align}
\lan \W_{\si}(\nv_l,t)\Om, \W_{\si}(\hv_j,t)\Om\ran= e^{-\fr{\|h_{l,j}^{\si} \|^2_2 }{2}}=:e^{-\h C_{l,j,\si}}.
\end{align}

\vspace{0.3cm}

\nin\bf Notation: \rm If there is no danger of confusion, we will use the following short-hand notation:
\beqa
& &\W_{\si}:=\W_{\si}(\hv_j,t), \quad\quad\quad \ \  \W_{\si}':=\W_{\si}(\nv_l,t),\quad  \quad a:= a(h_{l,j,t}^{\si} ),\quad \hW^{\hal}:=\hW_{\si,l,j}^{\hal}(\sn), \quad \\
& &\reta^*_{2,\si}:=\reta^*_{2,\si}(h^{(s)}_{2,j,t} e^{i\ga_{\si}(\hv_j,t)}), \quad \reta^{*\prime}_{2,\si}:=\reta^*_{2,\si}(h_{2,l,t}^{\prime (s)} e^{i\ga_{\si}(\nv_l,t)}).
\eeqa

\vspace{0.3cm}

\nin Similarly as in \cite{Pi05}, we will use the following observation:
%%%%%%%%%%%%%%%%%%%%%%%%%%%%%%%%%%%%%%%%%%%%%%%%%%%%%%%
\bel\label{basic-clustering-lemma} Let $\Psi, \Psi' \in \mcD$. Then
\beqa
\lan \W_{\si}'\Psi', \W_{\si}\Psi\ran=\lan \W_{\si}'\Om, \W_{\si}\Om\ran \, \lan \Psi', \Psi \ran
+ \int_0^1r^{\hal'}_{l,j,\si}(\sn)e^{-\fr{   \|h_{l,j}^{\si} \|^2_2  }{2}(1-(\hal')^2) } d\hal', \label{basic-clustering}
\eeqa
where
\beqa
r^{\hal}_{l,j,\si}(\sn)
:=-\lan (\hW^{\hal})^*\Psi', a \Psi\ran+\lan  a\Psi', 
   \hW^{\hal} \Psi\ran. 
\eeqa
\eel
%%%%%%%%%%%%%%%%%%%%%%%%%%%%%%%%%%%%%%%%%%%%%%%%%%%%%%%%
\proof With the help of relation~(\ref{Weyl-standard}) we show that $f(\hal,t):=\lan \W^{\hal}_{\si}(\nv_l,t)\Psi', \W_{\si}^{\hal}(\hv_j,t)\Psi\ran$ satisfies the following
differential equation
\beqa
\fr{ d f(\hal,t) }{d\hal}=-\hal  \|h_{l,j}^{\si} \|^2_2  f(\hal,t)  +r^{\hal}_{l,j,\si}(\sn), \label{M-equation}
\eeqa
whose solution is
\beqa
f(\hal,t)=e^{-\h  \|h_{l,j}^{\si} \|^2_2 \hal^2} f(0,t)+ \int_0^{\hal} r^{\hal'}_{ l,j,\si }(\sn)  e^{-\fr{ \|h_{l,j}^{\si} \|^2_2 }{2} (\hal^2-(\hal')^2) } d\hal'.
\label{solution}
\eeqa
By evaluating (\ref{solution}) at $\hal=1$ we obtain (\ref{basic-clustering}). \qed\\
%%%%%%%%%%%%%%%%%%%%%%%%%%%%%%%%%%%%%%%%%%%%%%%%%%%%%%%%%%%%%%%%%%%%%%%%
Lemma~\ref{basic-clustering-lemma} is useful if the rest term $r^{\hal}_{l,j,\si}(\sn)$  decays as $t\to\infty$. As noted in  \cite{Pi05}, this can be achieved  
exploiting the fact that single-particle vectors of the form $\nr^*_{1/2,\si}(h)\Om$ are annihilated by the asymptotic annihilation
operators of photons \cite{Pi05, FGS04}. For completeness, we provide an elementary proof of this fact, based on the Riemann-Lebesgue lemma, in Appendix~\ref{asymptotic-annihilation}.

%%%%%%%%%%%%%%%%%%%%%%%%%%%%%%
\bep \label{off-diagonal-thm-one}\cite{Pi05} The following equality holds true
\beqa
\lan \W_{\si}'  \reta^{*\prime}_{2,\si}\Om,  
\W_{\si} \reta^*_{2,\si}\Om\ran
= \lan\W_{\si}'\Om, \W_{\si}\Om\ran\, \lan  \reta^{*\prime}_{2,\si}\Om,  
\reta^*_{2,\si}\Om\ran 
+ o_{l,j,\si, \tn}(\sn),  \,\, \label{off-diagonal-one}
\eeqa
where $\lim_{\sn\to\infty}  o_{l,j,\si, \tn}(\sn)=0$. 
\eep
%%%%%%%%%%%%%%%%%%%%%%%%%%%%%%%%%%%%%%%%%%%%%
\proof We apply Lemma~\ref{basic-clustering-lemma} with
\beqa
\Psi=\reta^*_{2,\si}(h^{(\tn)}_{2,j,\sn} e^{i\ga_{\si}(\hv_j, \sn)})\Om, \quad
\Psi'=\reta^*_{2,\si}(h^{\prime (\tn)}_{2,l,\sn} e^{i\ga_{\si}(\nv_l, \sn)})\Om.
\eeqa
We note that {by Lemma~\ref{fiber-ground-states-lemma}}
\beqa
& &a(h_{l,j,t}^{\si})\Psi=a(h_{l,j,t}^{\si})\reta^*_{2,\si}(h^{(\tn)}_{2,j, \sn} e^{i\ga_{\si}(\hv_j, \sn)})\Om
=a(h_{l,j,t}^{\si}) e^{-iH_{\si} t} \reta^*_{2,\si}(h^{(\tn)}_{2,j} e^{i\ga_{\si}(\hv_j, \sn)})\Om, \label{asymptotic-vacuum}
\eeqa
and analogously for $\Psi'$. Since $\ga_{\si}(\hv_j, \sn)$ is constant for sufficiently large $t$ and fixed $\si$ (see (\ref{slow-cutoff})) and the single-particle state
$\reta^*_{2,\si}(h^{(\tn)}_{2,j} e^{i\ga_{\si}(\hv_j, \infty)})\Om$ is a vacuum of the asymptotic annihilation operator (see Appendix~\ref{asymptotic-annihilation}) we have
\beqa
\lim_{\sn\to\infty}|r^{\hal'}_{l,j,\sip,\si}(\tn,\sn)|=0.
\eeqa
This concludes the proof. \qed
%%%%%%%%%%%%%%%%%%%%%%%%%%%%%%%%%%%%%%%%%%%%%%%%%%%%%%%%%%%%%%%%%
\bet \label{off-diagonal-corr}\cite{Pi05} There holds the bound
\beqa
\lan  \W_{\si}(\nv_l,s) \reta^*_{2,\si}(h_{2,l,s}^{'(s)} e^{i\ga_{\si}(\nv_l,s)})\Om, \W_{\si}(\hv_j,s) 
\reta^*_{2,\si}(h_{2,j,s}^{(s)} e^{i\ga_{\si}(\hv_j,s)})\Om\ran\ph{444444444444444444444}\non\\
=\lan \W_{\si}(\nv_l)\Om, \W_{\si}(\hv_j)\Om\ran    \
\lan \reta^*_{2,\si}(h_{2,l}^{'(\tn)} e^{i\ga_{\si}(\nv_l,\infty)})\Om, \reta^*_{2,\si}(h_{2,j}^{(\tn)} e^{i\ga_{\si}(\hv_j,\infty)})\Om\ran
+O\bigg( \fr{1}{\si^{ \de_{\la_0} }}\bigg(\fr{1}{s^{7\peps}}\bigg)\bigg).
\label{off-diagonal-corr-main-zero}
\eeqa
\eet

%%%%%%%%%%%%%%%%%%%%%%%%%%%%%%%%%%%%%%%%%%%%%%%%%%%%%%%%%%%%%%%%%%%%%%%%%
\begin{remark} The power $6\peps$ (out of $7\peps$) in the error term on the r.h.s. of (\ref{off-diagonal-corr-main-zero})
will be needed for summation over the partition in the next section. 
\end{remark}
%%%%%%%%%%%%%%%%%%%%%%%%%%%%%%%%%%%%%%%%%%%%%%%%%%%%%%%%
\proof   Let us set
\beqa
\hat M_{l,j}(\si,  \tn, \sn)
:=\lan e^{iH_{\si} t} \W_{\si}(\nv_l,\sn)  \reta^*_{2,\si}(h_{2,l,\sn}^{'(\tn)} e^{i\ga_{\si}(\nv_l,\sn)})\Om, e^{iH_{\si} t}\W_{\si}(\hv_j,\sn) 
\reta^*_{2,\si}(h_{2,j,\sn}^{(\tn)} e^{i\ga_{\si}(\hv_j,\sn)})\Om\ran. \label{off-diag}
\eeqa
We note that  the first term  on the l.h.s. of (\ref{off-diagonal-corr-main-zero}) equals $\hat  M_{l,j}(\si, \tn, \tn)$.
We define
\beqa
& &\psi(s,t):=e^{iH_{\si} t}\W_{\si}(\hv_j,\sn) \reta^*_{2,\si}(h_{2,j,t}^{(s)} e^{i\ga_{\si}(\hv_j,t)})\Om,\\
& &\psi'(s,t):=e^{iH_{\si} t}\W_{\si}(\nv_l,\sn)\reta^*_{2,\si}(h_{2,l,t}^{'(s)} e^{i\ga_{\si}(\nv_l,t)})\Om.
\eeqa
Next, we write 
\beqa
\hat N_{l,j}(\si, s, t)
:=\lan \W_{\si}(\nv_l)\Om, \W_{\si}(\hv_j)\Om\ran \ 
\lan \reta^*_{2,\si}(h_{2,l}^{'(\tn)} e^{i\ga_{\sip}(\nv_l,t)})\Om, \reta^*_{2,\si}(h_{2,j}^{(\tn)} e^{i\ga_{\si}(\hv_j,t)})\Om\ran
\eeqa
and note that $\hat N_{l,j}(\si, s, \infty)$ is the first term on the r.h.s. of (\ref{off-diagonal-corr-main-zero}). Now we write for $t\geq s$
\beqa
|\hat M_{l,j}(\si, s,s) - \hat N_{l,j}(\si, s, t) |\ph{444444444444444444444444444444}& &\non\\
\leq |\hat M_{l,j}(\si, s, s)-\hat M_{l,j}(\si, s, t)|+ |\hat M_{l,j}(\si, s, t)-
\hat N_{l,j}(\si, s, t) |. \label{M-N-decomposition}
\eeqa
We consider the first term on the r.h.s. of (\ref{M-N-decomposition}):
\beqa
& &\hat M_{l,j}(\si, s,s) - \hat M_{l,j}(\si, s, t)=\int_s^t dt'\, \pa_{t'} \hat M_{l,j}(\si, s,t')\non\\
& &=\int_s^t dt'\, \bigg(\lan \pa_{t'} \psi'(s,t'), \psi(s,t')\ran+\lan \psi'(s,t'), \pa_{t'}\psi(s,t')\ran\bigg)=O\bigg( \fr{1}{\si^{\de_{\la_0}} }\fr{1}{s^{7\peps}}  \bigg).
\label{integral+derivative}
\eeqa
In the last step of (\ref{integral+derivative}) we used Corollary~\ref{Cook-corollary} and Lemma~\ref{fiber-ground-states-lemma} which give
\beqa\label{est-with-s}
\|\pa_{t'}  \psi'(s,t')\|, \ \ \|\pa_{t'} \psi(s,t')\| \leq \fr{c}{\si^{\de_{\la_0}} }\fr{1}{(t')^{1+7\peps}} \quad\textrm{ and }\quad  \|  \psi'(s,t')\|, \ \ \| \psi(s,t')\|\leq c,
\eeqa
respectively. By Proposition~\ref{off-diagonal-thm-one}
\beqa
 \lim_{t\to\infty}|\hat M_{l,j}(\si, s, t)-\hat N_{l,j}(\si, s, t) |=0.
\eeqa
Thus by taking the limit $t\to\infty$ in (\ref{M-N-decomposition}) we conclude the proof. \qed\\
\begin{remark}
{\color{black}We stress that the factor $\fr{1}{\si^{ \de_{\la_0} }}$ obtained here in the remainder term of (\ref{off-diagonal-corr-main-zero}) is an artefact of the estimate in (\ref{est-with-s}) but it is absent in \cite{Pi05}. We refrain from a more accurate estimate since from the clustering of the full expression in Corollary \ref{off-diagonal-corr-new-corr}  this factor will appear anyway in the error term. This is further explained in Remark \ref{rem-corr-off-diagonal-corr-new-corr}.}
\end{remark}
%%%%%%%%%%%%%%%%%%%%%%%%%%%%%%%%%%%%%%%%%%%%%%%%%%%%%%%%the 
\nin The following corollary is a straightforward consequence of Theorems~\ref{off-diagonal-corr-new} and \ref{off-diagonal-corr}.
%%%%%%%%%%%%%%%%%%%%%%%%%%%%%%%%%%%%%%%%%%%
\bec \label{off-diagonal-corr-new-corr} The following estimate  holds true
\beqa
& &\lan  \W_{\si}(\nv_l,s) \reta^*_{1, \si}(h'_{1,s})\reta^*_{2,\si}(h_{2,l,s}^{'(s)} e^{i\ga_{\si}(\nv_l,s)})\Om, \W_{\si}(\hv_j,s) 
\reta^*_{1, \si}(h_{1,s})\reta^*_{2,\si}(h_{2,j,s}^{(s)} e^{i\ga_{\si}(\hv_j,s)})\Om\ran\non\\
& &= \lan \reta^*_{1, \si}(h_{1,s}')\Om, \reta^*_{1, \si}(h_{1,s})\Om\ran \times \non\\
& &\ph{44}\times \lan \W_{\si}(\nv_l,s)\Om,  \W_{\si}(\hv_j,s) \Om\ran  \   \lan \reta^*_{2,\si}(h_{2,l,s}^{'(\tn)} e^{i\ga_{\si}(\nv_l,\infty)})\Om, \reta^*_{2,\si}(h_{2,j,s}^{(\tn)} e^{i\ga_{\si}(\hv_j,\infty)})\Om\ran
+ R^{W}_{l,j}(\si,s),\quad\quad\quad
\label{off-diagonal-corr-main-x-x}
\eeqa
where, for $0<\peps<1/15$,
\beqa
|R^{W}_{l,j}(\si,s)|\leq \fr{c_{\la_0}}{\si^{\de_{\la_0}}}\bigg\{ 
  \fr{1}{  s\si^{ 1/(8\tiga) } } +  \fr{1}{s^{7\peps}}   +\si^{\alf/(8\ga_0)} \bigg\}.
\eeqa
\eec

\begin{remark}\label{rem-corr-off-diagonal-corr-new-corr}
{\color{black} When we compare the estimate in Corollary  \ref{off-diagonal-corr-new-corr}  to analogous estimates (see \cite{Pi05})  for the scattering of one single charged particle (interacting with the quantized massless boson field), we observe that the atom-electron scattering (with the quantized massless boson field) studied here yields the extra factor $\fr{c_{\la_0}}{\si^{\de_{\la_0}}}$ that is related to the control of the norm of what  in the approximating vector (\ref{approximating-vectors}) is associated with the atom-electron system. We recall that $\delta_{\la_0}$ can be chosen arbitrarily small provided the coupling constant $\lambda_0$ is small enough. This makes still possible to use a telescopic argument as in (\ref{telescop}).}

\end{remark}

%%%%%%%%%%%%%%%%%%%%%%%%%%%%%%%%%%%%%%%%%%%%%%%%%%%%%%%%%%%%%%%%%%%%%%%%%
%%%%%%%%%%%%%%%%%%%%%%%%%%%%%%%%%%%%%%%%%%
\section{Convergence of the approximating vector}
\label{convergence-atom-electron}
%%%%%%%%%%%%%%%%%%%%%%%%%%%%%%%%%%%%%%%%%%
\setcounter{equation}{0}

We recall the definition of atom-electron scattering states approximants from Subsection~\ref{Results-subsection}: Let $h_1,h_2\in C_0^2(\PSS)$ have disjoint velocity supports (see (\ref{velocity-support})) which do not contain zero.  
For each cube $\Ga_j^{(t)}$ we write 
\beq
\Psi_{\si,j}(s,t)=e^{iHt}\W_{\si}(\nv_j,t)\reta^*_{1, \si}(h_{1,t})\reta^*_{2,\si}(h_{2,j,t}^{(s)} e^{i\ga_{\si}(\nv_j,t)})\Om, \label{approximating-vectors-one}
\eeq
where $h_{1,t}(p)=\e^{-i E_{1,p,\si}t}h_1(p)$, $h_{2,j,t}^{(s)}(p)=\e^{-i E_{2,p,\si}t}h_{2,j}^{(s)}(p)$ and  the two dispersion relations $p\mapsto E_{1,p,\si}$ and $p\mapsto E_{2,p,\si}$ may be different.
Now we set  $\si_t:=\kal/ t^{\gamma}$ and define
\beqa
\Psi_{h_1,h_2,t}=\sum_{j\in \Ga^{(t)}} \Psi_{\si_t,j}(t,t).
\eeqa
Let $N(t)=2^{3\np(t)}$ be the number of cells in the partition $\Ga^{(t)}$. We will write  for $t_2\geq t_1$
\beqa
& &\sum_{j\in \Ga^{(t)}}=\sum_{j=1}^{N(t)},\quad\quad
 \sum_{j=1}^{N(t_2)}=\sum_{j=1}^{N(t_1)}\sum_{l(j)},
\eeqa 
where $l(j)$ numbers the sub-cells of a given cell at $t_1$, so we have
\beqa
1\leq l(j)\leq \fr{N(t_2)}{N(t_1)}. \label{l-j-range}
\eeqa
The main result of this section  is Theorem~\ref{main-technical-result} below {\color{black} that uses various ingredients derived in Sects. \ref{partition}, \ref{Cook}, and \ref{variation-ir}}. It implies the existence of the electron-atom scattering states, stated in  Theorem~\ref{main-general-result}, via the telescopic argument. 
%%%%%%%%%%%%%%%%%%%%%%%%%%%%%%%%%%%%%%
\bet\label{main-technical-result}  We consider the difference
\beqa
\Psi_{h_1,h_2,t_2}-\Psi_{h_1,h_2,t_1}
\1&=&\1e^{iHt_2}\sum_{j=1}^{N(t_1)}\sum_{l(j)}\W_{\si_{t_2}}(\nv_{l(j)},t_2)\reta^*_{1, \si_{t_2}}(h_{1,t_2})\reta^*_{2,\si_{t_2}}(h_{2,l(j),t_2}^{(t_2)} 
e^{i\ga_{\si_{t_2}}(\nv_{l(j)},t_2)})\Om\non\\
\1& &\1-e^{iHt_1}\sum_{j=1}^{N(t_1)}\W_{\si_{t_1}}(\nv_j,t_1)\reta^*_{1, \si_{t_1}}(h_{1,t_1})\reta^*_{2,\si_{t_1}}(h_{2,j,t_1}^{(t_1)} e^{i\ga_{\si_{t_1}}(\nv_j,t_1)})\Om.   \label{total-diff} 
\eeqa
It satisfies, for some finite $M\in \nat$ and $\eta_i>2\eps_i\geq 0${\color{black},}
\beqa
\|\Psi_{h_1,h_2,t_2}-\Psi_{h_1,h_2,t_1}\|\leq  c_{\la_0} \sum_{i=1}^{M}\fr{t_2^{\eps_i}}{t_1^{\eta_i}}.   
\eeqa
\eet
%%%%%%%%%%%%%%%%%%%%%%%%%%%%%%%%%%%%%%%%%%%
\proof Follows from Propositions~\ref{change-of-partition}, \ref{cook-arg-prop}{\color{black},} and \ref{change-of-cut-off}. \qed
%%%%%%%%%%%%%%%%%%%%%%%%%%%%%%%%%%%%%%%%%%

\noindent
{\color{black} In order to estimate the difference on the r.h.s. of (\ref{total-diff}) we split it into subsequent steps that can be distinguished by identifying the various roles of the time variable $t_2$ appearing in the first term on the r.h.s. of (\ref{total-diff}), namely
\begin{itemize}
\item
the time variable $t_2$ underlined in 
\begin{equation}\label{shift-1}
e^{iHt_2}\sum_{j=1}^{N(\underline{t_2})}\sum_{l(j)}\W_{\si_{t_2}}(\nv_{l(j)},t_2)\reta^*_{1, \si_{t_2}}(h_{1, t_2})\reta^*_{2,\si_{t_2}}(h_{2,l(j),t_2}^{(\underline{t_2})} 
e^{i\ga_{\si_{t_2}}(\nv_{l(j)},t_2)})\Om 
\end{equation}
is associated with the cell partition (in expression (\ref{vec-s-t}) is represented by $s$);
\item
the time variable  $t_2$ underlined in
\begin{equation}\label{shift-2}
e^{iH\underline{t_2}}\sum_{j=1}^{N(t_2)}\W_{\si_{t_2}}(\nv_{l(j)},\underline{t_2})\reta^*_{1, \si_{t_2}}(h_{1,\underline{t_2}})\reta^*_{2,\si_{t_2}}(h_{2,l(j),\underline{t_2}}^{(t_2)} 
e^{i\ga_{\si_{t_2}}(\nv_{l(j)},\underline{t_2})})\Om 
\end{equation}
is associated with the evolution $e^{iHt_2}$ and the free
evolution (in expression (\ref{vec-s-t}) is represented by $t$);
\item
the time variable  $t_2$ underlined in
\begin{equation}\label{shift-3}
e^{iHt_2}\sum_{j=1}^{N(t_2)}\W_{\si_{\underline{t_2}}}(\nv_{l(j)},t_2)\reta^*_{1, \si_{\underline{t_2}}}(h_{1,t_2})\reta^*_{2,\si_{\underline{t_2}}}(h_{2,l(j),t_2}^{(t_2)} 
e^{i\ga_{\si_{\underline{t_2}}}(\nv_{l(j)},t_2)})\Om 
\end{equation}
is associated with the infrared cut-off (in expression (\ref{vec-s-t}) the cut-off is $\sigma$ with no time dependence).
\end{itemize}
In the three sections below we analyse the changes of the vector when the time variable is shifted from $t_1$ to $t_2$ in the three restricted contexts described in (\ref{shift-1}), (\ref{shift-2}), and (\ref{shift-3}), respectively.  The given ordering  is crucial to state our estimates.}
%associated with the change of the partition from $\Ga^{(t_2)}$ to $\Ga^{(t_1)}$ by keeping the infrared cut-off fixed at $\sigma_{t_2}$ and the time variable equal to $t_2$}
%%%%%%%%%%%%%%%%%%%%%%%%%%%%%%%%%%%%%%%%%
\subsection{Change of the partition}\label{partition}
%%%%%%%%%%%%%%%%%%%%%%%%%%%%%%%%%%%%%%%%%%
In this subsection we consider the change of the partition from $\Ga^{(t_1)}$ to $\Ga^{(t_2)}$ as time goes 
from $t_1$ to $t_2> t_1$. 
%%%%%%%%%%%%%%%%%%%%%%%%%%%%%%%%%%
\bep\label{change-of-partition} We consider the difference
\beqa
D0\1&:=&\1 e^{iHt_2}\sum_{j=1}^{N(t_1)}\sum_{l(j)}\W_{\si_{t_2}}(\nv_{l(j)},t_2)\reta^*_{1, \si_{t_2}}(h_{1,t_2})\reta^*_{2,\si_{t_2}}(h_{2,l(j),t_2}^{(t_2)} 
e^{i\ga_{\si_{t_2}}(\nv_{l(j)},t_2)})\Om \\
& &-e^{iHt_2}\sum_{j=1}^{N(t_1)}  \W_{\si_{t_2}}(\nv_{j},t_2)\reta^*_{1, \si_{t_2}}(h_{1,t_2})
\reta^*_{2,\si_{t_2}}(h_{2,j,t_2}^{(t_1)} e^{i\ga_{\si_{t_2}}(\nv_{{\mc j}},t_2)})\Om.
\label{partition-second}
\eeqa
It satisfies
\beqa
\|D0\|\leq \fr{c}{\si_{t_2}^{\de_{\la_0}}}\fr{1}{t_1^{\peps}}.
\eeqa
\eep
%%%%%%%%%%%%%%%%%%%%%%%%%%%%%%%%%
\proof Follows immediately from Lemmas~\ref{lemma-partition-one} and \ref{lemma-partition-two} below. \qed\\ 
%%%%%%%%%%%%%%%%%%%%%%%%%%%%%%%%%
In the following lemma we replace $h_{2,j,t_2}^{(t_1)}$ with $\sum_{l(j)}h_{2,l(j),t_2}^{ (t_2) }$ in (\ref{partition-second}) at a cost of an error term.
%%%%%%%%%%%%%%%%%%%%%%%%%%
\bel\label{lemma-partition-one} Let $D0$ be as defined above. Then 
\beqa
\|D0\|\1&=&\1\| \sum_{j=1}^{N(t_1)}\sum_{l(j)}\W_{\si_{t_2}}(\nv_{l(j)},t_2)\reta^*_{1, \si_{t_2}}(h_{1,t_2})\reta^*_{2,\si_{t_2}}(h_{2,l(j),t_2}^{(t_2)} 
e^{i\ga_{\si_{t_2}}(\nv_{l(j)},t_2)})\Om\non\\
& &-\sum_{j=1}^{N(t_1)} \sum_{l(j)}\W_{\si_{t_2}}(\nv_{j},t_2)\reta^*_{1, \si_{t_2}}(h_{1,t_2})
\reta^*_{2,\si_{t_2}}(h_{2,l(j),t_2}^{ (t_2) } 
e^{i\ga_{\si_{t_2}}(\nv_{j},t_2)})\Om\|+O\bigg(\fr{1}{\si_{t_2}^{\de_{\la_0}}}  \fr{1}{t_1^{2\peps}} \bigg).
\label{difference-under-norm}
\eeqa
\eel
%%%%%%%%%%%%%%%%%%%%%%%%%%
\begin{remark} In the case of a sharp partition the error term above would be zero.
\end{remark}
\proof {\color{black} We recall the definition of $\app_{\Ga^{(t)}_j}$ in (\ref{def-tilde-identity}), and we write}
\beqa
h_{2,j}^{(t_1)}(p)\1&:=&\1\app_{\Ga^{(t_1)}_j}(p)h_{2}(p)=(\app_{\Ga^{(t_1)}_j}(p)-\one_{\Ga^{(t_1)}_j}(p))h_{2}(p)+
\one_{ \Ga^{(t_1)}_j}(p)h_{2}(p)\non\\
\1&=&\1(\app_{\Ga^{(t_1)}_j}(p)-\one_{\Ga^{(t_1)}_j}(p))h_{2}(p)+\sum_{l(j)}\one_{\Ga^{(t_2)}_{l(j)}}(p)h_{2}(p)\non\\
\1&=&\1(\app_{\Ga^{(t_1)}_j}(p)-\one_{\Ga^{(t_1)}_j}(p))h_{2}(p)+\sum_{l(j)}( \one_{\Ga^{(t_2)}_{l(j)}}(p) -  \app_{\Ga^{(t_2)}_{l(j)}}(p) )h_{2}(p)
+\sum_{l(j)}   h_{2,l(j)}^{(t_2)}(p) \non\\
\1&=&\1 \sum_{l(j)}   h_{2,l(j)}^{ (t_2) }(p)+O\big( t_1^{-5\peps}\big)  +O\big(2^{3(\np(t_2)-\np(t_1)) } 
t_2^{-5\peps} \big)\non\\
\1&=&\1 \sum_{l(j)}   h_{2,l(j)}^{ (t_2) }(p)+O\big( t_1^{-5\peps}\big),
\eeqa
where we made use of Lemma~\ref{smooth-to-sharp}, relation~(\ref{l-j-range}), and the fact that
$2^{\np(t)}\leq t^{\peps}\leq 2^{\np(t)+1}$ which gives $2^{3(\np(t_2)-\np(t_1)) } t_2^{-5\peps}\leq 8 t_1^{-5\peps}$ since $t_1\leq t_2$ and  the rest term is in the $L^2$ norm.
Let us set
\beqa
\De h_{2,j}^{(t_1,t_2)}:=h_{2,j}^{(t_1)}-\sum_{l(j)}   h_{2,l(j)}^{ (t_2) }=O( t_1^{-5\peps}). \label{change-of-h}
\eeqa
Now we can write
\beqa
-(\ref{partition-second})\1&=&\1e^{iHt_2}\sum_{j=1}^{N(t_1)} \sum_{l(j)} \W_{\si_{t_2}}(\nv_j,t_2)\reta^*_{1, \si_{t_2}}(h_{1,t_2})\reta^*_{2,\si_{t_2}}(h_{2,l(j),t_2}^{ (t_2) } e^{i\ga_{\si_{t_2}}(\nv_j,t_2)})\Om\non\\
& &+e^{iHt_2}\sum_{j=1}^{N(t_1)}\W_{\si_{t_2}}(\nv_j,t_2)\reta^*_{1, \si_{t_2}}( h_{1,t_2})\reta^*_{2,\si_{t_2}}(\De h_{{\color{black}2,j}, {t_2} }^{(t_1,t_2)} e^{i\ga_{\si_{t_2}}(\nv_j,t_2)})\Om\non\\
\1&=&\1
e^{iHt_2}\sum_{j=1}^{N(t_1)} \sum_{l(j)} \W_{\si_{t_2}}(\nv_j,t_2)\reta^*_{1, \si_{t_2}}(h_{1,t_2})\reta^*_{2,\si_{t_2}}(h_{2,l(j),t_2}^{ (t_2) } e^{i\ga_{\si_{t_2}}(\nv_j,t_2)})\Om
+O\bigg(\fr{1}{\si_{t_2}^{\de_{\la_0}}}  \fr{1}{t_1^{2\peps}} \bigg),
\label{Delta-term}
\eeqa
where we made use of Proposition~\ref{Simple-L-2-bound}, (\ref{change-of-h}) and $N(t_1)\leq c t_1^{3\peps}$. \qed
%%%%%%%%%%%%%%%%%%%%%%%%%%%%%%%%%%%%%%%%%%%%%
\bel\label{lemma-partition-two} Let us denote by $D01$ the difference under the norm in (\ref{difference-under-norm}). 
Then 
\beqa
\|D01\|\leq \fr{c_{\la_0}}{\si_{t_2}^{ \de_{\la_0} }}\fr{1}{t_1^{\peps}}. 
\eeqa
\eel
%%%%%%%%%%%%%%%%%%%%%%%%%%%%%%%%%%%%%%%%%%%%%%%%%
\proof We write
\beqa
D01\1&=&\1\sum_{j=1}^{N(t_1)}\sum_{l(j)}\bigg(\W_{\si_{t_2}}(\nv_{l(j)},t_2)\Psi^{\nv_{l(j)}}_{l(j)}(t_2)-\W_{\si_{t_2}}(\nv_j,t_2)\Psi^{\nv_j}_{l(j)}(t_2)\bigg),\\
\Psi^{\nv}_{l(j)}(t_2)\1&:=&\1\reta^*_{1, \si_{t_2}}(h_{1,t_2})\reta^*_{2,\si_{t_2}}(h_{2,l(j),t_2}^{(t_2)} e^{i\ga_{\si_{t_2}}(\nv,t_2)})\Om,\\
\hW_{j', j}(t_2)\1&:=&\1\W_{\si_{t_2}}(\nv_{j'},t_2)^*\W_{\si_{t_2}}(\nv_{j},t_2).
\eeqa
Let us consider the norm squared of this expression 
\beqa
\|D01\|^2\1&=&\1\sum_{j',j=1}^{N(t_1)}\sum_{l'(j'), l(j)} \bigg(    \lan \Psi^{\nv_{l'(j')}}_{l'(j')}(t_2), \hW_{l'(j'), l(j)}(t_2)  \Psi^{\nv_{l(j)}}_{l(j)}(t_2) \ran+
\lan \Psi^{\nv_{j'}}_{l'(j')}(t_2), \hW_{j', j}(t_2)  \Psi^{\nv_{j}}_{l(j)}(t_2) \ran\non\\
\1& &\1-2\mathrm{Re}\, \lan \Psi^{\nv_{l'(j')}}_{l'(j')}(t_2), \hW_{l'(j'), j}(t_2)  \Psi^{\nv_{j} }_{l(j)}(t_2) \ran\bigg)\,{\color{black}.}
\eeqa

First, we look at the sum of the  off-diagonal terms (i.e.{\color{black},} terms where $j\neq j'$ or $l(j)\neq l'(j)$) which we denote $\|D01\|^2_{\mathrm{off-diag}}$. 
Here Corollary~\ref{off-diagonal-corr-new-corr}  and Lemma~\ref{fiber-ground-states-lemma} give
\beqa
\|D01\|^2_{\mathrm{off-diag}}\leq N(t_2)^2 \fr{c_{\la_0}}{\si_{t_2}^{\de_{\la_0}}}\bigg\{ 
  \fr{1}{  t_2\si_{t_2}^{ 1/(8\tiga) } } +  \fr{1}{t_2^{7\peps}}   +(\si_{t_2})^{\alf/(8\ga_0)} \bigg\}
\leq \fr{1}{\si_{t_2}^{\de_{\la_0}}} \fr{c_{\la_0}}{t_2^{\peps}}, \label{small-peps-term}
\eeqa
where we used that $N(t_2)^2\leq ct_2^{6\peps}$ and set $\peps$ sufficiently small. {\color{black}(We recall that
 $\peps$ is constrained, in particular, by the conditions (\ref{first-condition-peps})-(\ref{second-condition-peps}) in Corollary \ref{Cook-corollary}  that reflect the auxiliary role of $\peps$ -- coming from the cell partition in the approximating vector -- w.r.t. to the exponent $\gamma>4$ that is related to the physical decoupling between the photon cloud and the atom-electron system.)}
Now we consider the sum of the diagonal terms:
\beqa
\|D01\|^2_{\mathrm{diag}}\1&:=&\1\sum_{j=1}^{N(t_1)}\sum_{l(j)} \bigg(    \lan \Psi^{\nv_{l(j)}}_{l(j)}(t_2), \Psi^{\nv_{l(j)}}_{l(j)}(t_2) \ran+
\lan \Psi^{\nv_{j}}_{l(j)}(t_2), \Psi^{\nv_{j}}_{l(j)}(t_2) \ran\non\\
& &-2\mathrm{Re}\, \lan \Psi^{\nv_{l(j)}}_{l(j)}(t_2), \hW_{l(j), j}(t_2)  \Psi^{\nv_{j} }_{l(j)}(t_2) \ran\bigg). \label{diagonal-summation}
\eeqa 
Making use of   Corollary~\ref{off-diagonal-corr-new-corr},   for small $\peps$ we get similarly as in (\ref{small-peps-term})
\beqa
\lan \Psi^{\nv_j}_{l(j)}(t_2), \Psi^{\nv_j}_{l(j)}(t_2) \ran\1&=&\1\|h_1\|^2\| h_{2,l(j)}^{(t_2)}\|^2
+O\bigg(\fr{c_{\la_0}}{\si_{t_2}^{ \de_{\la_0} }} \fr{1}{t_2^{7\peps}}\bigg), \\
\label{change-of-partition-diagonal-terms}
\lan \Psi^{\nv_{l(j)}}_{l(j)}(t_2), \hW_{l(j), j}(t_2)  \Psi^{\nv_{j} }_{l(j)}(t_2) \ran\1&=&\1
e^{-\fr{C_{l(j),j,\si_{t_2} }}{2}}\|h_1\|_2^2\lan  h^{(t_2)}_{2,l(j)} e^{i\ga_{\si_{t_2}}(\nv_{l(j)}, \infty)}, h^{(t_2)}_{2,l(j)} e^{i\ga_{\si_{t_2}}(\nv_j, \infty)}\ran\non\\
& &\ph{4444444444444444444444}+O\bigg(\fr{c_{\la_0}}{\si_{t_2}^{ \de_{\la_0} }} \fr{1}{t_2^{7\peps}}\bigg),
\label{change-of-partition-off-diagonal-terms}
\eeqa
where  
\beqa \label{def-C-1}
C_{l(j),j,\si_{t_2}}:=\int |\vv_2^{\si_{t_2}}(k)|^2  |f_{\nv_j,\nv_{l(j)}}(\hatk)|^2 \fr{d^3k}{2|k|^2},\quad f_{\nv_j,\nv_{l(j)} }(\hatk):=\fr{ \hatk\cdot( \nv_{j}-\nv_{l(j)} )}{(1-\hatk\cdot  \nv_{j}) (1-\hatk\cdot \nv_{l(j)})}.
\eeqa
Since summation over the cells in (\ref{diagonal-summation}) gives a factor $t_2^{3\peps}$, it follows that 
\beqa
\|D01\|^2_{\mathrm{diag}}=2\|h_1\|_2^2\sum_{j=1}^{N(t_1)}\sum_{l(j)} \bigg( \| h_{2,l(j)}^{(t_2)}\|_2^2-e^{-\fr{C_{l(j),j,\si_{t_2} }}{2}}\mathrm{Re}\, 
\lan  h^{(t_2)}_{2,l(j)} e^{i\ga_{\si_{t_2}}(\nv_{l(j)}, \infty)}, h^{(t_2)}_{2,l(j)} e^{i\ga_{\si_{t_2}}(\nv_j, \infty)}\ran\bigg)\non\\+O\bigg(\fr{c_{\la_0}}{\si_{t_2}^{ \de_{\la_0} }} \fr{1}{t_2^{4\peps}}\bigg).
\eeqa
Now we will show that the leading terms above are  of order $O(|\log\,\si_{t_2}|^2 t_1^{-\peps})$, by exploiting the fact that  $\nv_{j}-\nv_{l(j)}$ is `small' and
therefore $C_{l(j),j,\si_{t_2}}$ {\color{black}(see (\ref{def-C-1}))} is close to zero, and $\ga_{\si_{t_2}}(\nv_{l(j)}, \infty)$ is close to $\ga_{\si_{t_2} }(\nv_j, \infty)$.
In fact, we have
\beqa
 |\nv_{l(j)}-\nv_j|\leq c|p_{l(j)}-p_j|\leq c\fr{\sqrt{3}}{2^{\ov{n}(t_1)+1} }\leq  \fr{c'}{t_1^{\peps} }, \label{velocity-shift}
\eeqa
where we used that $\nv_j:=\nabla E_{2,p_j,\si}$, where $p_j$ is the momentum in the center of the $j$-th cube, and that
the second derivatives of $p\mapsto E_{2,p_j,\si}$ are bounded uniformly in $\si$ (see second  estimate in (\ref{velocity-boundedness}) below).
Consequently,
\beqa
C_{l(j),j,\si_{t_2} }\leq \fr{c}{t_1^{2\peps}}|\log\,\si_{t_2}|, \quad\textrm{ hence } \quad \big(1-e^{-\fr{C_{l(j),j,\si_{t_2} }}{2}}\big)\leq  \fr{c}{t_1^{2\peps}}|\log\,\si_{t_2}|. \label{exp-C-shift}
\eeqa
Now  {\color{black}from Lemma~\ref{phase-shift} and formula (\ref{velocity-shift}) we get}:
\beqa
|\ga_{\si_{t_2}}(\nv_{l(j)},\infty)(p)- \ga_{\si_{t_2}}(\nv_j,\infty)(p)|\leq c|\log\,\si_{t_2}|^2 |\nv_{l(j)}-\nv_j|\leq \fr{c}{t_1^{\peps}}|\log\,\si_{t_2}|^2.
\eeqa
Hence
\beqa
|1-e^{i(\ga_{\si_{t_2}}(\nv_{l(j)},\infty)(p)- \ga_{\si_{t_2}}(\nv_j,\infty)(p)     )}| \leq \fr{c}{t_1^{\peps}}|\log\,\si_{t_2}|^2. \label{exp-phase-shift}
\eeqa
Making use of (\ref{exp-phase-shift}), (\ref{exp-C-shift}) and the fact that
\beqa
\sum_{j=1}^{N(t_1)}\sum_{l(j)} \| h_{2,l(j)}^{(t_2)}\|_2^2\leq \|h_2\|^2
\eeqa
we obtain that
\beqa
\|D01\|^2_{\mathrm{diag}}=O(|\log\,\si_{t_2}|^2 t_1^{-\peps})+O\bigg(\fr{c_{\la_0}}{\si_{t_2}^{ \de_{\la_0} }} \fr{1}{t_2^{4\peps}}\bigg),
\eeqa
which concludes the proof. \qed
%%%%%%%%%%%%%%%%%%%%%%%%%%%%%%%%%%%%%%%%%%%%%%%%
\bel\label{phase-shift} For $p\neq 0$ there holds the bound 
\beqa
|\ga_{\si}(\nv_i,\infty)(p)- \ga_{\si}(\nv_j,\infty)(p)|\leq {\color{black}\frac{c}{| \nabla E_{2,p,\si}|}}|\log\,\si|^2 |\nv_i-\nv_j|.
\eeqa
%{\color{red}\sout{where $c$ can be chosen uniformly for $p$ in  closed subsets of $S$ not
%containing zero.}} 
\eel
%%%%%%%%%%%%%%%%%%%%%%%%%%%%%%%%%%%%%%%%%%%%%%%%%%%
\proof We write
\beqa
&  &\ga_{\si}(\nv_i,\infty)(p)- \ga_{\si}(\nv_j,\infty)(p)\non\\
&  &= \mathrm{Re}\int_1^{ ({\kal /\si})^{1/\al}}d\tau \int_{\si}^{\si_{\tau}^\S} d|k|\, \vv_2^{\si}(k)^2(2|k|) e^{-i|k|\tau} \int d\Omm(\nee) 
 e^{i|k| \nabla E_{2,p,\si}\cdot \nee  \tau} (-)f_{\nv_i, \nv_j}(\nee),
\eeqa
where 
\beqa
f_{\nv_i, \nv_j}(\nee):=\fr{\nee\cdot (\nv_i - \nv_j)}{ (1-\nee\cdot \nv_i)(1-\nee\cdot \nv_j)}.
\eeqa
Now Lemma~\ref{two-angular-integration-lemma-x} gives
\beqa
|\int d\Omm(\nee)    e^{i \nabla E_{2,p,\si}\cdot \nee |k| \tau}  f_{\nv_i,\nv_j}(\nee)|\leq \fr{c}{{\color{black}| \nabla E_{2,p,\si}|}|k|\tau}\sup_{\nee }\sup_{\ell\in\{0,1\}}|\pa_{\theta}^{\ell}f_{\nv_i, \nv_j}(\nee)|,
\eeqa
where $(\theta, \phi)$ refers to spherical coordinates with $\nabla E_{2,p,\si}$ in the direction of the $z$-axis. In our case
\beqa
\sup_{\nee }\sup_{\ell\in\{0,1\}}|\pa_{\theta}^{\ell}f_{\nv_i,\nv_j}(\nee)|\leq c_{\nv}:=c|\nv_i-\nv_j|.
\eeqa
Thus we get
\begin{eqnarray}
|\ga_{\si}(\nv_i,\infty)(p)- \ga_{\si}(\nv_j,\infty)(p)|\!\!&\leq&\!\! \frac{c_{\nv}}{{\color{black}| \nabla E_{2,p,\si}|}}
\int_1^{({\kal/\si})^{1/\al}}\fr{d\tau}{\tau} \int_{\si}^{\si_{\tau}^\S}  \fr{ d|k|}{|k|}\\
\!\!&\leq &\!\! 
\frac{c_{\nv}}{{\color{black}| \nabla E_{2,p,\si}|}}\int_1^{({\kal/\si})^{1/\al}}\fr{d\tau}{\tau}  |\log\,\si|\leq \frac{{\color{black}c'_{\nv}}}{{\color{black}| \nabla E_{2,p,\si}|}}|\log\, \si|^2.\,
\end{eqnarray}
This concludes the proof. \qed

\bel\label{smooth-to-sharp} Let $h\in C_0^2(S)$. Then
\beqa
\|(\app_{\Ga^{(t)}_j} -  \mathbf{1}_{\Ga^{(t)}_j}  )h\|_2\leq c t^{-5\peps}.
\eeqa
\eel
%%%%%%%%%%%%%%%%%%%%%%%%%%%%%%%%
\proof Making use of the definition of $\app_{\Ga^{(t)}_j}$ in (\ref{def-tilde-identity}) we compute
\beqa
\|(\app_{\Ga^{(t)}_j} -  \mathbf{1}_{\Ga^{(t)}_j}  )h\|_2^2\1&=&\1\int d^3p\, \mathbf{1}_{\Ga^{(t)}_0}(p)|h(p+p_j)|^2 (\app_{\Ga^{(t)}_0}(p) -  \mathbf{1}_{\Ga^{(t)}_0}(p)  )^2 \non\\
&\leq & \int d^3p\, \mathbf{1}_{\Ga^{(t)}_0}(p)|h(p+p_j)|^2 
\chi\big(p\in [-a_{\np}, a_{\np}]^{\times 3}\backslash [-a_{\np}(1-\uneps), a_{\np}(1-\uneps)]^{\times 3} \big)\non\\
\1&\leq &\1 c a_{\np}^3 \uneps,
\eeqa
where $a_{\np}=2^{-(\np+1)}$, $a_{\np}^3\leq t^{-3\peps}$, $\uneps(t)=t^{-\3\peps}$ and $\Ga^{(t)}_0$ denotes a cube centered at zero. \qed 
%%%%%%%%%%%%%%%%%%%%%%%%%%%%%%%

%%%%%%%%%%%%%%%%%%%%%%%%%%%%%%%%%%%%%%%%%%%%%%%%%%%%%%%%%%%%%%
\subsection{Cook's argument}\label{Cook}
%%%%%%%%%%%%%%%%%%%%%%%%%%%%%%%%%%%%%%%%%%%%%%%%%%%%%%%%%%%%%%

%%%%%%%%%%%%%%%%%%%%%%%%%%%%%%%%%%%%%%%%%%%%%%%%%%%%%%%
\bep\label{cook-arg-prop} We consider the difference $D1$ which has the form
\beqa
D1\1&:=&\1e^{iHt_2}\sum_{j=1}^{N(t_1)}\W_{\si_{t_2}}(\nv_j,t_2)\reta^*_{1, \si_{t_2}}(h_{1,t_2})\reta^*_{2,\si_{t_2}}(h_{2,j,t_2}^{(t_1)} e^{i\ga_{\si_{t_2}}(\nv_j,t_2)})\Om\non\\
& &-e^{iHt_1}\sum_{j=1}^{N(t_1)}\W_{\si_{t_2}}(\nv_j,t_1)\reta^*_{1, \si_{t_2}}(h_{1,t_1})\reta^*_{2,\si_{t_2}}(h_{2,j,t_1}^{(t_1)} e^{i\ga_{\si_{t_2}}(\nv_j,t_1)})\Om.
\eeqa
Then {\color{black} for $7\peps\leq 3$} we have 
\beqa
\|D1\|\leq \fr{c_{\la_0}}{\si_{t_2}^{\de_{\la_0}}}\fr{1}{t_1^{4\peps}}.
\eeqa
\eep
%%%%%%%%%%%%%%%%%%%%%%%%%%%%%%%%%%%%%%%%%%%%%%%%%%%%%%%%%%%%%%%%%%%%
\proof Let $D1_j$ be the contribution to $D1$ from the $j$-th cell. We have by Corollary~\ref{Cook-corollary}
in the case $\sib=\emp$
\beqa
\|D1_j\|\1&\leq&\1 \int^{t_2}_{t_1} dt\, \bigg\|\pa_t\bigg\{ e^{iHt}\W_{\si_{t_2}}(\nv_j,t)\reta^*_{1, \si_{t_2}}(h_{1,t})\reta^*_{2,\si_{t_2}}(h_{2,j,t}^{(t_1)} e^{i\ga_{\si_{t_2}}(\nv_j,t)})\Om\bigg\}\bigg\|\non\\
\1&\leq&\1 \int^{t_2}_{t_1} dt\,\bigg(\fr{c_{\la_0}}{\si_{t_2}^{\de_{\la_0}}}\fr{1}{t^{1+7\peps}}+c\si_{t_2}^{1-\de_{\g_0}}\bigg)
\leq \fr{c_{\la_0}}{\si_{t_2}^{\de_{\la_0}}} \bigg(\fr{1}{t_1^{7\peps}}+\fr{1}{t_1^{\ga-1}}\bigg)\leq \fr{c'_{\la_0}}{\si_{t_2}^{\de_{\la_0}}} 
\fr{1}{t_1^{7\peps}},
\eeqa
%%%%%%%%%%%%%%%%%%%%%%%%%%%%%%%%%%%%%%%%%%%%%%%%%%%
where in the last step we used that $7\peps\leq 3\leq \ga-1$. Since $N(t_1)\leq t_1^{3\peps}$, the proof is complete. \qed.
%%%%%%%%%%%%%%%%%%%%%%%%%%%%%%%%%%%%%%%%%%%%%%%%%%%%%%%%

%%%%%%%%%%%%%%%%%%%%%%%%%%%%%%%%%%%%%%
\subsection{{\color{black}Shift} of the infrared cut-off}\label{variation-ir}

%%%%%%%%%%%%%%%%%%%%%%%%%%%%%%%%%%%%%%%%%%%%%%%%%%%%%
\bep\label{change-of-cut-off} We consider the difference:
\beqa
D2\1&:=&\1e^{iHt_1}\sum_{j=1}^{N(t_1)}\W_{\si_{t_2}}(\nv_j,t_1)\reta^*_{1, \si_{t_2}}(h_{1,t_1})\reta^*_{2,\si_{t_2}}(h_{2,j,t_1}^{(t_1)} e^{i\ga_{\si_{t_2}}(\nv_j,t_1)})\Om\non\\
& &-e^{iHt_1}\sum_{j=1}^{N(t_1)}\W_{\si_{t_1}}(\nv_j,t_1)
\reta^*_{1, \si_{t_1}}(h_{1,t_1})\reta^*_{2,\si_{t_1}}(h_{2,j,t_1}^{(t_1)} e^{i\ga_{\si_{t_1}}(\nv_j,t_1)})\Om.
\eeqa
Then we have
\beqa
\|D2\|\leq c_{\la_0}\sum_{i=1}^{12}\fr{t_2^{\ti\eps_i}}{t_1^{\ti\eta_i}}\,,
\eeqa
for some $\ti\eta_i>2\ti\eps_i\geq 0$.
\eep
%%%%%%%%%%%%%%%%%%%%%%%%%%%%%%%%%%%%%%%%%%%%%%%%%%%%%%
\proof  We abbreviate the notation as follows
\beqa
& &\Psi_{\si,j}:=\W_{\si}(\nv_j,t_1)\reta^*_{1, \si}(h_{1,t_1})
\reta^*_{2,\si}(h_{2,j,t_1}^{(t_1)} e^{i\ga_{\si}(\nv_j,t_1)})\Om, \quad \W_{\si}:=\W_{\si}(\nv_j,t_1),\\
& &\reta^*_{1,\si}:=\reta^*_{1, \si}(h_{1,t_1}), \quad \reta^*_{2,\si,j}:=\reta^*_{2,\si}(h_{2,j,t_1}^{(t_1)} e^{i\ga_{\si}(\nv_j,\infty)}),
\quad \reta_{2,\si,j}^{*(t_1)}:=\reta^*_{2,\si,j}(h_{2,j,t_1}^{(t_1)} e^{i\ga_{\si}(\nv_j,t_1)}), 
\eeqa
and denote $\sip:=\si_{t_1}$, $\si:=\si_{t_2}$. First, we note that
\beqa
\|D2\|^2\1&=&\1\sum_{l,j=1}^{N(t_1)}\bigg(\lan\Psi_{\sip,l},\Psi_{\sip,j}\ran
-\lan\Psi_{\sip,l},\Psi_{\si,j}\ran-\lan\Psi_{\si,l},\Psi_{\sip,j}\ran+\lan\Psi_{\si,l},\Psi_{\si,j}\ran\bigg). \label{first-difference}
\eeqa
Now we cluster each term above into physical particles according to Theorem~\ref{off-diagonal-corr-new}:
\beqa
& &\lan\Psi_{\sip,l},\Psi_{\sip,j}\ran
=\lan \reta^*_{1,\sip}\Om, \reta^*_{1,\sip} \Om\ran \ \lan \W_{\sip} \reta_{2,\sip,l}^{*(t_1)}\Om, \W_{\sip} \reta_{2,\sip,j}^{*(t_1)}\Om\ran
+R^{W}_{l,j}(\sip,\sip,t_1), \\
& &\lan\Psi_{\sip,l},\Psi_{\si,j}\ran
=\lan \reta^*_{1,\sip}\Om, \reta^*_{1,\si} \Om\ran \ \lan \W_{\sip} \reta_{2,\sip,l}^{*(t_1)}\Om, \W_{\si} \reta_{2,\si,j}^{*(t_1)}\Om\ran
+R^{W}_{l,j}(\sip,\si,t_1), \\
& &\lan\Psi_{\si,l},\Psi_{\sip,j}\ran
=\lan \reta^*_{1,\si}\Om, \reta^*_{1,\sip} \Om\ran \ \lan \W_{\si} \reta_{2,\si,l}^{*(t_1)}\Om, \W_{\sip} \reta_{2,\sip,j}^{*(t_1)}\Om\ran
+R^{W}_{l,j}(\sip,\si,t_1), \label{strange-sigmas} \\
& &\lan\Psi_{\si,l},\Psi_{\si,j}\ran
=\lan \reta^*_{1,\si}\Om, \reta^*_{1,\si} \Om\ran \ \lan \W_{\si} \reta_{2,\si,l}^{*(t_1)}\Om, \W_{\si} \reta_{2,\si,j}^{*(t_1)}\Om\ran
+R^{W}_{l,j}(\si,\si,t_1).
\label{scalar-products-cut-off-removal}
\eeqa
(We note that in Theorem~\ref{off-diagonal-corr-new}  always $\sip\geq \si$, therefore we have   $R^{W}_{l,j}(\sip,\si,t_1)$, rather than
$R^{W}_{l,j}(\si,\sip,t_1)$, in (\ref{strange-sigmas}). { Indeed, before applying  Theorem~\ref{off-diagonal-corr-new} we use 
$\lan\Psi_{\si,l},\Psi_{\sip,j}\ran= \ov{\lan\Psi_{\sip,j}, \Psi_{\si,l}\ran}$ which interchanged the order of $\sigma$ and $\sigma'$}). 
Thus we can rewrite the first difference in (\ref{first-difference}) as follows by adding and subtracting the term in (\ref{term-added})
\beqa
\lan\Psi_{\sip,l},\Psi_{\sip,j}\ran-\lan\Psi_{\sip,l},\Psi_{\si,j}\ran\1&=&\1\lan \reta^*_{1,\sip}\Om, \reta^*_{1,\sip} \Om\ran \ \lan \W_{\sip} \reta_{2,\sip,l}^{*(t_1)}\Om, \W_{\sip} \reta_{2,\sip,j}^{*(t_1)}\Om\ran \\
& &-\lan \reta^*_{1,\sip}\Om, \reta^*_{1,\si} \Om\ran \ \lan \W_{\sip} \reta_{2,\sip,l}^{*(t_1)}\Om, \W_{\sip} \reta_{2,\sip,j}^{*(t_1)}\Om\ran 
\label{term-added}\\
& &+\lan \reta^*_{1,\sip}\Om, \reta^*_{1,\si} \Om\ran \ \lan \W_{\sip} \reta_{2,\sip,l}^{*(t_1)}\Om, \W_{\sip} \reta_{2,\sip,j}^{*(t_1)}\Om\ran
\label{term-subtracted} \\
& &-\lan \reta^*_{1,\sip}\Om, \reta^*_{1,\si} \Om\ran \ \lan \W_{\sip} \reta_{2,\sip,l}^{*(t_1)}\Om, \W_{\si} \reta_{2,\si,j}^{*(t_1)}\Om\ran+R_{l,j}^{(1,2)},
\eeqa
where $R_{l,j}^{(1,2)}:=R^{W}_{l,j}(\sip,\sip,t_1)+R^{W}_{l,j}(\sip,\si,t_1)$. This can be rearranged as follows
\beqa
& &\lan\Psi_{\sip,l},\Psi_{\sip,j}\ran-\lan\Psi_{\sip,l},\Psi_{\si,j}\ran=\lan \reta^*_{1,\sip}\Om, (\reta^*_{1,\sip}\Om-\reta^*_{1,\si} \Om)\ran \ \lan \W_{\sip} \reta_{2,\sip,l}^{*(t_1)}\Om, \W_{\sip} \reta_{2,\sip,j}^{*(t_1)}\Om\ran  \\
& &\ph{444444444444444}+\lan \reta^*_{1,\sip}\Om, \reta^*_{1,\si} \Om\ran \ \lan \W_{\sip} \reta_{2,\sip,l}^{*(t_1)}\Om, (\W_{\sip} \reta_{2,\sip,j}^{*(t_1)}\Om - \W_{\si} \reta_{2,\si,j}^{*(t_1)}\Om) \ran +R_{l,j}^{(1,2)} {\color{black}.}\quad\quad\quad 
\eeqa
{Now taking for a moment for granted (\ref{cut-off-shifts})--(\ref{last-R-term})  below}, we estimate
\beqa
|\lan\Psi_{\sip,l},\Psi_{\sip,j}\ran-\lan\Psi_{\sip,l},\Psi_{\si,j}\ran|\leq c(\sip)^{\alf}+c\sum_{i=1}^3\fr{t_2^{\eps_i}}{t_1^{\eta_i}}
+c\fr{ {t_2^{\de_{\la_0}} }  }{t_1^{\rho}}+\fr{c}{t_1^{\eta}} \, {\leq}\, c\sum_{i=1}^6\fr{t_2^{\eps_i}}{t_1^{\eta_i}},
\eeqa
where all the terms can be estimated by $t_2^{\eps_i}/t_1^{\eta_i}$ for some $\eps_i,\eta_i>0$ s.t. $\eta_i> 2\eps_i$. 
Since the argument for the last two terms in (\ref{first-difference}) is analogous (related by the exchange $\si\leftrightarrow \sip$), we obtain
\beqa
\|D2\| \leq c\bigg(\sum_{i=1}^{12}\fr{t_2^{\eps_i}}{t_1^{\eta_i}}\bigg)^{1/2}\leq c \sum_{i=1}^{12}\fr{t_2^{\eps_i/2}}{t_1^{\eta_i/2}}.   
\eeqa
This completes the proof of the proposition, given  (\ref{cut-off-shifts})--(\ref{last-R-term})  below.

The estimates for the variation of the cut-off in the  single-electron and single-atom case  have the form:
\beqa
\bigg\|\sum_{j=1}^{N(t_1)}( \W_{\sip}\reta^{*(t_1)}_{2,\sip}\Om-\W_{\si}\reta^{*(t_1)}_{2,\si}\Om)\bigg\|\leq 
c\fr{  { t_2^{\de_{\la_0}} }  }{t_1^{\rho}}, \quad \|\reta^*_{1,\si}\Om-\reta^*_{1,\sip}\Om\|\leq
c(\sip)^{\alf}. \label{cut-off-shifts}
\eeqa
for some $\rho>0$. We recall that the former estimate was first proven in \cite{Pi05}, {actually for $t_2^{\de_{\la_0} }$ replaced with 
$|\ln(t_2)|^2$}.  We give a simplified proof in  
Appendix~\ref{Cut-off-variation} {and refer to the last lines of this appendix and to Lemma~\ref{auxiliary-commutator-lemma} for the origin of $t_2^{\de_{\la_0}}$
(as opposed to the logarithm)}. The latter estimate in (\ref{cut-off-shifts}) follows from Theorem~\ref{preliminaries-on-spectrum} and Lemma~\ref{fiber-ground-states-lemma}.

Finally, recalling that $\sip:=\si_{t_1}$, $\si:=\si_{t_2}$, setting $\peps$ sufficiently small and using $N(t_1)\leq t_1^{3\peps}$
we obtain from Theorem~\ref{off-diagonal-corr-new}
\beqa
\sum_{l,j}^{N(t_1)}|R^{W}_{l,j}(\si',\si,t_1)|\1&\leq&\1 c_{\la_0}\fr{t_1^{6\peps}}{\si^{\de_{\la_0}}}
\bigg\{ \fr{1}{  t_1\si^{ 1/(8\tiga) } } +  \fr{1}{t_1^{1-8\peps}}   +(\sip)^{\alf/(8\ga_0)} \bigg\}\label{zero-R-term} \\
\1&\leq&\1 c_{\la_0}\fr{t_2^{\ga   \de_{\la_0}+\fr{\ga}{8\ga_0}  }  }{t_1^{1-6\peps}}+c\fr{t_2^{\ga\de_{\la_0}}}{t_1^{1-14\peps}   }
+c\fr{t_2^{\ga\de_{\la_0} } }{t_1^{\fr{\alf\ga}{8\ga_0}-6\peps}}=c_{\la_0}\sum_{i=1}^3\fr{t_2^{\eps_i}}{t_1^{\eta_i}}, \label{first-R-term}\\
 \sum_{l,j}^{N(t_1)}|R^{W}_{l,j}(\sip,\sip,t_1)|\1&\leq&\1 c_{\la_0}\fr{t_1^{ \ga  \de_{\la_0}+\fr{\ga}{8\ga_0}  }  }{t_1^{1-6\peps}}+c\fr{t_1^{\ga\de_{\la_0}}}{t_1^{1-14\peps}   }
+c\fr{t_1^{\ga\de_{\la_0} } }{t_1^{\fr{\alf\ga}{8\ga_0}-6\peps}}=\fr{c_{\la_0}}{t_1^{\eta}},\\
\sum_{l,j}^{N(t_1)}|R^{W}_{l,j}(\si,\si,t_1)|\1&\leq&\1  c_{\la_0}\fr{t_1^{6\peps}}{\si^{\de_{\la_0}}}
\bigg\{ \fr{1}{  t_1\si^{ 1/(8\tiga) } } +  \fr{1}{t_1^{1-8\peps}}   +(\si)^{\alf/(8\ga_0)} \bigg\}\leq c_{\la_0}\sum_{i=1}^3\fr{t_2^{\eps_i}}{t_1^{\eta_i}}, \label{last-R-term}
\eeqa
where   $2\eps_i<\eta_i$ and in (\ref{last-R-term}) we used that $\si\leq \sip$ {and thus the first term on the r.h.s. of (\ref{last-R-term}) can be estimated from above by the first term on the r.h.s. of (\ref{zero-R-term}).}
%and thus we could estimate this term as in (\ref{first-R-term}) {\color{red}by replacing $\sigma'$ by $\sigma$.} 
\qed
%%%%%%%%%%%%%%%%%%%%%%%%%%%%%%%%%%%%%%%%%%%%%%%%%%%%%%%%%%%%%%%%%%%%%%%%%%

\appendix

\section{Spectral theory} \label{spectral-theory-app}
\setcounter{equation}{0}

In this appendix we restate our spectral results from \cite{DP12, DP16}. We refer to these latter references for
a detailed comparison with earlier literature and only mention here that the most innovative items are
(\ref{state-bound}) and (\ref{infrared-spectral-bound}) below for $|\be|=2$.

We do not write the index $1,2$ distinguishing the particles in this appendix, since the statement holds both 
in the infrared regular and infrared singular situation or it is clear from the context which situation is meant.

%%%%%%%%%%%%%%%%%%%%%%%%%%%%%%%%%%%%%%%%%%%%%%%
\begin{thm}\cite{DP12}\label{preliminaries-on-spectrum} Fix $0\leq \alf \leq 1/2$ and let $p_{\maxi}=1/3$. Then there exists $\g_0>0$ and $\ka \geq \ka_{\la_0}>0$ s.t. for all $|\g|\in (0, \g_0]$ and $p\in S:=\mcB_{\mc p_{\maxi}}$
  the following statements hold:
\begin{enumerate}[label = \textup{(\alph*)}, ref =\textup{(\alph*)},leftmargin=*]
\item\label{cut-off-part} For $\si\in (0,\ka_{\la_0}]$, $E_{p,\si}$ is a simple eigenvalue corresponding to a normalized eigenvector $\cpsi_{p,\si}$,
whose phase is fixed in Definition 5.3 of \cite{DP12}. $S\ni p\mapsto E_{p,\si}$ is analytic and strictly convex,   for all $\si\in (0,\ka_{\la_0}]$.
Moreover,  for some $0<\de_{\la_0}<1/4$, specified below
\beqa
& &|\pa_{p}^{\be_1}E_{p,\si}|\leq c,\quad  |\pa_{p}^{\be_2}E_{p,\si}|\leq c, \quad  |\pa_{p}^{\be_3}E_{p,\si}|\leq c/\si^{\de_{\la_0}}, \label{velocity-boundedness}\\
& &\|\pa^{\be}_p\cpsi_{p,\si}\|\leq  c /\si^{\de_{\g_0}} \label{state-bound}
\eeqa
for  multiindices $\be_i$, $\be$ s.t. $|\be_i|=i$ and $0<|\be|\leq 2$. 
\item\label{energy-part} For $\si\in (0,\ka_{\la_0}]$ the estimate 
\begin{equation}
|E_{p}-E_{p,\si}|\leq c\si \label{energy-convergence-bound}
\end{equation}
holds true. Moreover,  $S\ni p\mapsto E_p$ is twice continuously differentiable and strictly convex.
\item\label{eigenvectors-convergence} For $\alf>0$, $E_p$ is an eigenvalue corresponding to a normalized eigenvector $\cpsi_{p}$.
Moreover, for a suitable choice of the phase of $\cpsi_{p}$ and  $\si\in (0,\ka_{\la_0}]$
\beqa
\|\cpsi_p-\cpsi_{p,\si}\|\leq c  (\alf)^{-1}    \si^{\alf}.
\eeqa
\end{enumerate}
The constant $c$ above is independent of $\si$, $p$, $\g$, $\alf$ within the assumed restrictions. Clearly, all statements above
remain true after replacing $\g_0$ by some $\ti\g_0\in (0,\g_0]$. The resulting function
  $\ti\g_0\mapsto \de_{\ti\g_0}$ can be chosen s.t. $\lim_{\ti\g_0\to 0} \de_{\ti\g_0}=0$.
\end{thm}
%%%%%%%%%%%%%%%%%%%%%%%%%%%%%%%%%%%%%%%%%%%%%

{In Appendix~\ref{Cut-off-variation} we also use properties (\ref{additional-spectral-relation})--(\ref{res-formula-x}) below, which are not stated explicitly in the above theorem. 
Estimate~\ref{additional-spectral-relation} is also used in the discussion of approximate velocity supports in `Standing assumptions and conventions'. Estimate~(\ref{second-derivative-bound})
enters into the proof of Lemma~\ref{two-angular-integration-lemma}.

 Properties (\ref{additional-spectral-relation})--(\ref{holder-2-x})  date back to \cite{Pi03} and (\ref{res-formula-x}) even to \cite{Fr73, Fr74.1}.
 Estimate~(\ref{second-derivative-bound}), which is slightly stronger than strict convexity stated in (\ref{preliminaries-on-spectrum}), was shown in \cite{KM12, FP10} for different models.
  For the reader's convenience, we indicate below how to extract them from \cite{DP12, DP16}. 
%%%%%%%%%%%%%%%%%%%%%%%%%%%%%%%%%%%%%%%%%%%%%
\begin{prop} \label{App-A-additional-spectral-information} Under the assumptions of Theorem~\ref{preliminaries-on-spectrum}
\beqa
& & |\nabla E_{2,p,\si}-\nabla E_{2,p,\sip}|\leq c\sip^{1/4 }, \label{additional-spectral-relation}\\
& &\|\phi_{2,p,\si}-  \phi_{2,p,\sip}\|\leq c \sip^{1/4}, \label{phi-vector-x}\\
& &|\nabla E_{2,p,\si}-\nabla E_{2,p',\si}|\leq C|p-p'|,  \label{holder-1-x} \\ 
& &\|\phi_{2,p,\si}-  \phi_{2,p-k,\si}\|\leq c |k|^{\frac{1}{16}}, \label{holder-2-x}\\
&& \|b(k_1) \ldots b(k_m)\psi_{p,\sigma}\| \leq  (c\la)^m\, \frac{\chi_{[\sigma,\kappa]}(k_1)}{|k_1|^{3/2}}  \ldots  \frac{\chi_{[\sigma,\kappa]}(k_m)}{|k_m|^{3/2}},  
\label{res-formula-x}\\
& & \pa_{|p|}^2E_{|p|e, \si}\geq c>0,   \label{second-derivative-bound}
\eeqa
where $0<\si\leq \sip \leq \kal$, $p,p'\in S$. In (\ref{second-derivative-bound}) we used that $p\mapsto E_{p,\si}$ is rotation invariant and $e$ is an
arbitrary vector on the unit sphere.
\end{prop}
%%%%%%%%%%%%%%%%%%%%%%%%%%%%%%%%%%%
\proof For (\ref{additional-spectral-relation}) and (\ref{phi-vector-x}) we refer to relation~(C.41) and Corollary 5.6 (a) of  \cite{DP12}, respectively.
Estimate~(\ref{holder-1-x}) is a consequence of $\pa_p^{\be} E_{2,p\si}\leq c$ for $|\be|=2$ (cf. Theorem~\ref{preliminaries-on-spectrum})
via a Taylor expansion.  As for (\ref{holder-2-x}), formula (1.8) and estimate (1.3) of \cite{DP12} combined with definition (4.42) and estimate (4.44) of \cite{DP16}
(for $n=1$) give $\|\pa_p \phi_{2,p,\si}\|\leq  c/\si^{\de_{\la_0}}$. Now (\ref{holder-2-x}) follows from (\ref{phi-vector-x}) by a consideration analogous to (E.23) of \cite{DP12}.
Property (\ref{res-formula-x}) is shown in Appendix D of \cite{DP12}, where also the limiting procedure defining $b(k_1)\ldots b(k_m)\psi_{p,\sigma}$ is described.
Finally, (\ref{second-derivative-bound}) is shown in Appendix E of \cite{DP12}. \qed
%%%%%%%%%%%%%%%%%%%%%%%%%%%%%%%%%%%%%%%%%%%%%%%%%%%%%%%%%%%
}

Let us express $\psi_{\P,\si}$ in terms of its $m$-particle components in the Fock space: 
\beqa
\psi_{\P,\si}=\{f^m_{\P,\si}\}_{m\in \nat_0}.
\eeqa
Here $f^m_{\P,\si}\in L^2_{\sym}(\real^{3m}, d^{3m}k)$, i.e., each 
$f^m_{\P,\si}$   is a square-integrable function symmetric in $m$ variables from $\real^3$. Let us introduce the following auxiliary functions:
\beqa \label{functions g}
& & g_{1, \si}^m(k_1,\ldots, k_m):=\prod_{i=1}^m\fr{ c\g  \chi_{[\si,\kas)}(k_i) |k_i|^{\alf} }{|k_i|^{3/2}}, \label{def-g-ks-z}  \\
& &g_{2,\si}^m(k_1,\ldots, k_m):=\prod_{i=1}^m\fr{ c\g  \chi_{[\si,\kas)}(k_i)}{|k_i|^{3/2}}, \label{def-g-ks}
\eeqa
where $\kas:=(1-\eps_0)^{-1}\ka$,  $0<\eps_0<1$ {\mc appeared below (\ref{form-factor-definitions})} and 
$\tic$ is some positive constant independent of $m,\si, \P$ and $\g$ within the restrictions specified above. 
(We will also write $g^m_{\si}$ if there is no need to distinguish between the infrared regular and infrared singular situation).
Finally, we introduce the notation
\beqa
\mcA_{r_1,r_2}:=\{\, k\in \real^3\,|\, r_1<|k|<r_2 \,\}, \label{A-set}
\eeqa
where $0\leq r_1< r_2$. 
Now we are ready to state the required properties of the functions 
$f^m_{\P,\si}$:
%%%%%%%%%%%%%%%%%%%%%%%%%%%%%%%%%%%%%%%%%%%%%%%%%%%%%%%%%%%
%%%%%%%%%%%%%%%%%%%%%%%%%%%%%%%%%%%%%%%%%%%%%%%%
\begin{thm}\cite{DP16}\label{main-theorem-spectral} 
Fix $0\leq \alf \leq 1/2$  and let $\P_{\maxi}=1/3$. Then there exists  $\lambda_0>0$ and $\ka> \ka_{\la_0}>0$ s.t. for all $\P \in S=\mathcal{B}_{\P_{\maxi}}$,
$\lambda\in (0, \lambda_0]$ and $\si\in (0, \ka_{\la_0}]$ there holds:
\begin{enumerate}[label = \textup{(\alph*)}, ref =\textup{(\alph*)},leftmargin=*]

\item \label{f-m-support} Let  $\{f^{{\mc m}}_{\P,\sigma}\}_{{\mc m}\in\nat_0}$ be the ${\mc m}$-particle components of $\cpsi_{\P,\sigma}$
and let $\overline{\mathcal{A}}_{\sigma,\kappa}^{\times {\mc m}}$ be defined  as the Cartesian product of ${\mc m}$ copies of the closure of the set $\mathcal{A}_{\sigma,\kappa}$ introduced in (\ref{A-set}). Then,
 for any ${\mc \P}\in S$, the function $f^{{\mc m}}_{\P,\sigma}$ is supported in $\overline{\mathcal{A}}_{\sigma,\kappa}^{\times {\mc m}}$.

\item \label{f-m-smoothness} The  function
\begin{equation}
S\times \mathcal{A}_{\sigma,\infty}^{\times {\mc m}}\ni (\P; k_1,\dots, k_{\mc m}) \mapsto f^{{\mc m}}_{\P,\sigma}(k_1,\dots, k_{\mc m}) \label{momentum-wave-functions}
\end{equation}
is twice continuously differentiable and extends by continuity, together with its derivatives, to the set 
$S\times \overline{\mathcal{A}}_{\sigma,\infty}^{\times {\mc m}}$. 

\item\label{derivatives-bounds} For any multiindex $\beta$, $0\leq |\beta|\leq 2$, the function (\ref{momentum-wave-functions}) satisfies
\begin{eqnarray}
|\partial^{\beta}_{k_l}f^{{\mc m}}_{\P,\sigma}(k_1,\dots, k_{\mc m})|  \1&\leq&  \1 \frac{1}{\sqrt{{\mc m!}}} |k_l|^{-|\beta|}g^{{\mc m}}_{\sigma}(k_1,\dots, k_{\mc m}), \label{simple-spectral-bound}\\
|\partial^{\beta}_{\mc p}f^{{\mc m}}_{\P,\sigma}(k_1,\dots, k_{\mc m})|  \1 &\leq&  \1  \frac{1}{\sqrt{\mc m!}} \bigg( \frac{1}{\sigma^{\delta_{\lambda_0}}}\bigg)^{| \beta |}  g^{{\mc m}}_{\sigma}(k_1,\dots, k_{\mc m}), \label{infrared-spectral-bound}\\
|\partial_{\P^{i'}}\partial_{k_l^i}f^{{\mc m}}_{\P,\sigma}(k_1,\dots, k_{\mc m})|  \1 &\leq&  \1 \frac{1}{\sqrt{\mc m!}}\frac{1}{\sigma^{\delta_{\lambda_0}}}|k_l|^{-1}g^{{\mc m}}_{\sigma}(k_1,\dots, k_{\mc m}), \label{mixed-spectral-bound}
\end{eqnarray}
where the function $\tilde{\lambda}_0 \mapsto \delta_{\tilde{\lambda_0}}$ has the properties specified in Theorem~\ref{preliminaries-on-spectrum} and $g^{{\mc m}}_{\sigma}$ 
denotes $g^{{\mc m}}_{1,\sigma}$ or $g_{2, \sigma}^{{\mc m}}$ (cf. (\ref{def-g-ks-z}), (\ref{def-g-ks}) above) depending whether $f^{{\mc m}}_{\P,\sigma}$ is a wave function of an atom or of an electron.

\end{enumerate}
\end{thm}
%%%%%%%%%%%%%%%%%%%%%%%%%%%%%%%%%%

{ 
In the proof of Lemma~\ref{I-lemma-tx}  we will also need information about the wave functions of $\phi_{p,\si}$. Thus we  write $\phi_{p,\si}=\{f^{m}_{\mrm{w},p,\si} \}_{m\in\nat_0}$, 
and note the general relation
\beqa
f^{m}_{\mrm{w},p,\si}(k_1, \ldots, k_m)=\fr{1}{\sqrt{m!}} \lan \Om, b(k_1)\ldots b(k_m)\phi_{p,\si}\ran. \label{modified-wave-functions-I}
\eeqa
(The r.h.s. above is well defined by considerations from Appendix D of \cite{DP12} and by the simple relation between $\phi_{p,\si}$ and $\psi_{p,\si}$). 
We have the following:
%%%%%%%%%%%%%%%%%%%%%%%%%%%%%%%%%%
\begin{prop}\label{modified-wavefunctions-lemma} Under the assumptions of Theorem~\ref{main-theorem-spectral}, we have
\beqa
|f^m_{\mrm{w}, p,\si}(k)|\leq \fr{1}{\sqrt{m!}} g^m_{\si}(k), \quad   |\pa_{p^j}f^m_{\mrm{w}, p,\si}(k)|\leq \fr{1}{\si^{\de_{\la_0}}} \fr{1}{\sqrt{m!}}  g^m_{\si}(k). \label{modified-wave-functions}
\eeqa
\end{prop}
%%%%%%%%%%%%%%%%%%%%%%%%%%%%%%%%%%
\proof In formula (4.42) of \cite{DP16} we define
\beqa
\hat{f}^m_{p,\si}(k_1, \ldots, k_m)=W_{p,\si}^*b(k_1)\ldots b(k_n)\phi_{p,\si}. \label{functions-from-different-reference}
\eeqa
Then in Proposition 4.7 of  \cite{DP16} and the subsequent analysis in this reference we establish that
\beqa
\|\hat{f}^m_{p,\si}(k)\|\leq g^m_{\si}(k), \quad   \|\pa_{p^j}\hat{f}^m_{p,\si}(k)\|\leq \fr{1}{\si^{\de_{\la_0}}}g^m_{\si}(k). \label{spectral-info-I}
\eeqa
By comparing (\ref{modified-wave-functions-I}) with (\ref{functions-from-different-reference}) and recalling  that
$W_{p,\si}=e^{b^*(f_{p,\si}) -b(f_{p,\si}) }$, $f_{p,\si}(k):=\fr{\chi_{[\si,\ka)}(k) }{\sqrt{2|k|}|k|(1-\nabla E_{p,\si}\cdot e_k)}$, we obtain
\beqa
f^{m}_{\mrm{w},p,\si}(k)\1&=&\1\fr{1}{\sqrt{m!}} \lan W_{p,\si}^*\Om, \hat{f}^m_{p,\si}(k)\ran,\\
\pa_{p^j}  f^{m}_{\mrm{w},p,\si}(k)\1&=&\1- \fr{1}{\sqrt{m!}} \lan W_{p,\si}^*b^*(f_{p,\si})  \Om, \hat{f}^m_{p,\si}(k)\ran+\fr{1}{\sqrt{m!}} \lan W_{p,\si}^*\Om, \pa_{p^j}\hat{f}^m_{p,\si}(k)\ran.
\eeqa
Thus (\ref{spectral-info-I}) gives (\ref{modified-wave-functions}). \qed
}

%%%%%%%%%%%%%%%%%%%%%%%%%%%%%%%%%%%%%%%
\section{Domain questions} \label{domain-questions}
%%%%%%%%%%%%%%%%%%%%%%%%%%%%%%%%%%%%%%%

\setcounter{equation}{0}

%%%%%%%%%%%%%%%%%%%%%%
\bel\label{self-adjointness} 
There exist constants $0\leq a<1$  and $b\geq 0$ s.t. for any  $\Psi\in \mcC^{(n_1,n_2)}$ there holds the bound
\beqa
\|H_{\I}^{(n_1,n_2)}\Psi\|\leq a\|H_{\free}^{(n_1,n_2)}\Psi\|+b\|\Psi\|, \label{Kato-Relich}
\eeqa 
where $H_{\I}^{(n_1,n_2)}=H_{\I}|_{\mcC^{(n_1,n_2)}}$, $H_{\free}^{(n_1,n_2)}=H_{\free}|_{\mcC^{(n_1,n_2)}}$ 
and the constant $b$ may depend on $n$.
\eel
%%%%%%%%%%%%%%%%%%%%%
\proof Let us use the form of the interaction Hamiltonian appearing in formula (\ref{explicit-Hamiltonian}). We have
 $H_{\I}^{(n_1,n_2)}=H_{\I}^{\ain, (n_1,n_2)}+H_{\I}^{\cin, (n_1,n_2)}$, where 
\beqa
H_{\I}^{\ain, (n_1,n_2)}:=H_{\I,1}^{\ain, (n_1)}+H_{\I,2}^{\ain, (n_2)}
 :=\sum_{j=1}^{n_1}\int d^3k\, \vv_1(k)\, e^{ikx_{1,j}}a(k)+\sum_{j=1}^{n_2}\int d^3k\, \vv_2(k)\, e^{ikx_{2,j}}a(k)
\eeqa
and $H_{\I}^{\cin, (n_1,n_2)}:=(H_{\I}^{\ain, (n_1,n_2)})^*$, $H_{\I,1/2}^{\cin, (n_{1/2})}:=(H_{\I,1/2}^{\ain, (n_{1/2})})^*$. 
Let us set $C_{1/2}(k):= \sum_{j=1}^{n_{1/2}} e^{ikx_{1/2,j}}$ 
and compute for some $\Psi\in \mcC^{(n_1,n_2)}$ 
\beqa
\|H_{\I,1/2}^{\ain,(n_{1/2})}\Psi\|\1&\leq&\1 \int d^3k\,\vv_{1/2}(k)\|C_{1/2}(k) a(k)\Psi\|\leq n_{1/2} \int d^3k\,\vv_{1/2}(k)\|a(k)\Psi\|\non\\
\1&\leq&\1 n_{1/2}\|\om^{-1/2}\vv_{1/2}\|_2\lan \Psi,H_{\pho}\Psi\ran^{\h}\leq \fr{1}{8}\|H_{\free}\Psi\|+2n_{1/2}^2\|\om^{-1/2}\vv_{1/2}\|_2^2\|\Psi\|, \label{KR-bound}
\eeqa
where in the last step we anticipate that  (\ref{Kato-Relich}) should hold with $0<a<1$.

Let us now consider the creation part of $H_{\I}^{(n_1,n_2)}$. Making use of the canonical commutation relations, we get
\beqa
\|\ti{H}_{\I, 1/2}^{\cin,(n_{1/2})}\Psi\|^2\1&=&\1\int d^3k_1d^3k_2\,\vv_{1/2}(k_1)\vv_{1/2}(k_2)\lan C_{1/2}(k_1)^*a^*(k_1)\Psi,C_{1/2}(k_2)^*a^*(k_2)\Psi\ran\non\\
\1&\leq &\1 \|\ti{H}_{\I,1/2}^{\ain,(n_{1/2})}\Psi\|^2+  n_{1/2}^2 \|\vv_{1/2}\|_2^2   \|\Psi\|^2,
\eeqa
{where $\ti{H}_{\I,1/2}^{\ain,(n_{1/2})}$ differs from $H_{\I,1/2}^{\ain,(n_{1/2})}$ by an (inessential) replacement $x_{1/2,j}\to -x_{1/2,j}$}. Since  
$\|\ti{H}_{\I,1/2}^{\ain,(n_{1/2})}\Psi\|$ satisfies the bound (\ref{KR-bound}),  the proof is complete. \qed
%%%%%%%%%%%%%%%%%%%%%%%%%%%%%%%%%

%%%%%%%%%%%%%%%%%%%%%%%%%%%
\bel \label{ren-creation-operator} The domain $\mcD$, defined in (\ref{invariant-domain}), is contained in the
domains of  $H, H_{\at}, H_{\el}, H_{\pho}, H_{\I}^{\ain/\cin}$ and $\nr^*_{1/2,\si}(h)$, $h\in \B(\PSS)$. Moreover, these
operators leave $\mcD$ invariant.
\eel
%%%%%%%%%%%%%%%%%%%%%%%%%%%%%%
\proof Let $\Psi_{l_1,l_2}^{r_1,r_2}$ be a vector of the form (\ref{psi-l}). Then
\beqa
& &\nr^*_{1,\si}(h)\Psi_{l_1,l_2}^{r_1,r_2}\non\\
&=&\sum_{m,m'}\fr{1}{\sqrt{m'!}} \fr{1}{\sqrt{m!}} \int d^3p'\,  d^{3l_1}p_1\,d^{3l_2}p_2\,d^{3m'}k'\, d^{3m}k\, 
F'_{1,m'}(p';k')F_{l_1,l_2,m}(p_1,p_2;k)   \non\\
& &\ph{44444444444444444444444444}\times \nn_1^*(p') \nn_1^*(p_1)^{l_1} \nn_2^*(p_2)^{l_2}\bb^*(k')^{m'}\bb^*(k)^m\vac \non\\
&=& \sum_{\ti m=0}^{\infty}\fr{1}{\sqrt{\ti m!}}    
\int   d^{3(l_1+1)}\ti p_1\,d^{3l_2} p_2\, d^{3\ti m}\ti k \, \ti F_{\ti m}(\ti p_1, p_2;\ti k)  \nn_1^*(\ti p_1)^{l_1+1}\nn_2^*(p_2)^{l_2} \bb^*(\ti k)^{\ti m}\vac,    
\eeqa
where $F'_{1,m'}(p';k'):= h(p'+\unk') f^{m'}_{1, p'+\unk,\si}(k')$, $\ti m:=m+m'$, $\ti k:=(k,k')$, $\ti p_1:=(p_1,p')$ and
\beqa
\ti F_{\ti m}(\ti p_1,p_2;\ti k)=\sum_{m=0}^{\ti m} \fr{ \sqrt{\ti m!} }{   \sqrt{(\ti m-m)!} \sqrt{m!}  }(F'_{1,(\ti m-m) }F_{l_1,l_2,m})_{\sym}  (\ti p_1,p_2;\ti k),
\eeqa
where the symmetrization is performed in the $\ti p_1$, ${\mc p_2}$ and $\ti k$ variables separately. By Theorem~\ref{main-theorem-spectral},   $F'_{1,m'}$
satisfies the bound  (\ref{factorial-bound}), and therefore
\beqa
\|\ti F_{\ti m}\|_2\leq \fr{c^{\ti m}}{\sqrt{\ti m!}},
\eeqa
for some constant $c$. Hence $\nr^*_{1,\si}(h)\Psi_{l_1,l_2}^{r_1,r_2}$ is well defined and belongs to $\mcD$. The corresponding statements for
$\nr^*_{2,\si}(h)$ are proven analogously.

Next, we note that
\beqa
& &H_{\at}\Psi_{l_1,l_2}^{r_1,r_2}=\sum_{m=0}^{\infty}\fr{1}{\sqrt{m!}}\int d^{3l_1}p_1\, d^{3l_2}p_2\,d^{3m}k\,(\Omt(p_{1,1})+\cdots+\Omt(p_{1,l_1}) ) \non\\
& &\ph{4444444444444444444444444444444}\times F_{l_1,l_2,m}(p_1;p_2;k)\nn^*(p_1)^{l_1} \nn^*(p_2)^{l_2}   \bb^*(k)^m\vac,\\
& &H_{\el}\Psi_{l_1,l_2}^{r_1,r_2}=\sum_{m=0}^{\infty}\fr{1}{\sqrt{m!}}\int d^{3l_1}p_1\, d^{3l_2}p_2\, d^{3m}k\,(\Omt(p_{2,1})+\cdots+\Omt(p_{2,l_2}) ) \non\\
& &\ph{4444444444444444444444444444444}\times F_{l_1,l_2,m}(p_1;p_2;k)   \nn^*(p_1)^{l_1} \nn^*(p_2)^{l_2}\bb^*(k)^m\vac,\\
& &H_{\pho}\Psi_{l_1,l_2}^{r_1,r_2}=\sum_{m=0}^{\infty}\fr{1}{\sqrt{m!}}\int d^{3l_1}p_1\, d^{3l_2}p_2\, \,d^{3m}k\,(\om(k_1)+\cdots+\om(k_m) ) \non\\
& &\ph{4444444444444444444444444444444}\times   F_{l_1,l_2, m}(p_1;p_2;k)   \nn^*(p_1)^{l_1} \nn^*(p_2)^{l_2}\bb^*(k)^m\vac.
\eeqa
Due to the support properties of $F_{l_1,l_2,m}$ these vectors are well defined and belong to $\mcD$.

Finally, we consider the operators  $H_{\I}^{\ain/\cin}$. We recall that the interaction Hamiltonian restricted to $\hil^{(l_1,l_2)}$
has the form $H_{\I}^{(l_1,l_2)}=H_{\I}^{\ain, (l_1,l_2)}+H_{\I}^{\cin, (l_1,l_2)}$, where 
\beqa
H_{\I}^{\ain, (l_1,l_2)}:=H_{\I,1}^{\ain, (l_1)}+H_{\I,2}^{\ain, (l_2)}
 :=\sum_{j=1}^{l_1}\int d^3k\, \vv_1(k)\, e^{ikx_{1,j}}a(k)+\sum_{j=1}^{l_2}\int d^3k\, \vv_2(k)\, e^{ikx_{2,j}}a(k)
\eeqa
and $H_{\I}^{\cin, (l_1,l_2)}:=(H_{\I}^{\ain, (l_1,l_2)})^*$, $H_{\I,1/2}^{\cin, (l_{1/2})}:=(H_{\I,1/2}^{\ain, (l_{1/2})})^*$. 
Now we express $\Psi_{l_1,l_2}^{r_1,r_2}$ in terms of its $m$-photon components, i.e., 
\beqa
\{\Psi_{l_1,l_2}^{r_1,r_2}\}^{(l_1,l_2,m)}(p_1; p_2;k_1,\ldots,k_m)=F_{l_1,l_2,m}(p_{1,1},\ldots, p_{1,l_1}; p_{2,1},\ldots, p_{2,l_2}; k_1,\ldots, k_m).
\eeqa
We can write
\beqa
(H_{\I}^{\ain}\Psi_{l_1,l_2}^{r_1,r_2})^{(l_1,l_2,m)}(p_1; p_2; k_1,\ldots, k_m)\ph{4444444444444444444444444444444444444444444}\non\\
=\sqrt{m+1}\int d^3k\,\vv_1(k)\sum_{i=1}^{l_1} (\Psi_{l_1,l_2}^{r_1,r_2})^{(l_1,l_2,m+1)}(p_{1,1},\ldots, p_{1,i}-k,\ldots, p_{1,l_1}; p_2; k, k_1,\ldots, k_m)\non\\
+\sqrt{m+1}\int d^3k\,\vv_2(k)\sum_{i=1}^{l_2} (\Psi_{l_1,l_2}^{r_1,r_2})^{(l_1,l_2,m+1)}(p_1; p_{2,1},\ldots, p_{2,i}-k,\ldots, p_{2,l_2};  k, k_1,\ldots, k_m).
\eeqa
It is easy to see that for some constant $c$, independent of $m$,
\beqa
\|(H_{\I}^{\ain}\Psi_{l_1,l_2}^{r_1,r_2})^{(l_1,l_2,m)}\|_2\leq \fr{c^m}{\sqrt{m!}}. \label{c-m-bound}
\eeqa
Similarly, we obtain that
\beqa
& &(H_{\I}^{\cin}\Psi_{{\mc l_1,l_2}}^{r_1,r_2})^{(l_1, l_2, m)}(p_1; p_2; k_1,\ldots, k_m)\ph{444444444444444444444444444}\non\\
& &=\fr{1}{\sqrt{m}}\sum_{i=1}^m\sum_{j=1}^{l_1}\vv_1(k_i)(\Psi_{l_1,l_2}^{r_1,r_2})^{(l_1,l_2,m-1)}(p_{1,1},\ldots,p_{1,j}+k_i,\ldots, p_{1,l_1}; p_2; k_1, \ldots,k_{i_*},\ldots, k_m)\non\\
& &+\fr{1}{\sqrt{m}}\sum_{i=1}^m\sum_{j=1}^{l_2}\vv_2(k_i)(\Psi_{l_1,l_2}^{r_1,r_2})^{(l_1,l_2,m-1)}(p_1; p_{2,1},\ldots,p_{2,j}+k_i,\ldots, p_{2,l_2};  k_1,\ldots,k_{i_*},\ldots, k_m),\quad\quad
\eeqa
where $k_{i_*}$ means omission of the $i$-th variable.
This gives, again, a bound of the form (\ref{c-m-bound}) and concludes the proof. \qed 
%%%%%%%%%%%%%%%%%%%%%%%%%%%%%%%%%%%%%%%%%%%%%%%%%%%%%%%%%%%%%%%%%%%%%%%%%%%%%%%%%

%%%%%%%%%%%%%%%%%%%%%%%%%%%%%%%%%%%%%%%%%
\section{Fock space combinatorics}\label{Fock-space-combinatorics}
\setcounter{equation}{0}

%%%%%%%%%%%%%%%%%%%%%%%%%%%%%%%%%%%%%%%%%%
In Lemmas~\ref{norms-of-single-particle-states}, \ref{double-creation-lemma} and  \ref{norms-of-scattering-states-one}
below we verify the identities first for $G_{i,m}, G_{i,m}', F_{n,m}, F'_{n,m}$ of Schwartz class and then extend them to
square integrable functions using Theorem X.44 of \cite{RS2}.
%%%%%%%%%%%%%%%%%%%%%%%%%%%%%%%%%%%%%%%%%%
\bel\label{norms-of-single-particle-states} Let  $G_{i,m}, G'_{i',m}\in L^2(\real^3\times \real^{3m})$ be symmetric in their photon variables, 
see (\ref{notation-G-m}).
Let us define as operators on $\mcC$
\beqa
B_{i,m}^*(G_{i,m}):=\int d^3p\,d^{3m}k\, G_{i,m}(p;k) \neta_i^*(p-\unk)a^*(k)^m, \quad i=1,2
\eeqa
and $B_{i,m}(G_{i,m}):=(B_{i,m}^*(\ov{G}_{i,m}))^*$.
Then there holds the identity
\beqa
\lan\vac, B_{i',m}(G_{i',m}') B_{i,m}^*(G_{i,m}) \vac\ran=\de_{i,i'} m!\int d^3p  d^{3m}k\,  \ov  G_{i',m}'(p;  k) G_{i,m}(p;  k ) .
\eeqa
\eel
%%%%%%%%%%%%%%%%%%%%%%%%%%%%%%%%%%%%%%%%%%
\proof We compute
\beqa
& &\lan\vac, B_{i',m}(G_{i',m}) B_{i,m}^*(G_{i,m}) \vac\ran\non\\
\1&=&\1\int d^3p\,d^{3m}k\,\int d^3p'\,d^{3m}k'\, \ov G_{i',m}'(p'; k') G_{i,m}(p;k)                     
 \lan\vac, a(k')^m\neta_{i'}(p'-\unk') \neta_i^*(p-\unk)a^*(k)^m\vac\ran\non\\
\1&=&\1\de_{i,i'}\int d^3p\,d^{3m}k\,\int d^3p'\,d^{3m}k'\,    \ov G_{i',m}'(p'; k') G_{i,m}(p;k)                     
\de(p-\unk-p'+\unk') \lan\vac, a(k')^m a^*(k)^m\vac\ran\non\\
\1&=&\1\de_{i,i'}\int d^3p\,d^{3m}k\,\int\,d^{3m}k'\, \ov G_{i',m}'(p-\unk+\unk'; k') G_{i,m}(p;k)                      
 \lan\vac, a(k')^m a^*(k)^m\vac\ran\non\\
\1&=&\1\de_{i,i'} \int d^3p\,d^{3m}k\,\int\,d^{3m}k'\, \ov G_{i',m}'(p-\unk+\unk'; k') G_{i,m}(p;k)                      
 \sum_{\rho\in S_m}\prod_{j=1}^m\de(k_{\rho(j)}-k'_j) \non\\
\1&=&\1\de_{i,i'} m!\int d^3p\,d^{3m}k\, G_{i,m}(p;k)  \ov G_{i',m}'(p; k),                    
 \eeqa
where $S_m$ is the set of all permutations of an $m$-element set and in the last step we exploited the fact that $G_{m}'$
is symmetric in its photon variables. \qed
%%%%%%%%%%%%%%%%%%%%%%%%%%%%%%%%%%%%%%%%%

{In the following lemma we describe various contraction patterns which will appear in the expressions considered below in this appendix.
We refer to Section 3 of \cite{DP12.0} for graphical illustrations and simple examples of these contraction patterns.  }
%%%%%%%%%%%%%%%%%%%%%%%%%%%%%%%%%%%%%%%%%%
\bel\label{combinatorics} Let $n,m,\ti n,\ti m\in \nat_0$ be s.t. $n+m=\ti n+\ti m$.  Let us choose
\beqa
 r=(r_1,\ldots,r_n)\in \real^{3n},& & 
k=(k_1,\ldots, k_m)\in\real^{3m},  \\
\ti r=(\ti r_1,\ldots, \ti r_{\ti n})\in \real^{3\ti n}, & &
\ti k=(\ti k_1,\ldots,\ti k_{\ti m}  )\in\real^{3\ti m}
\eeqa
and define the sets
\beqa
& &C_n:=\{1,\ldots, n\}, \quad C_n':=\{n+1,\ldots,n+m\}, \\
& &C_{\ti n}:=\{1,\ldots,\ti n\}, \quad  C_{\ti n}':=\{\ti n+1,\ldots,\ti n+ \ti m\}. 
\eeqa
(Note that $C_n'$ is the complement of $C_n$ in $\{1,\ldots, n+m\}$. Similarly for $C_{\ti n}'$).
Let $S_{m+n}$ be the set of all permutations of an $m+n$ element set. 
For any $\sig\in S_{m+n}$ we introduce the following notation: 
\beqa
& &\hr:=(r_i)_{(i,\sig(i))\in C_n\times C_{\ti n}},\quad\quad \chr:=(r_i)_{(i,\sig(i))\in C_n\times C_{\ti n}' },\\
& &\hk:=(k_{i-n})_{(i,\sig(i))\in C_n'\times C_{\ti n}' },\quad \chk:=(k_{i-n})_{(i,\sig(i))\in C_n'\times C_{\ti n} },
\eeqa
so that $r=(\hr,\chr)$, $k=(\hk,\chk)$. Similarly,
\beqa
& &\hat{\ti r}:=(\ti{r}_{\sig(i)})_{(i,\sig(i))\in C_n\times C_{\ti n}},\quad\quad \check{\ti r}:=(\ti r_{\sig(i)})_{(i,\sig(i))\in C_n'\times C_{\ti n} },\\
& &\hat{\ti k}:=(\ti k_{\sig(i)-\ti n})_{(i,\sig(i))\in C_n'\times C_{\ti n}' },\quad \check{\ti k}:=(\ti k_{\sig(i)-\ti n })_{(i,\sig(i))\in C_n\times C_{\ti n}' },
\eeqa
so that $\ti r=(\hat{\ti r},  \check{\ti r})$ and $\ti k=(\hat{\ti k}, \check{\ti k})$. (If $\{\, i\,|\, (i,\sig(i))\in C_n\times C_{\ti n}\,\}=\emptyset$
then we say that $\hr$ is empty, and analogously for other collections of photon variables introduced above).
Finally, we define
\beqa
\de(\hr-\hat{\ti  r})\1&:=&\1\prod_{(i,\sig(i))\in C_n\times C_{\ti n} } \de(r_i-\ti r_{\sig(i)}),\\
\de(\chr-\check{\ti k})\1&:=&\1\prod_{(i,\sig(i))\in C_n\times C_{\ti n}' } \de(r_i-\ti k_{\sig(i)-\ti n} ),\\
\de(\chk-\check{\ti r})\1&:=&\1\prod_{(i,\sig(i))\in C_n'\times C_{\ti n} } \de(k_{i-n}-\ti r_{\sig(i)} ), \\
\de(\hk-\hat{\ti k} )\1&:=&\1\prod_{(i,\sig(i))\in C_n'\times C_{\ti n}' } \de(k_{i-n}-\ti k_{\sig(i)-\ti n} ).
\eeqa 
Then there holds, {as an equality of tempered distributions}
\beqa
\lan\vac, \bb(\ti r)^{\ti n} \bb(\ti k)^{\ti m} \bb^*(r)^{n} \bb^*(k)^{m}\vac\ran=\sum_{\sig\in S_{m+n}}
\de(\hr-\hat{\ti  r})\de(\chr-\check{\ti k})\de(\chk-\check{\ti r})\de(\hk-\hat{\ti k} ).
\eeqa
\eel
%%%%%%%%%%%%%%%%%%%%%%%%%%%%%%%%%%%%%%%%%%%%%%%%%%%%%%%%%%%%%%%%%
\proof Let $(v_1,\ldots, v_{n+m})=(r_1,\ldots,r_n,k_1,\ldots, k_m)$
and $(\ti v_1,\ldots,\ti v_{n+m})=(\ti r_1,\ldots,\ti r_{\ti n},\ti k_1,\ldots, \ti k_{\ti m})$. There holds
\beqa
& &\lan\vac, \bb(\ti r)^{\ti n} \bb(\ti k)^{\ti m} \bb^*(r)^{n} \bb^*(k)^{m}\vac\ran=\sum_{\sig\in S_{m+n}}\prod_{j=1}^{m+n}\de(v_{j}-\ti v_{\sig(j)})\non\\
& &=\sum_{\sig\in S_{m+n}} \de(r_1-\ti v_{\sig(1)} )\ldots  \de(r_n-\ti v_{\sig(n)})\de(k_1-\ti v_{\sig(n+1)})\ldots \de(k_m-\ti v_{\sig(n+m)})\\
& &=\sum_{\sig\in S_{m+n}}\bigg(\prod_{(i, \sig(i))\in C_n\times C_{\ti n}} \de(r_i-\ti r_{\sig(i)} )\bigg)
\bigg(\prod_{(i,\sig(i))\in C_n\times C_{\ti n}' } \de(r_i-\ti k_{\sig(i)-\ti n} )\bigg)\non\\
& &\quad\quad\quad \times\bigg(\prod_{(i,\sig(i))\in C_n'\times C_{\ti n} } \de(k_{i-n}-\ti r_{\sig(i)} )\bigg) 
\bigg(\prod_{(i,\sig(i))\in C_n'\times C_{\ti n}' } 
\de(k_{i-n}-\ti k_{\sig(i)-\ti n} )\bigg),
\eeqa
which concludes the proof. \qed
%%%%%%%%%%%%%%%%%%%%%%%%%%%%%%%%%%%%%%%%%
\bel\label{double-creation-lemma} Let $F_{n,m}, F'_{n,m} \in  L^2((\real^3\times\real^{3n})\times (\real^3\times\real^{3m}))$ be symmetric in the photon variables.  Let us introduce the following operators on $\mcC$
\beqa
B^*_{n,m}(F_{n,m}):=\int d^3q d^3p\int d^{3n}r  d^{3m}k\, F_{n,m}(q;r \ba p ; k) \bb^*(r)^{n}\bb^*(k)^m \nn_2^*(p-\unk)\nn_1^*(q-\unr)
\label{double-creation-two}
\eeqa
and set $B_{n,m}(F_{n,m}):=(B^*_{n,m}({\mc {F}_{n,m}}))^*$.  There holds
\beqa
& &\lan  B^{{\mc *}}_{\ti n,\ti m}(F'_{\ti n,\ti m})\vac, B_{n,m}^{{\mc *}}(F_{n,m})\vac\ran
=\sum_{\sig\in S_{m+n}}\int  d^3 q d^3p \int  d^{3n}r d^{3m}k\, F_{n,m}(q;r \ba p; k) \non\\
& &\ph{44444444444444444444444444}\times \ov F_{\ti n,\ti m}'(q+\uchk-\uchr; \hr, \chk \ba p-\uchk+\uchr; \hk, \chr) 
\eeqa
for $n+m=\ti n+\ti m$. Otherwise the expression on the l.h.s.  is zero. Here $S_{m+n}$ is the set of permutations of an $m+n$ element set
and the notation $\hk, \chk, \hr, \chr$ is explained in Lemma~\ref{combinatorics}.
\eel
%%%%%%%%%%%%%%%%%%%%%%%%%%%%%%%%%%%%%%%%%%
\proof We compute the expectation value
\beqa
 & &\lan  B_{\ti n,\ti m}^*(F'_{\ti n,\ti m})\vac, B_{n,m}^*(F_{n,m})\vac\ran \\
&=&\int d^3\ti q  d^3\ti p d^3 q d^3p \int d^{3\ti n}\ti r d^{3\ti m}\ti k   d^{3n}r d^{3m}k\,  
\ov F'_{\ti n,\ti m}(\ti q;\ti r \ba \ti p; \ti k) F_{n,m}(q;r \ba p; k)\non\\   
& &\times\de(\ti q- q-\ti{\unr}+\unr)\de(\ti p+\ti q-p-q)
\lan \vac, \bb(\ti r)^{\ti n} \bb(\ti k)^{\ti m} \bb^*(r)^{n} \bb^*(k)^{m}\vac\ran.       
\label{generalized-creation-operators-one}
\eeqa
The last factor is non-zero only if $\ti n+\ti m=n+m$.  
Then
\beqa
\lan\vac, \bb(\ti r)^{\ti n} \bb(\ti k)^{\ti m} \bb^*(r)^{n} \bb^*(k)^{m}\vac\ran
=\sum_{\sig\in S_{m+n}}\de(\hr-\hat{\ti  r})\de(\chr-\check{\ti k})\de(\chk-\check{\ti r})\de(\hk-\hat{\ti k} ),
\eeqa
where we made use of   Lemma~\ref{combinatorics}.
Thus the r.h.s. of (\ref{generalized-creation-operators-one}) is a sum over $\sig\in S_{m+n}$ of terms of the form:
\beqa
& &\int d^3\ti q  d^3\ti p d^3 q d^3p \int  d^{3n}r d^{3m}k\,  
 \ov F'_{\ti n,\ti m}( \ti q; \hr, \chk \ba \ti p; \hk, \chr)    F_{n,m}(q; r \ba p; k )    \non\\ 
 & &\ph{444444444444444}\times\de(\ti p+\ti q-p-q) \de(\ti q- q-\uchk+\uchr) \non\\
&=&\int  d^3 q d^3p \int  d^{3n}r d^{3m}k\,  
 F_{n,m}(q; r \ba p; k )   \ov F_{\ti n,\ti m}'( q+\uchk-\uchr; \hr, \chk \ba p-\uchk+\uchr; \hk, \chr ) 
\eeqa   
which concludes the proof. \qed

%%%%%%%%%%%%%%%%%%%%%%%%%%%%%%%%%%%%%%%%%%%
\bel\label{norms-of-scattering-states} Let $G_{1,m},G_{1,m}',G_{2,m}, G_{2,m}'\in L^2(\real^3\times \real^{3m})$ be symmetric in the photon variables. 
We define, as an operator on $\mcC$,
\beqa
B_{i,m}^*(G_{i,m}):=\int d^3p\,d^{3m}k\, G_{i,m}(p;k) \neta_i^*(p-\unk)a^*(k)^m \label{B-star-def-new}
\eeqa
and set $B_{i,m}(G_{i,m})=(B_{i,m}^*({\mc {G}_{i,m}}))^*$. There holds the identity
\beqa
\lan\vac, B_{1,\ti n}(G_{1,\ti n}' )B_{2,\ti m}(G_{2,\ti m}' )B_{1,n}^*(G_{1,n}) B_{2,m}^*(G_{2,m})\vac\ran\pha{444444444444444444444444444444}\non\\
=\sum_{\sig\in S_{m+n}}\int  d^3 q d^3p \int  d^{3n}r d^{3m}k\,  
G_{1,n}(q;  r) G_{2,m}(p;  k)\pha{44444444444444444444444}\non\\
 \times\bigg(\ov G_{1,\ti n}'( q+\uchk-\uchr ;  \hr, \chk)
\ov  G_{2,\ti m}'(p-\uchk+\uchr; \hk, \chr)\bigg)
\eeqa
for $n+m=\ti n+\ti m$. Otherwise the expression on the l.h.s.  is zero. Here $S_{m+n}$ is the set of permutations of an $m+n$ element set
and the notation $\hk, \chk, \hr, \chr$ is explained in Lemma~\ref{combinatorics}.
\eel
%%%%%%%%%%%%%%%%%%%%%%%%%%%%%%%%%%%%%%%%%%%
\proof Follows immediately from Lemma~\ref{double-creation-lemma}. \qed 
%%%%%%%%%%%%%%%%%%%%%%%%%%%%%%%%%%%%%%%%%%%

 We recall from (\ref{check-hamiltonian}) that
\beqa
\cH_{\I,\sipty}^{\cin}:=\sum_{i\in\{1,2\}}  \cH_{\I,\sipty,i}^{\cin}:=\sum_{i\in\{1,2\}}\int d^3p d^3k\,\cvv_i^{\sipty}(k)\nn_i^*(p-k)\bb^*(k)\nn_i(p), \label{check-hamiltonian-app}
\eeqa
where $\cvv_1^{\sipty}(k)=\g\fr{\mathbf{1}_{\mcB_\si}(k) |k|^{\alf} }{(2|k|)^{1/2}}$,  
$\cvv_2^{\sipty}(k)=\g\fr{\mathbf{1}_{\mcB_\si}(k)}{(2|k|)^{1/2}}$. 
%%%%%%%%%%%%%%%%%%%%%%%%%%%%%%%%%%%%%%%%%%%%%%%%%%%%%%%%%%%%%%%%%%%%%
\bel\label{norms-of-scattering-states-one-zero} Let $G_{2,m} \in L^2(\real^3\times \real^{3m})$ be supported in 
$\real^3\times \{\, k\in \real^3\,|\, |k|\geq \si\,\}^{\times m}$ and symmetric in its photon variables.  There holds the identity
\beqa
\lan\vac, B_{2,\tin}(\Gtin ) (\cH_{\I,\sipty,i'}^{\cin} )^*   \cH_{\I,\sipty,i}^{\cin}  B_{2,m}^*(\Hm)\vac\ran=
\|\cvv_2^{\sipty}\|_2^2 m!  \de_{m,\ti n} \int    d^3p \int   d^{3m}k\,  |\Hm(p; k)|^2.\,\,\,
\eeqa
\eel
%%%%%%%%%%%%%%%%%%%%%%%%%%%%%%%%%%%%%%%%%%
\proof We compute the expectation value
\beqa
& &\lan\vac, B_{2,\tin}(\Gtin ) (\cH_{\I,\sipty,i'}^{\cin} )^*   \cH_{\I,\sipty,i}^{\cin}  B_{2,m}^*(\Hm)\vac\ran\non\\  
& &=\int d^3\ti u d^3\ti w d^3u d^3w\int d^3\ti q   d^3p \int d^{3\tin}\ti r  d^{3m}k\, 
 \cvv_{i'}^{\sipty}(\ti w) \cvv_i^{\sipty}(w){\mc \ov{G}_{2, \ti n}}(\ti q;\ti r)  \Hm(p; k)\non\\ 
& &\ph{44}\times\lan \vac,  \neta_{2}(\ti q-\ti{\unr} )  \nn_{i'}^*(\ti u) \nn_{i'}(\ti u-\ti w)   
 \nn_i^*(u-w)\nn_i(u)\neta_2^*(p-\unk) \vac\ran\non\\
& &\ph{44}\times\lan \vac, \bb(\ti r)^{\tin}\bb(\ti w)  \bb^*(w) \bb^*(k)^{m}\vac\ran.       
\label{generalized-creation-operators-zero}
\eeqa
We note that
\beqa
& &\lan \vac,  \neta_2( \ti q-\ti{\unr}  )  \nn_{i'}^*(\ti u) \nn_{i'}(\ti u-\ti w)    \nn_i^*(u-w)\nn_i(u)    
 \neta_2^*(p-\unk) \vac\ran\non\\
& &=\de_{i',2}\de_{i,2}\de(\ti q-\ti{\unr} -\ti u)\de(p-\unk- u)\lan \vac,\nn_{i'}(\ti u-\ti w)  \nn_i^*(u-w) \vac\ran\non\\
& &=\de_{i',2}\de_{i,2}\de(\ti q-\ti{\unr}-\ti u)\de(p-\unk- u)   
\de(\ti u-\ti w-u+w).\label{deltas-with-insertions-zero}
\eeqa
In the following we can assume $i=2,i'=2$, as otherwise the expression is zero. 
Now we consider the expectation value of the photon creation operators:
\beqa
\lan \vac, \bb(\ti r)^{\ti n}  \bb(\ti w)\bb^*(w)   \bb^*(k)^{m}\vac\ran
=\de(w-\ti w)\lan \vac, \bb(\ti r)^{\ti n}   \bb^*(k)^{m}\vac\ran\label{many-deltas-zero},
\eeqa
for $\ti r,  k, w, \ti w$ in the supports of the respective functions. (Here we made use of the fact that $|\ti w|\leq \si$, 
whereas $|r_i|\geq \si$, $|k_j|\geq \si$).
Thus we have
\beqa
& &\!\!\!\!\!\!\!
\lan\vac, B_{2,\tin}(\Gtin) (\cH_{\I,\sipty,2}^{\cin}  )^*   \cH_{\I,\sipty,2}^{\cin}  B_{2,m}^*(\Hm)\vac\ran\non\\  
\1&=&\1\int d^3w\int d^3\ti q    d^3p \int d^{3\tin}\ti r  d^{3m}k\,  
\cvv_2^{\sipty}(\ti q-\ti{\unr}-p+\unk+w) \cvv_2^{\sipty}(w)\non\\
& &\times {\mc \ov{G}_{2,\ti n}}(\ti q;\ti r)  \Hm(p; k)
\lan \vac, \bb(\ti r)^{\tin}\bb(\ti q-\ti{\unr}-p+\unk+w)  \bb^*(w) \bb^*(k)^{m}\vac\ran \non\\
\1&=&\1 \int d^3w\int  d^3\ti q d^3p \int d^{3\tin}\ti r  d^{3m}k\,  
\cvv_2^{\sipty}(\ti q-\ti{\unr}-p+\unk+w) \cvv_2^{\sipty}(w)\non\\
& &\times {\mc \ov{G}_{2,\ti n}}(\ti q;\ti r)  \Hm(p; k)\de(\ti q-\ti{\unr}-p+\unk)
\lan \vac, \bb(\ti r)^{\tin} \bb^*(k)^{m}\vac\ran \non\\
\1&=&\1 \|\cvv_2^{\sipty}\|_2^2 m!  \de_{m,\ti n} \int    d^3p \int   d^{3m}k\,  \Hm(p; k)\ov{G}_{2,\ti n}(p;k),
\eeqa
where in the first step we substituted~(\ref{deltas-with-insertions-zero}) and integrated over $\ti u, \ti w, u$, in the second step we used (\ref{many-deltas-zero})
and in the last step we used Lemma~\ref{norms-of-single-particle-states}. This concludes the proof. \qed
%%%%%%%%%%%%%%%%%%%%%%%%%%%%%%%%%%

%%%%%%%%%%%%%%%%%%%%%%%%%%%%%%%%%%%%%%%%%%%%%%%%%%%%%%%%%%%%%%%%%%%%%
\bel\label{norms-of-scattering-states-one} Let $G_{1,m},G_{2,m} \in L^2(\real^3\times \real^{3m})$ be supported in 
$\real^3\times \{\, k\in \real^3\,|\, |k|\geq \si\,\}^{\times m}$ and symmetric in their photon variables.  
There holds the identity
\beqa
& &\lan\vac, B_{2,\tin}(\Gtin ) (\cH_{\I,\sipty}^{\cin} )^*  B_{1,\tim}(\Htim )B_{1,n}^*(\Gn) \cH_{\I,\sipty}^{\cin}  B_{2,m}^*(\Hm)\vac\ran\non\\
& &=\|\cvv^{\sipty}_2\|_2^2\sum_{\sig\in S_{m+n}}\int   d^3 q d^3p \int    d^{3n}r d^{3m}k\, \Gn(q;r) \Hm(p; k)\non\\
& &\pha{4444444444444444444} \times\ovGtin(q+\un{\chk}-\un{\chr};\hr,\chk) 
\ovHtim(p-\un{\chk}+\un{\chr}; \hk,\chr)   
\label{H-check-auxiliary}
\eeqa
for $m+n=\tim+\tin$, otherwise the l.h.s. is zero.  Here $S_{m+n}$ is the set of permutations of an $m+n$ element set
and the notation $\hk, \chk, \hr, \chr$ is explained in Lemma~\ref{combinatorics}.
\eel
%%%%%%%%%%%%%%%%%%%%%%%%%%%%%%%%%%%%%%%%%%
\proof We compute the expectation value
\beqa
& &\1\1\1\1\lan\vac, B_{2,\tin}(\Gtin ) (\cH_{\I,\sipty,i'}^{\cin} )^*  B_{1,\tim}(\Htim )B_{1,n}^*(\Gn)  \cH_{\I,\sipty,i}^{\cin}  B_{2,m}^*(\Hm)\vac\ran\non\\  
\1&=&\1\int d^3\ti u d^3\ti w d^3u d^3w\int d^3\ti q  d^3\ti p d^3 q d^3p \int d^{3\tin}\ti r d^{3\tim}\ti k   d^{3n}r d^{3m}k\, \non\\ 
& & \cvv_{i'}^{\sipty}(\ti w) \cvv_i^{\sipty}(w)\ovGtin(\ti q;\ti r) \ovHtim(\ti p; \ti k)  \Gn(q;r) \Hm(p; k)\non\\ 
& &\times\lan \vac,  \neta_{2}( \ti p-\ti{\unk}  )  \nn_{i'}^*(\ti u) \nn_{i'}(\ti u-\ti w)   \neta_1(    \ti q-\ti{\unr}) \neta_1^*(q-\unr)\nn_i^*(u-w)\nn_i(u)\neta_2^*(p-\unk) \vac\ran\non\\
& &\times\lan \vac, \bb(\ti r)^{\tin}\bb(\ti w) \bb(\ti k)^{\tim} \bb^*(r)^{n} \bb^*(w) \bb^*(k)^{m}\vac\ran.       
\label{generalized-creation-operators}
\eeqa
We note that
\beqa
& &\lan \vac,  \neta_2( \ti p-\ti{\unk}    )  \nn_{i'}^*(\ti u) \nn_{i'}(\ti u-\ti w)   \neta_1(  \ti q-\ti{\unr} )      \neta_1^*(q-\unr)  \nn_i^*(u-w)\nn_i(u)    
 \neta_2^*(p-\unk) \vac\ran\non\\
& &=\de_{i',2}\de_{i,2}\de(\ti p-\ti{\unk} -\ti u)\de(p-\unk- u)\lan \vac,\nn_{i'}(\ti u-\ti w) \neta_1( \ti q-\ti{\unr} ) \neta_1^*(q-\unr)\nn_i^*(u-w) \vac\ran\non\\
& &=\de_{i',2}\de_{i,2}\de(\ti p-\ti{\unk}-\ti u)\de(p-\unk- u)   
\de(\ti u-\ti w-u+w)\de(\ti q-\ti{\unr}-q+\unr).\label{deltas-with-insertions}
\eeqa
In the following we can assume $i=2,i'=2$, as otherwise the expression is zero. Now we consider the expectation value of the photon creation operators:
\beqa
\lan \vac, \bb(\ti r)^{\ti n} \bb(\ti k)^{\ti m} \bb(\ti w)\bb^*(w) \bb^*(r)^{n}  \bb^*(k)^{m}\vac\ran
=\de(w-\ti w)\lan \vac, \bb(\ti r)^{\ti n} \bb(\ti k)^{\ti m} \bb^*(r)^{n}  \bb^*(k)^{m}\vac\ran\label{many-deltas},
\eeqa
for $r,k,w, \ti r,\ti k,\ti w$ in the supports of the respective functions. (Here we made use of the fact that $|\ti w|\leq \si$, 
whereas $|r_i|\geq \si$, $|k_j|\geq \si$).
Thus we have
\beqa
& &\lan\vac, B_{2,\tin}(\Gtin) (\cH_{\I,\sipty,2}^{\cin}  )^*  B_{1,\tim}(\Htim )B_{1,n}^*(\Gn) \cH_{\I,\sipty,2}^{\cin}  B_{2,m}^*(\Hm)\vac\ran\non\\  
&=&\int d^3w\int d^3\ti p  d^3\ti q   d^3 q d^3p \int d^{3\tin}\ti r d^{3\tim}\ti k   d^{3n}r d^{3m}k\,  
\cvv_2^{\sipty}(\ti p-\ti{\unk}-p+\unk+w) \cvv_2^{\sipty}(w)\non\\
& &\times \ovGtin(\ti q;\ti r) \ovHtim(\ti p  ; \ti k)  \Gn(q;r) \Hm(p; k)\de(\ti q-\ti{\unr}-q+\unr)\de(\ti p-\ti{\unk}-p+\unk) \non\\
& &\times\lan \vac, \bb(\ti r)^{\tin} \bb(\ti k)^{\tim} \bb^*(r)^{n} \bb^*(k)^{m}\vac\ran \non\\
&=& \|\cvv_2^{\sipty}\|_2^2     \int   d^3 q d^3p \int d^{3\tin}\ti r d^{3\tim}\ti k   d^{3n}r d^{3m}k\,  
\non\\
& &\times \ovGtin(q+\ti{\un r}-\un{r}   ;\ti r) \ovHtim(p+\un{\ti k}-\un{k} ; \ti k)  \Gn(q;r) \Hm(p; k)\non\\
& &\times\lan \vac, \bb(\ti r)^{\tin}\bb(\ti k)^{\tim} \bb^*(r)^{n} \bb^*(k)^{m}\vac\ran \non\\
&=& \|\cvv_2^{\sipty}\|_2^2 \sum_{\sig\in S_{m+n}}   \int   d^3 q d^3p \int   d^{3n}r d^{3m}k\,  \Gn(q;r) \Hm(p; k) \non\\
& &\pha{44444444444444444444444444}\times  \ovGtin(q +\check{\un{k}}-\check{\un{r}};\hr,\chk) \ovHtim(p-\check{\un{k}} +\check{\un{r}}; \hk,\chr), \label{hat-check-controversy}
\eeqa
where in the first step we substituted~(\ref{deltas-with-insertions}), integrated over $\ti u, \ti w, u$, and used (\ref{many-deltas}).  In the last step  we made use  of Lemma~\ref{combinatorics}.

%%%%%%%%%%%%%%%%%%%%%%%%%%%%%%%%%%%%%%%%%%%
\section{Vacuum expectation values of renormalized creation operators} \label{Vacuum-exp-ren-creation}
%%%%%%%%%%%%%%%%%%%%%%%%%%%%%%%%%%%%%%%%%%%%
\setcounter{equation}{0}

%%%%%%%%%%%%%%%%%%%%%%%%%%%%%%%%%%%%%%%%%%
\subsection{$L^2$-bound}
%%%%%%%%%%%%%%%%%%%%%%%%%%%%%%%%%%%%%%%%%%
\bep \label{Simple-L-2-bound} Let $h_1, h_2\in C_0^2(\PSS)$. Then
\beqa
\|\reta^*_{1, \si}(h_{1})\reta^*_{2,\si}(h_{2})\Om\|\leq \fr{c}{\si^{\de_{\la_0}}}\|h_1\|_2 \|h_2\|_2.
\eeqa
\eep
%%%%%%%%%%%%%%%%%%%%%%%%%%%%%%%%%%%%%%%%%%%%%
\proof Let us set
\begin{align}
&G_{1,m}(q;k):=h_1(q)f^{m}_{1,q,\si}(k),\quad  G_{2,m}(q;k):=h_{2}(q)f^{m}_{2,q,\si}(k).
\label{G-def-new} 
\end{align}
Now we can write
\beqa
\|\reta^*_{1, \si}(h_{1})\reta^*_{2,\si}(h_{2})\Om\|^2=
\sum_{\substack{m,n,\ti m,\ti n\in\nat_0 \\ \ti m+\ti n=m+n } }\fr{1}{\sqrt{m!n!\ti m!\ti n!}} 
\lan\vac, B_{1,\ti n}(G_{1,\ti n} )B_{2,\ti m}(G_{2,\ti m} )B_{1,n}^*(G_{1,n}) B_{2,m}^*(G_{2,m})\vac\ran.\,\,
\label{scalar-product-formula-new}
\eeqa
Making use of Lemma~\ref{norms-of-scattering-states}, we obtain
\beqa
\lan\vac, B_{1,\ti n}(G_{1,\ti n} )B_{2,\ti m}(G_{2,\ti m} )B_{1,n}^*(G_{1,n}) B_{2,m}^*(G_{2,m})\vac\ran \pha{4444444444444444444444444444444}  \non\\
=\sum_{\sig\in S_{m+n}}\int  d^3 q d^3p \int  d^{3n}r d^{3m}k\,  
G_{1,n}(q;  r) G_{2,m}(p;  k)\pha{444444444444444444444444}\non\\
 \times\bigg(\ov G_{1,\ti n}( q+\un{\check{k}}-\un{\check{r}}; \hat r, \check k)\ov  G_{2,\ti m}(p-\un{\check{k}}+\un{\check{r}}; \hat{k}, \check r)
\bigg),
\label{B-expectation-value-new}
\eeqa
where the notation in (\ref{B-expectation-value-new}) is explained in Lemma~\ref{combinatorics}. Now by applying the Cauchy-Schwarz inequality to
the integrals w.r.t. $p$ and $q$, using definitions (\ref{G-def-new}) and applying the bound (\ref{infrared-spectral-bound}) for $\be=0$, we obtain 
\beqa
(\ref{B-expectation-value-new})\leq \|h_1\|^2_2\|h_2\|^2_2\fr{(m+n)!}{\sqrt{m!n!\ti m! \ti n!} }\|{g}_{2,\si}^n\|_2^2\|g_{2,\si}^m\|_2^2, 
\eeqa
{where we estimated $g_{1,\si}^n(k)\leq g_{2,\si}^n(k)$ in the above computation}.   
Now the claim follows from estimate (\ref{no-shifted-index}) below. \qed
%%%%%%%%%%%%%%%%%%%%%%%%%%%%%%%%%%%%%%%%%%%%%%%%%%%%%%%%%%%%%

%%%%%%%%%%%%%%%%%%%%%%%%%%%%%%%%%%%%%%%%%%%%%%%%%%%%%%%%%%%%%%%%%
\subsection{Double commutator}\label{double-commutator}
%%%%%%%%%%%%%%%%%%%%%%%%%%%%%%%%%%%%%%%%%%%%%%%%%%%%%%%%%%%%%%%%%

In this section we define
\beqa
G_{1,m}(q;k):=e^{-iE_{1,q,\si}t}h_1(q)f^{m}_{1,q,\si}(k),\quad  G_{2,m}(q;k):=e^{-iE_{2,q,\si}t}e^{i\ga_{\si}(\nv_j,t)(q)  }h_{2,j}^{(s)}(q)f^{m}_{2,q,\si}(k).
\label{zero-G-def-new} 
\eeqa
Moreover, we set  
\beqa
B_{i,m}^*(G_{i,m}):=\int d^3q d^{3m}k\, G_{i,m}(q;k) \bb^*(k)^m \nn_i^*(q-\unk). \label{B-star-def}
\eeqa

It turns out that the behaviour of the double commutator in (\ref{zero-double-comm}) is governed by
the decay of the functions
\beqa
F_{i,n,m}^{G_1,G_2}(q; r \ba p; k):=(n+1)\int d^3r_{n+1}\, \vv_i(r_{n+1})G_{1,n+1}(q+r_{n+1} ; r,r_{n+1}) G_{2,m}(p-r_{n+1} ; k), \label{FGH}
\eeqa
%%%%%%%%%%%%%%%%%%%%%%%%%%%%%%%%%%%%%%%%%%%%%%%%%%%%%%%%
where $q,p\in \real^3$, $r\in\real^{3n}$, $k\in\real^{3m}$.
For any such function we define an auxiliary operator
\beqa
B_{n,m}^*(F_{i,n,m}^{G_1,G_2}):=\int d^3q d^3p\int d^{3n}r  d^{3m}k\, F_{i,n,m}^{G_1,G_2}(q; r \ba p; k) \bb^*(r)^{n}\bb^*(k)^m\nn_1^*(q-\unr)\nn_2^*(p-\unk).
\label{double-creation}
\eeqa
We do not specify domains as it will act only on the vacuum below.
Key properties of expectation values of such operators are given in Lemma~\ref{double-creation-lemma} above. We write
\beqa
& &H_{\I,i}^{\ain}=\int d^3p d^3k\,\vv_i(k)\nn_i^*(p+k)\bb(k)\nn_i(p),\quad
H_{\I}^{\ain}:=\sum_{i\in\{1,2\}} H_{\I,i}^{\ain}. 
\eeqa
Now we  state and prove the estimate on the double commutator.
%%%%%%%%%%%%%%%%%%%%%%%%%%%%%%%%%%%%%%%%%%%%%%%%%%%%%%%%%%%%%%%%%
\bep\label{double-commutator-proposition}  For $1\leq s\leq t$ there holds the bound 
\beqa
\|[ [H_{\I}^{\ain}, \nr_{1,\si}^*(h_{1,t})  ], \nr_{2,\si}^*( h_{2,j,\sn}^{(\tn)} e^{i\ga_{\si}(\nv_j,\sn)}   )]\vac\|\leq  
\fr{c_{1,t} }{\si^{\de_{\g_0}}  }  \fr{1}{(\kal)^2}  \bigg(    \fr{\si_t^{\alf/(8\tiga)  } }{t}+\fr{1}{t^2\si_t^{1/(4\tiga)}}   \bigg),
\eeqa
where $c_{1,t}=c\bigg(t^{(1-\al)}+ ( \uneps(t) t^{-\peps} )^{-2} \bigg)$. (The same bound holds if $H_{\I}^{\ain}$
is replaced with $H_{\I,\si}^{\ain}$, since $F_{i,n,m}^{G_1,G_2}$, given by (\ref{FGH}), does not change if $\vv_i$ is
replaced with $\vv_i^{\si}$). 
\eep
%%%%%%%%%%%%%%%%%%%%%%%%%%%%%%%%%%%%%%%%%%%%%%%%%%%%%%%%%%%%%%%%%
\proof  We write $h_{2,t}^{\ga}(q):=e^{-iE_{2,q,\si}t}e^{i\ga_{\si}(\nv_j,t)(q)  }h_{2,j}^{(s)}(q)$ and compute:
\beqa
& &\lan [ [H_{\I}^{\ain}, \nr_{1,\si}^*(h_{1,t})  ], \nr_{2,\si}^*(h_{2,t}^{\ga})]\vac,   [ [H_{\I}^{\ain}, \nr_{1,\si}^*(h_{1,t})  ], \nr_{2,\si}^*(h_{2,t}^{\ga})]\vac\ran\non\\
&=&\sum_{\substack{ m,n,\ti m,\ti n\in\nat_0 \\   m+n=\ti m+\ti n} }
\fr{1}{\sqrt{m!n!\ti m!\ti n!}} \lan  [[H_{\I}^{\ain},  B_{1,\ti n}^*(G_{1,\ti n})], B_{2,\ti m}^*(G_{2,\ti m})]  \vac,  [[H_{\I}^{\ain},  B_{1,n}^*(G_{1,n})], B_{2,m}^*(G_{2,m})]\vac\ran\non\\
&=&\sum_{\substack{m,n,\ti m,\ti n \in \nat_0 \\  m+n=\ti m+\ti n} }
\fr{1}{\sqrt{m!n!\ti m!\ti n!}} \lan (B_{\ti n-1,\ti m}^*(F_{2,\ti n-1,\ti m}^{G_1,G_2})+B^*_{\ti m-1,\ti n}(F_{1,\ti m-1,\ti n}^{G_2,G_1}) ) \vac, \non\\
& &\pha{4444444444444444444444444444444} (B^*_{n-1,m}(F_{2,n-1,m}^{G_1,G_2})+B_{m-1,n}^*(F_{1,m-1,n}^{G_2,G_1}) )\vac\ran,
\label{scalar-product-formula-two}
\eeqa
where in the last step we made use of Lemma~\ref{double-comm-lemma} and the operators $B^*_{m,n}(\,\cdot\,)$ are defined in~(\ref{double-creation}).
(By convention, $B_{-1,m}^*(F_{i,-1,m}^{G_1,G_2})\vac=B_{-1,n}^*(F_{i,-1,n}^{G_2,G_1})\vac=0$).
With functions $G_{i_1,n}, G_{i_2,n}, G_{j_1,n}, G_{j_2,n}$ of the form~(\ref{zero-G-def-new}), with $i_1,i_2,j_1,j_2\in\{1,2\}$,
we define
\beqa
C_{i,i'}(G_{i_1},G_{i_2};G_{j_1},G_{j_2})\pha{444444444444444444444444444444444444444444444444}\non\\
:=\sum_{\substack{m, n, \ti m, \ti n\in \nat_0 \\  m+n=\ti m+\ti n} }
\fr{1}{\sqrt{m!(n+1)!\ti m!(\ti n+1)!}}  \lan B_{\ti n,\ti m}^*(F_{i,\ti n,\ti m}^{G_{i_1},G_{i_2}})\vac, B_{n,m}^*(F_{i',n,m}^{G_{j_1},G_{j_2}})\vac\ran.
\label{C-definition}
\eeqa
From (\ref{scalar-product-formula-two}) we obtain
\beqa
\|[ [H_{\I}^{\ain}, \nr_{1,\si}^*(h_{1,t})  ], \nr_{2,\si}^*(h_{2,t}^{\ga} )]\vac\|^2\1&=&\1C_{2,2}(G_{1},G_2;G_1,G_2)+C_{1,1}(G_2,G_1;G_2,G_1)\non\\
& &+C_{2,1}(G_1,G_2; G_2, G_1)+C_{1,2}(G_2,G_1; G_1, G_2) .
\eeqa
We will study in detail $C_{2,2}(G_{1},G_2;G_1,G_2)$ as the analysis of the remaining terms is analogous. 
We recall from Lemma~\ref{double-creation-lemma} that
\beqa
& &\lan  B_{\ti n,\ti m}^*(F^{G_1,G_2}_{2,\ti n,\ti m})\vac, B_{n,m}^*(F^{G_1,G_2}_{2,n,m})\vac\ran
=\sum_{\sig\in S_{m+n}}\int  d^3 q d^3p \int  d^{3n}r d^{3m}k\, F^{G_1,G_2}_{2,n,m}(q;r \ba p;k) \non\\
& &\ph{444444444444444444444444444}\times     
 \ov F_{2,\ti n,\ti m}^{G_1,G_2}(q+\uchk-\uchr; \hr, \chk \ba p-\uchk+\uchr;\hk, \chr). \label{double-commutator-exp}
\eeqa
The notation $\hk, \chk, \hr, \chr$ is explained  in Lemma~\ref{combinatorics}.  
Now  from Lemma~\ref{smooth-bounds} we obtain
\beqa
 |F^{G_1,G_2}_{2,n,m}(q; r \ba p; k)|\leq  \fr{c_{1,t}}{\si ^{\de_{\g_0}}  } \fr{1}{(\kal)^2} \bigg( \fr{\si_t^{\alf/(8\tiga)  } }{t}+\fr{1}{t^2\si_t^{1/(4\tiga)}} \bigg)
\fr{1}{\sqrt{m!n!}}  D(p,q)g^m_{2,\si}(k) g^n_{{2},\si}(r),  \label{FGH-bound-one}
\eeqa
where  $D\in C_0^{\infty}(\real^3\times \real^3)$ and $c_{1,t}= c \bigg(t^{2(1-\al)}+ ( \uneps(t) t^{-\peps} )^{-2} \bigg)$.
From this bound we get 
\beqa
& &|F^{G_1,G_2}_{2,n,m}(q; r \ba p ; k) \ov F_{2,\ti n,\ti m}^{G_1,G_2}(q+\uchk-\uchr; \hr, \chk \ba p-\uchk+\uchr; \hk, \chr)|\non\\
& &\pha{444}\leq  \fr{c_{1,t}^2 }{\si^{2\de_{\g_0}}  } \fr{1}{(\kal)^4}\bigg(    \fr{\si_t^{\alf/(8\tiga)  } }{t}+\fr{1}{t^2\si_t^{1/(4\tiga)}}   \bigg)^2D'(p,q)
  \fr{1}{\sqrt{m!n!\ti m!\ti n!} }    g^m_{2,\si}(k)^2 g^n_{2,\si}(r)^2,
\eeqa
where $D'$ is again a smooth, compactly supported function. 
Making use of (\ref{C-definition}), (\ref{double-commutator-exp}) and the last two bounds we get
\beqa
& &|C_{2,2}(G_1,G_2;G_1,G_2)|\non\\
& &\leq \fr{c_{1,t}^2 }{\si^{2\de_{\g_0}}  } \fr{1}{(\kal)^4}\bigg(    \fr{\si_t^{\alf/(8\tiga)  } }{t}+\fr{1}{t^2\si_t^{1/(4\tiga)}}   \bigg)^2
\sum_{\substack{m, n, \ti m, \ti n \in \nat_0 \\  m+n=\ti m+\ti n} }
\fr{(m+n)!}{\sqrt{m!(n+1)!\ti m!(\ti n+1)!}}\non\\
& &\pha{44444444444444444444444444444444444}\times \fr{1}{\sqrt{m!n!\ti m!\ti n!} }   \|g^m_{2,\si}\|_2^2 \|g^n_{2,\si}\|_2^2\non\\
&  &\leq \fr{c_{1,t}^2 }{\si^{2\de_{\g_0}}  }\fr{1}{(\kal)^4}\bigg(    \fr{\si_t^{\alf/(8\tiga)  } }{t}+\fr{1}{t^2\si_t^{1/(4\tiga)}}   \bigg)^2
\bigg(\fr{\kas}{\si}\bigg)^{4\g^2 c^2} 
 \leq  \fr{c_{1,t}^2 }{\si^{2\de_{\g_0}}  } \fr{1}{(\kal)^4}\bigg(    \fr{\si_t^{\alf/(8\tiga)  } }{t}+\fr{1}{t^2\si_t^{1/(4\tiga)}}   \bigg)^2, \,\,\,\,\,\,\,\,\,\,\,
\eeqa
where we made use of Lemma~\ref{summation} and definition~(\ref{def-g-ks}). This concludes the proof. \qed\\
%%%%%%%%%%%%%%%%%%%%%%%%%%%%%%%%%%%%%%%%%%%%%%%%%%%%%%%%%%%%%%%%
%%%%%%%%%%%%%%%%%%%%%%%%%%%%%%%%%%%%%%%%%%%%%%%%%%%%%%%%%%%%%%%
\bel\label{double-comm-lemma} 
 For any  $n,m\in \nat_0$ there holds
\beqa
[[H_{\I}^{\ain},  B_{1,n}^*(G_{1,n})],B_{2,m}^*(G_{2,m})]\vac=B_{n-1,m}^*(F_{2,n-1,m}^{G_1,G_2})\vac+B_{m-1,n}^*(F_{1,m-1,n}^{G_2,G_1})\vac,\label{double-comm-form}
\eeqa
where $F_{i,n-1,m}^{G_1,G_2}$ is defined in (\ref{FGH}) and  we set $B_{-1,m}^*(F_{i,-1,m}^{G_1,G_2})\vac=B_{-1,n}^*(F_{i,-1,n}^{G_2,G_1})\vac=0$. 
\eel
%%%%%%%%%%%%%%%%%%%%%%%%%%%%%%%%%%%%%%%%%%%%%%%%%%%%
\proof First we compute the inner commutator on  $\mcC$:
\beqa \label{commutator}
[H_{\I,i}^{\ain},  B_{1,n}^*(G_{1,n})]=\int d^3q d^{3n}r d^3u d^3w\,  G_{1,n}(q;r)\vv_i(w)[ \bb(w)\nn_i^*(u+w)\nn_i(u), \bb^*(r)^n \nn_1^*(q-\unr)].\,\,
\eeqa
We note that
\beqa
[\bb(w)\nn_i^*(u+w)\nn_i(u), \bb^*(r)^n \nn_1^*(q-\unr)]
\1&=&\1\sum_{i'=1}^n\de(w-r_{i'})\bb^*(r_{\ci})^{n-1}\nn_i^*(u+w)\nn_i(u)\nn_1^*(q-\unr)\non\\
& &+\bb(w)\bb^*(r)^n\de(u-q+\unr)\nn_i^*(u+w)\de_{i,1}, \label{commutator-two-terms}
\eeqa 
where $\bb^*(r_{\ci})^{n-1}=\bb^*(r_1)\ldots \bb^*(r_{i'-1})\bb^*(r_{i'+1})\ldots \bb^*(r_n)$ for $n\geq 1$ and $\bb^*(r_{\ci})^{n-1}=0$ for $n=0$. Since $G_{1,n}$ is symmetric in the photon variables,
the contributions to (\ref{commutator}) proportional to the first and the second term on the r.h.s. of (\ref{commutator-two-terms}) are
\beqa
& &\,[H_{\I,i}^{\ain},  B_{1,n}^*(G_{1,n})]_1\non\\
& &:= n\int d^3q d^{3(n-1)}r d^3r_n d^3u \,  G_{1,n}(q;r,r_n)\vv_i(r_n)\bb^*(r)^{n-1}\nn_i^*(u+r_n)\nn_i(u)\nn_1^*(q-\unr-r_n),\,\,\,\,\,\,\,\,\,
\eeqa
and
\beqa
 [H_{\I,i}^{\ain},  B_{1,n}^*(G_{1,n})]_2:=\int d^3q d^{3n}r d^3w\,  G_{1,n}(q;r)\vv_i(w) \bb(w)\bb^*(r)^n\nn_i^*(q-\unr+w)\de_{i,1},
\eeqa
respectively.
Now let us compute the first contribution to the double commutator:
\beqa
& &[[H_{\I,i}^{\ain},  B_{1,n}^*(G_{1,n})]_1, B_{2,m}^*(G_{2,m})]\vac\non\\
& &\pha{444444444}=n\int d^3q d^{3(n-1)}r d^3r_n d^3u\int d^3p d^{3m}k\,  G_{1,n}(q;r,r_n)\vv_i(r_n) G_{2,m}(p;k)\non\\ 
& &\pha{44444444444}\times [\bb^*(r)^{n-1}\nn_i^*(u+r_n)\nn_i(u)\nn_1^*(q-\unr-r_n),\bb^*(k)^m \nn_2^*(p-\unk)]\vac\non\\
& &\pha{444444444}=n\int d^3q d^3p  \int d^{3(n-1)}r d^3r_n d^{3m}k\,  G_{1,n}(q;r,r_n)\vv_2(r_n) G_{2,m}(p;k)\non\\ 
& &\pha{44444444444}\times \bb^*(r)^{n-1}\bb^*(k)^m\nn_2^*(p-\unk+r_n)\nn_1^*(q-\unr-r_n)\vac\,\de_{i,2}.  \label{double-commutator-one}
\eeqa
By changing variables $p\to p-r_n$ and $q\to q+r_n$, we get 
\beqa
[[H_{\I,i}^{\ain},  B_{1,n}^*(G_{1,n})]_1, B_{2,m}^*(G_{2,m})]\vac=\de_{i,2} B_{n-1,m}^*(F_{2,n-1,m}^{G_1,G_2})\vac .
\eeqa
The second contribution to the double commutator has the form:
\beqa
& &[[ H_{\I,i}^{\ain} ,  B_{1,n}^*(G_{1,n})]_2, B_{2,m}^*(G_{2,m})]\vac\non\\
& &\pha{444444444}=\int d^3q d^{3n}r d^3w \int d^3p d^{3m}k\,  G_{1,n}(q;r)\vv_i(w)G_{2,m}(p;k)\non\\ 
& &\pha{44444444444}\times [\bb(w)\bb^*(r)^n\nn_i^*(q-\unr+w), \bb^*(k)^m \nn_2^*(p-\unk)]\vac \de_{i,1}\non\\
& &\pha{444444444}=m\int d^3q d^3p  \int d^{3n}r  d^{3(m-1)}k d^3k_m\,  G_{1,n}(q;r)\vv_1(k_m)G_{2,m}(p;k,k_m)\non\\ 
& &\pha{44444444444}\times \bb^*(r)^n\bb^*(k)^{m-1}\nn_1^*(q-\unr+k_m)  \nn_2^*(p-\unk-k_m)\vac \de_{i,1}. \label{double-commutator-two}
\eeqa
By changing variables $q\to q-k_m$ and $p\to p+k_m$ we obtain
\beqa
[[ H_{\I,i}^{\ain},  B_{1,n}^*(G_{1,n})]_2, B_{2,m}^*(G_{2,m})]\vac=\de_{i,1}B_{m-1,n}^*(F_{1,m-1,n}^{G_2,G_1})\vac,
\eeqa
which concludes the proof. \qed 
%%%%%%%%%%%%%%%%%%%%%%%%%%%%%%%%%%%%%%%%%%%%%%%%%%%%%%%%%%%%%%%%%%%%%%%%%%%%%%%%%
%%%%%%%%%%%%%%%%%%%
\bel\label{summation}\cite{DP12.0} There hold the estimates
\beqa
& &\sum_{\substack{m,n,\ti m, \ti n\in\nat_0, \\ m+n=\ti m+\ti n} }  \fr{(\ti m+\ti n)!}{m!n! \ti m!\ti n! } \|g^m_{i,\si}\|_2^2 \,\|g^n_{i',\si}\|_2^2\leq \bigg(\fr{\kas}{\si}\bigg)^{4\la^2 c^2},\label{no-shifted-index}\\
& &\sum_{\substack{m, n, \ti m, \ti n\in\nat_0 \\ m+n=\ti m+\ti n }}  \fr{(\ti m+\ti n)!}{m!n! \ti m!\ti n! } \|g^{m-1}_{i,\si}\|_2^2 \,\|g^n_{i',\si}\|_2^2
\leq 2 \bigg(\fr{\kas}{\si}\bigg)^{4\la^2 c^2},   \label{shifted-index}\\
& &\sum_{\substack{m, n, \ti m, \ti n\in\nat_0 \\ m+n=\ti m+\ti n }}  \fr{(\ti m+\ti n)!}{m!n! \ti m!\ti n! } \|g^{m-1}_{i,\si}\|_2^2 \,\|g^{n-1}_{i',\si}\|_2^2
\leq 4 \bigg(\fr{\kas}{\si}\bigg)^{4\la^2 c^2}, \label{shifted-second-index}
\eeqa
where $g^m_{i,\si}, \kas$ are defined in (\ref{def-g-ks-z}), (\ref{def-g-ks}), $i,i'=1,2$, and we set by convention $g^{-1}_{i,\si}=0$.
\eel
%%%%%%%%%%%%%%%%%%%%%%%%%%%%%%%%%%%%%%%%%%%%%%%%%%%
%%%%%%%%%%%%%%%%%%%
\bel\label{summation-ell} There hold the estimates
\beqa
& &\sum_{\ell,\ellp=0}^{\infty}\fr{1}{\ell!\ellp!}\sum_{\substack{m\geq \ell, \ti m\geq \ellp,  n, \ti n\geq 0, \\ m+n=\ti m+\ti n} }  \fr{(\ti m+\ti n)!}{(m-\ell)!n! (\ti m-\ellp)!\ti n! } \|g^m_{i,\si}\|_2^2 \,\|g^n_{i',\si}\|_2^2
\leq \bigg(\fr{\kas}{\si}\bigg)^{16\la^2 c^2},  \label{no-shifted-index-ell} \\
& &\sum_{\ell,\ellp=0}^{\infty}\fr{1}{\ell!\ellp!} \sum_{\substack{m\geq \ell, \ti m\geq \ellp,  n, \ti n\geq 0, \\ m+n=\ti m+\ti n} }  \fr{(\ti m+\ti n)!}{(m-\ell)!n! (\ti m-\ellp)!\ti n! } \|g^{m-1}_{i,\si}\|_2^2 \,\|g^n_{i',\si}\|_2^2\leq {\mc 8} \bigg(\fr{\kas}{\si}\bigg)^{16\la^2 c^2},   \label{shifted-index-ell}\\
& &\sum_{\ell,\ellp=0}^{\infty} \fr{1}{\ell!\ellp!} \sum_{\substack{ m\geq \ell, \ti m\geq \ellp,  n, \ti n\geq 0  \\ m+n=\ti m+\ti n }}  \fr{(\ti m+\ti n)!}{(m-\ell)!n! (\ti m-\ellp)!\ti n! }      \|g^{m-1}_{i,\si}\|_2^2 \,\|g^{n-1}_{i',\si}\|_2^2
\leq  {\mc 32} \bigg(\fr{\kas}{\si}\bigg)^{4\la^2 c^2}, \label{shifted-second-index-ell}
\eeqa
where $g^m_{i,\si}, \kas$ are defined in (\ref{def-g-ks-z}), (\ref{def-g-ks}), $i,i'=1,2$, and we set by convention $g^{-1}_{{\mc i},\si}=0$.
\eel
%%%%%%%%%%%%%%%%%%%%
\begin{remark} Up to irrelevant numerical constants Lemma~\ref{summation} follows from Lemma~\ref{summation-ell}.
\end{remark}
%%%%%%%%%%%%%%%%%%%%%%%%%%%%%%
\proof  We first note that by definition of the functions $g^m_{i,\si}$
\beqa
\|g^m_{i,\si}\|_2^2\leq (\g  c)^{2m}(\log(\kas/\si))^m.
\eeqa
Now for any $a, b\in \{0,1\}$ we compute
\beqa
& &  \sum_{\ell,\ellp=0}^{\infty} \fr{1}{\ell!\ellp!} \sum_{\substack{m\geq \ell, \ti m\geq \ellp,  n, \ti n\geq 0, \\ m+n=\ti m+\ti n} }   
 \fr{(\ti m+\ti n)!}{(m-\ell)!n! (\ti m-\ellp)!\ti n! }    
  \|g^{m-a}_{i,\si}\|_2^2 \|g^{n-b}_{i',\si}\|_2^2\non\\
& &\ph{444444444444444}= \sum_{\ell,\ellp=0}^{\infty} \fr{1}{\ell!\ellp!} \sum_{\substack{m+a\geq \ell, \ti m\geq \ellp,  n+b, \ti n\geq 0, \\ m+n+a+b=\ti m+\ti n} }    
\fr{(\ti m+\ti n)!}{(m+a-\ell)!(n+b)! (\ti m-\ellp)!\ti n! }    
  \|g^{m}_{i,\si}\|_2^2 \|g^n_{i',\si}\|_2^2\non\\
& &\ph{444444444444444}\leq \sum_{\substack{m+a, \ti m, n+b, \ti n\geq 0 \\ m+n+a+b=\ti m+\ti n} } \sum_{\ell\leq m+a,\ellp\leq\ti m} {m+a \choose \ell} {\ti m \choose \ellp} \fr{(\ti m+\ti n)!}{m!n! \ti m!\ti n! }    
  \|g^m_{i,\si}\|_2^2 \|g^n_{i',\si}\|_2^2\non\\
& &\ph{444444444444444}=\sum_{\substack{m+a, \ti m, n+b, \ti n\geq 0 \\ m+n+a+b=\ti m+\ti n} } 2^{m+a+\ti m} \fr{(\ti m+\ti n)!}{m!n! \ti m!\ti n! }    
  \|g^m_{i,\si}\|_2^2 \|g^n_{i',\si}\|_2^2\non\\
& &\ph{444444444444444} \leq 2^{{\mc 2a+b}}\sum_{m+a,n+b\geq 0}\bigg(\sum_{ \substack{\ti m,\ti n\geq 0 \\ \ti m+\ti n=m+n+a+b}}   \fr{(\ti m+\ti n)!}{ \ti m!\ti n!}\bigg) \fr{ (2\g  c)^{2(m+n)} }{m!n!}   (\log(\kas/\si))^{m+n}\non\\
& &\ph{444444444444444} \leq 2^{{\mc 3a+2b}}\sum_{m,n\geq 0} \fr{ (2\sqrt{2}\la  c)^{2(m+n)} }{m!n!}   (\log(\kas/\si))^{m+n}
=2^{{\mc 3a+2b}}\bigg(\fr{\kas}{\si}\bigg)^{16\g^2 c^2},\,\,\,\,\,\,\,\quad
\eeqa
{\mc where in the fourth step we used $ 2^{m+a+\ti m}\leq 2^{m+a+\ti m+\ti n}\leq 2^{2(m+n)+2a+b}$}.
This concludes the proof. \qed
%%%%%%%%%%%%%%%%%%%%%%%%%%%%%%%%%%%%%%%%%
%%%%%%%%%%%%%%%%%%%%%%%%%%%%%%%%%%%%%%%%%%%%%%%%%%%%%%%%%%%%%%%%%%
\subsection{Contributions involving $\cH_{\I, \sipty}^{\cin}$ }\label{single-commutator-section}
%%%%%%%%%%%%%%%%%%%%%%%%%%%%%%%%%%%%%%%%%%%%%%%%%%%%%%%%%%%%%%%%%%
This subsection is devoted to terms involving $\cH_{\I, \si}^{\cin}$, appearing in  (\ref{zero-double-comm-rest-one}). The following
elementary proposition, which relies on Lemma~\ref{norms-of-scattering-states-one}, gives  contributions
 (\ref{H-check-contribution-one}), (\ref{H-check-contribution}) to expressions from Theorem~\ref{non-diagonal-theorem}. 
%%%%%%%%%%%%%%%%%%%%%%%%%%%
\bep\label{check-contribution} Let $\cH_{\I,\sipty}^{\cin}$ be  defined as in  (\ref{check-hamiltonian}). Then there hold the bounds 
\beqa
& &\| \cH_{\I,\sipty}^{\cin} \nr_{2,\si}^*(h^{\ga}_{2,t} ) \vac\|\leq c\si^{1-\de_{\g_0}}, 
\label{first-check-bound}\\
& &\|\nr_{1,\si}^*(h_{1,t}) \cH_{\I,\sipty}^{\cin} \nr_{2,\si}^*(h^{\ga}_{2,t} ) \vac\|\leq c\si^{1-\de_{\g_0}}, \label{second-check-bound}
\eeqa
where $h^{\ga}_{2,t}(p):=h_{2,j,t}^{(s)}(p)e^{ i\ga_{\si}(\nv_j,t)(p) }$.
\eep
%%%%%%%%%%%%%%%%%%%%%%%%%%%
\proof Let us show (\ref{first-check-bound}). We rewrite the expression from the statement of the proposition as follows:
\beqa
   \lan \cH_{\I,\sipty}^{\cin} \nr_{2,\si}^*(h_{2,t}^{\ga})\vac, \cH_{\I,\sipty}^{\cin} \nr_{2,\si}^*(h_{2,t}^\ga)  \vac\ran \pha{444444444444444444444444444}  \non\\
=\sum_{\substack{m\in\nat_0} }
\fr{1}{m!} \lan\vac, B_{2,m}(G_{2,m}) (\cH_{\I,\sipty}^{\cin} )^*  
 \cH_{\I,\sipty}^{\cin}  B_{2,m}^*(G_{2,m})\vac\ran.
\label{scalar-product-formula-one-zero}
\eeqa
Now Lemma~\ref{norms-of-scattering-states-one-zero} gives
\beqa
\lan\vac, B_{2,\ti n}(G_{2,\ti n} ) ( \cH_{\I,\sipty}^{\cin} )^*  
 \cH_{\I,\sipty}^{\cin}  B_{2,m}^*(G_{2,m})\vac\ran
=\|\cvv_2^{\sipty}\|_2^2 m!  \de_{m,\ti n} \int    d^3p \int   d^{3m}k\,  |\Hm(p; k)|^2.
\label{H-check-auxiliary-one-zero}
\eeqa
Making use of  Theorem~\ref{main-theorem-spectral}, and of definition~(\ref{zero-G-def-new}), we obtain the bound
\beqa
& &|G_{2,m}(p; k)|^2 \leq \fr{1}{m!} D(p)g_{2,\si}^{{\mc m}}(k)^2,
\eeqa
 where $p\mapsto D(p)$ is some smooth compactly supported function independent of $\si$, $t$, {\mc $\la$, $\la_0$}. 
Consequently, the r.h.s. of (\ref{H-check-auxiliary-one-zero}) can be estimated by
$c\|\cvv_2^{\sipty}\|_2^2 \|g_{2,\si}^m\|^2_2.$
Substituting this bound to (\ref{scalar-product-formula-one-zero}) and making use of
definition~(\ref{def-g-ks}) of functions $g_{2,\si}^m$, we get 
\beqa
 \|\cH_{\I,\sipty}^{\cin} \nr_{2,\si}^*(h_{2,t}) \vac\|^2\leq 
c\|\cvv_2^{\sipty}\|_2^2\bigg(\fr{\kas}{\si}\bigg)^{c'\la^2 }.
\eeqa
Exploiting the fact that $\|\cvv_2^{\sipty}\|_2\leq c\si$, we conclude the proof of (\ref{first-check-bound}).
%%%%%%%%%%%%%%%%%%%%%%%%%%%%%%%%%%%%%%%%%%%%%%%%%%%%%%%%%%

Let us now show  (\ref{second-check-bound}). We rewrite the expression from the statement of the proposition as follows:
\beqa
   \lan \nr_{1,\si}^*(h_{1,t})  \cH_{\I,\sipty}^{\cin} \nr_{2,\si}^*(h_{2,t}^{\ga})   \vac, \nr_{1,\si}^*(h_{1,t}) \cH_{\I,\sipty}^{\cin} \nr_{2,\si}^*(h_{2,t}^\ga)  \vac\ran \pha{444444444444444444444444444}  \non\\
=\sum_{\substack{m,n,\ti m,\ti n\in\nat_0 \\  m+n=\ti m+\ti n} }
\fr{1}{\sqrt{m!n!\ti m!\ti n!}} \lan\vac, B_{2,\ti n}(G_{2,\ti n}) (\cH_{\I,\sipty}^{\cin} )^*  B_{1,\ti m}(G_{1,\ti m} )
B_{1,n}^*(G_{1,n}) \cH_{\I,\sipty}^{\cin}  B_{2,m}^*(G_{2,m})\vac\ran.
\label{scalar-product-formula-one}
\eeqa
Now Lemma~\ref{norms-of-scattering-states-one} gives
\beqa
& &\lan\vac, B_{2,\ti n}(G_{2,\ti n} ) ( \cH_{\I,\sipty}^{\cin} )^*  
B_{1,\ti m}(G_{1,\ti m} )B_{1,n}^*(G_{1,n}) \cH_{\I,\sipty}^{\cin}  B_{2,m}^*(G_{2,m})\vac\ran\non\\
& &=\|\cvv_2^{\sipty}\|_2^2\sum_{\sig\in S_{m+n}}\int   d^3 q d^3p \int    d^{3n}r d^{3m}k\, G_{1,n}(q;r) G_{2,m}(p; k)\non\\
& &\pha{444444444444444444444}\times\ov G_{1,\ti m}(q+\un{\chk}-\un{\chr};\hr,\chk) \ov G_{2,\ti n}(p-\un{\chk}+\un{\chr}; \hk,\chr), 
\label{H-check-auxiliary-one}
\eeqa
where $S_{m+n}$ is the set of permutations of an $m+n$ element set.
Making use of  Theorem~\ref{main-theorem-spectral}, and of definition~(\ref{zero-G-def-new}), we obtain the bounds
\beqa
& &|G_{1,n}(q;r) G_{2,m}(p; k) \ov G_{1,\ti m}(q+\un{\chk}-\un{\chr};\hr,\chk) \ov G_{2,\ti n}(p-\un{\chk}+\un{\chr}; \hk,\chr)|\non\\
& &\pha{44444444444444444444444444444}\leq \fr{1}{\sqrt{m!n!\ti m!\ti n!}}    D(p,q) g_{2,\si}^n(r)^2g_{2,\si}^m(k)^2,
\eeqa
 where $(p,q)\mapsto D(p,q)$ is some smooth compactly supported function independent of $\si$, $t$, {\mc $\la$, $\la_0$} . 
Consequently, the r.h.s. of (\ref{H-check-auxiliary-one}) can be estimated by
\beqa
c\|\cvv_2^{\sipty}\|_2^2(m+n)! \fr{1}{\sqrt{m!n!\ti m!\ti n!}}   \|g_{2,\si}^n\|^2_2\|g_{2,\si}^m\|^2_2.
\eeqa
Substituting this bound to (\ref{scalar-product-formula-one}) and making use of Lemma~\ref{summation}, we get
\beqa
\|\nr_{1,\si}^*(h_{1,t}) \cH_{\I,\sipty}^{\cin} \nr_{2,\si}^*(h_{2,t}) \vac\|^2\leq c\|\cvv_2^{\sipty}\|_2^2\bigg(\fr{\kas}{\si}\bigg)^{4\la^2 c^2}.
\eeqa
Exploiting the fact that $\|\cvv_2^{\sipty}\|_2\leq c\si$, we conclude the proof of (\ref{second-check-bound}).  \qed 
%%%%%%%%%%%%%%%%%%%%%%%%%%%%%%%%%%%%%%%%%%%%%%%%%%%%%%%%%%%%%%%%%%%%%%%%%%%%%%%%%%%%%

%%%%%%%%%%%%%%%%%%%%%%%%%%%%%%%%%
\subsection{Clustering estimates} 
%%%%%%%%%%%%%%%%%%%%%%%%%%%%%%%%%

Consider the expressions
\beqa
& &\1\1\1M^W_{l,j}(\si',\si,s,t)\non\\
& &\1\1\1:=\lan\W_{\sip}(\nv_l,\sn)   \reta^*_{1, \si'}(h'_{1,t})\reta^*_{2,\si'}(h_{2,l,t}^{\prime (s)} e^{i\ga_{\si'}(\nv_l,t)})\Om,  \W_{\si}(\hv_j,\sn)  \reta^*_{1, \si}(h_{1,t})
\reta^*_{2,\si}(h^{(s)}_{2,j,t} e^{i\ga_{\si}(\hv_j,t)})\Om\ran, \label{M-W-def}  \\
& &\1\1\1N_{l,j}^W(\si',\si, s,t)\non\\
& &\1\1\1:=\lan   \reta^*_{1, \si'}(h'_{1,t})\Om,    \reta^*_{1, \si}(h_{1,t}) \Om \ran \ 
\lan \W_{\sip}(\nv_l,\sn) \reta^*_{2,\si'}(h_{2,l,t}^{\prime (s)} e^{i\ga_{\si'}(\nv_l,t)})\Om,   \W_{\si}(\hv_j,\sn)\reta^*_{2,\si}(h^{(s)}_{2,j,t} e^{i\ga_{\si}(\hv_j,t)})\Om\ran.  
\label{N-W-def} \quad \quad\quad
\eeqa
We want to estimate the difference $M^W_{l,j}(\si',\si,s,t)-N_{l,j}^W(\si',\si, s,t)$.
We  set for $\kal\geq \sip\geq \si>0$ consistently with (\ref{two-Weyl-operators})
\beqa
\hW_{\sip,\si,l,j}^{1}(\sn):=\W_{\sip}(\nv_l,t)^*\W_{\si}(\hv_j,t).
\eeqa
(We suppress the dependence of the l.h.s. on $\nv_l, \hv_j$ as it will be clear from the context). 
 We generalize some  definitions from Subsection~\ref{clustering-subsection}  and state two simple estimates 
\begin{align}
\hW_{\sip,\si,l,j}^{1}(\sn)&=\exp\big\{- \big(a(h_{l,j,t}^{\sip,\si})-a^*(h_{l,j,t}^{\sip,\si}) \big)    \big\}=e^{-a(h_{l,j,t}^{\sip,\si}) } e^{a^*(h_{l,j,t}^{\sip,\si}) }
e^{\h C_{l,j,\sip,\si}  }, \label{h-W-x} \\
h_{l,j,t}^{\sip,\si}(k)&:=\fr{\vv_2^{\si}(k) f^{\sip,\si}_{  \hat{\nv}_j, \nv_l  }(k)}{|k|}\e^{-i|k|t}, \quad   f^{\sip,\si}_{\hat{\nv}_j, \nv_l}(k):= 
\fr{ \hatk\cdot(\hv_j-\nv_l)}{(1-\hatk\cdot \hv_j) (1-\hatk\cdot \nv_l)}+\fr{\chi_{[\si,\sip]}(k) }{1-\hatk\cdot \nv_l},  \label{h-photon-def-app}\\
 C_{l,j,\sip,\si}&:=\|h_{l,j,t}^{\sip,\si} \|^2_2, \quad\quad\quad\quad\quad e^{\h C_{l,j,\sip,\si}  }=\mco(\si^{-\de_{\la_0}}), \label{C-e-C-item}\\
|h_{l,j,t}^{\sip,\si}(k)|&\leq c\la \fr{\chi_{[\si,\ka)}(k)}{|k|^{3/2}}. \label{photon-h-bounds}
\end{align}
Now we are ready to prove the following proposition:
%%%%%%%%%%%%%%%%%%%%%%%%%%%%%%%%%%%%
\bep\label{strong-clustering-proposition} We have for $\kal\geq \sip\geq \si>0$ and $M^W_{l,j}, N^W_{l,j}$ given by (\ref{M-W-def}), (\ref{N-W-def})
\beqa
M^{W}_{l,j}(\si',\si,s,t)=N^{W}_{l,j}(\si',\si,s,t)+R^{W}_{l,j}(\si',\si,s,t), \label{rest-term-W}
\eeqa
where the rest term satisfies
\beqa \label{formula-strong-clustering-proposition}
|R^{W}_{l,j}(\si',\si,s,t)|\leq \fr{c}{\si^{\de_{\la_0}}} \fr{1}{\kal}  \bigg\{ 
  \fr{1}{  t\si^{ 1/(8\tiga) } } +  \fr{1}{\uneps(s) t s^{-\peps}}   +(\sip)^{\alf/(8\ga_0)} \bigg\}.
\eeqa
\eep
%%%%%%%%%%%%%%%%%%%
\proof First,  we obtain from (\ref{h-W-x}) and (\ref{M-W-def})
\beqa
& &M^W_{l,j}(\si',\si,s,t) =e^{\h C_{l,j,\sip,\si}  }\sum_{\ellp, \ell=0}^{\infty}  \fr{(-1)^{\ellp} }{\ellp!\ell!} M^{\ellp,\ell}_{l,j}(\si',\si,s,t), \textrm{ where } \label{M-W-C}\\ 
& & M^{\ellp,\ell}_{l,j}(\si',\si,s,t)\non\\
& &:= \lan  \reta^*_{1, \si'}(h'_{1,t})   a^*(h_{l,j,t}^{\sip,\si})^{\ellp} \reta^*_{2,\si'}(h_{2,l,t}^{\prime (s)} e^{i\ga_{\si'}(\nv_l,t)})\Om,    
 \reta^*_{1, \si}(h_{1,t}) a^*(h_{l,j,t}^{\sip,\si})^{\ell} 
\reta^*_{2,\si}(h^{(s)}_{2,j,t} e^{i\ga_{\si}(\hv_j,t)})\Om\ran. \label{M-W-def-mod} \quad\quad\quad\quad
\eeqa
Similarly,
\beqa
& &N^W_{l,j}(\si',\si,s,t) =e^{\h C_{l,j,\sip,\si}  }\sum_{\ellp, \ell=0}^{\infty}  \fr{(-1)^{\ellp} }{\ellp!\ell!} N^{\ellp,\ell}_{l,j}(\si',\si,s,t), \textrm{ where } \\ 
& & N^{\ellp, \ell}_{l,j}(\si',\si,s,t):= \lan  \reta^*_{1, \si'}(h'_{1,t}) \Om,  \reta^*_{1, \si}(h_{1,t})\Om\ran \times \non \\   
& & \ph{44444444444444}  \times \lan  \Om,  a^*(h_{l,j,t}^{\sip,\si})^{\ellp} \reta^*_{2,\si'}(h_{2,l,t}^{\prime (s)} e^{i\ga_{\si'}(\nv_l,t)})\Om,      
a^*(h_{l,j,t}^{\sip,\si})^{\ell} \reta^*_{2,\si}(h^{(s)}_{2,j,t} e^{i\ga_{\si}(\hv_j,t)})\Om\ran. \label{N-W-def-mod} \quad\quad\quad\quad
\eeqa
Let us write consistently with our earlier notations
\beqa
& &h^{\ga}_{2,t}(p):=h_{2,j,t}^{(s)}(p)e^{ i\ga_{\si}(\hv_j,t)(p) }, \quad  h_t:=h_{l,j,t}^{\sip,\si}, \quad h_t^{\ell}(k_1, \ldots, k_{\ell}):=h_t(k_1)\ldots h_t(k_{\ell}), \\  
& &k_{\ell}:=(k_1,\ldots, k_{\ell}), \quad \unk_{\ell}:=k_1+\cdots+k_{\ell}, \quad  |\unk|_{\ell}:=|k_1|+\cdots+|k_{\ell}|.  
\eeqa
Now let us consider the expression
\beqa
& &a^*(h_{t})^{\ell} \reta^*_{2,\si}(h^{\ga}_{2,t})\Om=\sum_{m=0}^{\infty}\fr{1}{\sqrt{m!}}\int d^3p\,d^{3m}k d^{3\ell}k'\,h_{2,t}^{\ga}(p) f^{m}_{2,p,\si}(k)h^{\ell}_t(k') a^*(k')^{\ell}  \bb^*(k)^m \, \nn_{2}^*(p- \unk)\non\\
& &=\sum_{m=0}^{\infty}\fr{1}{\sqrt{m!}}\int d^3p\,d^{3m}k d^{3\ell}k'\,h_{2,t}^{\ga}(p-\unk') f^{m}_{2,p-\unk',\si}(k)h^{\ell}_t(k') a^*(k')^{\ell}  \bb^*(k)^m \, \nn_{2}^*(p- \unk-\unk')\Om\non\\
& &=\sum_{m'\geq \ell}^{\infty}\fr{1}{\sqrt{(m'-\ell)!}}\int d^3p\,d^{3m'}\ti k \,h_{2,t}^{\ga}(p-\ti{\unk}_{\ell}) (h^{\ell}_t f^{m'-\ell}_{2,p-\ti{\unk}_{\ell},\si})(\ti{k})  \bb^*(\ti{k})^{m'} \, \nn_{2}^*(p- \ti{\unk})\Om, \label{above-symmetrization}
\eeqa
where we set $m':=m+\ell$, $\ti k:=(k,k')$ and $(h^{\ell} f^{m'-\ell}_{2,q-\ti{\unk}_{\ell},\si} )(\ti k):=
[h(\ti k_1)\ldots h(\ti k_{\ell})f^{m'-\ell}_{2,q-\ti{\unk}_{\ell},\si}(\ti k_{\ell+1}, \ldots, \ti k_{m'})]_{{\mc \mrm{sym}}}$.  {\mc Here $\mrm{sym}$ denotes symmetrization in variables $k_1, \ldots, k_m$, i.e., 
for any function $F$ of $m$ variables
 \beqa
 [F(k_1, \ldots, k_m)]_{\mrm{sym}}=\fr{1}{m!} \sum_{\si\in S_m} F(k_{\si_1}, \ldots, k_{\si_m}).  \label{symmetrization-sum}
 \eeqa
 }
Thus we put $a^*(h_{t})^{\ell} \reta^*_{2,\si}(h^{\ga}_{2,t})$ as a sum of terms of  the form (\ref{B-star-def-new}) and we set accordingly
\begin{align}
&G_{1,m}(q;k):=e^{-iE_{1,q,\si}t}h_1(q)f^{m}_{1,q,\si}(k), \label{G-def-new-ell} \\
&G^{\ell}_{2,m}(q;k):=e^{-i(E_{2,q-\unk_{\ell},\si}+|\unk|_{\ell})  t}e^{i\ga_{\si}(\hv_j,t)(q-\unk_{\ell})  }h_{2,j}^{(s)}(q-\unk_{\ell})
(h^{\ell}f^{m-\ell}_{2,q-\unk_{\ell},\si} )(k),  \\
&G'_{1,m}(q;k):=e^{-iE_{1,q,\si'}t}h'_1(q)f^{m}_{1,q,\si'}(k), \\
& G^{\prime\ell'}_{2,m}(q;k):=e^{-i(E_{2,q-\unk_{\ellp},\si'}+|\unk|_{\ellp})  t}e^{i\ga_{\si'}(\nv_l,t)(q-\unk_{\ellp})  }h_{2,l}^{\prime (s)}(q-\unk_{\ellp}) (h^{\ellp}f^{m-\ellp}_{2,q-\unk_{\ellp},\si'} )(k),   
\label{G-prime-def-one-ell} 
\end{align}
%where $(f^{m-\ell}_{2,q,\si} h^{\ell})(k):=[h(k_1)\ldots h(k_{\ell})f^{m-\ell}_{2,q,\si}(k_{\ell+1}, \ldots, k_m)]_{{\mc\mrm{sym}}}$,
 Now we can write
\beqa
& &M^{\ellp,\ell}_{l,j}(\si',\si,s,t) =
\sum_{\substack{m\geq \ell, \ti m\geq \ellp  ,n,\ti n\geq 0  \\ \ti m+\ti n=m+n } } \fr{1}{\sqrt{(m-\ell)!n!(\ti m-\ellp)!\ti n!}} \times \non\\
& &\ph{444444444444444}\times\lan\vac, B_{1,\ti n}(G_{1,\ti n}' )B_{2,\ti m}(G_{2,\ti m}^{\prime \ell'} )B_{1,n}^*(G_{1,n}) 
B_{2,m}^*(G_{2,m}^{\ell})\vac\ran.\,\,
\label{scalar-product-formula-new-ell}
\eeqa
Making use of Lemma~\ref{norms-of-scattering-states}, we obtain
\beqa
\lan\vac, B_{1,\ti n}(G_{1,\ti n}' )B_{2,\ti m}(G_{2,\ti m}^{\prime\ell'} )B_{1,n}^*(G_{1,n}) B_{2,m}^*(G^{\ell}_{2,m})\vac\ran \pha{4444444444444444444444444444444}  \non\\
=\sum_{\sig\in S_{m+n}}\int  d^3 q d^3p \int  d^{3n}r d^{3m}k\,  
G_{1,n}(q;  r) G_{2,m}^{\ell}(p;  k)\pha{444444444444444444444444}\non\\
 \times\bigg(\ov G_{1,\ti n}'( q+\un{\check{k}}-\un{\check{r}}; \hat r, \check k)\ov  G_{2,\ti m}^{\prime\ell'}(p-\un{\check{k}}+\un{\check{r}}; \hat{k}, \check r)
\bigg),
\label{B-expectation-value-new-ell}
\eeqa
where the notation in (\ref{B-expectation-value-new-ell}) is explained in Lemma~\ref{combinatorics}.

 Let us denote the summands on the r.h.s. of (\ref{B-expectation-value-new-ell}) by $I^{\ellp,\ell,(1)}_{m,n,\ti m,\ti n}$.
Let $\check I^{\ellp,\ell,(1)}_{m,n,\ti m,\ti n}$ be such summands coming from permutations for which $\check k$
or $\check r$ are non-empty. They will give contributions to the rest term in (\ref{rest-term-W}). We note that there are $(m+n)!-m!n!$  such permutations. 
Let $M_{l,j}^{\ellp,\ell,(\check 1)}$
be the  contribution to (\ref{scalar-product-formula-new-ell}) involving all the summands $\check I^{\ellp,\ell,(1)}_{m,n,\ti m,\ti n}$.  
By Lemma~\ref{second-contribution-decay-new-old} we have
\beqa
 |M_{l,j}^{\ellp,\ell,(\check 1)}| \1&\leq&\1 \sum_{\substack{m\geq \ell, \ti m\geq \ellp  ,n,\ti n\geq 0  \\ \ti m+\ti n=m+n } }  \fr{(m+n)!}{\sqrt{(m-\ell)!n!(\ti m-\ellp)!\ti n!}}  |\check I^{\ellp,\ell,(1)}_{m,n,\ti m,\ti n}(\si,\sip,s,t)_{l,j}|\non\\
\1&\leq& \1 
\fr{c}{\kal}\bigg\{ 
  \fr{1}{  t\si^{ 1/(8\tiga) } } +  \fr{1}{\uneps(s) t s^{-\peps}}   +(\sip)^{\alf/(8\ga_0)}  
 \bigg\}\times \non\\
& &\ph{44444444}\times  \sum_{\substack{m\geq \ell, \ti m\geq \ellp  ,n,\ti n\geq 0  \\ \ti m+\ti n=m+n } } (m+n)!\fr{( \|g^{n-1}_{{2},\si}\|^2_2 +\|g^n_{{2},\si}\|^2_2) (\|g^{m-1}_{2,\si}\|^2_2+\|g^m_{2,\si}\|_2^2 )}{(m-\ell)!n!(\ti m-\ellp)! \ti n! }.
\label{I-two-final-formula-new-early}
\eeqa
Thus from  (\ref{M-W-C}), second relation in  (\ref{C-e-C-item}) and Lemma~\ref{summation-ell}  we obtain
\beqa
|R^{W}_{l,j}(\si',\si,s,t)|\leq \fr{c}{\si^{\de_{\la_0}}} \fr{1}{\kal} \bigg\{ 
  \fr{1}{  t\si^{ 1/(8\tiga) } } +  \fr{1}{\uneps(s) t s^{-\peps}}   +(\sip)^{\alf/(8\ga_0)} \bigg\}.
\eeqa

Now let $\hat I^{\ellp,\ell,(1)}_{m,n,\ti m,\ti n}$ denote the expressions $I^{\ellp,\ell, (1)}_{m,n,\ti m,\ti n}$ coming from permutations for which $\check k$ and $\check r$
are empty. (We note that there are $m!n!$ such permutations and that in this case $m=\ti m$, $n=\ti n$). We obtain from 
Lemma~\ref{norms-of-single-particle-states} that
\beqa
\hat I^{\ell',\ell,(1)}_{m,n,\ti m,\ti n}\1&=&\1\int  d^3 q  d^{3n}r \,  G_{1,n}(q;  r) \ov G_{1,\ti n}'(q;  r)  \int d^3p  d^{3m}k\,   
G_{2,m}^{\ell}(p;  k ) \ov  G_{2,\ti m}^{\prime \ell'}(p;  k)\non\\
\1&=&\1\fr{1}{n!}\fr{1}{m!}\lan \vac, B_{1,\ti n}(G_{1,\ti n}') B_{1,n}^*(G_{1,n})\vac\ran  \lan \vac,B_{2,\ti m}(G_{2,\ti m}^{\prime \ell'}) B_{2,m}^*(G_{2,m}^{\ell} )\vac\ran. 
\eeqa
Let $ M_{l,j}^{\ellp,\ell,(\hat 1)}$  be the contribution to $ M^{\ellp,\ell}_{l,j}$ involving all such  
$\hat I^{\ellp,\ell,(1)}_{m,n,\ti m,\ti n}$. Since the sum over permutations in (\ref{B-expectation-value-new-ell})  gives the compensating factor $m!n!$, we obtain
\beqa
M_{l,j}^{\ellp,\ell,(\hat 1)}(\si',\si,s,t)\1&=&\1\sum_{  \substack{m\geq \ell, \ti m\geq \ellp  ,n,\ti n\geq 0  \\ \ti m+\ti n=m+n }     }\fr{1}{\sqrt{(m-\ell)!n!(\ti m-\ellp)!\ti n!}} \times \non\\
& &\ph{44444444}\times \lan \vac, B_{1,\ti n}(G_{1,\ti n}') B_{1,n}^*(G_{1,n})\vac\ran  \lan \vac,B_{2,\ti m}(G_{2,\ti m}^{' \ell'}) B_{2,m}^*(G_{2,m}^{\ell})\vac\ran\non\\
\1&=&\1 N^{\ellp,\ell}_{l,j}(\si',\si,s,t),
\eeqa
where in the last step we compared  definition~(\ref{B-star-def-new}) of  $B_{i,n}^*(G_{i,n})$ with definition~(\ref{renormalized-new})
of the renormalized creation operator and with the definition of $N^{\ellp,\ell}_{l,j}(\si',\si,s,t)$ in (\ref{N-W-def-mod}). \qed
%%%%%%%%%%%%%%%%%%%%%%%%%%%%%%%%%%%%%%%%%%%%%%%%%%%%%%%%%
%%%%%%%%%%%%%%%%%%%%%%%%%%%%%%%%%%%%%%%%%%%%%%%%%%%%%%%%%%%%%%%%%%%%%%%%

\section{Non-stationary phase analysis} \label{non-stationary-app}
\setcounter{equation}{0}
%%%%%%%%%%%%%%%%%%%%%%%%%%%%%%%%%%
\subsection{Preliminaries concerning the partition and phases}
%%%%%%%%%%%%%%%%%%%%%%%%%%%%%%%%%%

As compared to the paper on scattering of two atoms \cite{DP12.0},
the non-stationary phase arguments must be modified due to the
presence of the phases and due to the time-dependent partition.

%%%%%%%%%%%%%%%%%%%%%%%%%%%%%%%%%%%%%%%%%%%%%
\bel\label{h-theta} 
 Recall from Definition~\ref{partition-def} that $h_{2,j}^{(s)}(p)=\app_{ \Ga_j^{(s)} }(p) h_{2}(p)$, where $\app_{\Ga_{j}^{(s)}}(p)=\hveta_{a_{\np},\uneps}(p-p_{j})$ and
\beqa
\hveta_{a_{\np},\uneps}(p)=\prod_{i=1}^3\veta\bigg(\fr{(p^i+a_{\np})}{a_{\np}\uneps}\bigg) \veta\bigg(-\fr{(p^i-a_{\np})}{a_{\np}\uneps}\bigg).
\eeqa
Here $\uneps=\uneps(s)=s^{-\3\peps}$, $a_{\np}:=2^{-(\np+1)}$, where $\np=\np(s)$ is restricted by  $2^{\np}\leq s^{\peps}< 2^{\np+1}$ and $\peps>0$ will be fixed a posteriori. We set $\theta:=a_{\np}\uneps$
and note that $\theta^{-1}\leq { 2}/(\uneps(s)s^{-\peps})\leq {2}/(\uneps(t)t^{-\peps})$. We have:
\begin{align}
|\pa_{p^i} h_{2,j}^{(s)}(p)|&\leq c \theta^{-1} (|h_{2}^j(p)|
+|(\pa_{p^i}h_{2})^j(p)|), \label{first-derivative-partition}\\ 
|\pa_{p^{i'}}\pa_{p^i} h_{2,j}^{(s)}(p)|&\leq c \theta^{-2} (|h_{2}^j(p)|
+|(\pa_{p^i}h_{2})^j(p)|+|(\pa_{p^{i'}}\pa_{p^i}h_{2})^j(p)| ), \label{second-derivative-partition}
\end{align}
where $h_{2}^j(p):= \mathbf{1}_{ \Ga_{j}^{(s)} }(p)h_{2}(p)$ (i.e.
upper index $j$ indicates restriction with a sharp characteristic function).
\eel
%%%%%%%%%%%%%%%%%%%%%%%%%%%%%%%%%%%%%%%%%%%%%%%
\proof We compute
\begin{align}
\pa_{p^i} h_{2,j}^{(s)}(p)&=(\pa_{p^i}\app_{\Ga_{j}^{(s)}}(p))h_{2}(p)
+\app_{\Ga_{j}^{(s)}}(p)\pa_{p^i} h_{2}(p),\\
\pa_{p^{i'}}\pa_{p^i} h_{2,j}^{(s)}(p)&=(\pa_{p^{i'}}\pa_{p^i}\app_{\Ga_{j}^{(s)}}(p))h_{2}(p)
+\pa_{p^{i'}}\app_{\Ga_{j}^{(s)}}(p)\pa_{p^i} h_{2}(p)\non\\
&\ph{44}+\pa_{p^{i}}\app_{\Ga_{j}^{(s)}}(p)\pa_{p^{i'}} h_{2}(p)+\app_{\Ga_{j}^{(s)}}(p)\pa_{p^{i'}}\pa_{p^i} h_{2}(p).
\end{align}
We have
\beqa
\pa_{p^1}\hveta_{a_{\np},\uneps}(p)=\fr{1}{a_{\np}\uneps}
\bigg( \veta'\bigg(\fr{(p^1+a_{\np})}{a_{\np}\uneps}\bigg) \veta\bigg(-\fr{(p^1-a_{\np})}{a_{\np}\uneps}\bigg)- \veta\bigg(\fr{(p^1+a_{\np})}{a_{\np}\uneps}\bigg) \veta'\bigg(-\fr{(p^1-a_{\np})}{a_{\np}\uneps}  \bigg)         \bigg)\times\non\\
\times\prod_{i=2}^3\veta\bigg(\fr{(p^i+a_{\np})}{a_{\np}\uneps}\bigg) \veta\bigg(-\fr{(p^i-a_{\np})}{a_{\np}\uneps}\bigg).
\eeqa
Thus we obtain
\beqa
|\pa_{p^1}\hveta_{a_{\np},\uneps}(p)|\leq 
\fr{c}{a_{\np}\uneps}\mathbf{1}_{[-a_{\np},a_{\np}]^{\times 3}}(p)
\eeqa
for a universal constant $c$ (depending on $\sup|\veta|$).
Consequently we get
\beqa
|\pa_{p^1}\app_{\Ga_{j}^{({\mc s})}}(p)|\leq \fr{c}{a_{\np}\uneps}
\mathbf{1}_{ \Ga_{j}^{({\mc s})} }(p),
\eeqa
which implies (\ref{first-derivative-partition}).
By an analogous argument we obtain
\beqa
| \pa_{p^{i'}}\pa_{p^i}\app_{\Ga_{j}^{({\mc s})}}(p)|\leq \fr{c}{(a_{\np}\uneps)^2}
\mathbf{1}_{ \Ga_{j}^{({\mc s})} }(p),
\eeqa
which concludes the proof. \qed

%%%%%%%%%%%%%%%%%%%%%%%%%%%%%%%%%%%%%%%%%%%%%%
\bel\label{phase-lemma} Let $1/2<\al<1$ and $\si^{\S}_{\tau}=\kal \tau^{-\al}$.   Set $t_{\si}:=\min(t, (\kal/\si)^{1/\al})$ and define as in (\ref{slow-cutoff})
\beqa
\ga_{\si}(\nv_j,t)(p)= -\int_1^{t_{\si}}\bigg\{\int_{\si}^{\si_{\tau}^\S}d|k| \int d\Omm(e_k)\, \vv_2^{\si}(k)^2(2|k|)
\bigg(\fr{\cos(k\cdot\nabla E_{2,p,\si}\tau-|k|\tau)}{1-\hatk\cdot \nv_j }\bigg)  \bigg\} d\tau.    \label{slow-cutoff-one}
\eeqa
Then, uniformly in $p\in S$,
\begin{align}
|\ga_{\si}(\nv_j,t)(p)|&\leq c, \label{phase-zero-derivative} \\
|\pa_{p^i} \ga_{\si}(\nv_j,t)(p)|&\leq c,  \label{phase-first-derivative} \\
|\pa_{p^{i'}}\pa_{p^i} \ga_{\si}(\nv_j,t)(p)|&\leq \fr{c}{  \si^{\de_{\g_0}}  }  t^{(1-\al)}. \label{modified-phase-second-derivative}
\end{align}
\eel
%%%%%%%%%%%%%%%%%%%%%%%%%%%%%%%%%%%%%%%%%%%%%%%
\proof
First, we note that $\ga_{\si}(\nv_j,t)(p)$ is the real part of
\beqa
\ti\ga_{\si}(\nv_j,t)(p)\1&=&\1 -\int_1^{t_{\si}}\bigg\{\int_{\si\tau}^{\si_{\tau}^\S\tau} d|k| \int d\Omm(e_k)  \, \chi_{[\si,\ka)}(k/\tau)^2
\bigg(\fr{\e^{i(k\cdot \nabla E_{2,p,\si}-|k|)} }{1-\hatk\cdot  \nv_j }\bigg)  \bigg\} \fr{d\tau}{\tau}\non\\
\1&=&\1-\int_1^{t_{\si}}\bigg\{\int_{\si\tau}^{\si_{\tau}^\S\tau} d|k| \int d\Omm(\nee)\,
\bigg(\fr{\e^{i|k| (\nee\cdot \nabla E_{2,p,\si}-1)} }{1-\nee \cdot \nv_j }\bigg) \bigg\} \fr{d\tau}{\tau}\non\\
\1&=&\1-\int_1^{t_{\si}}\bigg\{\int \fr{d\Omm(\nee)}{(1-\nee \cdot \nv_j) }
\bigg(\fr{ \e^{i \si_{\tau}^\S\tau (\nee\cdot \nabla E_{2,p,\si}-1)} -   \e^{i \si\tau  (\nee\cdot \nabla E_{2,p,\si}-1)}  }
{i(\nee\cdot \nabla E_{2,p,\si}-1)}\bigg)   \bigg\} \fr{d\tau}{\tau},
\eeqa
where we used that $\chi_{[\si,\ka)}(k/\tau)=1$ in the region of integration.
Let us introduce notation
\begin{align}
d\ti{\Omm}(\nee):=\fr{d\Omm(\nee)}{(1-\nee \cdot \nv_j)},\quad
f(p,\nee):= (\nee\cdot \nabla E_{2,p,\si}-1).
\end{align}
Then we can write
\beqa
\ti\ga_{\si}(\nv_j,t)(p)=i \int_1^{t_{\si}}\bigg\{\int d\ti{\Omm}(\nee)  f(p,\nee)^{-1}
\bigg( \e^{i \si_{\tau}^\S\tau f(p,\nee)} -   \e^{i \si\tau  f(p,\nee)}  \bigg)   \bigg\} \fr{d\tau}{\tau}.
\eeqa
Let us consider auxiliary integrals, depending on a parameter $a\in\real$
\beqa
I_{a}:=\int_1^{t_{\si} } \e^{i a\si\tau  f(p,\nee)} \fr{d\tau}{\tau}, \quad J_a:=\int_1^{t_{\si} } \e^{i a \kal \tau^{1-\al}  f(p,\nee)} \fr{d\tau}{\tau}.
\eeqa
We have
\begin{align}
\pa_a I_a&=
 \int_1^{t_{\si} } \e^{i a\si\tau  f(p,\nee)} i\si  f(p,\nee) d\tau= \fr{ \e^{i a\si t_{\si}  f(p,\nee)} -\e^{i a\si   f(p,\nee)} }{ a}, \\
\pa_a J_a&=\int_1^{t_{\si} } \e^{i a \kal \tau^{1-\al}  f(p,\nee)} i\kal \tau^{1-\al} f(p,\nee)\fr{d\tau}{\tau}=
\fr{1}{1-\al} \fr{ \e^{i a\kal t_{\si}^{1-\al}  f(p,\nee)} -\e^{i a\kal   f(p,\nee)} }{a}.
\end{align}
Thus we obtain
\begin{align}
I_1&=I_0+\int_0^1da \fr{ \e^{i a\si t_{\si}  f(p,\nee)} -\e^{i a\si   f(p,\nee)} }{ a},\\
J_1&=J_0+\fr{1}{1-\al}\int_0^1da  \fr{ \e^{i a\ka t_{\si}^{1-\al}  f(p,\nee)} -\e^{i a\ka   f(p,\nee)} }{a}.
\end{align}
The expressions we are interested in are $\mathrm{Re}(-iI_1)$, $\mathrm{Re}(iJ_1)$. Since $I_0$ and $J_0$ are real, we get
\beqa
\mathrm{Re}(-iI_1)\1&=&\1\int_0^1da \fr{ \sin( a\si t_{\si}  f(p,\nee)) -\sin( a\si   f(p,\nee))  }{ a}\non\\
\1&=&\1 \int_0^{\si t_{\si}  f(p,\nee)} dx\, \fr{ \sin(x)}{x} -\int_0^{\si   f(p,\nee)} dx \,\fr{ \sin(x) }{x}\non\\
\1&=&\1\Sin(\si t_{\si}  f(p,\nee))-\Sin(\si   f(p,\nee)).\\
\mathrm{Re}(iJ_1)\1&=&\1- \fr{1}{1-\al}\int_0^1da \fr{ \sin( a\kal t_{\si}^{1-\al}  f(p,\nee)) -\sin( a\kal   f(p,\nee))  }{ a}\non\\
\1&=&\1-\fr{1}{1-\al}\big(\Sin(\kal t_{\si}^{1-\al}  f(p,\nee))-\Sin(\kal   f(p,\nee))\big).
\eeqa
Thus we get 
\begin{align}
\ga_{\si}(\nv_j,t)(p)&= \int d\ti{\Omm}(\nee)  f(p,\nee)^{-1}\bigg(\Sin(\si t_{\si}  f(p,\nee))-\Sin(\si   f(p,\nee))\non\\
&\ph{44}-\fr{1}{1-\al}\Sin(\kal t_{\si}^{1-\al}  f(p,\nee))+\fr{1}{1-\al}\Sin(\kal   f(p,\nee))\bigg).
\end{align}
Now we note that for $x\in \real$
\beqa
|\Sin(x)|\leq c, \quad |\Sin'(x)|\leq \fr{c}{|x|}, \quad |\Sin''(x)|\leq \fr{c}{|x|}.
\eeqa
Hence
\begin{align}
|\ga_{\si}(\nv_j,t)(p)|&\leq c, \\
|\pa_{p^i} \ga_{\si}(\nv_j,t)(p)|&\leq c, \\
|\pa_{p^i}\pa_{p^{i'}} \ga_{\si}(\nv_j,t)(p)|&\leq \fr{c}{\si^{\de_{\g_0}} }(\si t_{\si}+\si+\kal t_{\si}^{1-\al}+\kal)\leq \fr{c}{\si^{\de_{\g_0}} } t^{1-\al},
\end{align}
where in the last step we estimated $\si t_{\si}\leq \si t_{\si}^{\al} t_{\si}^{1-\al}\leq t^{1-\al}$, since  $t_{\si}\leq (\kal)^{1/\al}\si^{-1/\al}$ and $t_{\si}\leq t$. 
We also noted that in the second derivative of $\pa_{p^i}\pa_{p^{i'}} \ga_{\si}(\nv_j,t)(p)$ the third derivative of $E_{2,p,\si}$ arises. \qed
%%%%%%%%%%%%%%%%%%%%%%%%%%%%%%%%%%%%%%%%%%%%%%%%%%%%%%%%%%%%%%%%%%%%%%%%%%%%%%%%%%%%%%%%%%%%%%%%%%%%

%%%%%%%%%%%%%%%%%%%%%%%%%%%%%%%%%%%%%%%%%%%%%%%%%
\subsection{Non-stationary phase arguments} \label{non-stationary-subsection}
%%%%%%%%%%%%%%%%%%%%%%%%%%%%%%%%%%%%%%%%%
\bel\label{smooth-bounds} Let $G_{i, m}$, $i\in \{1,2\}$, be as specified in  (\ref{zero-G-def-new}) and let $F^{G_1,G_2}_{2,n,m}$ be defined as in (\ref{FGH}) i.e., it has the form
\beqa
F^{G_1,G_2}_{2,n,m}(q;r \ba p;k)
\!=\!(n+1)\int d^3\ti r\, \vv_{{\mc 2}}(\ti r) e^{-i(E_{1, q+\ti r,\si }+E_{2,p-\ti r,\si} )t}h^{\ga}_2(p-\ti r )h_1(q+\ti r )
f^{n+1}_{1,q+\ti r ,\si}(r,\ti r ) f^{m}_{2,p-\ti r,\si}(k),\,\,\,\,\,\, \label{factor-n+1}
\eeqa
where $h_1,h_2\in C_0^{2}(\PSS)$ have disjoint {velocity} supports (cf. (\ref{velocity-support})) and $h^{\ga}_2(p)=h_{2,j'}^{(s)}(p)e^{i\ga_{\si}(\nv_{j'},t)(p)}$, $1\leq s\leq t$.
There holds the bound 
\beqa
 |F^{G_1,G_2}_{{\mc 2},n,m}(q; r \ba p; k)|\leq  \fr{c_{1,t}}{\si ^{\de_{\g_0}}  } \fr{1}{(\kal)^2} \bigg( \fr{\si_t^{\alf/(8\tiga)  } }{t}+\fr{1}{t^2\si_t^{1/(4\tiga)}} \bigg)
\fr{1}{\sqrt{m!n!}}  D(p,q)g^m_{2,\si}(k) g^n_{{2},\si}(r),  \label{FGH-bound}
\eeqa
where  $D\in C_0^{\infty}(\real^3\times \real^3)$, $c_{1,t}=c\bigg(t^{(1-\al)}+ ( \uneps(t) t^{-\peps} )^{-2} \bigg)$ and the estimate is uniform in $j'$. The same estimate holds for  $F^{G_2,G_1}_{1,n,m}$. 
\eel
%%%%%%%%%%%%%%%%%%%%%%%%%%%%
\proof 
For $0< \sigma\leq \kal$, we introduce the slow (time-dependent) cut-off $\si_{\s,t}:=\kal(\si_t/\kal)^{ 1/(8\tiga) }$,
where $\si_t=\kal/t^{\ga}$, $4<\ga\leq \ga_0$. 
Let $\chi\in C_0^{\infty}(\real^3)$, $0\leq \chi\leq 1$,  be supported in $\mcB_{1+\eps}$ (the ball of radius $1+\eps$ {\mc centered at zero}) 
for some  $0<\eps<1$ and be equal to one on $\mcB_{1}$. We set $\chi_1(\ti k):=\chi(\ti k/\si_{\s,t})$, $\chi_2(\ti k):=1-\chi_1(\ti k)$ and define
 \beqa
& &F^{G_1,G_2}_{j,2,n,m}(q; r \ba p; k)
:=\int d^3\ti r\, \vv_2(\ti r) \chi_j(\ti r) e^{-i(E_{1, q+\ti r,\si }+E_{2,p-\ti r,\si})t }h_2^{\ga}(p-\ti r )h_1(q+\ti r )\non\\
& &\pha{4444444444444444444444444444444444444444} \times f^{n+1}_{1,q+\ti r ,\si}(r,\ti r ) f^{m}_{2,p-\ti r,\si}(k).   \label{decomposition-of-FGH}
\eeqa
We set $\nF(q,p,\ti r):=E_{1, q+\ti r,\si }+E_{2,p-\ti r,\si }$ and note that by disjointness of the  approximate velocity
supports\footnote{Cf. `Standing assumptions and conventions'.} of $h_1$, $h_2$, the condition $h_2(p-\ti r )h_1(q+\ti r )\neq 0$, together with Theorem~\ref{preliminaries-on-spectrum}, implies that
\beqa
|\nabla_{\ti r}\nF(q,p,\ti r)|\geq \eps'>0
\eeqa
for some fixed $\eps'$, independent of $q,p, \ti r$ within the above restrictions. Thus we can write
\beqa
e^{-i\nF(q,p,\ti r)t}=\fr{\nabla_{\ti r} \nF(q,p,\ti r)\cdot \nabla_{\ti r} e^{-i\nF(q,p,\ti r) t} }{(-it) |\nabla_{\ti r}\nF(q,p,\ti r)|^2}.
\eeqa
Now we define the function
\beqa
J(q,p,\ti r):=\fr{\nabla_{\ti r} \nF(q,p,\ti r)}{ |\nabla_{\ti r}\nF(q,p,\ti r)|^2}  h_2^{\ga}(p-\ti r )h_1(q+\ti r )\chi^{\ka}(\ti r), \label{function-J}
\eeqa
where $\chi^{\ka}\in C_0^{\infty}(\real^3)$ is equal to one on $\mcB_{\ka}$  and vanishes outside of a slightly larger set.
We note that, by Theorem~\ref{preliminaries-on-spectrum}, Lemma~\ref{h-theta} and Lemma~\ref{phase-lemma},
 for any multiindex $\be$ s.t. $0\leq|\be|\leq 2$ 
\beqa
|\pa_{\ti r}^{\be} J(q,p,\ti r)|\leq D(q,p)c_{1,t} , \quad c_{1,t}=\fr{c}{\si^{\de_{\la_0}}}\big(t^{(1-\al)}+  (\uneps(t) t^{-\peps})^{-2}  \big),
\label{j-estimate}
\eeqa
where $(q,p)\mapsto D(q,p)$ is a smooth, compactly supported  function, independent of $j'$  
 and we exploited that $t\geq s$, {\mc where $s$ is the time-scale of the partition}. 
Moreover, for  $0\leq|\be|\leq 2$, 
\beqa
|\pa_{\ti r}^{\be} \vv_2(\ti r)|\leq \fr{\chi_3(\ti r) }{|\ti r|^{\h+|\be|}},\quad
|\pa_{\ti r}^{\be}\chi_j(\ti r)|\leq \fr{c}{(\si_{\s,t})^{|\be|}}, \label{derivative-bounds}
\eeqa
where  $\chi_3$ is a smooth, compactly supported  function, independent of $\si$.
In addition, for  $0\leq|\be|\leq 2$  we obtain from  Theorem~\ref{main-theorem-spectral} 
\beqa
& &|\pa^{\be}_{\ti r} f^{m}_{2,p-\ti r,\si}(k)|\leq \fr{1}{\sqrt{m!}} \fr{c}{\si^{\de_{\g_0}} } g^m_{2,\si}(k),\label{m-derivatives} \\
& &|\pa^{\be}_{\ti r} f^{n+1}_{1,q+\ti r,\si}(r,\ti r)|\leq \fr{1}{\sqrt{n!}}\fr{c}{\si^{\de_{\g_0}} } \fr{|\ti r|^{\alf}}{|\ti r|^{3/2+|\be|}}  g^{n}_{1,\si}(r).
\label{n+1-derivatives}
\eeqa

We note that, by the support properties of $\chi_1$, we have
\beqa
F^{G_1,G_2}_{1,2,n,m}(q;r \ba p;k)=\mathbf{1}_{\{ \si \leq (1+\eps)\si_{\s,t}  \}}(t) F^{G_1,G_2}_{1,2,n,m}(q;r \ba p;k).
\eeqa
Thus we can assume that  $\si \leq (1+\eps)\si_{\s,t}$.
Now using the Gauss Law we obtain from (\ref{decomposition-of-FGH}) 
\beqa
F^{G_1,G_2}_{1,2,n,m}(q;r \ba p;k)
=\fr{1}{it}\int_{ \si\leq |\ti r|\leq (1+\eps)\si_{\s,t}} d^3\ti r\,  e^{-i\nF(q,p,\ti r)t} \nabla_{\ti r}
 \cdot\bigg(J(q,p,\ti r)  \vv_2(\ti r) \chi_1(\ti r) f^{n+1}_{1,q+\ti r ,\si}(r,\ti r )
 f^{m}_{2,p-\ti r,\si}(k)\bigg)\non\\
 +\fr{\si^2}{it}\int d\Omm(\nee )\,  e^{-i\nF(q,p,\si \nee )t} \nee
 \cdot\bigg( J(q,p,\ti r)\vv_2(\ti r) \chi_1(\ti r) f^{n+1}_{1,q+\ti r ,\si}(r,\ti r ) f^{m}_{2,p-\ti r,\si}(k)\bigg)\bigg|_{\ti r=\si \nee}, \,\,\,\,
\label{FGH-bound-integration-by-parts}
\eeqa
where $\nee$ is the normal vector to the unit sphere and  $d\Omm(\nee  )$ is the spherical measure. 
(Notice that the contribution corresponding to the outer surface vanishes because of the smooth cut-off associated with $\chi_{1}$).

\noindent
Let us consider the first term on the r.h.s.
of (\ref{FGH-bound-integration-by-parts}). Let $I_1$ be the corresponding  integrand.
Making use of (\ref{j-estimate})--(\ref{n+1-derivatives}), we obtain
\beqa
|I_1|\1&\leq&\1\fr{1 }{\sqrt{m!n!}} \sum_{0\leq |\be_1|+|\be_2|+|\be_3|\leq 1}\fr{ c_{1,t}}{(\si_{\s,t})^{|\be_1|}}\fr{  D(q,p)  }{\si^{\de_{\g_0}} }
\fr{\chi_3(\ti r)|\ti r|^{\alf}   }{|\ti r|^{2+|\be_2|+|\be_3|}}   g^{n}_{1,\si}(r)g^m_{2,\si}(k)\non\\
\1&\leq&\1\fr{ c_{1,t}}{\sqrt{m!n!}} \fr{  D(q,p)    }{\si^{\de_{\g_0}} }\fr{\chi_3(\ti r)|\ti r|^{\alf}  }{|\ti r|^{3}}   g^{n}_{1,\si}(r)g^m_{2,\si}(k),
\label{I-one-bound}
\eeqa
where in the last step above we made use of the fact that $|\ti r|\leq (1+\eps)\si_{\s,t}$ in the region of integration.
Now let $I_2$ be the integrand in the boundary integral on the r.h.s. of (\ref{FGH-bound-integration-by-parts}). Making use, again, of 
bounds (\ref{j-estimate})--(\ref{n+1-derivatives}), we get
\beqa
|I_2|\1&\leq&\1\fr{c_{1,t}}{\sqrt{m!n!}} \fr{  D(q,p)}{\si^{\de_{\g_0}} }\fr{\chi_3(\ti r)|\ti r|^{\alf}   }{|\ti r|^{2} }   g^{n}_{1,\si}(r)g^m_{2,\si}(k)
\bigg|_{\ti r=\si\nee }\non\\
\1&=&\1\fr{c_{1,t}}{\sqrt{m!n!}} \fr{  D(q,p)}{\si^{\de_{\g_0}} }\fr{\chi_3(\si \nee) \si^{\alf}   }{\si^{2} }   g^{n}_{1,\si}(r)g^m_{2,\si}(k).
\label{I-two-bound}
\eeqa
Thus (\ref{FGH-bound-integration-by-parts}), (\ref{I-one-bound}), (\ref{I-two-bound}) give 
\beqa
|F^{G_1,G_2}_{1,2,n,m}(q;r \ba p;k)|\leq \fr{ c_{1,t}}{\sqrt{m!n!}}\fr{  D(q,p) }{\si^{\de_{\g_0}} } \fr{(\si_{\s,t})^{\alf} }{t}   g^{n}_{1,\si}(r)g^m_{2,\si}(k),
\eeqa
which is the first contribution to the bound in (\ref{FGH-bound}), where we also use $g^{n}_{1,\si}(r)\leq g^n_{2,\si}(r)$.

Now we proceed to $F^{G_1,G_2}_{2,2, n,m}$. By integrating  twice by parts in the defining expression, we get
\beqa
F^{G_1,G_2}_{2,2, n,m}(q; r \ba p; k)=\fr{1}{(it)^2}\int d^3\ti r\,  e^{-i\nF(q,p,\ti r)t}\cdot\pha{4444444444444444444444444444}\non\\
\times\nabla_{\ti r} \cdot\bigg( \fr{\nabla_{\ti r}\nF(q,p,\ti r)}{ |\nabla_{\ti r}\nF(q,p,\ti r)|^2}\nabla_{\ti r} 
\cdot\big( J(q,p,\ti r)  \vv_2(\ti r) \chi_2(\ti r) f^{n+1}_{1,q+\ti r ,\si}(r,\ti r )
 f^{m}_{2,p-\ti r,\si}(k)\big)\bigg). \label{second-derivative}
\eeqa
By Theorem~\ref{preliminaries-on-spectrum}, the function 
\beqa
(q,p,\ti r)\mapsto \fr{\nabla_{\ti r}\nF(q,p,\ti r)}{ |\nabla_{\ti r}\nF(q,p,\ti r)|^2}
\eeqa
is bounded by $c/\si^{\de_{\la_0}}$, together with its first  derivatives, on the support of 
$(q,p,\ti r)\mapsto h_2(p-\ti r )h_1(q+\ti r )\chi^{\ka}(\ti r)$. Thus we obtain from the bounds (\ref{j-estimate}), (\ref{derivative-bounds}), 
(\ref{m-derivatives}), (\ref{n+1-derivatives})  that the integrand $I$ in (\ref{second-derivative})
satisfies
\beqa
|I| \1&\leq&\1 \fr{1}{\sqrt{m!n!}}\sum_{0\leq |\be_1|+|\be_2|+|\be_3|\leq 2}\fr{c_{1,t}}{(\si_{\s,t})^{|\be_1|}}\fr{  D(q,p)}{\si^{\de_{\g_0}} }
\fr{\chi_3(\ti r)|\ti r|^{\alf}   }{|\ti r|^{2+|\be_2|+|\be_3|}}   g^{n}_{1,\si}(r)g^m_{2,\si}(k)\non\\
\1&\leq&\1 \fr{1}{\sqrt{m!n!}}   \fr{c_{1,t}}{(\si_{\s,t})^{2}}\fr{  D(q,p)}{\si^{\de_{\g_0}} }
\fr{\chi_3(\ti r)|\ti r|^{\alf}   } {|\ti r|^{2}}   g^{n}_{1,\si}(r)g^m_{2,\si}(k), 
\eeqa
where in the second step we made use of the fact that $\si_{\s,t}\leq |\ti r|$ in the region of integration.
Thus we get from (\ref{second-derivative}) that
\beqa
|F^{G_1,G_2}_{2,2, n,m}(q; r\ba p; k)|\leq \fr{1}{\sqrt{m!n!}}\fr{ c_{1,t} D(q,p)}{\si^{\de_{\g_0}} } \fr{1 }{t^2 (\si_{\s,t})^2}   g^{n}_{1,\si}(r)g^m_{2,\si}(k),
\eeqa
which gives the second contribution to (\ref{FGH-bound}). The factor $(n+1)$, appearing in (\ref{factor-n+1}), can be estimated by $2^n$
and incorporated into the constant appearing in the definition of $g^{n}_{1,\si}$. 

To estimate  $F^{G_2,G_1}_{1,n,m}$, one repeats analogous steps. The main difference is that one needs to exchange the indices $1\leftrightarrow 2$
between formulas~(\ref{m-derivatives}), (\ref{n+1-derivatives}). Consequently, the crucial factor $|\ti r|^{\alf}$ does not appear in any of these
inequalities. However, it appears in the first bound in (\ref{derivative-bounds}) as now we have the regular form-factor $v_1$. \qed
%%%%%%%%%%%%%%%%%%%%%%%%%%%%%%%%%%%%%%%%%%%%%%%%%%%%%%%%%%%

%%%%%%%%%%%%%%%%%%%%%%%%%%%%%%%%%%%%%%%%%%%%%%%%%
\bel\label{second-contribution-decay-new-old} Let $\check I^{\ell,\ellp,(1)}_{m,n,\ti m,\ti n}$ be defined as follows 
\beqa
\check I^{\ellp,\ell,(1)}_{m,n,\ti m,\ti n}\1&:=&\1\int  d^3 q d^3p \int  d^{3n}r d^{3m}k\,  
G_{1,n}(q;  r) G_{2,m}^{\ell}(p;  k)\non\\
& &\pha{4444444444444444444}\times\ov G_{1,\ti n}'( q+\un{\check{k}}-\un{\check{r}}; \hat r, \check k)
\ov  G_{2,\ti m}^{\prime\ellp}(p-\un{\check{k}}+\un{\check{r}}; \hat k, \check r), \label{definition-I}
\eeqa
where $G_{1,n}, G_{2,m}^{\ell}, G_{1,\ti n}', G_{2,\ti m}^{\prime\ellp}$  are defined in (\ref{G-def-new-ell})--(\ref{G-prime-def-one-ell}).  
The notation
$k=(\hat k, \check k)$, $r=(\hat r, \check r)$ is explained in Lemma~\ref{combinatorics}, and we consider a permutation
in (\ref{B-expectation-value-new-ell}) for which  $\check k$
or $\check r$ are non-empty. Then there holds for $0<\si\leq \sip\leq \kal$ and uniformly in $l,j$  and $\ellp,\ell$
\beqa
|\check I^{\ellp,\ell,(1)}_{m,n,\ti m,\ti n}(\si,\sip,s,t)_{l,j}|
\1&\leq&\1  
\fr{c}{\kal}\bigg\{ 
  \fr{1}{  t\si^{ 1/(8\tiga) } } +  \fr{1}{\uneps(s) t s^{-\peps}}   +(\sip)^{\alf/(8\ga_0)}  
 \bigg\}\non\\
& &\times \fr{( \|g^{n-1}_{{2},\si}\|^2_2 +\|g^n_{{2},\si}\|^2_2) (\|g^{m-1}_{2,\si}\|^2_2+\|g^m_{2,\si}\|_2^2 )}{\sqrt{(m-\ell)!n!(\ti m-\ellp)! \ti n!}}.
\label{I-two-final-formula-new}
\eeqa
We note that for $\check k$ (resp. $\check r$) non-empty we have $m\neq 0$ (resp. $n\neq 0$)
 and there always holds $m+n=\ti m+\ti n$. We set by convention $g^{-1}_{2,\si}=0$. 
\eel
%%%%%%%%%%%%%%%%%%%%%%%%%%%%%%%%%%%%%%%%%%%%%%%%%
\proof We will write $h_{2,j}(p):= h_{2,j}^{(s)}(p)$, 
$h'_{2,l}(p):= h_{2,l}^{\prime (s)}(p)$.  Making use of definitions  (\ref{G-def-new-ell})--(\ref{G-prime-def-one-ell}) we write
%%%%%%%%%%%%%%%%%%%%%%%%%%%%%%%%%%%%%
\begin{align}
&G_{1,n}(q;r):=e^{-iE_{1,q,\si}t}h_1(q)f^{n}_{1,q,\si}(r), \label{G-def-new-ell-new} \\
%%%%%%%%%%%%%%%%%%%%%%%%%%%%%%%%%%%%%%%%%%%%%
&G^{\ell}_{2,m}(p;k):=e^{-i(E_{2,p-\unk_{\ell},\si}+|\unk|_{\ell})  t}e^{i\ga_{\si}(\hv_j,t)(p-\unk_{\ell})  }
h_{2,j}(p-\unk_{\ell})(h^{\ell}f^{m-\ell}_{2,p-\unk_{\ell},\si} )(k),  \\
%%%%%%%%%%%%%%%%%%%%%%%%%%%%%%%%%%%%%%%%%%%%%%%%%
&G'_{1,\ti n}(q+\un{\check{k}}-\un{\check{r}}; \hat r, \check k):=e^{-iE_{1, q+\un{\check{k}}-\un{\check{r}} ,\si'}t}
h'_1(q+\un{\check{k}}-\un{\check{r}})f^{\ti n}_{1, q+\un{\check{k}}-\un{\check{r}} ,\si'}( \hat r, \check k), \\
%%%%%%%%%%%%%%%%%%%%%%%%%%%%%%%%%%%%%%%%%%%%%%%%%%
& G^{\prime\ell'}_{2,\ti m}(p-\un{\check{k}}+\un{\check{r}}; \hat k, \check r ):=
e^{-i(E_{2,p-\un{\check{k}}+\un{\check{r}}-(\hat{\un k}, \check{\un r})_{\ellp},\si'}+
| (\hat{\un k}, \check{\un r}) |_{\ellp})  t}e^{i\ga_{\si'}(\nv_l,t)(p-\un{\check{k}}+\un{\check{r}}- (\hat{\un k}, \check{\un r})_{\ellp})  }\non\\
&\ph{444444444444444444444444} \times h_{2,l}^{\prime}(p-\un{\check{k}}+\un{\check{r}}-(\hat{\un k}, \check{\un r})_{\ellp}) (h^{\ellp}f^{\ti m-\ellp}_{2,p-\un{\check{k}}+\un{\check{r}}-(\hat{\un{k}}, \check{\un{r}})_{\ellp},\si'} )(\hat k, \check r ).   
\label{G-prime-def-one-ell-new} 
\end{align}
%%%%%%%%%%%%%%%%%%%%%%%%%%%%%%%%%%%%
 By inserting these definitions, we obtain 
\beqa
\check I^{\ellp,\ell,(1)}_{m,n,\ti m,\ti n}
\1&=&\1 \int  d^3 q d^3p \int  d^{3n}r d^{3m}k\, G_{1,n}(q;  r) G^{\ell}_{2,m}(p;  k)\non\\
& &\pha{4444444444444444444}\times\ov G'_{1,\ti n}( q+\un{\check{k}}-\un{\check{r}}; \hat r, \check k)
\ov{G}^{\prime, \ellp }_{2,\ti m}(p-\un{\check{k}}+\un{\check{r}}; \hat k, \check r)\non\\
%%%%%%%%%%%%%%%%%%%%%%%%%%%%%%%%%%%%%%%%%%%%%%%%%%%%%%%%%%%
\1&=&\1 \int  d^3 q d^3p \int  d^{3n}r d^{3m}k\, e^{i( E_{1, q+\un{\check{k}}-\un{\check{r}} ,\si'}+ E_{2,p-\un{\check{k}}+\un{\check{r}}-(\hat{\un k}, \check{\un r})_{\ellp},\si'}+| (\hat{\un k}, \check{\un r}) |_{\ellp})  t}
e^{-i( E_{1,q,\si}+ E_{2,p-\unk_{\ell},\si}+|\unk|_{\ell})  t} \times\non\\
& &\times e^{i\ga_{\si}(\hv_j,t)(p-\unk_{\ell})  } e^{-i\ga_{\si'}(\nv_l,t)(p-\un{\check{k}}+\un{\check{r}}- (\hat{\un k}, \check{\un r})_{\ellp})}
 h_1(q)f^{n}_{1,q,\si}(r)  h_{2,j}(p-\unk_{\ell})(h^{\ell}f^{m-\ell}_{2,p-\unk_{\ell},\si} )(k)\times \non\\
& &\times \ov{h}'_1(q+\un{\check{k}}-\un{\check{r}})\ov{f}^{\ti n}_{1, q+\un{\check{k}}-\un{\check{r}} ,\si'}( \hat r, \check k) 
\ov{h}_{2,l}^{\prime}(p-\un{\check{k}}+\un{\check{r}}-(\hat{\un k}, \check{\un r})_{\ellp}) (\ov{h}^{\ellp}\ov{f}^{\ti m-\ellp}_{2,p-\un{\check{k}}+\un{\check{r}}-(\hat{\un{k}}, \check{\un{r}})_{\ellp},\si'} )(\hat k, \check r ). \quad \label{I-two-formula-new-new}
\eeqa
By making a change of variables $p\to p+\un{\check{k}}$ we arrive at our main formula
\beqa
\check I^{\ellp,\ell,(1)}_{m,n,\ti m,\ti n}\1&:=&\1 \int  d^3 q d^3p \int  d^{3n}r d^{3m}k\, 
e^{i( E_{1, q+\un{\check{k}}-\un{\check{r}} ,\si'}+ E_{2,p+\un{\check{r}}-(\hat{\un k}, \check{\un r})_{\ellp},\si'}+| (\hat{\un k}, \check{\un r}) |_{\ellp})  t}
e^{-i( E_{1,q,\si}+ E_{2,p+\un{\check{k}}-\unk_{\ell},\si}+|\unk|_{\ell})  t} \times\non\\
&\times&\!\!\!\!\!e^{i\ga_{\si}(\hv_j,t)(p+\un{\check{k}}-\unk_{\ell})  } e^{-i\ga_{\si'}(\nv_l,t)(p+\un{\check{r}}- (\hat{\un k}, \check{\un r})_{\ellp})}
 h_1(q)f^{n}_{1,q,\si}(r)  h_{2,j}(p+\un{\check{k}}-\unk_{\ell})(h^{\ell}f^{m-\ell}_{2,p+\un{\check{k}}-\unk_{\ell},\si} )(k)\times \non\\
&\times&\!\!\!\!\!\ov{h}'_1(q+\un{\check{k}}-\un{\check{r}})   \ov{f}^{\ti n}_{1, q+\un{\check{k}}-\un{\check{r}} ,\si'}( \hat r, \check k)
\ov{h}_{2,l}^{\prime}(p+\un{\check{r}}-(\hat{\un k}, \check{\un r})_{\ellp}) 
(\ov{h}^{\ellp}\ov{f}^{\ti m-\ellp}_{2,p+\un{\check{r}}-(\hat{\un{k}}, \check{\un{r}})_{\ellp},\si'} )(\hat k, \check r ). \quad\quad\quad 
\label{I-two-formula-two-new-main}
\eeqa
%%%%%%%%%%%%%%%%%%%%%%%%%%%%%%%%%%%%%%
{\mc We recall that the definition of the  expressions $(h^{\ell}f^{m-\ell}_{2,p+\un{\check{k}}-\unk_{\ell},\si} )(k)$, $ (\ov{h}^{\ellp}\ov{f}^{\ti m-\ellp}_{2,p+\un{\check{r}}-(\hat{\un{k}}, \check{\un{r}})_{\ellp},\si'} )(\hat k, \check r )$
 stated below (\ref{above-symmetrization}), involved symmetrization over all variables. As all terms in the resulting 
 sums over permutations are estimated analogously, we can redefine  $(h^{\ell}f^{m-\ell}_{2,p+\un{\check{k}}-\unk_{\ell},\si} )(k)$, $ (\ov{h}^{\ellp}\ov{f}^{\ti m-\ellp}_{2,p+\un{\check{r}}-(\hat{\un{k}}, \check{\un{r}})_{\ellp},\si'} )(\hat k, \check r )$ as
 some particular terms in these sums over permutations.   The expression (\ref{I-two-formula-two-new-main}) after this
 modification is still denoted $\check I^{\ellp,\ell,(1)}_{m,n,\ti m,\ti n}$.}

 %However,  different terms in the sum over all permutations from (\ref{symmetrization-sum}) give contributions to (\ref{I-two-formula-two-new-%main}) which have the same structure. (One can
%see this by renumbering the $k$-variables). Therefore, for the remaining part of the proof, we can redefine 
%$(h^{\ell}f^{m-\ell}_{2,p+\un{\check{k}}-\unk_{\ell},\si} )(k):=h(k_1)\ldots h(k_{\ell})
%f^{m-\ell}_{2, p+\un{\check{k}}-\unk_{\ell}  ,\si}(k_{\ell+1}, \ldots, k_{m})$.}\\
%$(h^{\ell}f^{m-\ell}_{2,p+\un{\check{k}}-\unk_{\ell},\si} )(k) $
%has the same form, as one can see by renumbering  can be estimated analogously
%put in the form (\ref{I-two-formula-two-new-main}), since  }
%%%%%%%%%%%%%%%%%%%%%%%%%%%%%%%%%%%%%%%
\nin\textbf{Case 1:} Either: In $(h^{\ell}f^{m-\ell}_{2,p+\un{\check{k}}-\unk_{\ell},\si} )(\hk, \chk)$ some components of $\chk$ are arguments of 
$f^{m-\ell}_{2,p+\un{\check{k}}-\unk_{\ell},\si}$. \\
\ph{4444444}Or: In  $(\ov{h}^{\ellp}\ov{f}^{\ti m-\ellp}_{2,p+\un{\check{r}}-(\hat{\un{k}}, \check{\un{r}})_{\ellp},\si'} )(\hat k, \check r )$
some components of $\chr$ are arguments of  $\ov{f}^{\ti m-\ellp}_{2,p+\un{\check{r}}-(\hat{\un{k}}, \check{\un{r}})_{\ellp},\si'}$. \vspace{0.4cm}

\nin In this case it suffices to consider the latter possibility, as the former one is analogous. Thus we suppose that $\chr_1$ (the first component of $\chr$) is an argument of  $\ov{f}^{\ti m-\ellp}_{2,p+\un{\check{r}}-(\hat{\un{k}}, \check{\un{r}})_{\ellp},\si'}$.   
Let us set $\sip_\s:=\kal (\sip/\kal)^{1/(8\tiga)}$, which clearly satisfies $\sip\leq\sip_\s\leq \kal$.
Let $\chi\in C^{\infty}(\real^3)$, $0\leq \chi\leq 1$, be supported in $\mcB_1$ (the unit ball) and be equal to one on $\mcB_{1-\epsilon}$ for
some  $0<\epsilon<1$.   We set $\chi_1(\chr_1  ):=\chi( \chr_1/\sip_\s)$, $\chi_2( \chr_1  ):= 1-\chi_1( \chr_1)$
and define $\check I^{\ellp,\ell,(1),(j')}_{m,n,\ti m,\ti n}$, for  $j'\in\{1,2\}$,  by multiplying the integrand in (\ref{I-two-formula-two-new-main}) with $\chi_j(\chr_1)$. Namely,
\beqa
\check I^{\ellp,\ell,(1),(j')}_{m,n,\ti m,\ti n}\1&:=&\1 \int  d^3 q d^3p \int  d^{3n}r d^{3m}k\, 
e^{i( E_{1, q+\un{\check{k}}-\un{\check{r}} ,\si'}+ E_{2,p+\un{\check{r}}-(\hat{\un k}, \check{\un r})_{\ellp},\si'}+| (\hat{\un k}, \check{\un r}) |_{\ellp})  t}
e^{-i( E_{1,q,\si}+ E_{2,p+\un{\check{k}}-\unk_{\ell},\si}+|\unk|_{\ell})  t} \times\non\\
&\times&\!\!\!\!\!e^{i\ga_{\si}(\hv_j,t)(p+\un{\check{k}}-\unk_{\ell})  } e^{-i\ga_{\si'}(\nv_l,t)(p+\un{\check{r}}- (\hat{\un k}, \check{\un r})_{\ellp})}
 h_1(q)(\chi_j(\chr_1)f^{n}_{1,q,\si}(r))  h_{2,j}(p+\un{\check{k}}-\unk_{\ell})(h^{\ell}f^{m-\ell}_{2,p+\un{\check{k}}-\unk_{\ell},\si} )(k)\times \non\\
&\times&\!\!\!\!\!\ov{h}'_1(q+\un{\check{k}}-\un{\check{r}})   \ov{f}^{\ti n}_{1, q+\un{\check{k}}-\un{\check{r}} ,\si'}( \hat r, \check k) 
\ov{h}_{2,l}^{\prime}(p+\un{\check{r}}-(\hat{\un k}, \check{\un r})_{\ellp}) 
(\ov{h}^{\ellp}\ov{f}^{\ti m-\ellp}_{2,p+\un{\check{r}}-(\hat{\un{k}}, \check{\un{r}})_{\ellp},\si'} )(\hat k, \check r ). \quad\quad\quad \label{I-two-formula-two-new}
\eeqa
Let us first consider (\ref{I-two-formula-two-new}) with $j'=1$. We conclude  from Theorem~\ref{main-theorem-spectral},  the definition
of the functions $g^n_{1/2,\si}$ in (\ref{def-g-ks-z})-(\ref{def-g-ks})  and estimate~(\ref{photon-h-bounds}) that
\beqa
& &|\chi_1(\chr_1)f^{n}_{1,q,\si}(r) (h^{\ell}f^{m-\ell}_{2,p+\un{\check{k}}-\unk_{\ell},\si} )(k) \ov{f}^{\ti n}_{1, q+\un{\check{k}}-\un{\check{r}} ,\si'}( \hat r, \check k) 
  (\ov{h}^{\ellp}\ov{f}^{\ti m-\ellp}_{2,p+\un{\check{r}}-(\hat{\un{k}}, \check{\un{r}})_{\ellp},\si'} )(\hat k, \check r )| \non\\
& &\pha{44444}\leq\fr{1}{\sqrt{(m-\ell)!n!(\ti m-\ellp)!\ti n!}}\chi_1(\chr_1)g^n_{1,\si}(r) g^{m}_{2,\si}(k)  g^{\ti n}_{1,\si}(\hat r, \check k)g^{\ti m}_{2,\si}(\hat k, \check r) \non\\
& &\pha{44444}\leq \fr{1}{\sqrt{(m-\ell)!n!(\ti m-\ellp)!\ti n!}}\fr{c\chi(\chr_1/\sip_\s)\chi_{[\si,\kas)}(\chr_1)^2 |\chr_1|^{\alf}    }{|\chr_1|^{3}} 
g^{n-1}_{1,\si}(r') g^{m}_{2,\si}(k)  g^{\ti n}_{1,\si}(\hat r, \chk)g^{\ti m-1}_{2,\si}(\hat k, \check r')\non\\
& &\pha{44444}\leq \fr{1}{\sqrt{(m-\ell)!n!(\ti m-\ellp)!\ti n!}}\fr{c\chi(\chr_1/\sip_\s)\chi_{[\si,\kas)}(\chr_1)^2  |\chr_1|^{\alf}    }{|\chr_1|^{3}}   
g^{n-1}_{2,\si}(r')^2 g^{m}_{2,\si}(k)^2, \label{pointwise-bounds-new}
\eeqa
where we decomposed $r=(\chr_1,r')$, $\chr=(\chr_1,\chr')$ and in the first step we used $\sip\geq \si$ and estimate~(\ref{photon-h-bounds}).  
Substituting~(\ref{pointwise-bounds-new}) to (\ref{I-two-formula-two-new}) and making use of the fact that (\ref{pointwise-bounds-new}) is independent
of $p$, $q$, (so we can apply the Cauchy-Schwarz inequality to the $p$, $q$ integration first) we get 
\beqa
|\check I^{\ellp,\ell,(1),(1)}_{m,n,\ti m,\ti n}|\leq (\sip_\s)^{\alf} \fr{c\|h_{2,j}\|_2 \|h'_{2,l}  \|_2  \|h_1\|_2\|h'_1\|_2   }{\sqrt{(m-\ell)!n!(\ti m-\ellp)!\ti n!}} \|g^{n-1}_{2,\si}\|^2_2 \|g^{m}_{2,\si}\|^2_2.
\label{below-slow-cut-off-part-new}
\eeqa

Let us now consider $\check I^{\ellp,\ell,(1), (2)}_{m,n,\ti m,\ti n}$ given by (\ref{I-two-formula-two-new})  for $j'=2$.  
For this purpose, we  set $w_1:=\un{\check{k}}-\un{\check{r}'}$, 
$w_2:=\un{\check{r}}'-(\hat{\un k}, \check{\un r})_{\ellp}$, $w=(w_1,w_2)$ and note that $\chr_1$ does not appear in $(\hat{\un k}, \check{\un r})_{\ellp}$,
since by assumption $\chr_1$ is not an argument of $h^{\ellp}$. We write
\beqa
& &\nF(\chr_1, p,q,w):=E_{1, q+\un{\check{k}}-\un{\check{r}} ,\si'}+ E_{2,p+\un{\check{r}}-(\hat{\un k}, \check{\un r})_{\ellp},\si'}=
E_{1, q-\chr_1+w_1, \sip}+E_{2, p+\chr_1+w_2,\sip}.
\eeqa
We note that, by disjointness of the velocity supports of $h_1$, $h_2$,  the condition 
$\ov{h}'_1(q-\chr_1+w_1 )\ov{h}'_{2,l}( p+\chr_1+w_2 )\neq 0$
implies that
\beqa
|\nabla_{\chr_1}  \nF( \chr_1, p, q, w)|\geq\eps'>0, \label{two-gradient-positivity}
\eeqa
for some $\eps'$ independent of $\chr_1$, $p$, $q$, $w$, $\si$, $\sip$ within the above restrictions. Thus we can write the following identity
\beqa
e^{i  \nF( \chr_1, p,q,w  ) t }=\fr{\nabla_{\chr_1} \nF(\chr_1, p,q,w   )\cdot \nabla_{\chr_1} e^{i  \nF(\chr_1, p,q,w )t} }{ i t|\nabla_{ \chr_1} \nF(\chr_1, p,q,w) |^2 }.
\eeqa
Now we define the function
\beqa
J(\chr_1, p,q,w ):=\fr{\nabla_{\chr_1}  \nF( \chr_1, p,q,w  ) }{ |\nabla_{\chr_1}  \nF( \chr_1, p,q,w  )|^2} 
 e^{i\ga_{\sip}(\nv_l,t)(p+\chr_1+w_2)  }  \ov{h}'_1(q-\chr_1+w_1)\ov{h}'_{2,l}(p+\chr_1+w_2). 
\eeqa
We note that, by Theorem~\ref{preliminaries-on-spectrum}, Lemma~\ref{h-theta} and Lemma~\ref{phase-lemma}
we have for $|\be|=0, 1$
\beqa
|\pa_{ \chr_1}^{\be} J(\chr_1, p,q,w)|\leq \theta^{-|\be|} C(\chr_1, p,q,w),\quad
|\pa_{  \chr_1 }^{\be}\chi_j( \chr_1  )|\leq \fr{c}{(\sip_\s)^{|\be|} },\label{derivative-bounds-one-two-new}
\eeqa
where 
\beqa
C(\chk_1, p,q,w):=c\sum_{\be_1,\be_2; 0\leq |\be_1|+|\be_2|\leq 1} |  \pa_{  \chr_1 }^{\be_1}h'_1( q-\chr_1+w_1 ) |\  
|(\pa_{  \chr_1 }^{\be_2} h_{2})^l( p+\chr_1+w_2 ) |.
\eeqa
Moreover, we obtain from  Theorem~\ref{main-theorem-spectral} and estimate~(\ref{photon-h-bounds}) that
\beqa
& &|\pa_{\chr_1}^{\be}(\chi_2(\check r_1)f^{n}_{1,q,\si}(r)) |\leq \fr{1}{\sqrt{n!}} \bigg(\fr{1}{\sip_\s}\bigg)^{|\be|} g^n_{1,\si}(r),\label{no-der-estimate-new}\\
& &| \pa_{\chr_1}^{\be} ( \ti{\chi}_2(\check r_1) (\ov{h}^{\ellp}\ov{f}^{\ti m-\ellp}_{2,p+\un{\check{r}}-(\hat{\un{k}}, \check{\un{r}})_{\ellp},\si'} )(\hat k, \check r ))|\leq \fr{1}{\sqrt{ (\ti m-\ellp)!}} \bigg(\fr{1}{(\mc \si')^{\de_{\la_0}}  }+\fr{1}{\sip_\s}\bigg)^{|\be|} g^{\ti m}_{2,\si}(\hat k, \check r ), \\
& &|  (h^{\ell}f^{m-\ell}_{2,p+\un{\check{k}}-\unk_{\ell},\si} )(k) |\leq 
\fr{1}{\sqrt{(m-\ell)!}}  g^{m}_{2,\si}( k ), \\
& &|\pa_{\chr_1}^{\be}\ov{f}^{\ti n}_{1, q+\un{\check{k}}-\un{\check{r}} ,\si'}( \hat r, \check k)|
\leq  \fr{1}{\sqrt{\ti n! }}\bigg(\fr{1}{({\mc \si'})^{\de_{\la_0}}  }\bigg)^{|\be|}  g^{\ti n}_{1,\sip}( \hat r, \check k  ),\label{der-estimate-two-new}
\eeqa
where the function $\ti\chi$ satisfies the same properties as $\chi$ and is s.t. $\chi\ti\chi=\chi$.
Now coming back to formula~(\ref{I-two-formula-two-new}) and integrating by parts we obtain
\beqa
\check I^{\ellp,\ell,(1),(2)}_{m,n,\ti m,\ti n}\1&:=&\1-\fr{1}{it} \int  d^3 q d^3p \int  d^{3n}r d^{3m}k\, e^{i( \nF( \chr_1, p,q,w  )t}
e^{i
 | (\hat{\un k}, \check{\un r}) |_{\ellp}  t}
e^{-i( E_{1,q,\si}+ E_{2,p+\un{\check{k}}-\unk_{\ell},\si}+|\unk|_{\ell})  t}
\times\non\\
\1&\times&\1 e^{i\ga_{\si}(\hv_j,t)(p+\un{\check{k}}-\unk_{\ell})  }   
h_1(q) h_{2,j}(p+\un{\check{k}}-\unk_{\ell})   (h^{\ell}f^{m-\ell}_{2,p+\un{\check{k}}-\unk_{\ell},\si} )(k) \times \non\\
\1&\times&\1\nabla_{\chr_1}\cdot\bigg( J(\chr_1, p,q,w )\chi_2(\check r_1)  \ov{f}^{\ti n}_{1, q+\un{\check{k}}-\un{\check{r}} ,\si'}( \hat r, \check k) f^{n}_{1,q,\si}(r)(\ov{h}^{\ellp}\ov{f}^{\ti m-\ellp}_{2,p+\un{\check{r}}-(\hat{\un{k}}, \check{\un{r}})_{\ellp},\si'} )(\hat k, \check r )
 \bigg).
\eeqa
Making use of the bounds (\ref{derivative-bounds-one-two-new}), (\ref{no-der-estimate-new})-(\ref{der-estimate-two-new})  we estimate
\beqa
|\check I^{\ellp,\ell,(1),(2)}_{m,n,\ti m,\ti n}|\1&\leq&\1 \fr{c}{t}\bigg(\fr{1}{(\sip)^{\de_{\la_0}}  }+\fr{1}{\sip_\s} +\theta^{-1}  \bigg)\non\\
& &\times \fr{\|h_1\|_2\|h_{2,j}\|_2 \sum_{\be_1,\be_2 ; 0\leq |\be_1|+|\be_2|\leq 1 } \|\pa^{\be_1}h_1'\|_2  
\|(\pa^{\be_2}h_2)^l \|_2 }{\sqrt{(m-\ell)!n!(\ti m-\ellp)!\ti n!}}  \| g^n_{ {2},\si}\|_2^2   \|g^m_{2,\si}\|_2^2.
 \label{above-slow-cut-off-part-new}
\eeqa
Exploiting (\ref{above-slow-cut-off-part-new}), (\ref{below-slow-cut-off-part-new}), the fact that $\sip_\s=\kal(\sip/\kal)^{1/(8\tiga)}$, and  $\de_{\g_0}\leq 1/(8\tiga)$ we obtain
\beqa
|\check I^{\ellp,\ell,(1)}_{m,n,\ti m,\ti n}(\si,\sip,s,t)|
\1&\leq&\1  
D_{l,j}(h,s)\fr{1}{\ka^{\la_0}}
\bigg( \fr{1}{t}\bigg(  \fr{1}{\si^{ 1/(8\tiga) } } +\theta^{-1} \bigg)+(\sip)^{\alf/(8\ga_0)}\bigg)  
 \non\\
& &\times \fr{( \|g^{n-1}_{{2},\si}\|^2_2 +\|g^n_{{2},\si}\|^2_2) (\|g^{m-1}_{2,\si}\|^2_2+\|g^m_{2,\si}\|_2^2 )}{\sqrt{(m-\ell)!n!(\ti m-\ellp)! \ti n!}},
\eeqa
where
\beqa
D_{l,j}(h,s)\1&:=&\1c\|h_1\|_2\|h_{2,j} \|_2 \sum_{\be_1,\be_2 ; 0\leq |\be_1|+|\be_2|\leq 1 } 
\|\pa^{\be_1}h_1'\|_2  \|(\pa^{\be_2}h_2)^l \|_2\non\\
\1&\leq&\1 c'\|h_1\|_2\|h_{2} \|_2 \sum_{\be_1,\be_2 ; 0\leq |\be_1|+|\be_2|\leq 1 } \|\pa^{\be_1}h_1'\|_2  \|(\pa^{\be_2}h_2) \|_2, \label{D-l-j}
\eeqa
and we estimated trivially
\beqa
 \|g^{n-1}_{{2},\si}\|^2_2 \|g^{m}_{2,\si}\|_2^2+\|g^n_{{2},\si}\|^2_2 \|g^m_{2,\si}\|_2^2 \leq  ( \|g^{n-1}_{{2},\si}\|^2_2 +\|g^n_{{2},\si}\|^2_2) 
(\|g^{m-1}_{2,\si}\|^2_2+\|g^m_{2,\si}\|_2^2 ) \label{m-n-symmetrization}
\eeqa
to obtain an expression in (\ref{I-two-final-formula-new}) which is symmetric under the substitution $m\leftrightarrow n$. 
 Since $\theta^{-1}\leq 1/(\uneps(s)s^{-\peps})$, this concludes the analysis of Case 1. \\\\
%%%%%%%%%%%%%%%%%%%%%%%%%%%%%%%%%%%%%%%%%%%%%%%
\nin\textbf{Case 2:} Either: In $(h^{\ell}f^{m-\ell}_{2,p+\un{\check{k}}-\unk_{\ell},\si} )(\hk, \chk)$ some components of $\chk$ are arguments of $h^{\ell}$.\\
\ph{4444444}Or: In  $(\ov{h}^{\ellp}\ov{f}^{\ti m-\ellp}_{2,p+\un{\check{r}}-(\hat{\un{k}}, \check{\un{r}})_{\ellp},\si'} )(\hat k, \check r )$
some components of $\chr$ are arguments of   $\ov{h}^{\ellp}$. \vspace{0.4cm}  
%%%%%%%%%%%%%%%%%%%%%%%

\nin We consider only the latter possibility, since the former one is analogous and simpler. We denote again by  $\chr_1$ the relevant variable.
We recall from (\ref{h-photon-def-app}) that 
\begin{align}
h(\chr_1):=h_{l,j}^{\sip,\si}(\chr_1)&:=\fr{\vv_2^{\si}(\chr_1) f^{\sip,\si}_{\hv_j,\nv_l}(\chr_1)}{|\chr_1|}, \\ 
f^{\sip,\si}_{\hv_j,\nv_l}(\chr_1)&:= 
\fr{ e_{ {\chr}_1}\cdot(\hv_j-\nv_l)}{(1-e_{{\chr}_1}\cdot \hv_j) (1-e_{\chr_1}\cdot \nv_l)}+\fr{\chi_{[\si,\sip]}(\chr_1) }{1-e_{ {\chr}_1}\cdot \nv_l}=:f_0(\chr_1)
+f_{[\si,\sip]}(\chr_1).
\end{align}
 Accordingly, we write the decomposition
\beqa
h(\chr_1)=h_0(\chr_1)+h_{[\si,\sip]}(\chr_1):=\fr{\la \chi_{[\si,\ka)}(\chr_1) f_0(\chr_1)}{\sqrt{2}|\chr_1|^{3/2}}
+\fr{\la\chi_{[\si,\ka)} (\chr_1) f_{[\si,\sip]}(\chr_1)}{\sqrt{2}|\chr_1|^{3/2}}.
\eeqa
For future reference, we note the bounds
\beqa
|h_0(\chr_1)|\leq c\la \fr{\chi_{[\si,\ka)}( \chr_1)}{| \chr_1|^{3/2}}, \quad |\pa_{\chr_1}h_0(\chr_1)|\leq \fr{c}{|\chr_1|}\la \fr{\chi_{[\si,\ka)}( \chr_1)}{| \chr_1 |^{3/2}}. \label{h-bound-def}
\eeqa
We note that $|\pa_{\chr_1}h_{[\si,\sip]}(\chr_1)|$ is problematic in the relevant region $\si\leq |\chr_1| \leq \ka$ due to the discontinuity at $|\chr_1|=\sip$.
Therefore, before we apply the non-stationary phase method, we have to treat the  $h_{[\si,\sip]}(\chr_1)$ contribution by a separate argument.
For this purpose, we recall the expression $(\ov{h}^{\ellp}\ov{f}^{\ti m-\ellp}_{2,p+\un{\check{r}}-(\hat{\un{k}}, \check{\un{r}})_{\ellp},\si'} )(\hat k, \check r )$ 
appearing in (\ref{I-two-formula-two-new-main}) and perform the corresponding decompositions
\beqa
(\ov{h}^{\ellp}\ov{f}^{\ti m-\ellp}_{2,p+\un{\check{r}}-(\hat{\un{k}}, \check{\un{r}})_{\ellp},\si'} )(\hat k, \check r )\1&:=&\1
(\ov{h}_0^{\ellp}\ov{f}^{\ti m-\ellp}_{2,p+\un{\check{r}}-(\hat{\un{k}}, \check{\un{r}})_{\ellp},\si'} )(\hat k, \check r )+
(\ov{h}_{[\si,\sip]}^{\ellp}\ov{f}^{\ti m-\ellp}_{2,p+\un{\check{r}}-(\hat{\un{k}}, \check{\un{r}})_{\ellp},\si'} )(\hat k, \check r ),\\
\check I^{\ellp,\ell,(1)}_{m,n,\ti m,\ti n}\1&:=&\1\check I^{\ellp,\ell,(1)}_{0,m,n,\ti m,\ti n}+\check I^{\ellp,\ell,(1)}_{[\si,\sip], m,n,\ti m,\ti n}, \label{I-decomp}
\eeqa
where only the factor involving variable $\chr_1$ was decomposed. We analyse first $\check I^{\ellp,\ell,(1)}_{[\si,\sip], m,n,\ti m,\ti n}$.
By (\ref{h-photon-def-app}) we have explicitly
\beqa
\check I^{\ellp,\ell,(1)}_{[\si,\sip],m,n,\ti m,\ti n}\1&:=&\1 \int  d^3 q d^3p \int  d^{3n}r d^{3m}k\, 
e^{i( E_{1, q+\un{\check{k}}-\un{\check{r}} ,\si'}+ E_{2,p+\un{\check{r}}-(\hat{\un k}, \check{\un r})_{\ellp},\si'}+| (\hat{\un k}, \check{\un r}) |_{\ellp})  t}
e^{-i( E_{1,q,\si}+ E_{2,p+\un{\check{k}}-\unk_{\ell},\si}+|\unk|_{\ell})  t} \times\non\\
&\times&\!\!\!\!\!e^{i\ga_{\si}(\hv_j,t)(p+\un{\check{k}}-\unk_{\ell})  } e^{-i\ga_{\si'}(\nv_l,t)(p+\un{\check{r}}- (\hat{\un k}, \check{\un r})_{\ellp})}
 h_1(q)f^{n}_{1,q,\si}(r)  h_{2,j}(p+\un{\check{k}}-\unk_{\ell})(h^{\ell}f^{m-\ell}_{2,p+\un{\check{k}}-\unk_{\ell},\si} )(k)\times \non\\
&\times&\!\!\!\!\!\ov{h}'_1(q+\un{\check{k}}-\un{\check{r}})   \ov{f}^{\ti n}_{1, q+\un{\check{k}}-\un{\check{r}} ,\si'}( \hat r, \check k)
\ov{h}_{2,l}^{\prime}(p+\un{\check{r}}-(\hat{\un k}, \check{\un r})_{\ellp}) 
(\ov{h}^{\ellp}_{[\si,\sip]}\ov{f}^{\ti m-\ellp}_{2,p+\un{\check{r}}-(\hat{\un{k}}, \check{\un{r}})_{\ellp},\si'} )(\hat k, \check r ). \quad\quad\quad 
\label{I-two-formula-two-new-main-sigmas}
\eeqa
Similarly as in  (\ref{pointwise-bounds-new}), we can write
\beqa
& &|f^{n}_{1,q,\si}(r) (h^{\ell}f^{m-\ell}_{2,p+\un{\check{k}}-\unk_{\ell},\si} )(k) \ov{f}^{\ti n}_{1, q+\un{\check{k}}-\un{\check{r}} ,\si'}( \hat r, \check k) 
  (\ov{h}^{\ellp}_{[\si,\sip]}\ov{f}^{\ti m-\ellp}_{2,p+\un{\check{r}}-(\hat{\un{k}}, \check{\un{r}})_{\ellp},\si'} )(\hat k , \check r )| \non\\
& &\pha{44444}\leq\fr{1}{\sqrt{(m-\ell)!n!(\ti m-\ellp)!\ti n!}}  \fr{c\chi_{[\si,\sip]} (\chr_1) }{ |\chr_1|^{3/2}}     g^n_{1,\si}(r) g^{m}_{2,\si}(k)  g^{\ti n}_{1,\si}(\hat r, \check k)g^{\ti m-1}_{2,\si}(\hat k, \check r') \non\\
& &\pha{44444}\leq \fr{1}{\sqrt{(m-\ell)!n!(\ti m-\ellp)!\ti n!}}\fr{c\chi_{[\si,\sip]}(\chr_1) |\chr_1|^{\alf}    }{|\chr_1|^{3}} 
g^{n-1}_{1,\si}(r') g^{m}_{2,\si}(k)  g^{\ti n}_{1,\si}(\hat r, \chk)g^{\ti m-1}_{2,\si}(\hat k, \check r')\non\\
& &\pha{44444}\leq \fr{1}{\sqrt{(m-\ell)!n!(\ti m-\ellp)!\ti n!}}\fr{c\chi_{[\si,\sip)}(\chr_1)  |\chr_1|^{\alf}    }{|\chr_1|^{3}}   
g^{n-1}_{{2},\si}(r')^2 g^{m}_{2,\si}(k)^2, \label{pointwise-bounds-new-sigmas}
\eeqa
where we decomposed $r=(\chr_1,r')$, $\chr=(\chr_1,\chr')$ and in the first step we used $\sip\geq \si$.  

Substituting~(\ref{pointwise-bounds-new-sigmas}) to (\ref{I-two-formula-two-new-main-sigmas}) and making use of the fact that (\ref{pointwise-bounds-new-sigmas}) is independent
of $p$, $q$, (so we can apply the Cauchy-Schwarz inequality  to the $p$, $q$ integration first) we get 
\beqa
|\check I^{\ellp,\ell,(1)}_{[\si,\sip], m,n,\ti m,\ti n}|\leq (\sip)^{\alf} \fr{c\|h_{2,j}\|_2 \|h'_{2,l}  \|_2  \|h_1\|_2\|h'_1\|_2   }{\sqrt{(m-\ell)!n!(\ti m-\ellp)!\ti n!}} \|g^{n-1}_{{2},\si}\|^2_2 \|g^{m}_{2,\si}\|^2_2.
\label{below-slow-cut-off-part-new-sigma-sigma-prime}
\eeqa
Now we proceed to the analysis of $\check I^{\ellp,\ell,(1)}_{0,m,n,\ti m,\ti n}$, appearing in (\ref{I-decomp}), which will be similar to the arguments
in Case 1. That is, we will use an auxiliary  infrared cut-off $\sip_\s:=\kal (\sip/\kal)^{1/(8\tiga)}$, which clearly satisfies $\sip\leq\sip_\s\leq \kal$.
Let $\chi\in C^{\infty}(\real^3)$, $0\leq \chi\leq 1$, be supported in $\mcB_1$  and be equal to one on $\mcB_{1-\epsilon}$ for
some  $0<\epsilon<1$.   We set $\chi_1(\chr_1  ):=\chi( \chr_1/\sip_\s)$, $\chi_2( \chr_1  ):= 1-\chi_1( \chr_1)$. Now we define $\check I^{\ellp,\ell,(1),(j')}_{0,m,n,\ti m,\ti n}$, for  $j'\in\{1,2\}$,  by 
\beqa
\check I^{\ellp,\ell,(1),(j)}_{0,m,n,\ti m,\ti n}\1&:=&\1 \int  d^3 q d^3p \int  d^{3n}r d^{3m}k\, 
e^{i( E_{1, q+\un{\check{k}}-\un{\check{r}} ,\si'}+ E_{2,p+\un{\check{r}}-(\hat{\un k}, \check{\un r})_{\ellp},\si'}+| (\hat{\un k}, \check{\un r}) |_{\ellp})  t}
e^{-i( E_{1,q,\si}+ E_{2,p+\un{\check{k}}-\unk_{\ell},\si}+|\unk|_{\ell})  t} \times\non\\
&\times&\!\!\!\!\!e^{i\ga_{\si}(\hv_j,t)(p+\un{\check{k}}-\unk_{\ell})  } e^{-i\ga_{\si'}(\nv_l,t)(p+\un{\check{r}}- (\hat{\un k}, \check{\un r})_{\ellp})}
 h_1(q)f^{n}_{1,q,\si}(r)  h_{2,j}(p+\un{\check{k}}-\unk_{\ell})(h^{\ell}f^{m-\ell}_{2,p+\un{\check{k}}-\unk_{\ell},\si} )(k)\times \non\\
&\times&\!\!\!\!\!\ov{h}'_1(q+\un{\check{k}}-\un{\check{r}})   \ov{f}^{\ti n}_{1, q+\un{\check{k}}-\un{\check{r}} ,\si'}( \hat r, \check k)
\ov{h}_{2,l}^{\prime}(p+\un{\check{r}}-(\hat{\un k}, \check{\un r})_{\ellp}) 
\chi_j(\chr_1  )(\ov{h}^{\ellp}_{0}\ov{f}^{\ti m-\ellp}_{2,p+\un{\check{r}}-(\hat{\un{k}}, \check{\un{r}})_{\ellp},\si'} )(\hat k, \check r ). \quad\quad\quad 
\label{I-two-formula-two-new-main-sigmas-two}
\eeqa
By repeating the arguments in (\ref{pointwise-bounds-new})-(\ref{below-slow-cut-off-part-new}), we obtain
\beqa
|\check I^{\ellp,\ell,(1),(1)}_{0, m,n,\ti m,\ti n}|\leq (\sip_{\s})^{\alf} \fr{c\|h_{2,j}\|_2 \|h'_{2,l}  \|_2  \|h_1\|_2\|h'_1\|_2   }{\sqrt{(m-\ell)!n!(\ti m-\ellp)!\ti n!}} \|g^{n-1}_{{2},\si}\|^2_2 \|g^{m}_{2,\si}\|^2_2.
\label{below-slow-cut-off-part-new-sigma-new}
\eeqa
\newcommand{\tiw}{w_1}
\newcommand{\tiV}{V_1}
\newcommand{\tiJ}{\ti J}
Now we proceed to the analysis of $\check I^{\ellp,\ell,(1),(2)}_{0,m,n,\ti m,\ti n}$. For this purpose, we recall the notation $w_1:= \un{\check{k}}-\un{\check{r}}' $, $w_2:=\un{\check{r}}'-(\hat{\un k}, \check{\un r})_{\ellp}$, $w:=(w_1,w_2)$ and write 
\beqa
\tiV(\chr_1,p,q, \tiw):=E_{1, q+\un{\check{k}}-\check{r}_1-\un{\check{r}}' ,\si'}+| (\hat{\un k}, \check{\un r}) |_{\ellp}=E_{1,q-\chr_1+\tiw, \sip}+| (\hat{\un k}, \check{\un r}) |_{\ellp}.
\eeqa
We note that with our assumptions $| (\hat{\un k}, \check{\un r}) |_{\ellp}$ contains a summand $|\chr_1|$. We also 
note that $E_{2,p+\un{\check{r}}-(\hat{\un k}, \check{\un r})_{\ellp},\si'}$, appearing in (\ref{I-two-formula-two-new-main-sigmas-two}),
is independent of $\chr_1$, because the contributions from $ \un{\check{r}}$ and $ (\hat{\un k}, \check{\un r})_{\ellp}$ cancel.
We will now apply a non-stationary phase argument exploiting that the velocity of the photon is always larger than the velocity
of the massive particle (in the relevant range of parameters). Namely,
\beqa
|\nabla_{\chr_1}  V_1(\chr_1,p,q,\tiw)|=\bigg|-(\nabla E)_{1, q-\chr_1+\tiw, \si'}+\fr{\chr_1}{|\chr_1|}\bigg|\geq \eps_0>0,  
\eeqa
uniformly in the arguments of $\tiV$ and in $\si$, $\sip$. (Recall from Proposition~3.2 of \cite{DP12} that $|\nabla E_{1,p,\si}|\leq 1/3+c|\la|$ for $p\in S$).
We define the function
\beqa
J_1(\chr_1, p,q,\tiw ):=\fr{\nabla_{\chk_1} \nF_1( \chr_1, p,q, \tiw  ) }{ |\nabla_{\chr_1}  \nF_1( \chr_1, p,q, \tiw  )|^2} 
 e^{i\ga_{\sip}(\nv_l,t)(p+\chr_1+w_2)  }  \ov{h}'_1(q-\chr_1+w_1)\ov{h}'_{2,l}(p+\chr_1+w_2). 
\eeqa
We note that, by Theorem~\ref{preliminaries-on-spectrum}, Lemma~\ref{h-theta} and Lemma~\ref{phase-lemma}
we have for $|\be|=0, 1$
\beqa
|\pa_{ \chr_1}^{\be} J_1(\chr_1, p,q,w_1)|\leq \theta^{-|\be|} C(\chr_1, p,q,w),\quad
|\pa_{  \chr_1 }^{\be}\chi_j( \chr_1  )|\leq \fr{c}{(\sip_\s)^{|\be|} },\label{derivative-bounds-one-two-new-photon}
\eeqa
where 
\beqa
C(\chk_1, p,q,w):=c\sum_{\be_1,\be_2; 0\leq |\be_1|+|\be_2|\leq 1} |  \pa_{  \chr_1 }^{\be_1}h'_1( q-\chr_1+w_1 ) |\  
|(\pa_{  \chr_1 }^{\be_2} h_{2})^l( p+\chr_1+w_2 ) |.
\eeqa
Moreover, we obtain from  Theorem~\ref{main-theorem-spectral} \ref{derivatives-bounds} and (\ref{h-bound-def})
\beqa
& &|\pa_{\chr_1}^{\be}(\chi_2(\check r_1)f^{n}_{1,q,\si}(r)) |\leq \fr{1}{\sqrt{n!}} \bigg(\fr{1}{\sip_\s}\bigg)^{|\be|} g^n_{1,\si}(r),\label{no-der-estimate-new-photon}\\
& &| \pa_{\chr_1}^{\be} ( \ti{\chi}_2(\check r_1) (\ov{h}_0^{\ellp}\ov{f}^{\ti m-\ellp}_{2,p+\un{\check{r}}-(\hat{\un{k}}, \check{\un{r}})_{\ellp},\si'} )(\hat k, \check r ))|\leq \fr{1}{\sqrt{ (\ti m-\ellp)!}} \bigg( \fr{1}{\sip_\s}\bigg)^{|\be|} g^{\ti m}_{2,\si}(\hat k, \check r ), \label{h_0-estimate}\\
& &|  (h^{\ell}f^{m-\ell}_{2,p+\un{\check{k}}-\unk_{\ell},\si} )(k) |\leq 
\fr{1}{\sqrt{(m-\ell)!}}  g^{m}_{2,\si}( k ), \\
& &|\pa_{\chr_1}^{\be}\ov{f}^{\ti n}_{1, q+\un{\check{k}}-\un{\check{r}} ,\si'}( \hat r, \check k)|
\leq  \fr{1}{\sqrt{\ti n! }}\bigg(\fr{1}{({\mc \si'})^{\de_{\la_0}}  }\bigg)^{|\be|}  g^{\ti n}_{1,\sip}( \hat r, \check k  ),\label{der-estimate-two-new-photon}
\eeqa
where the function $\ti\chi$ satisfies the same properties as $\chi$ and is s.t. $\chi\ti\chi=\chi$. These bounds are similar as in  (\ref{no-der-estimate-new})-(\ref{der-estimate-two-new}). The main difference is the appearance of $h_0(\chr_1)$  in (\ref{h_0-estimate}) which is differentiated and estimated with the help of (\ref{h-bound-def}) and the fact that $p+\un{\check{r}}-(\hat{\un{k}}, \check{\un{r}})_{\ellp}$ is independent of $\chr_1$ in the present case.

Now coming back to formula~(\ref{I-two-formula-two-new-main-sigmas-two}) and integrating by parts we obtain
\beqa
\check I^{\ellp,\ell,(1),(2)}_{0,m,n,\ti m,\ti n}\1&:=&\1-\fr{1}{it} \int  d^3 q d^3p \int  d^{3n}r d^{3m}k\, e^{i \nF_1( \chr_1, p,q,w_1)t} e^{i E_{2,p+\un{\check{r}}-(\hat{\un k}, \check{\un r})_{\ellp},\si'} t}
e^{-i( E_{1,q,\si}+ E_{2,p+\un{\check{k}}-\unk_{\ell},\si}+|\unk|_{\ell})  t}
\times\non\\
& &\times e^{i\ga_{\si}(\hv_j,t)(p+\un{\check{k}}-\unk_{\ell})  }   
h_1(q) h_{2,j}(p+\un{\check{k}}-\unk_{\ell})   (h^{\ell}f^{m-\ell}_{2,p+\un{\check{k}}-\unk_{\ell},\si} )(k) \times \non\\
 & &\times \nabla_{\chr_1}\cdot\bigg( J(\chr_1, p,q,w )\chi_2(\check r_1)  \ov{f}^{\ti n}_{1, q+\un{\check{k}}-\un{\check{r}} ,\si'}( \hat r, \check k) f^{n}_{1,q,\si}(r)(\ov{h}^{\ellp}\ov{f}^{\ti m-\ellp}_{2,p+\un{\check{r}}-(\hat{\un{k}}, \check{\un{r}})_{\ellp},\si'} )(\hat k, \check r )
 \bigg).
\eeqa
Now we conclude the argument as in Case 1: Making use of the bounds (\ref{derivative-bounds-one-two-new-photon}), (\ref{no-der-estimate-new-photon})-(\ref{der-estimate-two-new-photon}),  we estimate
\beqa
|\check I^{\ellp,\ell,(1),(2)}_{m,n,\ti m,\ti n}|\1&\leq&\1 \fr{c}{t}\bigg(\fr{1}{(\sip)^{\de_{\la_0}}  }+\fr{1}{\sip_\s} +\theta^{-1}  \bigg)\non\\
& &\times \fr{\|h_1\|_2\|h_{2,j}\|_2 \sum_{\be_1,\be_2 ; 0\leq |\be_1|+|\be_2|\leq 1 } \|\pa^{\be_1}h_1'\|_2  
\|(\pa^{\be_2}h_2)^l \|_2 }{\sqrt{(m-\ell)!n!(\ti m-\ellp)!\ti n!}}  \| g^n_{{\bc 2},\si}\|_2^2   \|g^m_{2,\si}\|_2^2.
 \label{above-slow-cut-off-part-new-photon}
\eeqa
Exploiting (\ref{above-slow-cut-off-part-new-photon}), (\ref{below-slow-cut-off-part-new-sigma-new}), (\ref{below-slow-cut-off-part-new-sigma-sigma-prime}) and  the fact that $\sip_\s=\kal(\sip/\kal)^{1/(8\tiga)}$, and  $\de_{\g_0}\leq 1/(8\tiga)$ we obtain
\beqa
|\check I^{\ellp,\ell,(1)}_{m,n,\ti m,\ti n}(\si,\sip,s,t)|
\1&\leq&\1  
D_{l,j}(h,s)\fr{1}{\ka^{\la_0}}
\bigg( \fr{1}{t}\bigg(  \fr{1}{\si^{ 1/(8\tiga) } } +\theta^{-1} \bigg)+(\sip)^{\alf/(8\ga_0)}\bigg)  
 \non\\
& &\times \fr{( \|g^{n-1}_{{2},\si}\|^2_2 +\|g^n_{{2},\si}\|^2_2) (\|g^{m-1}_{2,\si}\|^2_2+\|g^m_{2,\si}\|_2^2 )}{\sqrt{(m-\ell)!n!(\ti m-\ellp)! \ti n!}},
\eeqa
where  $D_{l,j}(h,s)$ is given by  (\ref{D-l-j}) and we used (\ref{m-n-symmetrization}). Since $\theta^{-1}\leq 1/(\uneps(s)s^{-\peps})$, this concludes the proof. \qed
%%%%%%%%%%%%%%%%%%%%%%%%%%%%%%%%%%%%%%%%%%
\section{Analysis of the phase I} \label{analysis-of-phase-I}
%%%%%%%%%%%%%%%%%%%%%%%%%%%%%%%%%%%%%%%%%%
\setcounter{equation}{0}

In this section we will often write $h_{2,t}^{\ga}:=h_{2,j,\sn}^{(\tn)} e^{i\ga_{\si}(\hv_j,\sn)}$. We note that $h_{2,t}^{\ga}$ depends on $j$ and $s$ 
although it is hidden in the notation. 
%%%%%%%%%%%%%%%%%%%%%%%%%%%%%%%%%%%%%%%%%%%%%%%%%%%%%%%%%%%%%%
\bep\label{two-Regular-phases-proposition} We set  
\begin{align}
A_{2, \si}&:=\int d^3p d^3k\,\vv_2^{\si}(k) \nn_2^*(p+k) F_{\si,\sn}(k) \nn_2(p),\\
F_{\si,\sn}(k)&:=
\fr{\vv_2^{\si}(k)}{|k|} \bigg(  \fr{e^{-i|k|\sn} }{(1-\hatk\cdot \hv_j ) }+\fr{e^{i|k|\sn} }{(1+\hatk\cdot \hv_j ) } \bigg):=f_{\si,\sn}(k)+\ov{f}_{\si,\sn}(-k).
\end{align}
Then, for any $0<\de<(\al^{-1}-1)$, $1\leq s\leq t$,
\begin{align}
\,[A_{2,\si}, \reta^*_{2,\si}(h_{2,j,\sn}^{(\tn)} e^{i\ga_{\si}(\hv_j,\sn)})]\Om&=\reta^*_{2,\si}(-i \dot\ga_{\si}(\hv_j,\sn)  h_{2,j,\sn}^{(\tn)} e^{i\ga_{\si}(\hv_j,\sn)})\Om+R_j(\si, t, s), 
\label{two-Regular-result-single}\\
\reta^*_{1, \si}(h_{1,\sn})[A_{2,\si}, \reta^*_{2,\si}(h_{2,j,\sn}^{(\tn)} e^{i\ga_{\si}(\hv_j,\sn)})]\Om&=\reta^*_{1, \si}(h_{1,\sn})
\reta^*_{2,\si}(-i \dot\ga_{\si}(\hv_j,\sn)  h_{2,j,\sn}^{(\tn)} e^{i\ga_{\si}(\hv_j,\sn)})\Om+R_j(\si, t, s), 
\label{two-Regular-result}
\end{align}
where {\mc the rest terms in (\ref{two-Regular-result-single}) and (\ref{two-Regular-result}) may be different, but they both satisfy}
\beqa
\|R_j(\si, t, s)\|\leq \fr{c_{t}^2}{\si^{\de_{\la_0}}}\bigg( t(\si^{\S}_t)^{3}+  \fr{(\si^{\S}_t)^{\de}}{t}+\fr{1}{(\si^{\S}_t)^{1+\de} t^2}  \bigg)
\|h_{1}\|_2, \quad c_{t}:=c(\uneps(t) t^{-\peps})^{-1}.\label{two-phases-estimate}
\eeqa
\eep
%%%%%%%%%%%%%%%%%%%%%%%%%%%%%%%%%%%%%%%%%%%%%%%%%%%%%%%%%%
\proof We consider only  (\ref{two-Regular-result}) as the proof of (\ref{two-Regular-result-single}) is analogous and simpler.  We recall that $h_{2,t}^{\ga}:=h_{2,j,t}^{(\tn)} e^{i\ga_{\si}(\nv_j,\sn)}$ and compute
\beqa
& & [A_{2,\si}, \reta^*_{2, \si}(h^{\ga}_{2,\sn})]\non\\
\1&=&\1[\int d^3p \int d^3k\,\nn_2^*(p+k) F_{\si,\sn}(k)\vv_2^{\si}(k)\nn_2(p), \reta^*_{2, \si}(h^{\ga}_{2,\sn})   ]\non\\
\1&=&\1\sum_{m=0}^{\infty}\fr{1}{\sqrt{m!}}\int d^{3m}k' d^3p \int d^3k  d^3q\, \nn_2^*(p+k)h^{\ga}_{2,\sn}(q)F_{\si,\sn}(k)\vv_2^{\si}(k)\de(p-q+\un{k'}) \bb^*(k')^m\,f^{m}_{2,q,\si}(k'^m)\non\\
\1&=&\1\sum_{m=0}^{\infty}\fr{1}{\sqrt{m!}}\int d^{3m}k' d^3p \int d^3k \, \nn_2^*(p+k)h^{\ga}_{2,\sn} (p+\un{k'})F_{\si,\sn}(k)\vv_2^{\si}(k) \bb^*(k')^m\, 
f^{m}_{2,p+\un{k}',\si }(k'^m)\non\\
\1&=&\1\sum_{m=0}^{\infty}\fr{1}{\sqrt{m!}}\int d^{3m}k' d^3p 
\bigg(\int d^3k \, h^{\ga}_{2,\sn}(p-k)F_{\si,\sn}(k)\vv_2^{\si}(k)f^{m}_{2,p-k,\si}(k'^m)\bigg)
\nn_2^*(p-\un{k}') \bb^*(k')^m. \label{two-full-formula-with-ca}
\eeqa
We obtain
\beqa
& &\int d^3k\, h^{\ga}_{2,\sn}(p-k)f^{m}_{2,p-k,\si}(k'^m)\big(f_{\si,\sn}(k)+\bar f_{\si,\sn}(-k)\big)\vv^{\si}_{2}(k)\non\\
\1&=&\1\int d^3k\, \big( e^{-iE_{2,p-k,\si} \sn}  h^{\ga}_{2}(p-k)f^{m}_{2,p-k,\si}(k'^m)-e^{-iE_{2,p,\si}\sn}e^{i\nabla E_{2,p,\si}\cdot k \sn } h^{\ga}_2(p)f^{m}_{2,p,\si}(k'^m) \big)\times\non\\ & &\ph{44444444444444444444444444444444444444}\times\big( f_{\si,\sn}(k)+\bar f_{\si,\sn}(-k)  \big)\vv_2^{\si}(k)\non\\
& &+\int  d^3k\,e^{-iE_{2,p,\si} \sn}e^{i\nabla E_{2,p,\si}\cdot k\sn}h^{\ga}_2(p)f^{m}_{2,p,\si}(k'^m)\bigg( \fr{e^{-i|k|\sn}}{ 
|k| (1-\hatk \cdot\hv_j)}+\fr{e^{i|k|\sn}}{ |k|(1+\hatk\cdot \hv_j)   }   \bigg) \vv^\si_2(k)^2.\label{two-leading-term/rest-term}
\eeqa
By a change of variables, the last term in (\ref{two-leading-term/rest-term}) can be rewritten as follows
\beqa
& &\int d^3k\,e^{-iE_{2,p,\si} \sn} h_2^\ga(p)f^{m}_{2,p,\si}(k'^m)  \bigg( \fr{e^{-i(|k|\sn-\nabla E_{2,p,\si}\cdot k\sn)}  }{|k|(1-\hatk\cdot \hv_j)}
+\fr{e^{i(|k|\sn-\nabla E_{2,p,\si}\cdot k\sn)}}{ |k|(1-\hatk\cdot \hv_j)   }   \bigg) \vv^\si_2(k)^2   \non\\
& &=h_2^\ga (p)e^{-iE_{2,p,\si}\sn}  f^{m}_{2,p,\si}(k'^m)2\int d^3k\, \vv^\si_2(k)^2 \fr{1}{|k|(1-\hatk\cdot \hv_j) }\cos( (\nabla E_{2,p,\si}\cdot  k
-|k|)\sn).    
\eeqa
We  define
\beqa
-\fr{1}{2}\dot{\ga}_{2,\si}(\hv_j,\sn)(p):=\int d^3k\, \vv^\si_2(k)^2 \fr{1}{|k|(1-\hatk\cdot \hv_j) }\cos( (\nabla E_{2,p,\si} \cdot k-|k|)\sn). \label{two-phase-derivatives}
\eeqa
We divide this expression into two parts:
\begin{align}
-\h\dot{\ga}^{{\color{black}(1)}}_{2,\si}(\hv_j,\sn)(p)&:=\int_{\si\leq |k|\leq \si_{\S,t}} d^3k\, \vv^\si_2(k)^2 \fr{1}{|k|(1-\hatk\cdot \hv_j) }\cos( (\nabla E_{2,p,\si}\cdot  k-|k|)\sn),
\label{gamma-first-part}\\
-\h\dot{\ga}^{{\color{black}(2)}}_{2,\si}(\hv_j,\sn)(p)&:=\int_{|k|\geq \si_{\S,t} } d^3k\, \vv^\si_2(k)^2 \fr{1}{|k|(1-\hatk\cdot \hv_j) }\cos( (\nabla E_{2,p,\si}\cdot  k-|k|)\sn),
\label{gamma-second-part}
\end{align}
where $\si_{\S,t}:=\max\{ \si^{\S}_t, \si\, \}$. Comparing with (\ref{slow-cutoff}), we see that 
$\dot{\ga}^{{\color{black}(1)}}_{2,\si}(\hv_j,\sn)(p)=\dot{\ga}_{\si}(\hv_j,\sn)(p)$. 

Let us denote by $[A_{2,\si}, \reta^*_{2, \si}(h^{\ga}_{2,\sn})]_2$ the contribution to 
$[A_{2,\si}, \reta^*_{2, \si}(h^{\ga}_{2,\sn})]$ coming from the last term on the r.h.s. of (\ref{two-leading-term/rest-term}).
By inserting the decomposition from (\ref{gamma-first-part}), (\ref{gamma-second-part}), we obtain
\begin{align}
\reta^*_{1, \si}(h_{1,\sn})[A_{2,\si}, \reta^*_{2,\si}(h^{\ga}_{2,\sn})]_2\Om&= \reta^*_{1, \si}(h_{1,\sn})\reta^*_{2, \si}( -\dot{\ga}_{\si}(\hv_j,\sn) h^{\ga}_{2,\sn})\Om\label{phase-cancellation-term}\\
&\ph{44} +\reta^*_{1, \si}(h_{1,\sn})\reta^*_{2, \si}( -\dot{\ga}^{{\color{black}(2)}}_{2,\si}(\hv_j,\sn) h^{\ga}_{2,\sn})\Om. \label{leading-term}
\end{align}
Clearly, (\ref{phase-cancellation-term}) gives the first term on the r.h.s. of (\ref{two-Regular-result}) while (\ref{leading-term}) contributes to the
rest term. In fact,
\beqa
\|\reta^*_{1, \si}(h_{1,\sn})\reta^*_{2, \si}( -\dot{\ga}^{{\color{black}(2)}}_{2,\si}(\hv_j,\sn) h^{\ga}_{2,\sn})\Om\|\leq
 \fr{  c }{\si^{\de_{\la_0}}} \bigg(  \fr{1 }{t^2\si^{\S}_{t} }\bigg) \| h_{1}\|_2  \|h_{2,j}^{(s)}\|_2,
\eeqa
where we made use of Proposition~\ref{two-phase-vector-lemma} (a).

Now let us denote  the contribution to $[A_{2,\si}, \reta^*_{2, \si}(h^{\ga}_{2,\sn})]$ coming from the first term on the r.h.s. of (\ref{two-leading-term/rest-term})
by $[A_{2,\si}, \reta^*_{2, \si}(h^{\ga}_{2,\sn})]_1$. By Proposition~\ref{two-phase-vector-lemma} (b), we have
\beqa
\| \reta^*_{1, \si}(h_{1,\sn})   [A_{2,\si}, \reta^*_{2, \si}(h^{\ga}_{2,\sn})]_1 \Om\|\leq 
\fr{c_{t}^2}{\si^{\de_{\la_0}}}\bigg(t(\si^{\S}_t)^{3}+  \fr{(\si^{\S}_t)^{\de}}{t}+\fr{1}{(\si^{\S}_t)^{1+\de} t^2}  \bigg)\|h_{1}\|_2,
\eeqa
where $c_{t}:=c(\uneps(t) t^{-\peps})^{-1}$. This completes the proof. \qed
%%%%%%%%%%%%%%%%%%%%%%%%%%%%%%%%%%%%%%%%%%%%%%
\bep\label{two-phase-vector-lemma}(a) Let us define as in (\ref{gamma-second-part})
\beqa
-\h\dot{\ga}^{{\color{black}(2)}}_{2,\si}(\hv_j,\sn)(p):=\int_{|k|\geq \si_{\S,t}} d^3k\, \vv^\si_2(k)^2 \fr{1}{|k|(1-\hatk \cdot \hv_j) }\cos( (\nabla E_{2,p,\si}\cdot k-|k|)\sn).
\eeqa
Then 
\beqa
 M_{j}^{\ga}(\si,s,t)^{1/2}:= 
\|\reta^*_{1, \si}(h_{1,\sn})\reta^*_{2, \si}( -\dot{\ga}^{{\color{black}(2)}}_{2,\si}(\hv_j,\sn) h^{\ga}_{2,\sn})\Om\|\leq
 \fr{  c }{\si^{\de_{\la_0}}} \bigg(  \fr{1 }{t^2\si^{\S}_{t} }\bigg) \| h_{1}\|_2  \|h_{2,j}^{(s)}\|_2.
\eeqa
(b) Let us define as in (\ref{two-leading-term/rest-term})
\beqa
& &F_{m}(p;k'):=\int d^3k\, 
\big( e^{-iE_{2,p-k,\si} \sn}  h^{\ga}_{2}(p-k)f^{m}_{2,p-k,\si}(k'^m)-e^{-iE_{2,p,\si}\sn}e^{i\nabla E_{2,p,\si}k \sn } h^{\ga}_2(p)f^{m}_{2,p,\si}(k'^m) \big)\times \non\\
& &\ph{444444444444444444444444444444444444444444444444444}\times F_{\si,\sn}(k)  \vv_2^{\si}(k).
\eeqa
and set 
\beqa
 [A_{2,\si}, \reta^*_{2, \si}(h^{\ga}_{2,\sn})]_1:=\sum_{m=0}^{\infty}\fr{1}{\sqrt{m!}}\int d^{3m}k' d^3p 
 F_{m}(p;k') 
\nn_2^*(p-\un{k}') \bb^*(k')^m.
\eeqa
Then, for any $0<\de<(\al^{-1}-1)$,  
\begin{align}
R_{j}^{\ga}(\si,s,t)^{1/2}&:=\| \reta^*_{1, \si}(h_{1,\sn})   [A_{2,\si}, \reta^*_{2, \si}(h^{\ga}_{2,\sn})]_1 \Om\|
\leq 
\fr{c_{t}^2}{\si^{\de_{\la_0}}}\bigg(t(\si^{\S}_t)^{3}+  \fr{(\si^{\S}_t)^{\de}}{t}+\fr{1}{(\si^{\S}_t)^{1+\de} t^2}  \bigg)\|h_{1}\|_2,
\end{align}
where $c_{t}:=c (\uneps(t) t^{-\peps})^{-1}$. 
\eep
%%%%%%%%%%%%%%%%%%%%%%%%%%%%%%%%%%%%%%%%%%%%
\proof To prove (a), let us set
\begin{align}
G_{1,m}(q;k)&:=e^{-iE_{1,q,\si} t}h_1(q)f^{m}_{1,q,\si}(k),\\  
G_{2,m}(q;k)&:=e^{-iE_{2,q,\si}t} h_{2}^{\ga}(q)(-\dot{\ga}_{2,\si}^{{\color{black}(2)}} )(\hv_j,t)(q)   f^{m}_{2,q,\si}(k),  \label{two-G-def-new-new}  \\
G'_{1,m}(q;k)&:=G_{1,m}(q;k), \\
G'_{2,m}(q;k)&:=G_{2,m}(q;k).  \label{two-G-prime-def-one-new}
\end{align}
Now, {\color{black}recalling (\ref{B-star-def})},  we can write
\beqa
 M_{j}^{\ga}(\si,s,t) =
\sum_{\substack{m,n,\ti m,\ti n\in\nat_0 \\ \ti m+\ti n=m+n } }\fr{1}{\sqrt{m!n!\ti m!\ti n!}} 
\lan\vac, B_{1,\ti n}(G_{1,\ti n}' )B_{2,\ti m}(G_{2,\ti m}' )B_{1,n}^*(G_{1,n}) B_{2,m}^*(G_{2,m})\vac\ran.\,\,
\label{two-scalar-product-formula-new}
\eeqa
Making use of Lemma~\ref{norms-of-scattering-states}, we obtain
\beqa
|\lan\vac, B_{1,\ti n}(G_{1,\ti n}' )B_{2,\ti m}(G_{2,\ti m}' )B_{1,n}^*(G_{1,n}) B_{2,m}^*(G_{2,m})\vac\ran| \pha{4444444444444444444444444444444}  \non\\
\leq\sum_{\sig\in S_{m+n}}\int  d^3 q d^3p \int  d^{3n}r d^{3m}k\,  
\big|G_{1,n}(q;  r) G_{2,m}(p;  k)\pha{444444444444444444444444}\non\\
 \times\bigg(\ov G_{1,\ti n}'( q+\un{\check{k}}-\un{\check{r}}; \hat r, \check k)\ov  G_{2,\ti m}'(p-\un{\check{k}}+\un{\check{r}}; \hat{k}, \check r)
\bigg)\big|\label{strange-notation}\\
\leq\sum_{\sig\in S_{m+n}} \int  d^{3n}r d^{3m}k\,  
\big( \|G_{1,n}(\,\cdot\,;  r)\|_2 \|G_{2,m}(\,\cdot\,;  k)\|_2\big)
\big(\| G_{1,\ti n}'(\,\cdot\,; \hat r, \check k)\|_2 \| G_{2,\ti m}'(\,\cdot\, ; \hat{k}, \check r)\|_2\big)\label{Cauchy-Schwarz-explanation}\\
\leq (m+n)!  \|G_{1,n}\|_2 \|G_{2,m}\|_2
\| G_{1,\ti n}'\|_2 \| G_{2,\ti m}'\|_2,
\label{two-B-expectation-value-new-new}
\eeqa
where the notation in (\ref{strange-notation}) is explained in Lemma~\ref{combinatorics}, and we applied the Cauchy-Schwarz inequality
in variables $r, k$ to factorize the two brackets in (\ref{Cauchy-Schwarz-explanation}).
Making use of Theorem~\ref{main-theorem-spectral}, we get
\beqa
\|G_{2,m}\|_2 \leq \frac{1}{\sqrt{m!}} \|\dot{\ga}_{2,\si}^{{\color{black}(2)}}(\hv_j,t)h_{2,j}^{(s)}\|_2 \|g_{2,\si}^m\|_2, \quad
\|G_{1,n}\|_2 \leq  \frac{1}{\sqrt{n!}} \|h_{1} \|_2 \|g_{1,\si}^{n}\|_2,
\eeqa
{\mc and analogous bounds hold for $\|G_{2,\ti m}'\|_2$ and $\|G_{1,\ti n}'\|_2$, respectively}.
Thus we obtain from~(\ref{two-B-expectation-value-new-new})
\beqa
& &|\lan\vac, B_{1,\ti n}(G_{1,\ti n}' )B_{2,\ti m}(G_{2,\ti m}' )B_{1,n}^*(G_{1,n}) B_{2,m}^*(G_{2,m})\vac\ran|\non\\
& &\leq c \frac{(m+n)!}{\sqrt{m!n!\ti m!\ti n!}} \| h_1\|^2_2  \|\dot{\ga}^{{\color{black}(2)}}_{2,\si}(\hv_j,t)h_{2,j}^{(s)}\|_2^2 \|g_{2,\si}^m\|_2 \|g_{2,\si}^{\ti m}\|_2 \|g_{1,\si}^{n}\|_2\|g_{1,\si}^{\ti n}\|_2\non\\
& &\leq c' \frac{(m+n)!}{\sqrt{m!n!\ti m!\ti n!}} \|h_1\|_2^2  \|\dot{\ga}_{2,\si}^{{\color{black}(2)}}(\hv_j,t)h_{2,j}^{(s)}\|_2^2 \|g_{2,\si}^{m}\|^2_2  \|g_{2,\si}^{n}\|^2_2, \label{two-auxiliary-bound-B}
\eeqa
where in the last step we made use of the facts that  $g_{1,\si}^{m}\leq  g_{2,\si}^m$, {\mc the functions} $g_{2,\si}^m$ have a  product structure and $m+n=\ti m+\ti n$.
Substituting (\ref{two-auxiliary-bound-B}) to (\ref{two-scalar-product-formula-new}) and making use of Lemma~\ref{summation}, we get
\beqa
 M_{j}^{\ga}(\si,s,t)
\leq \fr{c}{\si^{\de_{\la_0}}}  \|h_1\|^2_2  \|\dot{\ga}_{2,\si}^{{\color{black}(2)}}(\hv_j,t)h_{2,j}^{(s)}\|_2^2. \label{two-phases-last-step}
\eeqa
Making use of Lemma \ref{two-phase-upper-part} we conclude the proof of (a). 

To prove (b), let us set 
\begin{align}
G_{1,m}(q;k)&:=e^{-iE_{1,q,\si}t}  h_{1}(q)f^{m}_{1,q,\si}(k),\\  
G_{2,m}(q;k)&:=F_{m}(q;k),  \label{two-G-def-new-new-x}  \\    
G'_{1,m}(q;k)&:=G_{1,m}(q;k), \\
G'_{2,m}(q;k)&:=G_{2,m}(q;k).  \label{two-G-prime-def-one-new-x}
\end{align}
Making use of Theorem~\ref{main-theorem-spectral} and Lemma~\ref{two-rest-terms-new-lemma} we get
\beqa
& &\|G_{1,m}\|_2, \|G'_{1,m}\|_2\leq \frac{1}{\sqrt{m!}} \|h_{1}\|_2 \|g_{1,\si}^m\|_2,\\
& &\|G_{2,m}\|_2, \|G'_{2,m}\|_2 \leq  \fr{c_{t}^2}{\si^{\de_{\la_0}}}\bigg(t(\si^{\S}_t)^{3}+  \fr{(\si^{\S}_t)^{\de}}{t}+\fr{1}{(\si^{\S}_t)^{1+\de} t^2}  \bigg)\fr{1}{\sqrt{m!}} \|g^m_{2,\si}\|_2.
\eeqa
Thus repeating the steps (\ref{two-scalar-product-formula-new})--(\ref{two-phases-last-step}) above, we obtain for any  $0<\de<(\al^{-1}-1)$
\beqa
R_{j}^{\ga}(\si,s,t)\leq  
\fr{c_{t}^4}{\si^{\de_{\la_0}}}\bigg(t(\si^{\S}_t)^{3}+  \fr{(\si^{\S}_t)^{\de}}{t}+\fr{1}{(\si^{\S}_t)^{1+\de} t^2}  \bigg)^2\|h_{1}\|_2^2,
\eeqa
which concludes the proof. \qed
%%%%%%%%%%%%%%%%%%%%%%%%%%%%%%%%%%%%%%%%%
\subsection{Auxiliary results for Proposition~\ref{two-phase-vector-lemma} (a)}
%%%%%%%%%%%%%%%%%%%%%%%%%%%%%%%%%%%%%%%%%
\bel\label{two-phase-upper-part} Let us set $ \si_{\S,t}:=\max\{ \si^{\S}_t, \si\, \} $ and assume that $0<|\nabla E_{2,p,\si}|<1$. Then the expression
\beqa
-\h\dot{\ga}^{{\mc (2)}}_{2,\si}(\hv_j,\sn)(p):=\int_{|k|\geq \si_{\S,t}  } d^3k\, \vv^\si_2(k)^2 \fr{1}{|k|(1-\hatk\cdot \hv_j) }
\cos( (\nabla E_{2,p,\si}\cdot k-|k|)\sn)
\eeqa
satisfies
\beqa
|\dot{\ga}^{{\mc (2)}}_{2,\si}(\hv_j,\sn)(p)|\leq \fr{c}{ t^2 \si^{\S}_{t}}.\label{two-phase-bound}
\eeqa
\eel
%%%%%%%%%%%%%%%%%%%%%%%%%%%%%%%%%%%%%%%%%
\proof We consider the following expression whose real part is $-\h\dot{\ga}^{{\mc (2)}}_{2,\si}(\hv_j,\sn)(p)$ 
\beqa
I\1&:=&\1\int_{|k|\geq \si_{\S,t}  } d^3k\, \vv^\si_2(k)^2 \fr{1}{|k|(1-\hatk\cdot \hv_j) }e^{ i(\nabla E_{2,p,\si} k-|k|)\sn}\non\\
\1&=&\1\fr{\la}{2}\int d\Omm(\nee)\int_{|k|\geq \si_{\S,t}  } d|k| \, e^{ i(\nabla E_{2,p,\si}\cdot \nee-1)|k|\sn}  \fr{ \chit_{[\si,\ka)}^2(|k|) }{(1-\nee\cdot \hv_j) }\non\\
\1&=&\1\fr{\la}{2it}\int d\Omm(\nee) \, e^{ i(\nabla E_{2,p,\si}\cdot \nee-1)|k|\sn} 
 \fr{ \chit_{[\si,\ka)}^2(|k|) }{(\nabla E_{2,p,\si}\cdot \nee-1)(1-\nee\cdot \hv_j)}\bigg|_{|k|=\si_{\S,t}  }\label{two-dot-gamm-one}\\
& &-\fr{\la}{2it}\int d\Omm(\nee)\int_{|k|\geq \si_{\S,t}} d|k|\, e^{ i(\nabla E_{2,p,\si}\cdot \nee-1)|k|\sn} 
 \fr{\pa_{|k|}( \chit_{[\si,\ka)}^2(|k|) ) }{(\nabla E_{2,p,\si}\cdot \nee-1)(1-\nee\cdot \hv_j)}, \label{two-dot-gamm-two}
\eeqa
where $\ti\chi_{[\si, \ka)}$ was introduced below (\ref{form-factor-definitions-x}).
Let us first consider (\ref{two-dot-gamm-one}). We introduce the function $f_{|k|}(\nee):=\fr{ \chit_{[\si,\ka)}^2(|k|) }{(\nabla E_{2,p,\si}\cdot \nee-1)(1-\nee\cdot \hv_j)}$
for $|k|=\si_{\S,t}$. It satisfies
\beqa
|f_{|k|}(\nee)|\leq c, \quad |\pa_{\theta} f_{|k|}(\nee)|\leq c
\eeqa
w.r.t. spherical coordinates $(\theta, \phi)$ chosen with $z$-axis in the direction of $\nabla E_{2,p,\si}$.
Thus Lemma~\ref{two-angular-integration-lemma-x} gives
\beqa
|(\ref{two-dot-gamm-one})|\leq  \fr{c}{ \si_{\S,t} t^2}\leq \fr{c}{ \si^{\S}_t t^2}.
\eeqa
Now we estimate (\ref{two-dot-gamm-two}). We set $f'_{|k|}(\nee):=\fr{\pa_{|k|}( \chit_{[\si,\ka)}^2(|k|)) }{(\nabla E_{1,p,\si}\cdot \nee-1)(1-\nee\cdot \hv_j)}$
and note that
\beqa
|f'_{|k|}(\nee)|\leq c, \quad |\pa_{\theta}f'_{|k|}(\nee)|\leq c.
\eeqa
Thus we get from Lemma~\ref{two-angular-integration-lemma-x}
\beqa
|(\ref{two-dot-gamm-two})|\leq \int_{\si_{S,t}}^{\kappa}d|k|\,  \fr{c}{|k|t^2}\leq \fr{c|\log (\si^{\S}_t)|  }{t^2},
\eeqa
which completes the proof. \qed
%%%%%%%%%%%%%%%%%%%%%%%%%%%%%%%%%%%%%%%%%%%%%%%%%%%%%%%%%%%%%%%%%%%%%%%%
%%%%%%%%%%%%%%%%%%%%%%%%%%%%%%%%%%%%%%%%%%%%%%%%%%%%%%%%%%%%%%
\bel\label{two-angular-integration-lemma-x} Let $e=e(\theta,\phi)$ be a unit vector normal to the unit sphere and  consider a smooth function $f_{|k|}(\nee)=f_{|k|}(\theta, \phi)$
on a unit sphere depending on a parameter $\kappa\geq |k|>0$. For $ \nabla E_{2,p,\si} \neq 0$  we have 
\beqa
|\int d\Omm(\nee)    e^{i \nabla E_{2,p,\si}\cdot \nee |k| t}  f_{|k|}(\nee)|\leq \fr{c}{{\color{black}|\nabla E_{2,p,\si}|}|k|t}\sup_{\nee }\sup_{\ell\in\{0,1\}}|\pa_{\theta}^{\ell}f_{|k|}(\nee)|.
\eeqa 
(The spherical coordinates $(\theta,\phi)$ are chosen with $z$-axis in the direction of $\nabla E_{2,p,\si}$).
\eel
%%%%%%%%%%%%%%%%%%%%%%%%%%%%%%%%%%%%%%%%%%%%%%%
\proof  We consider the integral
\beqa
I_1=\int d\Omm(\nee)   e^{i \nabla E_{2,p,\si}\cdot \nee |k|  t}  f_{|k|}(\nee)=\int_0^{2\pi}  d\phi \int_{0}^{\pi} d\theta \,  e^{i t|k| |\nabla E_{2,p,\si}|\cos(\theta)} 
\sin(\theta) f_{|k|}(\theta,\phi).
 \eeqa
%We introduce a small parameter $0< \eps < 1$ and write
%\beqa
%I_1=\int_0^{2\pi}  d\phi \int_{\eps}^{\pi-\eps} d\theta \,   e^{i t|k| |\nabla E_{2,p,\si}|\cos(\theta)} \sin(\theta)f_{|k|}(\theta,\phi)+R. 
%\label{two-leading-term-R-x}
%\eeqa
%Clearly,
%\beqa
%|R|\leq c\eps \sup_{\nee}|f_{|k|}(\nee)|. \label{two-rest-term-theta-x}
%\eeqa
Now we note that
\beqa
\pa_{\theta} e^{i t|k| |\nabla E_{2,p,\si}|\cos(\theta)} = -it|k| |\nabla E_{2,p,\si}|\sin(\theta)e^{i t|k| |\nabla E_{2,p,\si}|\cos(\theta)}.\label{two-theta-int-x}
\eeqa
%We can estimate in the region of integration of the first term on the r.h.s. of (\ref{two-leading-term-R-x})
%\beqa
%t|k| |\nabla E_{2,p,\si}|\, |\sin(\theta)|\geq c_0t|k| |\nabla E_{2,p,\si}|\eps, \quad c_0>0.
%\eeqa
% Denoting by $I_{1}$ the first integral on the r.h.s. of  (\ref{two-leading-term-R-x}), 
Hence, we obtain from (\ref{two-theta-int-x}) by integration by parts
\beqa
I_{1}\1&:=&\1\fr{1}{(-i)|\nabla E_{2,p,\si}||k| t}\int_0^{2\pi}  d\phi   \,  
e^{i t|k| |\nabla E_{2,p,\si}|\cos(\theta)} 
   f_{|k|}(\theta,\phi)  \bigg|_{0}^{\pi}\non\\
& &-\fr{1}{(-i)|\nabla E_{2,p,\si}||k| t }\int_0^{2\pi}  d\phi \int_{0}^{\pi}d\theta   \, 
e^{i t|k| |\nabla E_{2,p,\si}|\cos(\theta)}
 \pa_{\theta}  f_{|k|}(\theta,\phi).
\eeqa
Thus we get
\beqa
|I_{1}|\leq \fr{c}{ { |\nabla E_{2,p,\si}|}  |k| t}\sup_{\nee}\sup_{\ell\in\{0,1\}}|\pa_{\theta}^{\ell}f_{|k|}(\nee)|,
\eeqa
%where the constant $c$ is actually independent of $\eps$. Thus we can take the limit $\eps\to 0$ in which we obtain
%\beqa
%|I_1|\leq \fr{c}{ {\bc |\nabla E_{2,p,\si}|}  |k| t}\sup_{\nee }\sup_{\ell\in\{0,1\}}|\pa_{\theta}^{\ell}f_{|k|}(\nee)|,
%\eeqa
which completes the proof. \qed
%%%%%%%%%%%%%%%%%%%%%%%%%%%%%%%%%%%%%%%%%%%%%%%%%%%%%%%%%%
\subsection{Auxiliary results for Proposition~\ref{two-phase-vector-lemma} (b) } 
%%%%%%%%%%%%%%%%%%%%%%%%%%%%%%%%%%%%%%%%%%%%%%%%%%%%%%%%%%
\bel\label{two-rest-terms-new-lemma} Let us define
\beqa
F_{m}(p;k'):=\int d^3k\, \bigg( e^{-iE_{2,p-k,\si} \sn}  h^{\ga}_{2}(p-k)f^{m}_{2,p-k,\si}(k'^m)-e^{-iE_{2,p,\si}\sn}
e^{i\nabla E_{2,p,\si}\cdot k \sn } h^{\ga}_2(p)f^{m}_{2,p,\si}(k'^m) \bigg)\times \non\\
 \ph{444444444444444444444444444444444444444} \times F_{\si,\sn}(k)  \vv_2(k),
\label{two-rest-term-difference}
\eeqa
where
\beqa
F_{\si,\sn}(k):=
\fr{\vv_2^{\si}(k)}{|k|} \bigg(  \fr{e^{-i|k|\sn} }{(1-\hatk\cdot \hv_j ) }+\fr{e^{i|k|\sn} }{(1+\hatk\cdot \hv_j ) } \bigg):=f_{\si,\sn}(k)+\ov{f}_{\si,\sn}(-k).
\eeqa
Then, for any $0<\de<(\al^{-1}-1)$,
\beqa
\|F_{m}\|_2\1&\leq&\1 
\fr{c_{t}^2}{\si^{\de_{\la_0}}}\bigg(t(\si^{\S}_t)^{3}+  \fr{(\si^{\S}_t)^{\de}}{t}+\fr{1}{(\si^{\S}_t)^{1+\de} t^2}  \bigg)\fr{1}{\sqrt{m!}} \|g^m_{2,\si}\|_2.
\eeqa
where  $c_{t}:=c (\uneps(t) t^{-\peps})^{-1}$. 
\eel
%%%%%%%%%%%%%%%%%%%%%%%%%%%%%%%%%%%%%%%%%%%%%%%%%%%%%%%%%%%
\proof Since $f_{\si,\sn}(k)$ and $\ov{f}_{\si,\sn}(-k)$ give two contributions to $F_{m}(p;k')$ whose analysis is analogous,
we consider only the contribution of $f_{\si,\sn}(k)$. Therefore we define 
\beqa
& &\hat F_{m}(p;k'):=\int d^3k\, \big( e^{-iE_{2,p-k,\si} \sn}  h_{2}^{\ga}(p-k)f^{m}_{2,p-k,\si}(k'^m)-e^{-iE_{2,p,\si}\sn}e^{i\nabla E_{2,p,\si}\cdot k \sn } h_2^{\ga}(p)f^{m}_{2,p,\si}(k'^m) \big)\times \non\\
& &\ph{4444444444444444444444444444444444444444444444444444444}\times f_{\si,\sn}(k) \vv_2(k).
\eeqa
Now we divide this expression into two parts:
\beqa
\hat F_{1,m}(p;k')\1&:=&\1\int d^3k\, \big( e^{-iE_{2,p-k,\si} \sn}-e^{-iE_{2,p,\si}\sn}e^{i\nabla E_{2,p,\si}\cdot k \sn } )h_2^{\ga}(p)f^{m}_{2,p,\si}(k'^m) f_{\si,\sn}(k) \vv_2(k),\label{F-83}\\
\hat F_{2,m}(p;k')\1&:=&\1\int d^3k\, e^{-iE_{2,p-k,\si} \sn}  \big(h_{2}^{\ga}(p-k)f^{m}_{2,p-k,\si}(k'^m)-h_{2}^{\ga}(p)f^{m}_{2,p,\si}(k'^m)\big)  f_{\si,\sn}(k) \vv_2(k).\label{F-84}
\eeqa
We will consider (\ref{F-83}), (\ref{F-84}) below and above the slow cut-off. More precisely, we set $\si_{\S,t}:=\max\{\si^\S_t,\si\}$
and define
\beqa
\hat F^{1}_{1,m}(p;k')\1&:=&\1\int_{\si\leq |k|\leq \si_{\S,t}} d^3k\, \big( e^{-iE_{2,p-k,\si} \sn}-e^{-iE_{2,p,\si}\sn}e^{i\nabla E_{2,p,\si}\cdot k \sn } )h^{\ga}_2(p)f^{m}_{2,p,\si}(k'^m) f_{\si,\sn}(k) \vv_2(k),\\
\hat F^{2}_{1,m}(p;k')\1&:=&\1\int_{ |k|\geq \si_{\S,t}} d^3k\, \big( e^{-iE_{2,p-k,\si} \sn}-e^{-iE_{2,p,\si}\sn}e^{i\nabla E_{2,p,\si}\cdot k \sn } )h^{\ga}_2(p)f^{m}_{2,p,\si}(k'^m) f_{\si,\sn}(k) \vv_2(k),\\
\hat F_{2,m}^1(p;k')\1&:=&\1\int_{\si\leq |k|\leq \si_{\S,t}} d^3k\, e^{-iE_{2,p-k,\si} \sn}  \big(h^{\ga}_{2}(p-k)f^{m}_{2,p-k,\si}(k'^m)-h^{\ga}_{2}(p)f^{m}_{2,p,\si}(k'^m)\big)  f_{\si,\sn}(k) \vv_2(k),\quad\quad\\
\hat F_{2,m}^2(p;k')\1&:=&\1\int_{  |k|\geq \si_{\S,t} }  d^3k\, e^{-iE_{2,p-k,\si} \sn}  \big(h_{2}^{\ga}(p-k)f^{m}_{2,p-k,\si}(k'^m)-h^{\ga}_{2}(p)f^{m}_{2,p,\si}(k'^m)\big)  f_{\si,\sn}(k) \vv_2(k).
\eeqa
As for $\hat F_{1,m}^{1}$, we note that
\beqa
\big| e^{-iE_{2,p-k,\si} \sn}-e^{-iE_{2,p,\si}\sn}e^{i\nabla E_{2,p,\si}\cdot k \sn } \big|\leq  t|E_{2,p-k,\si}- E_{2,p,\si}+\nabla E_{2,p,\si}\cdot k|\leq  c tk^2,
\eeqa
where we made use of the fact that the second derivative of $p\mapsto E_{2,p,\si}$ exists and is bounded uniformly in the cut-off (cf. Theorem~\ref{preliminaries-on-spectrum}). Thus we get
\beqa
|\hat F_{1,m}^{1}(p;k')|\leq |h^{\ga}_2(p)|\fr{c}{\sqrt{m!}} t g^m_{2,\si}(k') \int_{\si\leq |k|\leq \si_{\S,t}} d|k|\,|k|^2 \leq  |h^{\ga}_2(p)|\fr{c}{\sqrt{m!}} t(\si^{\S}_t)^{3} g^m_{2,\si}(k').
\eeqa
Now we consider $\hat F_{1,m}^{2}$.  We write
\begin{align}
\hat F_{1,m}^{2}(p;k')&=h^{\ga}_2(p)f^{m}_{2,p,\si}(k'^m) e^{-iE_{2,p,\si}t }\hat G_{1}^{2}(p), \label{two-F-two-m} \\
\hat G_{1}^{2}(p)&:=\int_{|k|\geq \si_{\S,t}} d^3k\, \big( e^{i(E_{2,p,\si}-E_{2,p-k,\si}-|k|) \sn}-e^{i(\nabla E_{2,p,\si}\cdot k-|k| )\sn } ) \fr{\vv_2^{\si}(k)^2}{|k|(1-\hatk\cdot \hv_j ) } . \label{two-G-two-m}
\end{align}
Let us denote by $\hat G_{1}^{2,1}(p)$ the term in (\ref{two-G-two-m}) involving $e^{i(E_{2,p,\si}-E_{2,p-k,\si}-|k|) \sn}$ and by
$\hat G_{1}^{2,2}(p)$ the term involving $e^{i(\nabla E_{2,p,\si}\cdot k-|k| )\sn }$. By  (the proof of) Lemma~\ref{two-phase-upper-part},
we get 
\beqa
|\hat G_{1}^{2,2}(p)|\leq  \fr{c }{t^2\si^{\S}_{t} }.
\eeqa
Similarly, from Lemma~\ref{two-angular-integration-lemma-one}  we have for any $0<\de<(\al^{-1}-1)$ 
\beqa
|\hat G_{1}^{2,1}(p)|\leq  \fr{c}{\si^{\de_{\la_0}}}\bigg( \fr{(\si^{\S}_t)^{\de}}{t}+\fr{1}{(\si^{\S}_t)^{1+\de} t^2}\bigg).   
\eeqa
Therefore 
\beqa
|\hat F_{1,m}^{2}(p;k')|\leq \fr{c}{\si^{\de_{\la_0}}}   \bigg(  \fr{(\si^{\S}_t)^{\de}}{t}+\fr{1}{(\si^{\S}_t)^{1+\de} t^2}+   \fr{1 }{t^2\si^{\S}_{t} } \bigg)\fr{1}{\sqrt{m!}} g^m_{2,\si}(k')|h^{\ga}_2(p)|.  
\eeqa
Consequently
\beqa
|\hat F_{1,m}(p;k')|
\leq \fr{c'}{\si^{\de_{\la_0}}}\bigg(t(\si^{\S}_t)^{3}+  \fr{(\si^{\S}_t)^{\de}}{t}+\fr{1}{(\si^{\S}_t)^{1+\de} t^2}  \bigg)\fr{1}{\sqrt{m!}} g^m_{2,\si}(k')|h^{\ga}_2(p)|. \label{two-sub-final-one}
\eeqa
Now we consider $\hat F_{2,m}$. From Lemmas~\ref{two-second-contribution-phase-rest} and \ref{two-second-contribution-phase-rest-x} we get for $\de$ as above
\begin{align}
|\hat F_{2,m}^1(p;k')|&\leq  \fr{c_t}{\si^{\de_{\la_0}}}\fr{\si^{\S}_t}{t}\fr{1}{\sqrt{m!}} g_{2,\si}^m(k'),  \\
|\hat F_{2,m}^2(p;k')|&\leq   \fr{c_{t}^2}{ \si^{\de_{\la_0} }} \bigg( \fr{(\si^{\S}_t)^{\de}}{t}+\fr{1}{(\si^{\S}_t)^{1+\de} t^2}\bigg)\fr{1}{\sqrt{m!}} g^m_{2,\si}(k'),  
\end{align}
where $c_t:=(\uneps(t) t^{-\peps})^{-1}$. Therefore
\beqa
|\hat F_{2,m}(p;k')|\leq   \fr{c_{t}^2 }{ \si^{\de_{\la_0} }} \bigg(\fr{(\si^{\S}_t)^{\de}}{t}+\fr{1}{(\si^{\S}_t)^{1+\de} t^2}\bigg)\fr{1}{\sqrt{m!}} g^m_{2,\si}(k').
\label{two-sub-final-two}
\eeqa
We note that by definition $\hat F_{2,m}(p;k')$ is compactly supported in $p$ thus we can multiply the r.h.s. of the above bound by
a characteristic function of a sufficiently large set before computing the $L^2$ norm.  Making use of (\ref{two-sub-final-one}) and (\ref{two-sub-final-two}) we conclude the proof.  \qed
%%%%%%%%%%%%%%%%%%%%%%%%%%%%%%%%%%%%%%%%%%%%%%%%%%%%%%%%%%%%%%%%%
%%%%%%%%%%%%%%%%%%%%%%%%%%%%%%%%%%%%%%%%%%%%%%%%%%%%%%%%%%%%%%%%%
\bel\label{partition-lemma} Let us recall that 
\beqa
h_{2}^{\ga}(p):=e^{i\ga_{\si}(\hv_j,\sn)(p)}  h_{2,j}^{(\tn)}(p)= e^{i\ga_{\si}(\hv_j,\sn)(p)} \app_{\Ga^{(s)}_j  }(p) h_{2}(p).
\eeqa
Then
\beqa
|\pa_{p^i} (h_{2}^{\ga})(p)|\leq  c_t, \quad  |\pa_{p^i}\pa_{p^{j}} (h_{2}^{\ga})(p)|
\leq  \fr{c_{1,t}}{\si^{\de_{\la_0} }  },
\eeqa
where $c_t:=c(\uneps(t) t^{-\peps})^{-1}$ and $c_{1,t}:=c((\uneps(t) t^{-\peps})^{-2}+t^{1-\al})$.
\eel
%%%%%%%%%%%%%%%%%%%%%%%%%%%%%%%%%%%%%%%%%%%%%%%%%%
\proof Follows immediately from Lemmas~\ref{h-theta} and \ref{phase-lemma}. \qed  
%%%%%%%%%%%%%%%%%%%%%%%%%%%%%%%%%%%%%%%%%%%%%%%%%%%%%%
\bel\label{two-second-contribution-phase-rest} Consider the expression
\beqa
\hat F_{2,m}^1(p;k'):=\int_{\si\leq |k|\leq \si_{\S,t}} d^3k\, e^{-i(E_{2,p-k,\si}+|k|) \sn}  \big(h_{2}^\ga(p-k)f^{m}_{2,p-k,\si}(k'^m)-h_{2}^\ga(p)f^{m}_{2,p,\si}(k'^m)\big)   \fr{ \vv_2^{\si}(k)^2    }{ |k| (1-\hatk\cdot \hv_j ) }.\non\\ \label{two-second-contribution-phase-statement}
\eeqa
There holds the bound
\beqa
|\hat F_{2,m}^1(p;k')|\leq \fr{c_t}{\si^{\de_{\la_0}}}\fr{\si^{\S}_t}{t}\fr{1}{\sqrt{m!}} g_{2,\si}^m(k'),  \label{two-first-bound-rest-terms}
\eeqa
where $c_{t}:=c(\uneps(t) t^{-\peps})^{-1}$.
\eel 
%%%%%%%%%%%%%%%%%%%%%%%%%%%%%%%%%%%%%%%%%%%%%%%%%%%%%%%
\proof Let us define the functions
\beqa
F^m(p,k,k'):=\big(h_{2}^{\ga}(p-k)f^{m}_{2,p-k,\si}(k'^m)-h_{2}^{\ga}(p)f^{m}_{2,p,\si}(k'^m)\big)   \fr{ \vv_2^{\si}(k)^2    }{ |k| (1-\hatk\cdot \hv_j ) }
\label{F-m}
\eeqa
and note that for any $0\leq |\be|\leq 1$,
\beqa
& &|\pa^{\be}_k F^m(p,k,k')|\leq \fr{c_{t}}{(\si^{\de_{\la_0}})^{|\be|} |k|^{1+|\be|}} \fr{1}{\sqrt{m!}}g_{2,\si}^m(k').
\label{two-F-general-bound}
\eeqa
In fact, we rewrite (\ref{F-m}) as 
\beqa
F^m(p,k,k')=\bigg(\fr{h_{2}^{\ga}(p-k)-h_2^{\ga}(p)}{|k|}f^{m}_{2,p-k,\si}(k'^m)+h_2^{\ga}(p)\fr{f^{m}_{2,p-k,\si}(k'^m)
-f^{m}_{2,p,\si}(k'^m)}{|k|}\bigg)   \fr{ \vv_2^{\si}(k)^2    }{  (1-\hatk\cdot \hv_j ) } \ \
\eeqa
and differentiate. Taylor expansion, Lemma~\ref{partition-lemma} and Theorem~\ref{main-theorem-spectral} give (\ref{two-F-general-bound}).

Now we define $\Om_p(k):=E_{2,p-k,\si}+|k|$ and note that
\beqa
|\nabla_k \Om_p(k)|=|-\nabla E_{2,p-k,\si}+k/|k| |\geq \eps>0,
\eeqa
independently of $p$, $k$ within the assumed restrictions. We write
\beqa
e^{-i\Om_p(k) t}=\fr{\nabla_k  \Om_p(k)\cdot \nabla_k e^{-i\Om_p(k) t}  }{(-it) |\nabla_k \Om_p(k)|^2 }.\label{two-non-stat-trick}
\eeqa
By substituting this equality to (\ref{two-second-contribution-phase-statement}), and integrating by parts we get
\beqa
& &\int_{\si\leq |k|\leq \si_{\S,t}} d^3k\, e^{-i(E_{2,p-k,\si}+|k|) \sn} F^m(p,k,k')\non\\
& &\ph{444444444444444}=
\int_{\si\leq |k|\leq \si_{\S,t}} d^3k\, \fr{\nabla_k  \Om_p(k)\cdot \nabla_k e^{-i\Om_p(k) t}  }{(-it) |\nabla_k \Om_p(k)|^2 }    F^m(p,k,k')\non\\
& &\ph{444444444444444}=\fr{1}{(-it)}\int d\Omm(\nee)\,|k|^2 \nee\cdot \fr{\nabla_k  \Om_p(k) e^{-i\Om_p(k) t}  }{ |\nabla_k \Om_p(k)|^2 }   
 F^m(p,k,k')|^{k=\si_{\S,t}e}_{k=\si \nee} \label{two-rest-term-boundary-phases} \\
& &\ph{444444444444444}-\fr{1}{(-it)}\int_{\si\leq |k|\leq \si_{\S,t}} d^3k\,  e^{-i\Om_p(k) t} \nabla_k\cdot\bigg( \fr{\nabla_k  \Om_p(k)   }{ |\nabla_k \Om_p(k)|^2 }    F^m(p,k,k')\bigg).\label{two-rest-term-boundary-phases-one-x}
\eeqa
Making use of (\ref{two-F-general-bound}) we obtain
\beqa
|(\ref{two-rest-term-boundary-phases})|\leq  c_{t}\fr{\si^{\S}_t}{t} \fr{1}{\sqrt{m!}} g_{2,\si}^m(k').
\eeqa
Keeping in mind (\ref{two-F-general-bound}) and the fact  that $|{\mc \pa_{k^i}\pa_{k^j}} \Om_p(k)|\leq c/|k|$, we also obtain
\beqa
|(\ref{two-rest-term-boundary-phases-one-x})|\leq  c_{t} \fr{1}{\si^{\de_{\la}}}\fr{\si^{\S}_t}{t}\fr{1}{\sqrt{m!}} g_{2,\si}^m(k').
\eeqa 
This proves (\ref{two-first-bound-rest-terms}). \qed
%%%%%%%%%%%%%%%%%%%%%%%%%%%%%%%%%%%%%%%%%%%%%%%%%%%%%%%%%%%%%%%%%%%%%%%%%%%%%%%

%%%%%%%%%%%%%%%%%%%%%%%%%%%%%%%%%%%%%%%%%%%%%%%%%%%%%%
\bel\label{two-second-contribution-phase-rest-x} Consider the expression
\beqa
\hat F_{2,m}^2(p;k'):=\int_{ |k|\geq \si_{\S,t}} d^3k\, e^{-i(E_{2,p-k,\si}+|k|) \sn}  
\big(h_{2}^{\ga}(p-k)f^{m}_{2,p-k,\si}(k'^m)-h_{2}^{\ga}(p)f^{m}_{2,p,\si}(k'^m)\big)   \fr{ \vv_2^{\si}(k)^2    }{ |k| (1-\hatk\cdot \hv_j ) }. \non\\ 
\label{two-last-contribution-phase-statement}
\eeqa
There holds the bound for any $0<\de<(\al^{-1}-1)$
\beqa
|\hat F_{2,m}^2(p;k')|\leq \fr{c_{t}^2}{ \si^{\de_{\la_0} }} \bigg( \fr{(\si^{\S}_t)^{\de}}{t}+\fr{1}{(\si^{\S}_t)^{1+\de} t^2}\bigg)\fr{1}{\sqrt{m!}} g^m_{2,\si}(k'), \label{similar-bound-phase}
\eeqa
where $c_{t}:=c (\uneps(t) t^{-\peps})^{-1}$. 
\eel 
%%%%%%%%%%%%%%%%%%%%%%%%%%%%%%%%%%%%%%%%%%%%%%%%%%%%%%%
\proof We express $k$ in spherical coordinates choosing the $z$-axis in the direction of $p$, and define functions
\beqa
F^m(p,k',|k|,\nee):=\big(h_{2}^{\ga}(p-|k|\nee)f^{m}_{2,p-|k|\nee,\si}(k'^m)-h_{2}^{\ga}(p)f^{m}_{2,p,\si}(k'^m)\big)   
\fr{ \chit_{[\si,\ka)}(|k|)^2 }{ 2 (1-\nee\cdot \hv_j ) }.
\eeqa
We also set $V(|k|,\nee):=E_{2,p-|k|\nee,\si}+|k|$ and note that $\pa_{|k|}V(|k|,\nee)=-\nee\cdot \nabla E_{2,p-|k|\nee,\si}+1$ and 
thus $|\pa_{|k|}V(|k|,\nee)|\geq \eps_1>0$ for some $\eps_1{\mc >}0$. We have
\beqa
\fr{i\pa_{|k|}e^{-iV(|k|,\nee)t} }{ \pa_{|k|}V(|k|,\nee) t } = e^{-iV(|k|,\nee)t}.
\eeqa
Now we integrate by parts:
\begin{align}
\hat F_{2,m}^2(p;k')&=\int d\Omm(\nee)\int_{ |k|\geq \si_{\S,t}} d|k|\, e^{-iV(|k|,\nee)  \sn}  F^m(p,k',|k|,\nee)\non\\
&=\fr{i}{t}\int d\Omm(\nee)\, e^{-iV(|k|,\nee)  \sn} \bigg( \fr{ F^m(p,k',|k|,\nee)}{\pa_{|k|}V(|k|,\nee)}  \bigg)\bigg|_{|k|=\si_{\S,t}} \label{two-difference-term-one} \\
&\ph{44} -\fr{i}{t}\int d\Omm(\nee)\int_{ |k|\geq \si_{\S,t}} d|k|\, e^{-iV(|k|,\nee)  \sn} \pa_{|k|}  \bigg( \fr{ F^m(p,k',|k|,\nee)}{\pa_{|k|}V(|k|,\nee)}  \bigg).
\label{two-difference-term-two}
\end{align}
Let us estimate (\ref{two-difference-term-one}). We write $f^{\si_{\S,t}}(\nee):= \fr{F^m(p,k',|k|,\nee)}{\pa_{|k|}V(|k|,\nee)}\big|_{|k|=\si_{\S,t}}$ and note that by Theorem~\ref{main-theorem-spectral} and Lemma~\ref{partition-lemma}
\beqa
 |f^{\si_{\S,t}}(\nee)|\leq \fr{c}{\sqrt{m!}} g^m_{2,\si}(k'), \quad |\pa_{\theta}f^{\si_{\S,t}}(\nee)|\leq \fr{c_t}{ \si^{\de_{\la_0} }}\fr{1}{\sqrt{m!}} g^m_{2,\si}(k'),
 \label{bounds-on-derivatives-part-one}
\eeqa
where $c_t= (\uneps(t) t^{-\peps})^{-1}$.  Thus we get from Lemma~\ref{two-angular-integration-lemma} 
\beqa
|(\ref{two-difference-term-one})|\leq c_t\bigg(\fr{\eps}{t}+\fr{1}{\si_{\S,t} t^2\eps}\bigg)\fr{1}{\sqrt{m!}} g^m_{2,\si}(k')\leq
 c_t\bigg( \fr{(\si^{\S}_t)^{\de}}{t}+\fr{1}{(\si^{\S}_t)^{1+\de} t^2}\bigg)\fr{1}{\sqrt{m!}} g^m_{2,\si}(k'),
\eeqa
where we set $\eps=(\si^{\S}_t)^{\de}$, $0<\de<(\al^{-1}-1)$  so that $\al(1+\de)<1$. 

Now we consider (\ref{two-difference-term-two}): We set $f_{|k|}(\nee):=\pa_{|k|}  \bigg( \fr{ F^m(p,k',|k|,\nee)}{\pa_{|k|}V(|k|,\nee)}\bigg)$
and note that, by Theorem~\ref{main-theorem-spectral} and Lemma~\ref{partition-lemma}
\beqa
|f_{|k|}(\nee)|\leq  \fr{1}{\si^{\de_{\la_0}}}\fr{c_t}{\sqrt{m!}} g^m_{2,\si}(k'),\quad |\pa_{\theta}f_{|k|}(\nee)|\leq  \fr{1}{\si^{\de_{\la_0}}}\fr{c_{1,t}}{\sqrt{m!}} g^m_{2,\si}(k'),
\label{bounds-on-derivatives-part-two}
\eeqa
where $c_{1,t}:=c((\uneps(t) t^{-\peps})^{-2}+t^{1-\al})$.
Consequently, by Lemma~\ref{two-angular-integration-lemma}, we obtain
\begin{align}
|(\ref{two-difference-term-two})|&\leq \fr{1}{ \si^{\de_{\la_0}} } 
\int_{\kappa\geq  |k|\geq \si_{\S,t}}d|k|\, \bigg(\fr{\eps}{t}c_t+\fr{1}{|k| t^2\eps} c_{1,t}\bigg) \fr{1}{\sqrt{m!}} g^m_{2,\si}(k')\non\\
&\leq \fr{1}{ \si^{\de_{\la_0} }}\bigg(\fr{\eps}{t}c_{t}+\fr{|\log(\si_{\S,t}) | }{ t^2 \eps}c_{1,t}\bigg)\fr{1}{\sqrt{m!}} g^m_{2,\si}(k')\non\\
&\leq  \fr{c_t^2}{ \si^{\de_{\la_0} }} \bigg( \fr{(\si^{\S}_t)^{\de}}{t}+\fr{1}{(\si^{\S}_t)^{1+\de} t^2} \bigg)\fr{1}{\sqrt{m!}} 
g^m_{2,\si}(k'),
\end{align}
where we set $\eps=(\si^{\S}_t)^{\de}$, $0<\de<(\al^{-1}-1)$ so that $\al(1+2\de)<1$. In the last step we observed
that $|\log(\si_{\S,t}) |\leq |\log (\si)|\leq c/\si^{\de_{\la_0}}$ and noted that $\si_t^\S c_{1,t}\leq c_t^2$ since $1>\al>1/2$.
This concludes the proof. \qed
%%%%%%

%%%%%%%%%%%%%%%%%%%%%%%%%%%%%%%%%%%%%%%%
\bel\label{two-angular-integration-lemma-one} Suppose that $0<|\nabla E_{2,p,\si}|<1$ (in particular $p\neq 0$). Consider the expression
\beqa
\check G_{1}^{2}(p)=\int_{|k|\geq \si_{\S,t}} d^3k\,  e^{-i(E_{2,p-k,\si}+|k|) \sn}  \fr{\vv_2^{\si}(k)^2}{|k|(1-\hatk\cdot \hv_j ) }. 
\eeqa
There holds for any  $0<\de< (\al^{-1}-1)$
\beqa
|\check G_{1}^{2}(p)|\leq \fr{c}{\si^{\de_{\la_0}}}\bigg( \fr{(\si^{\S}_t)^{\de}}{t}+\fr{1}{(\si^{\S}_t)^{1+\de} t^2}\bigg).
\eeqa
\eel
%%%%%%%%%%%%%%%%%%%%%%%%%%%%%%%%%%%%%%%%%%%%%%%%
\proof We choose spherical coordinates with the $z$-axis in the direction of $p$ and  write
\beqa
\check G_{1}^{2}(p):=\int d\Omm(\nee)\int_{|k|\geq \si_{\S,t}} d|k|\,  e^{-i(E_{2,p-|k|\nee,\si}+|k|) \sn}  \fr{\chi_{[\si,\kappa)}(k)^2  }{ {\mc 2}(1-\nee\cdot \hv_j ) }. 
\label{two-G-one-two-rewritten}
\eeqa
We set $V(|k|,\nee)=E_{2,p-|k|\nee,\si}+|k|$ and note that $\pa_{|k|}V(|k|,\nee)=-\nee\cdot \nabla E_{2,p-|k|\nee,\si}+1$, hence $|\pa_{|k|}V(|k|,\nee)|\geq\eps_1>0$. We have
\beqa
\fr{i\pa_{|k|}e^{-iV(|k|,\nee)t} }{ \pa_{|k|}V(|k|,\nee) t } = e^{-iV(|k|,\nee)t}.
\eeqa
Now we integrate by parts in (\ref{two-G-one-two-rewritten}):
\beqa
\check G_{1}^{2}(p)\1&=&\1\fr{1}{t}\int\fr{d\Omm(\nee)}{2(1-\nee\cdot \hv_j )}\int_{|k|\geq \si_{\S,t}} d|k|\,   \fr{i\pa_{|k|}e^{-iV(|k|,\nee)t} }{ \pa_{|k|}V(|k|,\nee)  }  
 \chit_{[\si,\kappa)}(|k|)^2  \non\\
\1&=&\1\fr{i \chit_{[\si,\kappa)}( \si_{\S,t})^2   }{t}\int\fr{d\Omm(\nee)}{2(1-\nee\cdot \hv_j )}  \fr{  e^{-i (E_{2,p- \si_{\S,t} \nee,\si}+  \si_{\S,t}) t} }
{    (-\nee\cdot \nabla E_{2,p- \si_{\S,t}\nee,\si}+1)  }  \label{two-first-term-estimate-theta}\\
& &-   \fr{1}{t}\int\fr{d\Omm(\nee)}{2(1-\nee\cdot \hv_j )}\int_{|k|\geq \si_{\S,t}} d|k|\,   ie^{-iV(|k|,\nee)t} 
  \pa_{|k|}\bigg(\fr{ \chit_{[\si,\kappa)}(|k|)^2   }{\pa_{|k|}V(|k|,\nee)}\bigg), \label{two-second-term-estimate-theta}
\eeqa
where $\ti\chi_{[\si,\kappa)}( |k|)=\chi_{[\si,\kappa)}(k)$.
To estimate (\ref{two-first-term-estimate-theta}), we write $f_{\si_{\S,t}}(\nee):=(1-\nee\cdot \hv_j )^{-1} (-\nee\cdot \nabla E_{2,p- \si_{\S,t}\nee,\si}+1)^{-1}$ 
and obtain from Lemma~\ref{two-angular-integration-lemma}
\beqa
|(\ref{two-first-term-estimate-theta})|\leq c\bigg(\fr{\eps}{t}+\fr{1}{\si_{\S,t}t^2 \eps}\bigg)\leq
 c\bigg( \fr{(\si^{\S}_t)^{\de}}{t}+\fr{1}{(\si^{\S}_t)^{1+\de} t^2}\bigg),
\eeqa
where we set $\eps=(\si^{\S}_t)^{\de}$, $0<\de< (\al^{-1}-1)$.

Now we analyse (\ref{two-second-term-estimate-theta}). We recall that $\pa_{|k|}V(|k|,\nee)=-\nee\cdot \nabla E_{2,p-|k|\nee,\si}+1$ and  define the function
\beqa
f_{|k|}(\nee)= \fr{1}{(1-\nee\cdot \hv_j )}\pa_{|k|}\bigg(\fr{ \chit_{[\si,\kappa)}(|k|)^2  }{-\nee\cdot \nabla_p E_{2,p-|k|\nee,\si}+1   }\bigg).
\eeqa
We note that 
\beqa
|f_{|k|}(\nee)|\leq c, \quad |\pa_{\theta}f_{|k|}(\nee)|\leq \fr{c}{\si^{\la_0}}.
\eeqa
Thus Lemma~\ref{two-angular-integration-lemma} gives
\beqa
| (\ref{two-second-term-estimate-theta})|\1&=&\1\bigg|\fr{1}{t}\int_{|k|\geq \si_{\S,t}} d|k| e^{-i|k| t}\int d\Omm(\nee) \, e^{-iE_{2,p-|k|\nee,\si} t} f_{|k|}(\nee)\bigg|\non\\
\1&\leq&\1 \fr{c}{\si^{\la_0}}  \fr{1}{t} \int_{\kappa\geq |k|\geq \si_{\S,t}} d|k|\bigg(\eps+\fr{1}{|k|t \eps}\bigg)\non\\
\1&\leq&\1  \fr{c}{\si^{\la_0}}  \bigg(\fr{\eps}{t}+\fr{|\log(\si_{\S,t}) |}{ t^2 \eps}\bigg)\non\\
\1&\leq &\1 \fr{c}{\si^{\la_0}}   \bigg( \fr{(\si^{\S}_t)^{\de}}{t}+\fr{1}{(\si^{\S}_t)^{1+\de} t^2}\bigg),
\eeqa
where we set $\eps=(\si^{\S}_t)^{\de}$, $\de$ as above and estimated $|\log(\si_{\S,t}) |\leq |\log (\si)|\leq c/\si^{\de_{\la_0}}$. We also 
estimated trivially $((\si^{\S}_t)^{\de})^{-1}\leq ((\si^{\S}_t)^{\de+1})^{-1}$ to obtain a bound similar to (\ref{similar-bound-phase}).  \qed

%%%%%%%%%%%%%%%%%%%%%%%%%%%%%%%%%%%%%%%%%%%%%%%%%%%%%%%%%%%%%%
\bel\label{two-angular-integration-lemma} Let $\nee=\nee(\theta,\phi)$ be a unit vector normal to the unit sphere and  consider a smooth function $f_{|k|}(\nee)=f_{|k|}(\theta, \phi)$
on a unit sphere depending on a parameter $\kappa\geq |k|>0$. For $p\neq 0$ and any $0<\eps\leq 1$
\beqa
|\int d\Omm(\nee)  \,  e^{-i (E_{2,p- |k| \nee,\si}) t}  f_{|k|}(\nee)|\leq c\bigg(\eps \sup_{\nee'}|f_{|k|}(\nee')| +\fr{1}{|k|t \eps}\sup_{\nee' }\sup_{\ell\in\{0,1\}}|\pa_{\theta}^{\ell}
f_{|k|}(\nee')|\bigg)
\eeqa 
(The spherical coordinates $(\theta,\phi)$ are chosen with $z$-axis in the direction of $p$).
\eel
%%%%%%%%%%%%%%%%%%%%%%%%%%%%%%%%%%%%%%%%%%%%%%%
\proof We consider the integral
\beqa
I_1=\int d\Omm(\nee)    e^{-i (E_{2,p- |k| \nee,\si}) t}  f_{|k|}(\nee)=\int_0^{2\pi}  d\phi \int_0^{\pi} d\theta \,  e^{-i (E_{2,p- |k| \nee(\theta,\phi),\si }) t}  
\sin(\theta)f_{|k|}(\theta,\phi).
 \eeqa
We introduce a parameter $0< \eps \leq 1$ and set
\beqa
I_1=\int_0^{2\pi}  d\phi \int_{\eps}^{\pi-\eps} d\theta \,  e^{-i (E_{2,p- |k| \nee(\theta,\phi),\si }) t}\sin(\theta)f_{|k|}(\theta,\phi)+R. \label{two-leading-term-R}
\eeqa
Clearly,
\beqa
|R|\leq c\eps \sup_{\nee'}|f_{|k|}(\nee')|. \label{two-rest-term-theta}
\eeqa
Now we note that
\beqa
\pa_{\theta} e^{-i (E_{2,p- |k| \nee(\theta,\phi),\si}) t}
= -ie^{-i (E_{2,p- |k| \nee(\theta,\phi),\si }) t} \big(\nabla E_{2,p- |k| \nee(\theta,\phi),\si}\cdot \pa_{\theta} \nee(\theta,\phi)\big) |k|t. \label{two-theta-int}
\eeqa
Before we integrate by parts w.r.t. $\theta$ in (\ref{two-leading-term-R}), we choose the direction of the $z$-axis parallel to $p$. By rotational invariance we can write $E_{2,q,\si}=\ti E(|q|)$ and $\nabla E_{2,q,\si}=\pa_{|q|} \ti E(|q|) q/|q|$,
with $|\nabla E_{2,q,\si}|=|\pa_{|q|} \ti E(|q|)|$. Thus we get
\beqa
\nabla E_{2,p- |k| \nee(\theta,\phi),\si}\cdot \pa_{\theta} \nee(\theta,\phi)=|\nabla E_{2,p- |k|\nee(\theta,\phi),\si }|
\fr{-|p| \sin(\theta)}{|p-|k|\nee(\theta,\phi)|}. \label{two-new-theta-int}
\eeqa
Now consider the function $g(|q|):=|\nabla E_{2,q,\si}|/|q|=|\pa_{|q|} \ti E(|q|) |/|q|$. 
Since $\nabla \ti E(0)=0$, we have
\beqa
g(|q|)=|\pa_{|\ti q|}^2\ti E(|\ti q|)|_{\ti q=q'}|  
\eeqa
where $q'$ is some point on the interval from $0$ to $q$.  
By (\ref{second-derivative-bound}), we have
(in the relevant region of momenta) 
$\pa^2_{|\ti q|}\ti E(|\ti q|)\geq \eps_1>0$ and therefore  $g(|q|)=\pa_{|q|} \ti E(|q|)/|q|$ (i.e. we could drop the absolute value from $\pa_{|q|} \ti E(|q|)$) and
\beqa
& &c_2\geq \fr{|\nabla E_{2,p- |k|\nee(\theta,\phi),\si }|}{|p- |k|\nee(\theta,\phi)|}\geq c_1, \quad c_1, c_2> 0, \label{c-one-c-two}\\
& &|\nabla E_{2,p- |k| \nee(\theta,\phi),\si}\cdot \pa_{\theta} \nee(\theta,\phi)|\geq c\eps
\eeqa
for  $\ka\geq |k|\geq 0 $ and $0<\eps\leq 1$ as above.  Denoting by $I_{1,1}$ the first integral on the r.h.s. of  (\ref{two-leading-term-R}), 
we obtain from (\ref{two-theta-int}), (\ref{two-new-theta-int}):
\begin{align}
I_{1,1}&:=\fr{1}{i|p||k| t}\int_0^{2\pi}  d\phi   \,  e^{-i (E_{2,p- |k| \nee(\theta,\phi),\si }) t}  
  \fr{ |p-|k|\nee(\theta,\phi)| f_{|k|}(\theta,\phi) }{ |\nabla E_{2,p- |k| \nee(\theta,\phi),\si }|    } \bigg|_{\theta=\eps}^{\theta=\pi-\eps}\label{two-g-function-one}\\
&\ph{44}-\fr{1}{i|p||k| t  }\int_0^{2\pi}  d\phi \int_{\theta=\eps}^{\theta=\pi-\eps}  d\theta \,  e^{-i (E_{2,p- |k| \nee(\theta,\phi),\si }) t}  
 \pa_{\theta}\bigg(\fr{ |p-|k|\nee(\theta,\phi)| f_{|k|}(\theta,\phi) }{ |\nabla E_{2,p- |k| \nee(\theta,\phi),\si}|    } \bigg). \label{two-g-function-two}
\end{align}
Thus we obtain from (\ref{c-one-c-two}) 
\beqa
|(\ref{two-g-function-one}) |\leq \fr{c}{|k| t }  \sup_{\nee'}  |f_{|k|}(\nee')|.
\eeqa
Term (\ref{two-g-function-two}) requires more careful analysis: We write $\pa:=\pa_{|q|}$ and  note that
\beqa
\pa (g(|q|))^{-1}=\fr{1}{\pa \ti E(|q|) }\bigg(1-\fr{|q|  }{\pa \ti E(|q|)}\pa^2 \ti E(|q|)  \bigg).
\eeqa
We recall from (\ref{c-one-c-two}) that $\pa \ti E(|q|)\geq c|q|$ for some $c>0$. We set $q=p- |k| \nee(\theta,\phi)$ and note that
\beqa
|q|^2=(|p|-|k|\cos(\theta))^2+|k|^2\sin^2(\theta)\geq |k|^2\sin^2(\theta).
\eeqa
Consequently,
\beqa
|\pa_{\theta}(g(|q|))^{-1}|=|k|\bigg|\pa (g(|q|))^{-1} \fr{q}{|q|}\cdot \pa_{\theta}\nee(\theta,\phi)\bigg|\leq \fr{c}{\sin(\theta)}.
\eeqa
This gives
\beqa
|(\ref{two-g-function-two})|\leq \fr{c}{|k| t \eps}  \sup_{\nee' }\sup_{\ell\in\{0,1\}}|\pa_{\theta}^{\ell}f_{|k|}(\nee')|,
\eeqa
which completes the proof. \qed.
%%%%%%%%%%%%%%%%%%%%%%%%%%%%%%%%%%%%%%%%%%%%%%%%%%%%%%%%

%%%%%%%%%%%%%%%%%%%%%%%%%%%%%%%%%%%%%%%%%%%%%%%%%%%%%%%%%
\section{Analysis of the phase II} \label{analysis-of-phase-II}
%%%%%%%%%%%%%%%%%%%%%%%%%%%%%%%%%%%%%%%%%%%%%%%%%%%%%%%%%%
\setcounter{equation}{0}

%%%%%%%%%%%%%%%%%%%%%%%%%%%%%%%%%%%%%%%%%%%%%%%%%%%%%%%%%%%%%%
\bep\label{Regular-phases-proposition} We set  
\begin{align}
A_{1,\si}&:=\int d^3p d^3k\,\vv_1^{\si}(k) \nn_1^*(p+k) F_{\si,\sn}(k) \nn_1(p),\\
F_{\si,\sn}(k)&:=
\fr{\vv_2^{\si}(k)}{|k|} \bigg(  \fr{e^{-i|k|\sn} }{(1-\hatk\cdot \hv_j ) }+\fr{e^{i|k|\sn} }{(1+\hatk\cdot \hv_j ) } \bigg):=f_{\si,\sn}(k)+\ov{f}_{\si,\sn}(-k).
\end{align}
Then, for any $0<\de<(\al^{-1}-1)$, $1\leq s\leq t$,
\beqa
\| [A_{1,\si}, \reta^*_{1, \si}(h_{1,\sn})]\reta^*_{2,\si}(h_{2,j,\sn}^{(\tn)} e^{i\ga_{\si}(\hv_j,\sn)})\Om\|^2\leq  
\fr{c}{\si^{\de_{\la_0}}}\bigg((\si^{\S}_{t})^{1+\alf}+t(\si^{\S}_t)^{3}+  \fr{(\si^{\S}_t)^{\de}}{t}+\fr{1}{(\si^{\S}_t)^{1+\de} t^2}  \bigg)^2\|h_{2,j}^{(s)}\|_2^2.\label{phases-estimate}
\eeqa
\eep
%%%%%%%%%%%%%%%%%%%%%%%%%%%%%%%%%%%%%%%%%%%%%%%%%%%%%%%%%%
\proof The argument follows very similar lines as the proof of  formula~(\ref{two-Regular-result})
so we only briefly explain the differences: First, we note that $h_{1,t}$ and 
$h_{2,j,\sn}^{(\tn)} e^{i\ga_{\si}(\hv_j,\sn)}$ are interchanged in (\ref{phases-estimate}) as compared to (\ref{two-Regular-result}). As a consequence, derivatives of $h_{2,j,\sn}^{(\tn)} e^{i\ga_{\si}(\hv_j,\sn)}$, responsible for the
time-dependence of $c_{t}$ in the error term of (\ref{two-Regular-result}), do not appear here and we simply obtain a constant $c$ on the r.h.s. of (\ref{phases-estimate}). Second, we have to justify the appearance
of the term $(\si^{\S}_{t})^{1+\alf}$ on the r.h.s. of (\ref{phases-estimate}). Its origin is the estimate
\beqa
\|  \reta^*_{1, \si}( \dot{\ga}^{{\mc (1)}}_{1,\si}(\hv_j,\sn)h_{1,\sn}   )  \reta^*_{2,\si}(h_{2,j,\sn}^{(\tn)} e^{i\ga_{\si}(\hv_j,\sn)})\Om\|  
\leq   \fr{  c }{\si^{\de_{\la}}} (\si^{\S}_{t})^{1+\alf}  \| h_{1}\|_2  \|h_{2,j}^{(s)}\|_2, \label{phase-non-cancellation}
\eeqa
where
\beqa
-\h\dot{\ga}^{{\mc (1)}}_{1,\si}(\hv_j,\sn)(p):=\int_{\si\leq |k|\leq \si_{\S,t}} d^3k\, \vv^\si_2(k) \vv_1^{\si}(k) \fr{1}{|k|(1-\hatk \cdot \hv_j) }\cos( (\nabla E_{1,p,\si}\cdot k-|k|)\sn),
\eeqa
and $ \si_{\S,t}:=\max\{ \si^{\S}_t, \si\, \}$.  (We note that the vector on the l.h.s of (\ref{phase-non-cancellation})
corresponds to the one on the r.h.s. of (\ref{phase-cancellation-term}). It is not estimated in the proof of Proposition~\ref{two-Regular-phases-proposition} as it is used for cancellation of the phase in the proof of
Theorem~\ref{non-diagonal-theorem}). The bound~(\ref{phase-non-cancellation}) follows from
\begin{align}
|\dot{\ga}^{{\mc (1)}}_{1,\si}(\hv_j,\sn)(p)|\leq c\int_{\si}^{ \si_{\S,t}  } d|k|\,  |k|^{\alf} & \leq c\bigg( ( \si_{\S,t})^{1+\alf}  -(\si)^{1+\alf}\bigg) \non\\
&\leq c\bigg( (\si^{\S}_{t} )^{(1+\alf)}-(\si)^{1+\alf}\bigg)\chi(\si^{\S}_{t} \geq \si)\leq c(\si^{\S}_{t} )^{(1+\alf)},
\end{align}
where $\chi$ is the characteristic function. \qed
%%%%%%%%%%%%%%%%%%%%%%%%%%%%%%%%%%%%%%%%%%%%%  
%%%%%%%%%%%%%%%%%%%%%%%%%%%%%%%%%%%%%%%%%
\section{Single particle states as asymptotic vacua of photons} \label{asymptotic-annihilation}
%%%%%%%%%%%%%%%%%%%%%%%%%%%%%%%%%%%%%%%%%%%
\setcounter{equation}{0}

%%%%%%%%%%%%%%%%%%%%%%%%%%%%%%%%%%%%%%%%%%%%%%%
\bep\label{Riemann-Lebesgue-proposition} Let $h\in L^2(\real^3)$, $\supp\, h\subset S$, and consider the single-particle states 
\beqa
\psi_{1/2,h,\si}=\Pi^*\int^{\oplus}d^3p\,  h(p) \psi_{1/2,p,\si}
\eeqa
for $\si>0$. Then, for any $g\in L^2(\real^3)$, supported outside of zero, we have
\beqa
\lim_{t\to\infty}a(g_t)e^{-iH_{\si}t}\psi_{1/2,h,\si}=0,
\eeqa
where $g_t(k):=e^{-i|k|t}g(k)$.
\eep
%%%%%%%%%%%%%%%%%%%%%%%%%%%%%%%%%%%%%%%%%%%%%%%%%
\proof We consider only the single-electron case, as the single-atom case is analogous. For convenience, we write
$E_{2,\si}(p):=E_{2,p,\si}$. First, we note that
\beqa
a(g_t)e^{-iH_{\si}t}\psi_{2,h,\si}=a(g_t)e^{-iE_{2,\si}(P)t}\psi_{2,h,\si},
\eeqa
where $P$ are the total momentum operators. Let us now decompose $\psi_{2,h,\si}$ into components with fixed photon number $n$, which we will denote by $\psi_{n}$. Their wave-functions are $\psi_n=\psi_n(k_1,\ldots,k_n;p)$, where $p$ is the bare electron's momentum. We have by a straigtforward computation
\beqa
\|a(g_t)\e^{-i H_{\si}t}\psi_{2,h,\si} \|^2=\sum_{n=1}^{\infty} n \int d^{3n}\la_n
\big|\int d^3k\, \e^{i(|k|-E_{2,\si}(k+\un\la_n))t}\bar g(k) \psi_n(k, \la_n)\big|^2. \label{Riemann-Lebesgue}
\eeqa
Here $\la_n=(k_2,\ldots,k_n;p)$ and $\un\la_n=k_2+\cdots+k_n+p$. We set $\Om_{\xi}(k):=|k|-E_{2,\si}(k+\xi)$. Since
$\supp\, h\subset S$ and $\supp\, g$ does not contain zero, it suffices to consider $\Om_{\xi}$ on the set 
$\De_{\xi}=\{\, k\in \real^3\,|\, k\neq 0, k+\xi\in S   \,\}$.   
We have e.g.  by Proposition 3.1 of \cite{DP12} 
\beqa
|\nabla\Om_{\xi}(k)|=\bigg|\fr{k}{|k|}-\nabla E_{2,\si}(k+\xi)\bigg|\geq \eps_1>0 \ \ \textrm{ for } \ \ k\in \De_{\xi},
\eeqa
since the velocity of the physical electron is strictly smaller than the velocity of light in the relevant region of parameters.
Thus, by the implicit function theorem, around every point  $k_0\in\De_{\un\la_n}$ we can find an open neighbourhood $U_{k_0}^n$ s.t. the map
\beqa
\phi_n(k^1,k^2, k^{3})=(q^1,q^2, q^{3})=(\Om_{\un\la_n}(k),k^2,k^3),
\eeqa
is invertible (and smooth) on $U_{k_0}^n$. Such sets $U_{k_0}^n$ form a covering of the compact set $\De_{\un\la_n}$ from which we can choose a finite sub-covering $U^n_1,\ldots, U^n_{m_n}$. Thus we can divide $\De_{\un{\la_n}}$ into a finite number of disjoint Borel sets  $K_i^n$, each of which is contained in some $U^n_j$. Hence, 
\beqa
(\ref{Riemann-Lebesgue})\1&=&\1\sum_{n=1}^{\infty} n \int d^{3n}\la_n \big|\sum_i\int_{K_i^n} d^3k\, \e^{i \Om_{\un\la_n}(k) t}\bar g(k) \psi_n(k, \la_n)\big|^2\non\\
\1&=&\1\sum_{n=1}^{\infty} n \int d^{3n}\la_n \big|\sum_i\int_{\phi_n(K_i^n)} d^3q\, \e^{i q_1t}\big|\mathrm{det}\fr{\pa \phi_n^{-1}(q)}{\pa q}\big|  \bar g(\phi_n^{-1}(q)) \psi_n(\phi_n^{-1}(q), \la_n)\big|^2.
\eeqa  
The determinant of the Jacobi matrix  is a bounded function by definition of $\De_{\un\la_n}$. The integrand is in $L^1(\real^{3})$ as can
be checked  by transforming back.  Thus the integral over $q$ tends to zero as $t\to\infty$ by the Riemann-Lebesgue lemma.
To enter with the limit under the summation and integration over $\la_n$, we use the dominated convergence theorem and the bound
\beqa
(\ref{Riemann-Lebesgue}) \1&\leq&\1\sum_{n=1}^{\infty} n \int d^{3n}\la_n \big(\int d^3k\, | g(k) \psi_n(k, \la_n)|\big)^2\non\\
\1&\leq&\1\sum_{n=1}^{\infty}n \int d^3k' |g(k')|^2\int d^3kd^{3n}\la_n |\psi_n(k, \la_n)|^2
\leq  \|g\|_2^2\lan \psi_{2,h,\si} ,N_{\pho}  \psi_{2,h,\si}\ran,
\eeqa
where $N_{\pho}$ is the photon number operator. Since $\si>0$, the last expression is finite as one can easily see from (\ref{simple-spectral-bound}).  \qed
%%%%%%%%%%%%%%%%%%%%%%%%%%%%%%%%%%%%%%%%%%%%%%%%%%%%%
\section{Shifting the infrared cut-off in the single-electron case}\label{Cut-off-variation}
%%%%%%%%%%%%%%%%%%%%%%%%%%%%%%%%%%%%%%%%%%%%%%%%%%%%%%

%%%%%%%%%%%%%%%%%%%%%%%%%%%%%%%%%%%%%%%%%%%%%%%%%%%%%%%%%%%%%%%%%%%%%%%%%%%%%%
\setcounter{equation}{0}

{In this appendix we provide the proof of the first estimate in (\ref{cut-off-shifts}). 
This estimate was first proven in  \cite{Pi05}. Here we adapt the presentation from \cite{CFP07}, where the argument from \cite{Pi05} was streamlined,
but a different model was used. For some steps we give a more detailed discussion than in \cite{CFP07}, especially in Step c) below. }

{\bc Specifically, we want to prove the estimate}
\begin{equation}\label{eq:IRchange-ineq-1}
	\| \, D2_{\mathrm{el}} \, \|
	\, \leq \, \mathcal{O} \big({\mc t_2^{\de_{\la_0}} } /t_1^{\rho}\big)  \,
\end{equation}
for some $\rho>0$, where
\beqa
D2_{\mathrm{el}}\1&:=&\1e^{iHt_1}\sum_{j=1}^{N(t_1)}\W_{\si_{t_2}}(\nv_j,t_1)\reta^*_{2,\si_{t_2}}(h_{2,j,t_1}^{(t_1)} e^{i\ga_{\si_{t_2}}(\nv_j,t_1)})\Om\label{first-vector-variation-cut-off-0}\\
\1& &\1-e^{iHt_1}\sum_{j=1}^{N(t_1)}\W_{\si_{t_1}}(\nv_j,t_1)
\reta^*_{2,\si_{t_1}}(h_{2,j,t_1}^{(t_1)} e^{i\ga_{\si_{t_1}}(\nv_j,t_1)})\Om.
\eeqa
{The proof of  estimate (\ref{eq:IRchange-ineq-1}) will require extensive use of the following spectral information, which we recall from Proposition~\ref{App-A-additional-spectral-information}: } 
\beqa
& & |\nabla E_{2,p,\si}-\nabla E_{2,p,\sip}|\leq c\sip^{1/4 }, \label{gradient-estimate-shift-cut-off}\\
& &\|\phi_{2,p,\si}-  \phi_{2,p,\sip}\|\leq c \sip^{1/4}, \label{phi-vector}\\
& &|\nabla E_{2,p,\si}-\nabla E_{2,p',\si}|\leq C|p-p'|,\label{holder-1}\\
& &\|\phi_{2,p,\si}-  \phi_{2,p-k,\si}\|\leq c |k|^{\frac{1}{16}}, \label{holder-2}\\
& & {\, \|b(k_1) \ldots b(k_m)\psi_{p,\sigma}\| \leq  (c\la)^m\, \frac{\chi_{[\sigma,\kappa]}(k_1)}{|k_1|^{3/2}}  \ldots  \frac{\chi_{[\sigma,\kappa]}(k_m)}{|k_m|^{3/2}},}
\label{res-formula}
\eeqa
{where $p,p'\in S$ and $0<\si\leq\sip \leq\kal$.}

For convenience of the notation, in the following discussion which involves only the electron, we drop the index $2$ in the electron wave- and energy function, i.e., $h_{2,j,t_1}^{(t_1)}\equiv e^{-iE_{p,\sigma_{t_2}}t_1} h_{j}^{(t_1)}$,  and we write
\begin{equation}\label{eq-II-86-x}
    	\reta^*_{2,\si}(h_{2,j,t}^{(t)} )\Om\,=
	{\Pi^*\int^{\oplus} \, d^3p\,  e^{-iE_{p,\sigma}t}  h_{j}^{(t)}(p) \, \psi_{p,\sigma}} =:e^{-iE_{p,\sigma}t}\psi_{j,\sigma}^{(t)},
\end{equation}
{where $\Pi:=Fe^{iP_{\pho}x}$,  cf. Lemma~\ref{fiber-ground-states-lemma}}. Furthermore, we will {often} write the electron form factor $\vv_2^{\si}(k):=\la \fr{\chi_{[\si,\kappa)}(k) }{(2|k|)^{1/2}}$ explicitly in the integrals, specifying the region of integration. { Also, we introduce the following dressing transformation
\beqa
W_{\si}(\nabla E_{p,\si}):=\Pi^* \int^{\oplus}d^3p'\, W_{p',\si} \Pi=\e^{-\int d^3k \, \vv_2^{\si}(k)  \fr{( e^{ik\cdot x} a(k)-  e^{-ik\cdot x}a^*(k))}{|k|(1-\hatk\cdot\nabla E_{p,\si} )}  }, 
\label{dressing-trafo-I}
\eeqa
where $W_{p',\si}$ was defined in (\ref{dressing-trafo-intro}). We note that in formula (\ref{dressing-trafo-I}) $p'\in \real^3$ denotes an integration variable,
while $p$ denotes the total momentum operator of the electron-photon system. This common notation will be used below without further notice.
 }

%%%%%%%%%%%%%%%%%%%%%%%%%%%%%%%%%%%%%%%%

The starting idea is to rewrite the term in (\ref{first-vector-variation-cut-off-0}),
\begin{equation}\label{eq-II-83}
	e^{iHt_1}\sum_{j=1}^{N(t_1)}\mathcal{W}_{\sigma_{t_2}}(\nv_j,t_1)
	\, e^{i\gamma_{\sigma_{t_2}}(\nv_j,t_1)}
	\, e^{-iE_{p,\sigma_{t_2}}t_1} \, \psi_{j,\sigma_{t_2}}^{(t_1)} \, ,
\end{equation}
as
\begin{equation}\label{eq-II-84}
	e^{iHt_1}\sum_{j=1}^{N(t_1)}\mathcal{W}_{\sigma_{t_2}}(\nv_j,t_1)
	\, W_{\sigma_{t_2}}^{*}(\nabla E_{p,\sigma_{t_2}})
	\, W_{\sigma_{t_2}}(\nabla E_{p,\sigma_{t_2}})
	\, e^{i\gamma_{\sigma_{t_2}}(\nv_j,t_1)}
	\, e^{-iE_{p,\sigma_{t_2}}t_1}
	\, \psi_{j,\sigma_{t_2}}^{(t_1)}\,,
\end{equation}
and to group the terms appearing in (\ref{eq-II-84}) in such a way that, cell by cell,
we consider the new {\em dressing operator}
\begin{equation}\label{eq-II-85}
    	e^{iHt_1} \, \mathcal{W}_{\sigma_{t_2}}(\nv_j,t_1)
	\, W_{\sigma_{t_2}}^{*}(\nabla E_{p,\sigma_{t_2}})
	\, e^{-iE_{p,\sigma_{t_2}}t_1}\,,
\end{equation}
which acts on
\begin{equation}\label{eq-II-86}
    	\Phi^{(t_1)}_{j,\sigma_{t_2}} \, := { \,\Pi^*
	\int^{\oplus}} h_{j}^{(t_1)}(p) \, \phi_{p,\sigma_{t_2}} \, d^3p \, ,
\end{equation}
where ${\phi_{p,\sigma}=W_{p,\sigma}\psi_{p,\si}}$. 
The key advantage is that the vector
$\Phi_{j,\sigma_{t_2}}^{(t_1)}$ inherits  the H\"older regularity
of $\phi_{p, \sigma}$; see (\ref{phi-vector}).
We will refer to (\ref{eq-II-86}) as an {\em infrared-regular vector}.

Accordingly, (\ref{eq-II-84}) now reads
\begin{equation}\label{eq-II-87}
    	e^{iHt_1}\sum_{j=1}^{N(t_1)}\mathcal{W}_{\sigma_{t_2}}(\nv_j,t_1)
	\, W_{\sigma_{t_2}}^{*}(\nabla E_{p,\sigma_{t_2}})
	\, e^{i\gamma_{\sigma_{t_2}}(\nv_j,t_1)}
	\, e^{-iE_{p,\sigma_{t_2}}t_1}
	\, \Phi_{j,\sigma_{t_2}}^{(t_1)}\,,
\end{equation}
and we proceed as follows.

%%%%%%%%%%%%%%%%%%%%%%%%%%%%%%%%%%%%%%%%%%%%%%%%%%
\subsection{Shifting the IR cutoff in the infrared-regular vector}
%%%%%%%%%%%%%%%%%%%%%%%%%%%%%%%%%%%%%%%%%%%%%%%%%%
%\\
%\\
First, we substitute
\begin{eqnarray}\label{eq-II-88}
    	\lefteqn{e^{iHt_1}\sum_{j=1}^{N(t_1)}\mathcal{W}_{\sigma_{t_2}}(\nv_j,t_1)W_{\sigma_{t_2}}^{*}(\nabla
    	E_{p,\sigma_{t_2}}) \,
	\underline{e^{i\gamma_{\sigma_{t_2}}(\nv_j,t_1)}
	e^{-iE_{p,\sigma_{t_2}}t_1}\Phi_{j,\sigma_{t_2}}^{(t_1)}}}\\
    	&\longrightarrow & e^{iHt_1}\sum_{j=1}^{N(t_1)}\mathcal{W}_{\sigma_{t_2}}(\nv_j,t_1)
	W_{\sigma_{t_2}}^{*}(\nabla E_{p,\sigma_{t_2}}) \,
	\underline{e^{i\gamma_{\sigma_{t_1}}(\nv_j,t_1)}
	e^{-iE_{p,\sigma_{t_1}}t_1}\Phi_{j,\sigma_{t_1}}^{(t_1)}} \, ,
	\nonumber\,
\end{eqnarray}
where $\sigma_{t_2}$ is replaced by $\sigma_{t_1}$ in the underlined terms.
We prove that the norm difference of these two vectors is bounded by the r.h.s. of (\ref{eq:IRchange-ineq-1}).
The necessary ingredients are:
\begin{itemize}
\item[1)]
The convergence, as $\sigma\to 0$, stated in (\ref{phi-vector}).
\item[2)]
The estimate
\begin{eqnarray} 
    |\gamma_{\sigma_{t_2}}(\nv_j,t_1)-\gamma_{\sigma_{t_1}}(\nv_j,t_1)| 
	\, \leq \,
    \mathcal{O}\big(\, \sigma_{t_1}^{{{   \frac{1}{4} } }}\,t_1^{2(1-\alpha)} \, \big)
	\, + \, \mathcal{O}\big( \, t_1 \,\sigma_{t_1} \,\big)  \, ,
    \nonumber
\end{eqnarray}
for $t_2>t_1\gg1$, proven in Lemma \ref{phases-shift-cut-off}. 
\item[3)]
The cell partition  $\Ga^{(t_1)}$ depends on $t_1<t_2$.
\item[4)]
The parameter $\gamma$ can be chosen arbitrarily large, independently of $\peps$,
so that the infrared cutoff $\sigma_{t_1}={\kal}t_1^{-\gamma}$ can be made as small as one wishes {for a given $t_1>
 1$}.
\end{itemize}
First of all, it is clear that the norm
difference of the two vectors in (\ref{eq-II-88}) is bounded by
the norm difference of the two underlined vectors, summed over all $N(t_1)$ cells.
Using 1) and 2), one straightforwardly derives that the norm difference between
the two underlined vectors in (\ref{eq-II-88}) is bounded from above by
\begin{equation}
    \cO(\,t_1\,\sigma_{t_1}^{\frac{1}{4}}\,t_1^{-\frac{3}{2}\peps}\,) \, ,
\end{equation}
where the last factor, $t_1^{-\frac{3}{2}\peps}$, accounts for the volume of an individual cell
in $\Ga^{(t_1)}$, by 3).
The sum over all cells in $\Ga^{(t_1)}$ yields a bound
\begin{equation}  \label{eq-II-89}
	\cO( \, N(t_1)\,\sigma_{t_1}^{\frac{1}{4}}\,t_1^{1-\frac{3}{2}\peps} \, )
\end{equation}
where $N(t_1) \leq t_1^{3\peps}$, by 3).
Picking $\gamma$ sufficiently large, by 4),  we find that the norm difference of the two
vectors in (\ref{eq-II-88})
is bounded by $t_1^{-\eta}$, for some $\eta>0$. This agrees with the bound stated in (\ref{eq:IRchange-ineq-1}).

%%%%%%%%%%%%%%%%%%%%%%%%%%%%%%%%%%%%%%%%%%%%%%%%%%%
\subsection{Shifting the IR cutoff in the dressing operator}
%%%%%%%%%%%%%%%%%%%%%%%%%%%%%%%%%%%%%%%%%%%%%%%%%%%

Subsequently to (\ref{eq-II-88}), we substitute
\begin{eqnarray}\label{eq-II-89bis}
    & &
    e^{iHt_1}\sum_{j=1}^{N(t_1)}\underline{\mathcal{W}_{\sigma_{t_2}}(\nv_j,t_1)W_{\sigma_{t_2}}^{*}(\nabla
    E_{p,\sigma_{t_2}})} e^{i\gamma_{\sigma_{t_1}}(\nv_j,t_1)}e^{-iE_{p,\sigma_{t_1}}t_1}\Phi_{j,\sigma_{t_1}}^{(t_1)}
    \\
    &\longrightarrow & e^{iHt_1}\sum_{j=1}^{N(t_1)}
    \underline{\mathcal{W}_{\sigma_{t_1}}(\nv_j,t_1)W_{\sigma_{t_1}}^{*}(\nabla
    E_{p,\sigma_{t_1}}}) e^{i\gamma_{\sigma_{t_1}}(\nv_j,t_1)}e^{-iE_{p,\sigma_{t_1}}t_1}\Phi_{j,\sigma_{t_1}}^{(t_1)}
    \nonumber\, ,
\end{eqnarray}
where $\sigma_{t_2}\rightarrow\sigma_{t_1}$ in the underlined operators.
A crucial point in our argument is that when $\sigma_{t_1}(>\sigma_{t_2})$
tends to $0$, the H\"older continuity of
$\phi_{p,\sigma_{t_1}}$ in $p$ offsets the (logarithmic)
divergence in $t_2$ which arises from the dressing operator.

We subdivide the shift $\sigma_{t_2}\rightarrow\sigma_{t_1}$ in
\begin{equation}\label{eq-II-90}
    \mathcal{W}_{\sigma_{t_2}}(\nv_j,t_1)W_{\sigma_{t_2}}^{*}(\nabla E_{p,\sigma_{t_2}})
    \, \longrightarrow \, \mathcal{W}_{\sigma_{t_1}}(\nv_j,t_1)W_{\sigma_{t_1}}^{*}(\nabla
    E_{p,\sigma_{t_1}})\,
\end{equation}
into the following three intermediate steps,
where the operators modified in each step are underlined:
\begin{itemize}
\item[]\underline{\em Step a)}
\begin{eqnarray}\label{eq-II-91}
    & &\underline{\mathcal{W}_{\sigma_{t_2}}(\nv_j,t_1)W_{\sigma_{t_2}}^{*}(\nv_j)}
    W_{\sigma_{t_2}}(\nv_j)W_{\sigma_{t_2}}^{*}(\nabla E_{p,\sigma_{t_2}})\\
    & &\longrightarrow \underline{\mathcal{W}_{\sigma_{t_1}}(\nv_j,t_1)W_{\sigma_{t_1}}^{*}(\nv_j)}
    W_{\sigma_{t_2}}(\nv_j)
    W_{\sigma_{t_2}}^{*}(\nabla E_{p,\sigma_{t_2}}) \nonumber\,,
\end{eqnarray}
\item[]\underline{\em Step b)}
\begin{eqnarray}\label{eq-II-92}
    & &\mathcal{W}_{\sigma_{t_1}}(\nv_j,t_1)W_{\sigma_{t_1}}^{*}(\nv_j)W_{\sigma_{t_2}}(\nv_j)
    \underline{W_{\sigma_{t_2}}^{*}(\nabla E_{p,\sigma_{t_2}})} \\
    & &\longrightarrow \mathcal{W}_{\sigma_{t_1}}(\nv_j,t_1)
    W_{\sigma_{t_1}}^{*}(\nv_j)W_{\sigma_{t_2}}(\nv_j)\underline{W_{\sigma_{t_2}}^{*}
    (\nabla E_{p,\sigma_{t_1}})}\nonumber\,,
\end{eqnarray}
\item[]\underline{\em Step c)}
\begin{eqnarray}\label{eq-II-93}
    & &\mathcal{W}_{\sigma_{t_1}}(\nv_j,t_1)W_{\sigma_{t_1}}^{*}(\nv_j)
    \underline{W_{\sigma_{t_2}}(\nv_j)W_{\sigma_{t_2}}^{*}(\nabla
    E_{p,\sigma_{t_1}})} \\
    & &\longrightarrow \mathcal{W}_{\sigma_{t_1}}(\nv_j,t_1)W_{\sigma_{t_1}}^{*}(\nv_j)
    \underline{W_{\sigma_{t_1}}(\nv_j)W_{\sigma_{t_1}}^{*}
    (\nabla E_{p,\sigma_{t_1}})}\nonumber\,.
\end{eqnarray}
\end{itemize}

\noindent
%%%%%%%%%%%%%%%%%%%%%%%%%%%%%%%%%%%%%%%%%%%%%%%%%%%
\emph{\underline{Analysis of Step a)}}
%%%%%%%%%%%%%%%%%%%%%%%%%%%%%%%%%%%%%%%%%%%%%%%%%%%
$\;$\\

\noindent
In step {\em a)}, we analyze the difference between the vectors
\begin{equation}\label{eq-II-93bis}
    e^{iHt_1}\underline{\mathcal{W}_{\sigma_{t_2}}(\nv_j,t_1)
    W_{\sigma_{t_2}}^{*}(\nv_j)}W_{\sigma_{t_2}}(\nv_j)W_{\sigma_{t_2}}^{*}(\nabla
    E_{p,\sigma_{t_2}})e^{i\gamma_{\sigma_{t_1}}(\nv_j,t_1)}
    e^{-iE_{p,\sigma_{t_1}}t_1}\Phi_{j,\sigma_{t_1}}^{(t_1)}
\end{equation}
and
\begin{equation}\label{eq-II-93bisbis}
    e^{iHt_1}\underline{\mathcal{W}_{\sigma_{t_1}}(\nv_j,t_1)W_{\sigma_{t_1}}^{*}(\nv_j)}
    W_{\sigma_{t_2}}(\nv_j)W_{\sigma_{t_2}}^{*}(\nabla E_{p,\sigma_{t_2}})
    e^{i\gamma_{\sigma_{t_1}}(\nv_j,t_1)}
    e^{-iE_{p,\sigma_{t_1}}t_1}\Phi_{j,\sigma_{t_1}}^{(t_1)}\, ,
\end{equation}
for each cell  in $\Ga^{(t_1)}$. 
Our goal is to prove that
\begin{equation}\label{eq-II-95bis}
    \| \, (\ref{eq-II-93bis}) - (\ref{eq-II-93bisbis}) \, \|
	\, \leq \, {c}\,  \ln(t_2) \,  P(t_1,t_2) \,,
\end{equation} 
where
\begin{eqnarray}\label{eq-II-96}
    P(t_1,t_2)\1& := &\1 \sup_{k \in\mathcal{B}_{\sigma_{t_1}}}\Big\| \, (e^{-i(|k|t_1-k\cdot x)}-1)
    W_{\sigma_{t_2}}(\nv_j)W_{\sigma_{t_2}}^{*}(\nabla
    E_{p,\sigma_{t_2}})) \, \times
   \\
    &&\quad\quad\quad\quad \times \, e^{i\gamma_{\sigma_{t_1}}(\nv_j,t_1)}e^{-iE_{p,\sigma_{t_1}}t_1}\Phi_{j,\sigma_{t_1}}^{(t_1)}
    \, \Big\|\, \leq \, \cO( \, t_1^{-\eta}\ln t_2 \, )
    \nonumber
\end{eqnarray}
as $t_1\to+\infty$, for some $\eta>0$,
and for $\gamma$ large enough.

Using the identity 
\begin{eqnarray}\label{eq-II-94}
    \lefteqn{
	\mathcal{W}_{\sigma_{t_2}}(\nv_j,t_1)W_{\sigma_{t_2}}^{*}(\nv_j)
    \, = \, \mathcal{W}_{\sigma_{t_1}}(\nv_j,t_1)W_{\sigma_{t_1}}^{*}(\nv_j) \, \times
	}
	\nonumber\\
    & &\ph{4444444444}\times \,
    \exp\Big(\frac{i\lambda^2}{2}\,\int_{\mathcal{B}_{\sigma_{t_1}}\setminus\mathcal{B}_{\sigma_{t_2}}}
    d^3k \,
    \frac{\sin(k\cdot x-|k|t_1)}
    {|k|^3(1-e_{k}\cdot\nv_j)^{2}}\Big)\times
	\quad\quad\quad\label{eq-II-95}\\
    & &\ph{4444444444}\times\,
    \exp\Big(\lambda\int_{\mathcal{B}_{\sigma_{t_1}}\setminus
    \mathcal{B}_{\sigma_{t_2}}} \frac{d^3k}{\sqrt{{2}|k|}} \frac{
	{ \{ a^*(k) e^{-ik\cdot x} }(e^{-i(|k|t_1-k\cdot x)}-1)-\,\mrm{h.c.}\,\rbrace}{|k|(1-e_{k}\cdot\nv_j)
      }\Big) \, ,\nonumber\,
\end{eqnarray} 
the difference between (\ref{eq-II-93bis}) and (\ref{eq-II-93bisbis}) is given by
\begin{eqnarray}
    \lefteqn{
    e^{iHt_1}\mathcal{W}_{\sigma_{t_1}}(\nv_j,t_1)W_{\sigma_{t_1}}^{*}(\nv_j)\,
    \,
    \exp\Big(\frac{i\lambda^2}{2}\,\int_{\mathcal{B}_{\sigma_{t_1}}\setminus\mathcal{B}_{\sigma_{t_2}}}
    d^3k \,
    \frac{\sin(k \cdot x-|k|t_1)}
    {|k|^3(1-e_{k}\cdot\nv_j)^{2} }\Big) \, \times
    }
    \nonumber\\
    & & \quad \quad \times \, \Big[ 
    \exp\Big(\lambda\int_{\mathcal{B}_{\sigma_{t_1}}\setminus
    \mathcal{B}_{\sigma_{t_2}}} \frac{d^3k}{\sqrt{{2}|k|}} \frac{  { a^*(k)e^{-ik\cdot x} }(e^{-i(|k|t_1-k\cdot x)}-1)-\mrm{h.c.}\,}{|k|(1-e_{k}\cdot\nv_j)}
    \Big)
    \, - \, {1} \Big]\times
    \nonumber \\
    & & \quad \quad \quad \quad \quad \quad \times \,
    W_{\sigma_{t_2}}(\nv_j)W_{\sigma_{t_2}}^{*}(\nabla
    E_{p,\sigma_{t_2}})\,e^{i\gamma_{\sigma_{t_1}}(\nv_j,t_1)}
    e^{-iE_{p,\sigma_{t_1}}t_1}\Phi_{j,\sigma_{t_1}}^{(t_1)}\,  \label{eq-II-96.1}
    \\
    & + &
    e^{iHt_1}\mathcal{W}_{\sigma_{t_1}}(\nv_j,t_1)W_{\sigma_{t_1}}^{*}(\nv_j)\,
    \Big[ \,
    \exp\Big(\frac{i\lambda}{2}\,\int_{\mathcal{B}_{\sigma_{t_1}}\setminus\mathcal{B}_{\sigma_{t_2}}}
    d^3k \,
    \frac{\sin(k\cdot x-|k|t_1)}
    {|k|^3(1-e_{k}\cdot\nv_j)}\Big) \,
    -1\Big]\times \nonumber \\
    & &
    \quad \quad \quad \quad \quad \quad \times \,
    W_{\sigma_{t_2}}(\nv_j)W_{\sigma_{t_2}}^{*}(\nabla
    E_{p, \sigma_{t_2}})\,e^{i\gamma_{\sigma_{t_1}}(\nv_j,t_1)}
    e^{-iE_{p,\sigma_{t_1}}t_1}\Phi_{j,\sigma_{t_1}}^{(t_1)}. \label{eq-II-96.1bis}
\end{eqnarray}

The norm of the vector (\ref{eq-II-96.1}) equals
\begin{eqnarray} 
    \Big\|\,\Big[
    \exp\big(\lambda\int_{\mathcal{B}_{\sigma_{t_1}}\setminus
    \mathcal{B}_{\sigma_{t_2}}} \frac{d^3k}{\sqrt{2|k|}} \frac{
    {a^*(k)e^{-ik\cdot x} } (e^{-i(|k|t_1-k\cdot x)}-\charf)-\,\mrm{h.c.}\,}{|k|(1-e_{k}\cdot\nv_j)}
    \big)-\charf \Big]\times 
    \nonumber \\ 
    \times \,W_{\sigma_{t_2}}(\nv_j)W_{\sigma_{t_2}}^{*}(\nabla E_{p,\sigma_{t_2}})
    e^{i\gamma_{\sigma_{t_1}}(\nv_j,t_1)}
    e^{-iE_{p,\sigma_{t_1}}t_1}\Phi_{j,\sigma_{t_1}}^{(t_1)}\Big\|\,.
    \quad\quad\quad
    \label{eq-II-96.2}
\end{eqnarray}

\noindent
\newcommand{\bks}{{a^*(k)e^{-ik\cdot x} }}
\newcommand{\bk}{{a(k)e^{ik\cdot x} }}

We now observe that
\begin{itemize}
\item
for $k\in\mathcal{B}_{\sigma_{t_1}}$,
\beqa
& & \bk\,W_{\sigma_{t_2}}(\nv_j)W_{\sigma_{t_2}}^{*}(\nabla
    E_{p,\sigma_{t_2}}) \,\non\\
 & &   = W_{\sigma_{t_2}}(\nv_j)W_{\sigma_{t_2}}^{*}(\nabla
    E_{p,\sigma_{t_2}}) \, \bk
    +\, W_{\sigma_{t_2}}(\nv_j)W_{\sigma_{t_2}}^{*}(\nabla
    E_{p,\sigma_{t_2}}) \, f_{k}(\nv_j,p)\,,
\eeqa
where
\begin{equation}
    \int_{\mathcal{B}_{\sigma_{t_1}}\setminus
    \mathcal{B}_{\sigma_{t_2}}}\,d^3k\,|f_{k}(\nv_j,p)|^2 \, \leq \, \cO(|\ln\sigma_{t_2}|)
\end{equation}
uniformly in $\nv_j$, and in $p\in {\mc S}$, and where $j$ enumerates the cells.
\item
for $k\in\mathcal{B}_{\sigma_{t_1}}$,
\begin{equation}
    \bk\,e^{i\gamma_{\sigma_{t_1}}(\nv_j,t_1)}e^{-iE_{p,\sigma_{t_1}}t_1}\Phi_{j,\sigma_{t_1}}^{(t_1)}=0\,,
\end{equation}
because of the infrared properties of $\Phi_{j,\sigma_{t_1}}^{(t_1)}$.
\end{itemize}

From the Schwarz inequality, we therefore get
\begin{equation} 
    (\ref{eq-II-96.2}) 
	\, \leq \,
	 c \, |\ln \sigma_{t_2}| \, P(t_1,t_2)
\end{equation}
for some finite constant $c$ as claimed in (\ref{eq-II-95bis}), where 
\begin{equation}\label{eq-II.96.2.1} 
    P(t_1,t_2) \, = \, \sup_{k\in\mathcal{B}_{\sigma_{t_1}}}
    \Big\| \, (e^{-i(|k|t_1-k\cdot x)}-1) \, 
    W_{\sigma_{t_2}}(\nv_j)
    W_{\sigma_{t_2}}^{*}(\nabla E_{p,\sigma_{t_2}})
    e^{i\gamma_{\sigma_{t_1}}(\nv_j,t_1)}e^{-iE_{p,\sigma_{t_1}}t_1}\Phi_{j,\sigma_{t_1}}^{(t_1)}
    \Big\|\,, 
\end{equation} 
as defined in (\ref{eq-II-96}).
To estimate $P(t_1,t_2)$, we regroup the terms inside the norm into
\begin{eqnarray}
    \lefteqn{ (e^{-i(|k|t_1-k\cdot x)}-\charf)
    W_{\sigma_{t_2}}(\nv_j)
    \, W_{\sigma_{t_2}}^{*}(\nabla E_{p,\sigma_{t_2}})
    \, e^{i\gamma_{\sigma_{t_1}}(\nv_j,t_1)}
    \, e^{-iE_{p,\sigma_{t_1}}t_1}
    \, \Phi_{j,\sigma_{t_1}}^{(t_1)}
    }
    \nonumber\\
    & &\1 = \1\; \;
    W_{\sigma_{t_2}}(\nv_j)
    \, W_{\sigma_{t_2}}^{*}(\nabla E_{p-k,\sigma_{t_2}})
    \, (e^{-i(|k|t_1-k\cdot x)}-\charf) 
    \label{eq-II-96.3} 
    \,e^{i\gamma_{\sigma_{t_1}}(\nv_j,t_1)}
    \, e^{-iE_{p,\sigma_{t_1}}t_1}
    \, \Phi_{j,\sigma_{t_1}}^{(t_1)} \quad 
	\\
    & & \; \; + \; W_{\sigma_{t_2}}(\nv_j)
    \, W_{\sigma_{t_2}}^{*}(\nabla E_{p-k,\sigma_{t_2}})
    \, e^{i\gamma_{\sigma_{t_1}}(\nv_j,t_1)}
    \, e^{-iE_{p,\sigma_{t_1}}t_1}
    \, \Phi_{j,\sigma_{t_1}}^{(t_1)}
    \label{eq-II-96.4}\\
    &  & \; \; - \; W_{\sigma_{t_2}}(\nv_j)
    \, W_{\sigma_{t_2}}^{*}(\nabla E_{p,\sigma_{t_2}})
    \, e^{i\gamma_{\sigma_{t_1}}(\nv_j,t_1)}
    \, e^{-iE_{p,\sigma_{t_1}}t_1}
    \, \Phi_{j,\sigma_{t_1}}^{(t_1)}\,.
    \label{eq-II-96.5}
\end{eqnarray}
In order to discuss the quantities above, it is important to recall the $p-$dependence of the phase factor through the operator $\nabla E_{p,\sigma}$, and it is useful to make our notation more trasparent by introducing the symbol
\begin{equation}
  \gamma_{\sigma}(\nv_j,\nabla E_{p,\sigma}, t)\equiv \gamma_{\sigma}(\nv_j,t)\,.
\end{equation}
Then, we prove that
\begin{equation}\label{eq-II-97}
    \|(\ref{eq-II-96.3})\|
    \; , \; \;
    \|(\ref{eq-II-96.4}){\mc +}(\ref{eq-II-96.5})\|
    \, \leq \, \cO( \, (\sigma_{t_1})^{\rho} \, t_1\, \ln t_2  \, )
\end{equation}
for some $\rho>0$, thanks to the following ingredients:
\begin{itemize}
\item[i)]
The H\"older regularity of $\phi_{p,\sigma}$ and $\nabla E_{p,\sigma}$
described under conditions (\ref{holder-1}) and (\ref{holder-2}).
\item[ii)]
The regularity of the phase function
\begin{equation}
    \gamma_{\sigma_{t_1}}(\nv_j,\nabla E_{p,\sigma_{t_1}}, t_1)
\end{equation}
with respect to $p\in \supp\, h \subset S$ expressed in the following estimate, which
is the content of Lemma~\ref{phases-shift-cut-off}: For $k \in \mathcal{B}_{\sigma_{t_1}}$ and $t_1$ large enough
\begin{equation}\label{eq-A-3bis}
    \big| \, \gamma_{\sigma_{t_1}}(\nv_j,\nabla E_{p,\sigma_{t_1}}, t_1)
    \, - \,
    \gamma_{\sigma_{t_1}}(\nv_j,\nabla E_{p-k,\sigma_{t_1}}t_1) \, \big|
    \, \leq \, \cO( |k| ) \,.
\end{equation}
\item[iii)]
The estimate
\begin{equation}\label{eq-II-97.1}
	\| \, b(k)\psi_{p,\sigma} \, \| \, \leq \, C \, \frac{\mathbf{1}_{[\sigma,\kappa]}(k)}{|k|^{3/2}}
\end{equation}
 from (\ref{res-formula}) for $p\in S$,
which implies
\begin{equation}\label{eq-II-97.2}
	\| \, N_{\pho}^{1/2} \, \psi_{p,\sigma} \, \|
	\, = \,
	\big( \, \int d^3k \, \| \,
	b(k) \, \psi_{p,\sigma} \, \|^2 \big)^{1/2}
	\, \leq \, C \, | \, \ln \sigma \, |^{1/2} \,. 
\end{equation}
Likewise,
\begin{eqnarray}\label{eq-II-97.3}
	\| \, N_{\pho}^{1/2} \, \phi_{p,\sigma} \, \|
	&\1 = \1&
	\big( \, \int_{\cB_{\kappa}\setminus\cB_\sigma} d^3k \, \big\| \,
	\big( \, b(k) \, + \, \cO( |k|^{-3/2}) \, \big) \,
	\psi_{p,\sigma} \, \big\|^2 \big)^{1/2}
	\nonumber\\
	&\1 \leq \1 & C \, | \, \ln \sigma \, |^{1/2} \,,
\end{eqnarray}
which controls the expected photon number in the states
$\lbrace\phi_{p,\sigma_{t_1}}\rbrace$.
As a side remark, we note that the true size is in fact $\cO(1)$, uniformly in $\sigma$,
but the logarithmically divergent bound here is sufficient for our purposes.
\item[iv)]
The cell decomposition $\Ga^{(t_1)}$ is determined by
$1<t_1<t_2$. Moreover, since $\gamma(>4)$ can be chosen
arbitrarily large and independent of $\peps$, the cut-off $\sigma_{t_1}=\kal t_1^{-\gamma}$ can be made
as small as desired.
\end{itemize}

We first prove the bound on $\|(\ref{eq-II-96.3})\|$ stated
in  (\ref{eq-II-97}). To this end, we use
\begin{eqnarray}
    \lefteqn{
    \big( \, e^{-i(|k|t_1-k\cdot x)}-\charf \, \big)
    \, e^{i\gamma_{\sigma_{t_1}}(\nv_j, \nabla E_{p,\sigma_{t_1}}t_1)}
    \, e^{-iE_{p,\sigma_{t_1}}t_1}
    \, \Phi_{j,\sigma_{t_1}}^{(t_1)}
    \,
    }
    \\
    \1& &\1= 
    e^{i\gamma_{\sigma_{t_2}}(\nv_j, \nabla E_{p-k, \sigma_{t_1}},  t_1)}
    \, e^{-iE_{p-k, \sigma_{t_1}}t_1}(e^{-i(|k|t_1-k\cdot x)}-\charf)
    \, \Phi_{j,\sigma_{t_1}}^{(t_1)}
    \label{eq-II-97.1.1}\\
    & & \; \; + \;
    e^{i\gamma_{\sigma_{t_2}}(\nv_j,\nabla E_{p-k,\sigma_{t_1}},t_1)}
    \, e^{-iE_{p-k, \sigma_{t_1}}t_1}
    \, \Phi_{j,\sigma_{t_1}}^{(t_1)}
    \label{eq-II-97.1.2}\\
    & & \; \; - \;
    e^{i\gamma_{\sigma_{t_2}}(\nv_j,\nabla E_{p,\sigma_{t_1}},t_1)}
    \, e^{-iE_{p,\sigma_{t_1}}t_1}
    \, \Phi_{j,\sigma_{t_1}}^{(t_1)} \,.
    \label{eq-II-97.1.3}
\end{eqnarray}
{\bc Lemma~\ref{I-lemma-tx} }  yields
\begin{equation}
    	\| \, (\ref{eq-II-97.1.1}) \, \| \,
	\leq \, \cO( \,t_1\,\sigma_{t_1} \,t_1^{-\frac{3\peps}{2}} \,).
\end{equation}

The H\"older continuity of $E_{p,\sigma_{t_1}}$ and
$\nabla E_{p,\sigma_{t_1}}$, again from i), combined
with  ii), %  {\bc\cancel{ with $\alpha$ sufficiently close to $1$,}}  
implies that, with $k\in\cB_{\sigma_{t_1}}$,
\begin{equation}
    \| \, (\ref{eq-II-97.1.2}){\mc +}(\ref{eq-II-97.1.3}) \, \| \, \leq \,
    \cO( \, t_1\,\sigma_{t_1}   t_1^{-\frac{3\peps}{2}} \, )\,,
\end{equation}
as desired.
\\

To prove the bound on $\| \, (\ref{eq-II-96.4}){\mc +}(\ref{eq-II-96.5}) \, \|$
asserted in  (\ref{eq-II-97}), we write
\begin{equation}
    W_{\sigma_{t_2}}^{*}(\nabla
    E_{p-k,\sigma_{t_2}})-W_{\sigma_{t_2}}^{*}(\nabla
    E_{p,\sigma_{t_2}}) \, = \, W_{\sigma_{t_2}}^{*}(\nabla
    E_{p,\sigma_{t_2}})(W_{\sigma_{t_2}}^{*}(\nabla
    E_{p,-k, \sigma_{t_2}} \, ; \, \nabla
    E_{p, \sigma_{t_2}})-\charf )\,,
\end{equation}
where
\begin{eqnarray*}
	W^*_{\sigma_{t_2}}(\nabla E_{p-k,\sigma_{t_2}} \, ; \,  \nabla
	E_{p,\sigma_{t_2}})  
	\, := \, W_{\sigma_{t_2}}(\nabla E_{p, \sigma_{t_2}}) \,  W^*_{\sigma_{t_2}}(\nabla
	E_{p-k,\sigma_{t_2}}) \,,
\end{eqnarray*}
and apply the Schwarz inequality in the form
\begin{eqnarray}\label{eq-97bisbisbis}
	\lefteqn{
	\Big\| \, (W_{\sigma_{t_2}}^{*}(\nabla
    E_{{\mc p-k},\sigma_{t_2}} \, ; \, \nabla
    E_{p,\sigma_{t_2}})-\charf ) \, \widetilde \Phi \, \Big\|
    }
    \\
    &\leq& c \, \bigg( \, \int_{ { \cB_{\ka} } \backslash \cB_{\sigma_{t_2}}} \frac{ d^3 q }{|q|^3}\, \bigg)^{1/2}
   \,\sup_{p\in \supp\, h,\,k \in \mathcal{B}_{\sigma_{t_1}}} \big| \, \nabla E_{p-k, \sigma_{t_2}}-\nabla E_{p,\sigma_{t_2}} \, \big|
    \, \| \, N_{\pho}^{1/2} \widetilde \Phi\, \|,
    \nonumber
\end{eqnarray}
where in our case, $\widetilde\Phi\equiv e^{i\gamma_{\sigma_{t_1}}(\nv_j,\nabla
    E_{p,\sigma_{t_1}},t_1)}e^{-iE_{p,\sigma_{t_1}}t_1}
    \, \Phi_{j,\sigma_{t_1}}^{(t_1)}$.
We have
\begin{equation}\label{eq-97bisbis}
    \| N_{\pho}^{1/2} \,{\phi_{p, \sigma_{t_1}} }\|\,\leq\, c \,|\ln\sigma_{t_1}|^{1/2}
    \, \leq \, c' \, ( \, \ln t_1 \, )^{1/2} \,,
\end{equation}
as a consequence of iii). Due to i),
\begin{equation}
    \sup_{p\in \supp\,h,\, k \in \mathcal{B}_{\sigma_{t_1}}} \,
    \big| \, \nabla E_{p-k,\sigma_{t_2}} \, - \, \nabla E_{p,\sigma_{t_2}}  \, \big|
    \, \leq \, \cO( \, \sigma_{t_1}  \, )\,.
\end{equation}
Therefore,
$$
    \sup_{k\in\mathcal{B}_{\sigma_{t_1}}}\,
    \big\| \, (\ref{eq-II-96.4}){\mc +}(\ref{eq-II-96.5}) \, \big\|\,
    \leq \, \cO((\ln t_{2}) \,(\sigma_{t_1})^{\rho'})
$$
for some $\rho'>0$ which does not depend on $\peps$ (recalling that ${1<t_1}<t_2$). {This concludes the
analysis of (\ref{eq-II-96.1}).  Contribution~(\ref{eq-II-96.1bis}) can be treated in a similar  way. }
\\

We may now return to (\ref{eq-II-95bis}). From iv), and the fact that
the number of cells is $N(t_1) \leq t_1^{3\peps}$, summation over all cells yields
\begin{equation}
    \sum_{j=1}^{N(t_1)}\big\| \, (\ref{eq-II-93bis}) \, - \, (\ref{eq-II-93bisbis}) \, \big\|
    \, \leq \, \cO(\frac{{\mc |\ln(t_2)|^2}}{t_1^\rho})
    \label{eq-II-98}
\end{equation}
for some $\rho>0$, provided that $\gamma$ is sufficiently large.
This agrees with (\ref{eq:IRchange-ineq-1}). \\

\noindent
%%%%%%%%%%%%%%%%%%%%%%%%%%%%%%%%%%%%%%%%%%%%%%%%%%%
\emph{\underline{Analysis of Step b)}}
%%%%%%%%%%%%%%%%%%%%%%%%%%%%%%%%%%%%%%%%%%%%%%%%%%%
$\;$\\
\\
To show that the norm difference of the two vectors
corresponding to the change (\ref{eq-II-92}) in (\ref{eq-II-89bis})
is bounded by the r.h.s. of  (\ref{eq:IRchange-ineq-1}), we argue similarly as for step {\em a)},
and we shall not reiterate the details.
One again uses properties i) -- iv) as in step {\em a)}. 
\\

\noindent
%%%%%%%%%%%%%%%%%%%%%%%%%%%%%%%%%%%%%%%%%%%%%%%%%%%
\emph{\underline{Analysis of Step c)}}
%%%%%%%%%%%%%%%%%%%%%%%%%%%%%%%%%%%%%%%%%%%%%%%%%%%
$\;$\\
\\
Finally, we prove that the difference of the vectors corresponding to (\ref{eq-II-93}) satisfies 
\begin{eqnarray}\label{eq-II-99}
	&&\Big\| \, \sum_{j=1}^{N(t_1)}\mathcal{W}_{\sigma_{t_1}}(\nv_j,t_1)
	\big[W|_{\sigma_{t_2}}^{\sigma_{t_1}}(\nv_j) W^{*}|_{\sigma_{t_2}}^{\sigma_{t_1}}
	(\nabla E_{p,\sigma_{t_1}})-\charf \big] 
	\, e^{i\gamma_{\sigma_{t_1}}(\nv_j,\nabla
  	E_{p,\sigma_{t_1}},t_1)}
	e^{-iE_{p,\sigma_{t_1}}t_1} { \psi_{j,\sigma_{t_1}}^{(t_1)} } \, \Big\|^2
	\nonumber\\
	&&\quad\quad\quad\quad\quad\quad\, \leq \, \cO\big( {\mc t_2^{\de_{\la_0}}} /t_1^{\rho}\big) \,,
\end{eqnarray}
where we define
\begin{eqnarray}
    W|_{\sigma_{t_2}}^{\sigma_{t_1}}(\nv_j)
    \, := \, W^{*}_{\sigma_{t_1}}(\nv_j)W_{\sigma_{t_2}}(\nv_j)\,,
    \label{eq-II-100}
\end{eqnarray}
and likewise,
\begin{eqnarray}
    W^{*}|_{\sigma_{t_2}}^{\sigma_{t_1}}(\nabla E_{p,\sigma_{t_1}})
    \, := \,
    W_{\sigma_{t_2}}^{*}(\nabla E_{p,\sigma_{t_1}})
    \, W_{\sigma_{t_1}}(\nabla E_{p,\sigma_{t_1}})\,.
    \label{eq-II-101}
\end{eqnarray}
We separately discuss the diagonal and off-diagonal contributions
to  (\ref{eq-II-99}) from the sum over cells in $\Ga^{(t_1)}$.
\\

\noindent
$\bullet$ {\em The diagonal terms in  (\ref{eq-II-99}).}
\\
\\
To bound the diagonal terms in (\ref{eq-II-99}),  
we use that,
with $\nv_j\equiv \nabla E_{p,\sigma_{t_1}}|_{p=p_{j(\ov{n}) } }$, {where $p_{j(\ov{n})}$ is the position of the center of the $j$-th cell in the $\ov{n}$-th partition, 
the expression} 
\begin{equation}
    W|_{\sigma_{t_2}}^{\sigma_{t_1}}(\nv_j \ ; \, \nabla E_{p,\sigma_{t_1}})
    \, := \,
    W|_{\sigma_{t_2}}^{\sigma_{t_1}}(\nv_j)
    W^{*}|_{\sigma_{t_2}}^{\sigma_{t_1}}(\nabla
    E_{p,\sigma_{t_1}})  \, \label{W-W-combination-I}
\end{equation}
allows for an estimate similar to (\ref{eq-97bisbisbis}),
where we now use that
\begin{equation}\label{eq-II-102}
    \sup_{p\in \Ga^{(t_1)}_{j}}|\nabla E_{p,\sigma_{t_1}}-\nv_j|
    \, \leq \, \cO( \, t_1^{-{\peps} } \, ) \,.
\end{equation}
The latter follows from the H\"older regularity of $\nabla E_{p,\sigma}$,
due to  (\ref{holder-1}). 
Moreover, we use (\ref{eq-97bisbis}) to bound the expected photon number in the states
{$\psi_{j, \sigma_{t_1}}^{(t_1)}$}. 

Hereby, we find that the sum of diagonal terms can be bounded by
\begin{equation}
    \cO( \, N(t_1)\,\|h_{j}^{(t_1)}\|_2^2 \, (\ln t_2)
    \, {\mc \big(t_1^{-\peps}\big)^2} \, )
    \, \leq \, \cO( \,  t_1^{-\rho} \, \ln t_2\, )
\end{equation}
for some $\rho>0$, using $N(t_1)=\cO(t_1^{3\peps})$,
and $\|h_{j}^{(t_1)}\|^2_2=\cO( t_1^{-3\peps})$.
\\

\noindent
$\bullet$ {\em The off-diagonal terms in  (\ref{eq-II-99}).}
\\
\\
Next, we bound the off-diagonal terms in (\ref{eq-II-99}), corresponding
to the inner product of vectors supported on cells $j\neq l$
of the partition $\Ga^{(t_1)}$.
Those are similar to the off-diagonal terms $\hat{M}_{l,j}(\si, s,t)$
in (\ref{off-diag}) that were discussed in detail previously.
Correspondingly,
we can apply the methods discussed in Section {\ref{clustering-subsection}}, up to
some modifications which we explain now.

Our goal is to prove the asymptotic orthogonality of the off-diagonal terms in  (\ref{eq-II-99}). 

 First of all we prove the auxiliary result, {for  functions $h^{\si_{t_1}}_{l,j,s}$ defined in (\ref{h-photon-def}), and supported at $|k|\geq \si_{t_1}$,} 
\begin{equation}\label{eq-II-103}
    \lim_{s\to+\infty}\| \, a({ {h}^{\si_{t_1}}_{l,j,s} } )\,
    W^{*}|_{\sigma_{t_2}}^{\sigma_{t_1}}(\nabla
    E_{p,\sigma_{t_1}})e^{-iH_{\sigma_{t_1}}s   }\psi_{j,\sigma_{t_1}}^{(t_1)} \, \| \, = \, 0\,.
\end{equation}
To this end, we compare
\begin{equation}\label{eq-II-104}
    W^{*}|_{\sigma_{t_2}}^{\sigma_{t_1}}(\nabla
    E_{p,\sigma_{t_1}})\mathbf{1}_{\Ga^{(t_1)}_j}(p) \,
\end{equation}
(where $\mathbf{1}_{\Ga^{(t_1)}_j}$ is the characteristic function of the cell
$\Ga^{(t_1)}_j$) to its discretization: 
\begin{itemize}
\item[{\em 1.}]
We pick $\bar{t}$ large enough such that
$\Ga^{(\bar{t})}$ is a sub-partition of $\Ga^{{\mc (t_1)}}$; in
particular,
$\Ga^{{\mc (t_1)}}_j=\sum_{ {l(j)}=1}^{M}\Ga^{(\bar{t})}_{l(j)}$,
where $M=\frac{N(\bar{t})}{N({\mc t_1})}$.
\item[{\em 2.}]
Furthermore, defining $u_{l(j)}\, :=\, \nabla
E_{p_{l  (j)(\ov{n}(\bar t))}, \sigma_{t_1}   }$, where { $p_{l(j)(\ov{n}(\bar t)) } $} is the center
of the cell $\Ga^{(\bar{t})}_{l(j)}$, we have, for $p\in \Ga^{(\bar{t})}_{l(j)}$,
\begin{equation}\label{eq-II-105}
    |u_{l(j)}-\nabla E_{p,\sigma_{t_1}}| \, \leq \,
    c \,\bigg(\frac{1}{\bar{t}}\bigg)^{\peps}\, ,
\end{equation}
where $c$ is uniform in $t_1$.  
\item[{\em 3.}]
We define
\begin{equation}\label{eq-II-106}
    \mathbb{W}_{\sigma_{t_2}}^{\sigma_{t_1}}(M)\,:=\,
    \sum_{l(j)=1}^{M}W^{*}|_{\sigma_{t_2}}^{\sigma_{t_1}}(u_{l(j)})\,
    \one_{\Ga^{(\bar{t})}_{l(j)}}(p)\,
\end{equation}
\noindent
and rewrite the vector
\begin{equation}\label{eq-II-107}
    a(h^{\si_{t_1}}_{l,j,s}  )\,W^{*}|_{\sigma_{t_2}}^{\sigma_{t_1}}
    (\nabla E_{p,\sigma_{t_1}})e^{-iH_{\sigma_{t_1}}s}\psi_{j,\sigma_{t_1}}^{(t_1)}
\end{equation}
in (\ref{eq-II-103}) as 
\begin{eqnarray}
    & &\int_{\mathcal{B}_{{\ka} }\setminus\mathcal{B}_{\sigma_{t_1}}}
    d^3k \, e^{-i k\cdot x}\,\big[W^{*}|_{\sigma_{t_2}}^{\sigma_{t_1}}
    (\nabla E_{p,\sigma_{t_1}})-\mathbb{W}_{\sigma_{t_2}}^{\sigma_{t_1}}(M)\big] \times\quad\quad\quad\quad\non\\
    & &\quad\quad\quad\quad\quad\quad\times \, \ov{h}^{\si_{t_1}}_{l,j}(k) {e^{ikx}}  a(k)  e^{i|k|s} \, e^{-iE_{p,\sigma_{t_1}}s}\psi_{j,\sigma_{t_1}}^{(t_1)}
    \label{eq-II-108}  \\
    \!&+&\1\sum_{m(j)=1}^{M}W^{*}|_{\sigma_{t_2}}^{\sigma_{t_1}}(u_{m(j)})
    \int_{\mathcal{B}_{{\ka}}\setminus\mathcal{B}_{\sigma_{t_1}}}\,
    d^3k \, \ov{h}^{\si_{t_1}}_{l,j}(k)a(k)e^{i|k|s} % \, \times
   % \nonumber\\
   % & &\quad\quad\quad\quad\quad\quad \times \,
    e^{-iE_{p,\sigma_{t_1}}s}\psi_{l(j),\sigma_{t_1}}^{(\bar{t})}. \quad  \quad\,\label{eq-II-109}
\end{eqnarray}
\end{itemize}
{The reason for this last step is that $a(h^{\si_{t_1}}_{l,j,s} )$ does not commute with   $W^{*}|_{\sigma_{t_2}}^{\sigma_{t_1}} (\nabla E_{p,\sigma_{t_1}})$, due to the presence of the total momentum operator $p$, but it does commute with   $  W^{*}|_{\sigma_{t_2}}^{\sigma_{t_1}}(u_{l(j)})$.  We also note
that the vector $\psi_{l(j),\sigma_{t_1}}^{(\bar{t})}$ in (\ref{eq-II-109}) slightly differs from definition~(\ref{eq-II-86-x}), since we have used a sharp sub-partition here. 
Clearly, expression (\ref{eq-II-109}) tends to zero, as $s\to\infty$, at any fixed $\bar{t}$,  by Proposition~\ref{Riemann-Lebesgue-proposition}. Thus, to conclude
the proof of (\ref{eq-II-103}), it suffices to show that (\ref{eq-II-108}) tends to zero as $\bar{t}\to \infty$ uniformly in $s$, at fixed   $t_1, t_2$. To this end one 
first estimates
\beqa
\|(\ref{eq-II-108})\|\1&\leq&\1 \int_{\mathcal{B}_{{\ka} }\setminus\mathcal{B}_{\sigma_{t_1}}}
    d^3k \, \|\sum_{l(j)=1}^{M} \, \one_{\Ga^{(\bar{t})}_{l(j)}}(p)  \big[W^{*}|_{\sigma_{t_2}}^{\sigma_{t_1}}
   (\nabla E_{p,\sigma_{t_1}})-  W^{*}|_{\sigma_{t_2}}^{\sigma_{t_1}}(u_{l(j)})    \big]   \times\quad\quad\quad\quad\non\\
    & &\quad\quad\quad\quad\quad\quad\times \, \ov{h}^{\si_{t_1}}_{l,j}(k) {e^{ikx}}{a(k)} e^{i|k|s} \, e^{-iE_{p,\sigma_{t_1}}s}
    \psi_{l(j),\sigma_{t_1}}^{(\bar{t})}\|.
\eeqa
It is important for further analysis that the summands under the norm above are orthogonal vectors. Thus we can estimate the square of this
norm as  follows
\beqa
\|\ldots\|^2\1&\leq &\1\sum_{l(j)=1}^{M}2 \| (a(h_{\nabla E_{p,\sigma_{t_1}}, u_{l(j)}  }) 
\ov{h}^{\si_{t_1}}_{l,j}(k) {e^{ikx}}{a(k)} e^{i|k|s} \, e^{-iE_{p,\sigma_{t_1}}s}
    \psi_{l(j),\sigma_{t_1}}^{(\bar{t})}\|^2\label{first-additional-comp-I}\\
\1&  &\1+ \sum_{l(j)=1}^{M} \|  h_{\nabla E_{p,\sigma_{t_1}}, u_{l(j)}  }\|^2_2\|  \ov{h}^{\si_{t_1}}_{l,j}(k) {e^{ikx}}{a(k) } e^{i|k|s} \, e^{-iE_{p,\sigma_{t_1}}s}
    \psi_{l(j),\sigma_{t_1}}^{(\bar{t})}\|^2, \label{second-additional-comp-I}   
     \eeqa
where   
 \begin{align}
h_{\nabla E_{p,\sigma_{t_1}}, u_{l(j)}  }(k):=\fr{\chi_{[\si_{t_2}, \si_{t_1}]}(k) f_{ \nabla E_{p,\sigma_{t_1}} ,   u_{l(j)}   }(e_k)}{\sqrt{2|k|}|k|}e^{-ik\cdot x}, \quad
f_{  \nabla E_{p,\sigma_{t_1}}   ,  u_{l(j)}   }(e_k):= 
\fr{ \hatk\cdot( \nabla E_{p,\sigma_{t_1}}  -   u_{l(j)}   )}{(1-\hatk\cdot  \nabla E_{p,\sigma_{t_1}} ) (1-\hatk\cdot   u_{l(j)}   )}.
\end{align}
As for~(\ref{first-additional-comp-I}), we can write
\beqa
 (\ref{first-additional-comp-I})  \1&\leq&\1 c\sum_{l(j)=1}^{M} \|h^{(\bar{t})}_{l(j)}\|_2^2 \bigg(\int d^3k' \, |h_{\nabla E_{p,\sigma_{t_1}}, u_{l(j)}  }(k')|\, 
 \fr{\chi_{[\si_{t_2}, \ka]} (k') }{|k'|^{3/2}} |\ov{h}^{\si_{t_1}}_{l,j}(k)|   \fr{\chi_{[\si_{t_2}, \ka]} (k) }{|k|^{3/2}} \bigg)^2 \non\\
 \1 & \leq & \1 c (\ov{t})^{-2\peps} (\si_{t_2}^{-6} \si_{t_1}^3)^2, \label{t-bar-decay}
\eeqa  
where in the first step we used (\ref{res-formula}) and in the last step we exploited (\ref{eq-II-105}). We also noted that $M\leq \ov{t}^{3\peps}$ is
compensated by $\|h^{(\bar{t})}_{l(j)}\|_2^2\leq c t^{-3\peps}$. (Here $h^{(\bar{t})}_{l(j)}$ involves a sharp sub-partition).  It is manifest that (\ref{t-bar-decay})
tends to zero as $\bar{t}\to \infty$ and $t_1,t_2$ are kept fixed. (\ref{second-additional-comp-I})
 is analysed analogously. } This proves (\ref{eq-II-103}).

The main difference between  (\ref{eq-II-99}) and the similar expression 
in Proposition \ref{off-diagonal-thm-one} 
is the operator
\begin{equation}\label{eq-II-111}
    W|_{\sigma_{t_2}}^{\sigma_{t_1}}(\nv_j)W^{*}|_{\sigma_{t_2}}^{\sigma_{t_1}}(\nabla E_{p,\sigma_{t_1}}) \,.
\end{equation}
 To control it,
we first note that the Hamiltonian 
\begin{equation}\label{eq-II-112}
\widehat{H}_{\sigma_{t_1}}:={\Pi^*}\bigg(\int^{\oplus}\widehat{H}_{p,\sigma_{t_1}}d^3p\bigg)\Pi,
\end{equation}
where
\begin{equation}\label{eq-II-113}
\widehat{H}_{p,\sigma_{t_1}}\, := \,
\frac{\big(p- {P_{\pho}|^{\ka}_{\sigma_{t_1} }  }\big)^2}{2}  + H_{\pho}|^{\ka}_{\sigma_{t_1}} \; +\la\;\int_{\mathbb{R}^3\setminus\mathcal{B}_{\sigma_{t_1}}} d^3k\, \fr{\chi_{\kappa}(k) }{(2|k|)^{1/2}}\, (b(k)+b^*(k) ) 
\end{equation}
with
\begin{equation}\label{eq-II-114}
    P_{\pho}  |^{\ka}_{\sigma_{t_1}}:=\int_{\mathbb{R}^3\setminus\mathcal{B}_{\sigma_{t_1}}}k\,
    b^*(k)b(k)\,d^3k \,,\quad  H_{\pho}|^{\ka}_{\sigma_{t_1}}:=\int_{\mathbb{R}^3\setminus\mathcal{B}_{\sigma_{t_1}}}|k|\,
    b^*(k)b(k)\,d^3k \,
    \end{equation}
satisfies
\begin{equation}\label{eq-II-116}
	\widehat{H}_{p,\sigma_{t_1}}\,\psi_{p,\sigma_{t_1}}
	\,=\,E_{p,\sigma_{t_1}}\,\psi_{p,\sigma_{t_1}},
\end{equation}
and
\begin{equation}\label{eq-II-117}
[W|_{\sigma_{t_2}}^{\sigma_{t_1}}(\nv_j)W^{*}|_{\sigma_{t_2}}^{\sigma_{t_1}}(\nabla
E_{p,\sigma_{t_1}})\,,\,\widehat{H}_{\sigma_{t_1}}]=0.
\end{equation}
{ We consider the vector in {\mc (\ref{eq-II-99}) } corresponding to the $j$-th cell. The part of this vector involving 
 $W|_{\sigma_{t_2}}^{\sigma_{t_1}}(\nv_j)W^{*}|_{\sigma_{t_2}}^{\sigma_{t_1}}(\nabla E_{p,\sigma_{t_1}})$ has the form
 \beqa
\Psi_{j, \si,\sip}(t,s):=     \mathcal{W}_{\sip }(\nv_j,s)
	\, e^{i\gamma_{\sip}(\nv_j,s)}
	\, e^{-i\widehat{H}_{\sip}s}
  	\, W|_{\si}^{\sip}(\nv_j \, ; \, \nabla E_{p,\sip })
	\, \psi_{j,\sip}^{(t)}\,,
 \eeqa
for $t=t_1,s=t_1$, $\sip=\si_{t_1}$, $\si=\si_{t_2}$. {\mc (We used here (\ref{eq-II-117}))}. We will {\mc study these vectors more generally for $s\geq t_1$}. We  consider the scalar product
\beqa
M_{l,j,\si,\sip}(t,s):=\lan \Psi_{l, \si,\sip}(t,s),\Psi_{j, \si,\sip}(t,s)\ran= \lan e^{i \widehat{H}_{\sip}s}\Psi_{l, \si,\sip}(t,s),
e^{i \widehat{H}_{\sip}s}\Psi_{j, \si,\sip}(t,s)\ran \label{two-representations-of-M}
\eeqa
and write
\beqa
M_{l,j,\si,\sip}(t,t)\1&=&\1(M_{l,j,\si,\sip}(t,t)-M_{l,j,\si,\sip}(t,s))+M_{l,j,\si,\sip}(t,s)\non\\
\1&=&-\1 \int_{t}^s ds'\, \pa_{s'}M_{l,j,\si,\sip}(t,s')+ M_{l,j,\si,\sip}(t,s). \label{two-terms}
\eeqa
For the integrand in (\ref{two-terms}) we use the second representation of  $M_{l,j,\si,\sip}(t,s')$ in (\ref{two-representations-of-M}) and consider the expression:
\beqa
\pa_s ( e^{i \widehat{H}_{\sip}s}\Psi_{j, \si,\sip}(t,s))
\1&=&\1
    i \, e^{i\widehat{H}_{\sip  }s}
    \, \mathcal{W}_{\sip}(\nv_j,s)
    \, \lambda^2\int_{\mathbb{R}^3\setminus\mathcal{B}_{\sip  }} 
   d^3k \,\chi_{\kappa}(k)^2
    \frac{\cos(k \cdot x-|k|s)}
    {2|k|^{2}(1-e_{k}\cdot\nv_j)} \times\nonumber\\
    & &\quad\quad\quad\quad\quad\quad\quad\quad\quad
    \times \,
    e^{-iE_{p,\sip}s}
    \, e^{i\gamma_{\sip}(\nv_j,s)}
    \, W|_{\si}^{\sip}(\nv_j \, ; \, \nabla E_{p,\sip })
    \, \psi_{j,\sip}^{(t)}\label{tail}\\
  & &+
    \, i \, e^{i\widehat{H}_{\sip}s}
    \, \mathcal{W}_{\sip}(\nv_j,s)
    \, \frac{d\gamma_{\sip}(\nv_j,\nabla E_{p,\sip},s)}{ds} \, \times
    \nonumber\\
    & & \quad\quad\quad\quad\quad\quad\quad\quad\quad
    \times
    \, e^{i\gamma_{\sip}(\nv_j,s)}
    \, e^{-iE_{p,\sip}s}
    \, W|_{\si}^{\sip}(\nv_j \, ; \, \nabla E_{p,\sip  })
    \, \psi_{j,\sip}^{(t)}\,.
\eeqa
By commuting $W|_{\si}^{\sip}(\nv_j \, ; \, \nabla E_{p,\sigma_{t_1}})$ to the left we can write 
\beqa
& &\|\pa_s ( e^{i \widehat{H}_{\sigma_{t_1}}s}\Psi_{j, \si,\sip}(t,s))\| \leq \|\pa_s  \widehat{\psi}^{(\sip)}_{2,\sip,j}(t,s)\| \non\\
& &\ph{44}+\bigg\|  \big[\la^2\int_{\mathbb{R}^3\setminus\mathcal{B}_{\sip  }} 
   d^3k \,\chi_{\kappa}(k)^2
    \frac{\cos(k \cdot x-|k|s)}
    {2|k|^2(1-e_{k}\cdot\nv_j)},   W|_{\si}^{\sip}(\nv_j \, ; \, \nabla E_{p,\sip  })\big]  
      e^{i\gamma_{\sip}(\nv_j,s)} e^{-iE_{p,\sip}s}\psi_{j,\sip}^{(t)}\|,
            \eeqa
where we used the notation from (\ref{single-particle-vector-for-I}) with an added hat to indicate that now the Hamiltonian $\wh{H}_{\sip}$ is used.
By Theorem~\ref{non-diagonal-theorem} and Lemma~\ref{auxiliary-commutator-lemma} we obtain for $\peps$ sufficiently small
\beqa
   \|\pa_s ( e^{i \widehat{H}_{\sigma_{t_1}}s}\Psi_{j, \si,\sip}(t,s))\|\leq     \fr{c_{s}^2}{{\mc \si}^{\de_{\la_0}}}\bigg(s(\si^{\S}_{s})^{3}+  \fr{(\si^{\S}_{s})^{\de}}{s}+\fr{1}{(\si^{\S}_{s})^{1+\de} {s}^2}  \bigg)\leq \fr{c}{{\mc \si}^{\de_{\la_0}} s^{\eta+1}}
\eeqa
for some $\eta>0$ independent of $\la_0$. Hence the first term in (\ref{two-terms}) can be bounded by $\fr{c}{{\mc \si}^{\de_{\la_0}} t^{\eta}}$, uniformly in $s$. The
last term on the r.h.s. of (\ref{two-terms}) converges to zero as $s\to \infty$ as one can conclude from Lemma~\ref{basic-clustering-lemma}, off-diagonality of the considered expression
and property (\ref{eq-II-103}). By setting back $t=t_1$, ${\si=\si_{t_2}}$ and choosing $\la_0$ sufficiently small we complete the proof of (\ref{eq-II-99}) and hence also of (\ref{eq:IRchange-ineq-1}).
}

%%%%%%%%%%%%%%%%%%%%%%%%%%%%%%%%%%%%%%%%
{\subsection{Auxiliary lemmas}}
%%%%%%%%%%%%%%%%%%%%%%%%%%%%%%%%%%%%%%%%
%%%%%%%%%%%%%%%%%%%%%%%%%%%%%%%%%%%%%%%%%
\bel\label{phases-shift-cut-off} Let $1/2<\al<1$ and $\si^{\S}_{\tau}=\kal \tau^{-\al}$.   Set $t_{\si}:=\min(t, (\kal/\si)^{1/\al})$ and define as in (\ref{slow-cutoff})
\beqa
\ga_{\si}(\nv_j,\nabla E_{2,p,\si}, t)= -\int_1^{t_{\si}}\bigg\{\int_{\si}^{\si_{\tau}^\S}d|k| \int d\Omm(e_k)\, \vv_2^{\si}(k)^2(2|k|)
\bigg(\fr{\cos(k\cdot\nabla E_{2,p,\si}\tau-|k|\tau)}{1-\hatk\cdot \nv_j }\bigg)  \bigg\} d\tau.    \label{slow-cutoff-one-x}
\eeqa
Then, for ${\mc p, p-\ti{k}\in S}$ and $\si_{t_1}=:\sip>\si:=\si_{t_2}$ and $t:=t_1$, the following holds
\beqa
& &|\ga_{\si'}(\nv_j, \nabla E_{2,p,\sip}, t)   t)(p)-\ga_{\si}(\nv_j, \nabla E_{2,p,\si}, t)  t)(p)|\leq c\sip t+c(\sip)^{1/4}  t^{2(1-\al)},\label{App-I-gamma-one} \\
  & &  \big| \, \gamma_{\sip}(\nv_j,\nabla E_{2, p,\sip}, t)  -  \gamma_{\sip}(\nv_j,\nabla E_{2,p-\ti{k},\sip}, t) \, \big|
     \leq  c|\ti{k}|. \label{App-I-gamma-two}
    \eeqa
\eel
%%%%%%%%%%%%%%%%%%%%%%%%%%%%%%%%%%%%%%%%%
\proof As for (\ref{App-I-gamma-one}), let us define the expression
\beqa
F(\si,\tau)=\int  d\Omm(\hatk) d|k|\, \ti\chi_{[\si, \si_{\tau}^\S ]}(|k|)^2 
\bigg(\fr{\cos(|k|(\hatk\cdot\nabla E_{2,p,\si}\tau-\tau))}{1-\hatk\cdot \nv_j }\bigg)
\eeqa
and note that  the difference on the l.h.s. of (\ref{App-I-gamma-one})  has the following general structure
\beqa
\int_1^{t_{\si}}d\tau\, F(\si,\tau)-\int_1^{t_{\sip}}d\tau\, F(\sip,\tau)=
 \int_{t_{\sip}}^{t_{\si}}d\tau\, F(\si,\tau)+\int_1^{t_{\sip}}d\tau\, (F(\si,\tau)- F(\sip,\tau)). \label{F-sigma}
\eeqa
Under our assumptions $t\leq (\kal/ \si)^{1/\al}$ and $t\leq (\kal/\sip)^{1/\al}$ so $t_{\si}=t_{\sip}=t$ and the first term on the r.h.s. of (\ref{F-sigma})  disappears. 
We consider the difference
\beqa
& &|F(\si,\tau)-F(\sip,\tau)|\leq\int  d\Omm(\hatk) d|k|\, \ti\chi_{[\si, \sip]}(|k|)^2 
\bigg(\fr{\cos(|k|(\hatk\cdot\nabla E_{2,p,\si}\tau-\tau))}{1-\hatk\cdot \nv_j }\bigg) \non\\
& &+\int  d\Omm(\hatk) d|k|\, \ti\chi_{[\sip, \si_{\tau}^\S ]}(|k|)^2 
\bigg(\fr{\cos(|k|(\hatk\cdot\nabla E_{2,p,\si}\tau-\tau))}{1-\hatk\cdot \nv_j }-\fr{\cos(|k|(\hatk\cdot\nabla E_{2,p,\sip}\tau-\tau))}{1-\hatk\cdot \nv_j } \bigg).\ \ 
\eeqa
This gives, by (\ref{gradient-estimate-shift-cut-off}),
\beqa
|F(\si,\tau)-F(\sip,\tau)|\1&\leq&\1 c\sip+ c(\sip)^{1/4} \int_0^{\si_{\tau}^{\S}} d|k|\, |k|\tau\non\\
\1&\leq&\1 c\sip+ c(\sip)^{1/4} \tau (\si_{\tau}^{\S})^2=c\sip+c(\sip)^{1/4}\tau^{1-2\al}. 
\eeqa
Thus we have
\beqa
|\ga_{\si'}(\nv_j,t)(p)-\ga_{\si}(\nv_j,t)(p)|\leq  c\sip t+c(\sip)^{1/4} t^{2(1-\al)},
\eeqa
which concludes the proof of (\ref{App-I-gamma-one}).

As for (\ref{App-I-gamma-two}),  it follows from (\ref{phase-first-derivative}) via the Taylor expansion. \qed
%%%%%%%%%%%%%%%%%%%%%%%%%%%%%%%%%%%%%%
\bel \label{I-lemma-tx}Consider the following expression for $|\ti k|\leq \ka$ and $0<\si\leq \kal$.
\beqa
I:=(e^{-i(|\ti k|t-\ti{k}\cdot x)}-\charf) \, \Phi_{j,\sigma}^{(t)}, \quad   \Phi_{j,\sigma}^{(t)}:=\Pi^*\int^{\oplus} \, d^3p\,  e^{-iE_{p,\sigma}t}  h_{j}^{(t)}(p) \, \phi_{p,\sigma}.
\eeqa
Then, for $\peps$ sufficiently small 
\beqa
\| \, I \,  \|  
\leq  \fr{c}{\si^{\de_{\la_0}}}|\ti k|  t^{1-\fr{3}{2}\peps }. \label{I-estimate-I}
\eeqa
\eel
%%%%%%%%%%%%%%%%%%
\proof We decompose this expression as follows
\beqa
I=  (e^{-i|\ti k|t}-1)e^{i\ti{k}\cdot x} \Phi_{j,\sigma}^{(t)}+   (e^{i\ti{k}\cdot x}   -1)\Phi_{j,\sigma}^{(t)}=:I_1+I_2. 
\eeqa 
By obvious estimates, including the volume of the cube in the partition, we get
\beqa
\|I_1\|\leq c|\ti k|  t^{1-\fr{3}{2}\peps }.
\eeqa

Now for  $\phi_{p,\si}=\{f^{m}_{\mrm{w},p,\si} \}_{m\in\nat_0}$ we define, analogously as is (\ref{renormalized})
\beqa
\nr_{\mrm{w},\si}^*(h_j^{(t)}):=\sum_{m=0}^{\infty}\fr{1}{\sqrt{m!}}\int d^3p\,d^{3m}k\,h_j^{(t)}(p) f^{m}_{\mrm{w},p,\si}(k)\bb^*(k)^m \, \nn^*(p- \unk).
\eeqa
We  write, considering that $e^{i\ti{k}\cdot x} \eta^*(p)\Om=\eta^*(p+\ti{k})\Om$,
\beqa
I_2\1&=&\1(e^{i\ti{k}\cdot x}-\charf) \nr_{\mrm{w},\si}^*(h_j^{(t)})\Om\non\\
 \1&=&\1 \sum_{m=0}^{\infty}\fr{1}{\sqrt{m!}}\int d^3p\,d^{3m}k\,\,   \chi_j^{(t)} (p)\,\bigg(h_j^{(t)}(p-\ti k) f^{m}_{\mrm{w},p-\ti{k},\si}(k)-  h_j^{(t)}(p) 
 f^{m}_{\mrm{w},p,\si}(k)\bigg)  \bb^*(k)^m \, \nn^*(p- \unk)\Om\,\,\ \ \ \  \ \ \ \ \ \ \ \\
 \1&=&\1 \sum_{m=0}^{\infty}\fr{1}{\sqrt{m!}}\int d^3p\,d^{3m}k\,\, \chi^{(t)}_j(p)\,\ti k\cdot  \nabla_{\ti k'}\bigg(   h_j^{(t)}(p-\ti k') f^{m}_{\mrm{w},p-\ti{k}',\si}(k)
 \bigg)  \bb^*(k)^m \, \nn^*(p- \unk)\Om,
 \eeqa
where  $\chi_j^{(t)}(p)=\mathbf{1}_{\Ga^j\cup \Ga^j+\ti k}(p)$ and  in the last step we used the Taylor expansion and $\ti k'$ is some vector between $0$ and $\ti k$. Using Proposition~\ref{modified-wavefunctions-lemma} and Lemma~\ref{h-theta},
we get
\beqa
|\nabla_{\ti k'}\big(   h_j^{(t)}(p-\ti k') f^{m}_{\mrm{w},p-\ti{k}',\si}(k)  \big)|\leq \si^{-\de_{\la_0}} \theta^{-1} \fr{c}{\sqrt{m!}} g^m_{\si}(k)\leq   \fr{ct^{8\peps}}{\si^{\de_{\la_0}}}   \fr{1}{\sqrt{m!}} g^m_{\si}(k).
\eeqa
Now Lemma~\ref{norms-of-single-particle-states} gives
\beqa
\|\, I_2\,\|\leq |\ti k|\fr{c}{\si^{\de_{\la_0}}} t^{8\peps} t^{-\fr{3}{2}\peps }. 
\eeqa
Thus, for $\peps$ sufficiently small, we get
\beqa
\| \, I \, \|\leq  c|\ti k|  t^{1-\fr{3}{2}\peps }+ |\ti k|\fr{c}{\si^{\de_{\la_0}}} t^{8\peps} t^{-\fr{3}{2}\peps }\leq  \fr{c}{\si^{\de_{\la_0}}}|\ti k|  t^{1-\fr{3}{2}\peps }.
\eeqa 
which concludes the proof. \qed

%%%%%%%%%%%%%%%%%%%
\bel\label{auxiliary-commutator-lemma} For $0<\si\leq \sip\leq \kal$ we have for any $0<\de<(\al^{-1}-1)$
\beqa
& &\bigg\|  \int_{\mathbb{R}^3\setminus\mathcal{B}_{\sip  }} 
   d^3k \,\chi_{\kappa}(k)^2\,
   \bigg[ \frac{\cos(k \cdot x-|k|s)}
    {|k|^2(1- e_k\cdot\nv_j)},   W|_{\si}^{\sip}(\nv_j \, ; \, \nabla E_{p,\sip  })\bigg]    \reta^*_{\sip}(h_{j,s}^{(t)} e^{i\ga_{\sip}(\nv_j,s)})\Om\|\non\\
    & &\ph{4444444444444444444444444444444444444}\leq \fr{c_{s}^2}{ {\mc \si}^{\de_{\la_0} }} \bigg( \fr{(\si^{\S}_s)^{\de}}{s}+\fr{1}{(\si^{\S}_s)^{1+\de} s^2}\bigg),
\eeqa
    where $W|_{\si}^{\sip}(\nv_j \, ; \, \nabla E_{p,\sip  })$ was defined in (\ref{W-W-combination-I}) and $ c_{s}:=c(\uneps(s) s^{-\peps})^{-1}=cs^{8\peps}$.
\eel
%%%%%%%%%%%%%%%%%%%
\proof We define the following functions
\beqa
f(k):=\fr{ \chi^2_{[\sip, \ka)}(k) }{|k|^2(1- e_k\cdot\nv_j)}, \quad F_p(k'):=\la\fr{\chi_{[\si,\sip]}(k')   }{\sqrt{2|k'|} |k'|} 
\bigg( \fr{1}{  (1-e_{k'}\cdot \nv_j)}- \fr{1}{(1-e_{k'}\cdot \nabla E_{p, \sip})}  \bigg),
\eeqa
so that setting $F_{p,x}(k):=e^{-ikx}F_p$, we have
\beqa
W|_{\si}^{\sip}(\nv_j \, ; \, \nabla E_{p,\sip  }) =   e^{a^*(F_{p,x})-a(F_{p,x})}=:W(F_{p,x}).
\eeqa
Clearly, it suffices to consider the expression
\beqa
& &\int d^3k \, f(k)e^{i|k|s} [e^{ik\cdot x},    W(F_{p,x})] \reta^*_{\si}(h_{j,s}^{(s)} e^{i\ga_{\sip}(\nv_j,s)})\Om\non\\
& &= \int d^3k \, f(k)e^{i|k|s} (W(F_{p-k,x})  - W(F_{p,x})) e^{ik\cdot x}\reta^*_{\sip}(h_{j,s}^{(t)} e^{i\ga_{\sip}(\nv_j,s)})\Om\non\\
& &=W(F_{p,x})  \int d^3k \, f(k)e^{i|k|s} (W(\De F_{p,k,x})  - 1) e^{ik\cdot x}\reta^*_{\sip}(h_{j,s}^{(t)} e^{i\ga_{\sip}(\nv_j,s)})\Om, \label{F-F-expression}
\eeqa
where we set $\De F_{p,k,x}:=F_{p-k,x}-F_{p,x}$ and we will also write $\De F_{p,k}:=F_{p-k}-F_{p}$. Using that $F_p$ is supported
in $\si\leq |k|\leq \sip$,  setting $h_{j,s}^{(t),\ga}:=h_{j,s}^{(t)} e^{i\ga_{\si}(\nv_j,s)}$, and recalling  definition~(\ref{renormalized}), we can write
\beqa
(\ref{F-F-expression})\1&=&\1 W(F_{p,x})  \sum_{\ell=1}^{\infty}\fr{1}{\ell !}  \int d^3k \, f(k) e^{-\h \|\De F_{p,k} \|^2  }  e^{i|k|s} 
a^*(\De F_{p,k,x})^{\ell} e^{ik\cdot x}\reta^*_{\sip}(h_{j,s}^{(t), \ga })\Om. \label{app-I-commutator-expression} 
\eeqa
Now using again definition~(\ref{renormalized}) and making use of the fact that $e^{ik\cdot x} \eta^*(p)\Om=\eta^*(p+k)\Om$
\beq
e^{ik\cdot x}\nr_{\si}^*(h_{j,s}^{(t), \ga } )\Om=\sum_{m=0}^{\infty}\fr{1}{\sqrt{m!}}\int d^3q\,d^{3m}r\,  h_{j,s}^{(t), \ga }(q-k)   f^{m}_{q-k,\si}(r)\bb^*(r)^m \, 
\nn^*(q- \un{r})\Om.
\eeq 
Furthermore, it is easy to see that
\beqa
& &a^*(\De F_{p,k,x})^{\ell} e^{ik\cdot x}\nr_{\sip}^*(h_{j,s}^{(t), \ga } )\Om\non\\
& &=\sum_{m=0}^{\infty}\fr{1}{\sqrt{m!}}\int d^3q\,d^{3(m+\ell)}\ti{r}\,  h_{j,s}^{(t), \ga }(q-k)  (\De F^{\ell}_{q,k} f^{m}_{q-k,\sip})(\ti{r})\bb^*(\ti{r})^{m+\ell} \, \nn^*(q- \un{\ti{r}})\Om \\
& &=\sum_{m'= \ell}^{\infty}\fr{1}{\sqrt{(m'-\ell)!}}\int d^3q\,d^{3m'}\ti{r}\,  h_{j,s}^{(t), \ga }(q-k)  (\De F^{\ell}_{q,k} f^{m'-\ell}_{q-k,\sip})(\ti{r})\bb^*(\ti{r})^{m'} \, \nn^*(q- \un{\ti{r}})\Om,
\eeqa
where $\ti{r}:=(r' , r)$,  and we set  
\beqa
(\De F^{\ell}_{q,k} f^{m'-\ell}_{q-k,\sip})(\ti{r}):=[\De F_{q,k}(r'_1)\ldots \De F_{q,k}(r'_{\ell}) f^{m'-\ell}_{q-k,\sip}(r)]_{\mc \mrm{sym}},
\eeqa
{\mc where $\mrm{sym}$ denotes symmetrisation over all variables.}  Coming back to (\ref{app-I-commutator-expression}), we can write
\beqa
(\ref{app-I-commutator-expression})= W(F_{p,x})\sum_{\ell=1}^{\infty} \fr{1}{\ell!} \sum_{m'= \ell}^{\infty}\fr{1}{\sqrt{(m'-\ell)!}} \int d^3q\,d^{3m'}\ti{r}\, G_{m'}^{\ell}(q, \ti{r})
    \bb^*(\ti{r})^{m'} \, \nn^*(q- \un{\ti{r}})\Om,
 \eeqa
where 
\beqa
G_{m'}^{\ell}(q, \ti{r}):=\int d^3k\, f(k)e^{i|k|s}  e^{-\h \|\De F_{q,k} \|^2  }   h_{j,s}^{(t), \ga }(q-k)  (\De F^{\ell}_{q,k} f^{m'-\ell}_{q-k,\sip})(\ti{r}). \label{G-expression}
\eeqa
Thus, by Lemma~\ref{norms-of-single-particle-states}, we have 
\beqa
\|(\ref{app-I-commutator-expression})\|^2\1&=&\1 \sum_{\ell=1}^{\infty} \fr{1}{\ell!}\sum_{\ell'=1}^{\infty} \fr{1}{\ell'!} \sum_{m= \ell}^{\infty}\fr{1}{\sqrt{(m-\ell)!}} \sum_{m'= \ell'}^{\infty}\fr{1}{\sqrt{(m'-\ell')!}}\non\\
\1& &\1\times \de_{m,m'} m! \int d^3 q\, d^{3m} \, \ti{r}\,   \ov{G}_{m}^{\ell'}(q, \ti{r})   G_{m}^{\ell}(q, \ti{r}). \label{main-L-2-formula}
\eeqa
To conclude, we need to establish suitable decay of (\ref{G-expression}) in $s$. In more explicit notation, we have
\beqa
G_{m}^{\ell}(q, \ti{r})\1&=&\1\int d^3k\, e^{i (|k| -  E_{q-k,\sip}) s}  
\fr{ \chi^2_{[\sip, \ka)}(k) }{|k|^2(1- e_k\cdot\nv_j)}  e^{-\h \|\De F_{q,k} \|^2  } h_{j}^{(t),\ga}(q-k)   (\De F^{\ell}_{q,k} f^{m-\ell}_{q-k,\sip})(\ti{r}). 
\eeqa
We set
\beqa
\ti{F}^{m,\ell}(q,k,\ti r):= \fr{ \chi^2_{[\sip, \ka)}(k) }{|k|^2(1- e_k\cdot\nv_j)}  e^{-\h \|\De F_{q,k} \|^2  } h_{j}^{(t),\ga}(q-k)   (\De F^{\ell}_{q,k} f^{m-\ell}_{q-k,\sip})(\ti{r}) \label{ti-F-end}
\eeqa
and decompose  $G_{m}^{\ell}$ into the sum of the following two parts
\beqa
& &G_{m}^{\ell,1}(q, \ti{r}):= \int_{ \sip\leq |k|\leq  \si_{\S,s} } d^3k \, e^{i (|k| -  E_{q-k,\sip}) s} \ti{F}^{m,\ell}(q,k,\ti r), \\
& &G_{m}^{\ell,2}(q, \ti{r}):= \int_{ |k| \geq \si_{\S,s}  } d^3k\, e^{i (|k| -  E_{q-k,\sip}) s} \ti{F}^{m,\ell}(q,k,\ti r), 
\eeqa
where $ \si_{\S,t}:=\max\{ \si^{\S}_t, \sip\, \}$.

First, we consider $G_{m}^{\ell,1}(q, \ti{r})$, following the steps of the proof of Lemma~\ref{two-second-contribution-phase-rest}.
We note that $\ti F^{m, \ell}$ defined in (\ref{ti-F-end}) plays the role of  $F^m$ stated in (\ref{F-m}).  We will check that $\ti F^m$ satisfies 
{\mc a similar} bound as given in (\ref{two-F-general-bound}) for $F^m$, namely  for any $0\leq |\be|\leq 1$,
\beqa
& &|\pa^{\be}_k \ti{F}^{m,\ell}(q,k,\ti r)|\leq \fr{c_{s}}{((\sip)^{\de_{\la_0}})^{|\be|} |k|^{1+|\be|}} \fr{1}{\sqrt{\mc (m-\ell)!}}g_{{\mc \si}}^m(\ti r),
\label{ti-F-general-bound}
\eeqa
where $c_{s}:=c(\uneps(s) s^{-\peps})^{-1}$. In fact, we have by the boundedness of the second derivative of $E_{q,\sip}$
w.r.t. $q$ (cf.~Theorem~\ref{preliminaries-on-spectrum})
\beqa
\De F_{q,k}(r'_i)\1&:=&\1\la\fr{\chi_{[\si,\sip]}(r'_i)   }{\sqrt{2|r'_i|} |r'_i|} 
\bigg(   \fr{ e_{r'_i}\cdot (\nabla E_{q,\si} - \nabla E_{q-k,\si} )  }{ (1-e_{r'_i}\cdot \nabla E_{q-k, \sip})  (1-e_{r'_i}\cdot \nabla E_{q, \sip})} \bigg)=
\la\fr{\chi_{[\si,\sip]}(r'_i)   }{\sqrt{2|r'_i|} |r'_i|} O(|k| ),  \quad \\
\pa_k^{\be} (\De F_{q,k})(r'_i) \1&=&\1 \la\fr{\chi_{[\si,\sip]}(r'_i)   }{\sqrt{2|r'_i|} |r'_i|} O(1 ).
\eeqa
Thus by standard arguments (see Theorem~\ref{main-theorem-spectral} and Lemma~\ref{partition-lemma}) we obtain (\ref{ti-F-general-bound}). 
Now the arguments from the proof of Lemma~\ref{two-second-contribution-phase-rest} give
\beqa
|G_{m}^{\ell,1}(q, \ti{r})|\leq   \fr{c_t}{(\sip)^{\de_{\la_0}}}\fr{\si^{\S}_s}{s}\fr{1}{\sqrt{\mc (m-\ell)!}} g_{{\mc\si}}^m(\ti r). \label{I-last-lemma-bound-one}
\eeqa
Next, we consider $G_{m}^{\ell,2}$ following the arguments from Lemma~\ref{two-second-contribution-phase-rest-x}. We define
\beqa
\ti{F}^{m,\ell}(q,\ti r, |k|, e):= \fr{ \ti{\chi}_{[\sip, \ka)}(|k|)^2 }{(1- e\cdot\nv_j)}  e^{-\h \|\De F_{q,|k|e} \|^2  } h_{j}^{(t),\ga}(q-|k|e) 
(\De F^{\ell}_{q,|k|e} f^{m-\ell}_{q-|k|e,\sip})(\ti{r}), \label{ti-F-m}
\eeqa
%{\mc SYMMETRIZATION?}
where we chose the spherical coordinates with the $z$-axis in the direction of $q$, 
and set $\pa_{|k|}V(|k|,\nee):=\nee\cdot \nabla E_{q-|k|\nee,\sip}+1$.  We introduce
\beqa
\ti{f}^{\si_{\S,s}}(\nee):= \fr{\ti{F}^{m,\ell}(q,\ti r,|k|,\nee)}{\pa_{|k|}V(|k|,\nee)}\big|_{|k|=\si_{\S,s}}, \quad \ti{f}_{|k|}(\nee):=\pa_{|k|}  
\bigg( \fr{ \ti{F}^{m,\ell}(q,\ti{r},|k|,\nee)}{\pa_{|k|}V(|k|,\nee)}\bigg).
\eeqa
It is easy to verify the following bounds which correspond to (\ref{bounds-on-derivatives-part-one}), (\ref{bounds-on-derivatives-part-two}) 
%{\mc     $(m-\ell)!?$, $\sip\to \si$?}
\beqa
& & |\ti f^{\si_{\S,s}}(\nee)|\leq \fr{c}{\sqrt{{\mc (m-\ell)}!}} g^m_{{\mc \si}}(\ti r), \quad\quad |\pa_{\theta}\ti f^{\si_{\S,s}}(\nee)|\leq \fr{c_s}{ (\sip)^{\de_{\la_0} }}\fr{1}{\sqrt{  {\mc (m-\ell)}   }!}  g^m_{{\mc \si}}(\ti r),\\
& &|\ti f_{|k|}(\nee)|\leq  \fr{1}{(\sip)^{\de_{\la_0}}}\fr{c_s}{\sqrt{\mc (m-\ell)!}} g^m_{{\mc \si}}(\ti r),\quad |\pa_{\theta}\ti f_{|k|}(\nee)|\leq  \fr{1}{(\sip)^{\de_{\la_0}}}\fr{c_{1,s}}{\sqrt{\mc (m-\ell)!}} g^m_{{\mc \si}}(\ti r),
\eeqa
where $c_{1,s}:=c((\uneps(s) s^{-\peps})^{-2}+s^{1-\al})$. Thus the arguments from the proof of Lemma~\ref{two-second-contribution-phase-rest-x}
give 
\beqa
|G_{m}^{\ell,2}(q, \ti{r})|\leq \fr{c_{s}^2}{ (\sip)^{\de_{\la_0} }} \bigg( \fr{(\si^{\S}_s)^{\de}}{s}+\fr{1}{(\si^{\S}_s)^{1+\de} s^2}\bigg)
\fr{1}{\sqrt{\mc (m-\ell)!}} g^m_{{\mc \si}}(\ti r).
\label{I-last-lemma-bound-two}
\eeqa
Coming back to (\ref{main-L-2-formula}) and using (\ref{I-last-lemma-bound-two}), (\ref{I-last-lemma-bound-one}) and the fact that $G_{m}^{\ell}(q, \ti{r})$
are compactly supported in $q$, we obtain using Lemma~\ref{summation-ell}
\beqa
\|(\ref{app-I-commutator-expression})\|\leq \fr{c_{s}^2}{ {\mc \si}^{\de_{\la_0} }} \bigg( \fr{(\si^{\S}_s)^{\de}}{s}+\fr{1}{(\si^{\S}_s)^{1+\de} s^2}\bigg)
\eeqa
for any $0<\de<(\al^{-1}-1)$. {\mc (We note that $(\sip)^{\de_{\la_0}}$ turned into $\si^{\de_{\la_0}}$ when we applied  Lemma~\ref{summation-ell})}. 
This concludes the proof. \qed

%%%%%%%%%%%%%%%%%%%%%%%%%%%%%%%%%%%%%%%%%%%%%%%%%%%%%%%%%%%%%%%%%%%%%%%%%%%%%%%%


\begin{thebibliography}{LNT2}


%%%%%%%%%%%%%%%%%%%%%%  A  %%%%%%%%%%%%%%%%%%%%%%
\bibitem[AD15]{AD15} S. Alazzawi and W. Dybalski. \emph{Compton scattering in the Buchholz-Roberts framework of relativistic QED}.
Lett. Math. Phys.  \textbf{107},   (2017)  81--106.


%\bibitem[AH12]{AH12} A. Abdesselam and D. Hasler, 
%\emph{Analyticity of the ground state energy for massless Nelson models}.
%Commun. Math. Phys. \bf 310\rm, (2012) 511--536.

\bibitem[Al73]{Al73} S. Albeverio. \emph{Scattering theory in a model of quantum fields. I.} J. Math. Phys.  \bf 14\rm, (1973) 1800--1816.

\bibitem[Al72]{Al72} S. Albeverio. \emph{Scattering theory in a model of quantum fields II.} Helv. Phys. Acta  \bf 45\rm, (1972) 303--321.



%%%%%%%%%%%%%%%%%%%%%%  B  %%%%%%%%%%%%%%%%%%%%%%
%\bibitem[BCFS07]{BCFS07}  V. Bach, T. Chen, J. Fr\"ohlich, and I. M. Sigal, \emph{The renormalized electron mass in non-relativistic quantum electrodynamics}. J. Funct. Anal.  \bf 243\rm, (2007) 426-535.  
%%%%%%%%%%%%%%%%%%%%%%  C  %%%%%%%%%%%%%%%%%%%%%%

\bibitem[Bu77]{Bu77} D. Buchholz. \emph{Collision theory for massless bosons}.
Commun. Math. Phys. \bf 52\rm, (1977) 147--173.


\bibitem[CFP07]{CFP07}  T. Chen, J. Fr\"ohlich and A. Pizzo. \emph{Infraparticle scattering states in non-relativistic QED: 
I. The Bloch-Nordsieck paradigm.} Commun. Math. Phys. \bf 294\rm, (2010) 761--825.

\bibitem[CFP09]{CFP09}  T. Chen, J. Fr\"ohlich and A. Pizzo. \emph{Infraparticle scattering states in non-relativistic QED: II. Mass shell properties.} 
J. Math. Phys.  \bf 50\rm, (2009) 012103--012134. 
%%%%%%%%%%%%%%%%%%%%%%  D  %%%%%%%%%%%%%%%%%%%%%%

\bibitem[Da18]{Da18} T.N. Dam. \emph{Non-existence of ground states in the translation invariant Nelson model}. arXiv:1808.00088. 


%\bibitem[DG99]{DG99}  J.~Derezi{\'n}ski and C.~G{\'e}rard,
% \emph{Asymptotic completeness in quantum
%  field theory. Massive {Pauli-Fierz} {Hamiltonians}}, Rev. Math. Phys.
 % \textbf{11}, (1999) 383--450.


%\bibitem[DG04]{DG04}  J.~Derezi{\'n}ski and  C. G{\'e}rard,
%\emph{Scattering theory of infrared divergent Pauli-Fierz Hamiltonians}.
%Ann. Henri Poincar{\'e} \textbf{5}, (2004) 523--577.

\bibitem[Du15]{Du15} P. Duch and A. Herdegen. \emph{Massless asymptotic fields and Haag-Ruelle theory}.  Lett. Math. Phys. \textbf{105},  (2015) 245--277.

\bibitem[Du17]{Du17} M. Duell. \emph{Strengthened Reeh-Schlieder property and scattering in quantum field theory without mass gaps}.  Commun. Math. Phys. \textbf{352}, (2017) 935--966.

\bibitem[Dy05]{Dy05} W. Dybalski. \emph{Haag-Ruelle scattering theory in presence of
massless particles}.  Lett. Math. Phys. \bf 72\rm, (2005) 27--38.

\bibitem[Dy17]{Dy17} W. Dybalski.   \emph{From Faddeev-Kulish to LSZ. Towards a non-perturbative description of colliding electrons}.  Nuclear Physics B \textbf{925}, (2017) 455--469. 

\bibitem[DP13.1]{DP12.0} W. Dybalski and A. Pizzo.
\emph{Coulomb scattering in the massless Nelson model I. Foundations of two-electron scattering.}   J. Stat. Phys. \textbf{154}, (2014) 543--587.
%Preprint: 
%arXiv:1302.5001. To appear in J. Stat. Phys.


\bibitem[DP13.2]{DP12} W. Dybalski and A. Pizzo.
\emph{Coulomb scattering in the massless Nelson model II. Regularity of ground states.}  \emph{Rev. Math. Phys}. \textbf{31}, (2019) 1950010. %Preprint: arXiv:1302.5012. 

\bibitem[DP17.1]{DP16} W. Dybalski and A. Pizzo. \emph{Coulomb scattering in the massless Nelson model III. 
Ground state wave functions and non-commutative recurrence relations}.  \emph{Ann. Henri Poincar\'e} \textbf{19}, (2018) 463-514.

%%%%%%%%%%%%%%%%%%%%%%  E  %%%%%%%%%%%%%%%%%%%%%%

%%%%%%%%%%%%%%%%%%%%%%  F  %%%%%%%%%%%%%%%%%%%%%%
\bibitem[Fr73]{Fr73} J. Fr\"ohlich. \emph{On the infrared problem in a model of scalar electrons and massless, scalar bosons.} %(French summary)
Ann. Inst. H. Poincar\'e Sect. A (N.S.) \bf 19\rm, (1973) 1--103. 

\bibitem[Fr]{Fr73.1} J. Fr\"ohlich, unpublished notes.


\bibitem[Fr74]{Fr74.1} J. Fr\"ohlich. \emph{Existence of dressed one electron states in a class of persistent models}. 
Fortschr. Phys.   \bf 22\rm, (1974) 158--198. 

\bibitem[FGS01]{FGS01} J.~Fr{\"o}hlich, M.~Griesemer and B.~Schlein.
  \emph{Asymptotic electromagnetic fields in models of quantum mechanical matter interacting with the quantized radiation field}.
  Adv. Math.  \textbf{164}, (2001) 349--398.



%\bibitem[FGS02]{FGS02} J.~Fr{\"o}hlich, M.~Griesemer and B.~Schlein,
 % \emph{Asymptotic completeness for Rayleigh scattering}.
%  Ann. Henri Poincar{\'e}  \textbf{3}, (2002) 107--170.

\bibitem[FGS04]{FGS04} J.~Fr{\"o}hlich, M.~Griesemer and B.~Schlein.
  \emph{Asymptotic completeness for {C}ompton scattering}.
  Commun. Math. Phys. \textbf{252}, (2004) 415--476.







\bibitem[FP10]{FP10} J. Fr\"ohlich and A. Pizzo. \emph{Renormalized electron mass in non-relativistic QED}. 
Commun. Math. Phys. \bf 294\rm,  (2010) 439--470.

%%%%%%%%%%%%%%%%%%%%%%  G  %%%%%%%%%%%%%%%%%%%%%%

%%%%%%%%%%%%%%%%%%%%%%  H  %%%%%%%%%%%%%%%%%%%%%%
%\bibitem[Ha58]{Ha58} R. Haag, \emph{Quantum field theories with composite particles
%and asymptotic conditions}.  Phys. Rev. \bf 112\rm, (1958) 669--673.

\bibitem[HH08]{HH08} D. Hasler and I. Herbst. 
\emph{Absence of ground states for a class of translation invariant models of non-relativistic QED.}
Commun. Math. Phys. \bf 279\rm, (2008) 769--787.


\bibitem[HS95]{HS95}  M.~H{\"u}bner and H.~Spohn.
\emph{Radiative decay: nonperturbative approaches}.
Rev. Math. Phys. \textbf{7}, (1995) 363--387.


\bibitem[He14]{He14} A.~Herdegen. \emph{Infraparticle problem, asymptotic fields and Haag-Ruelle theory}.  Ann. Henri Poincar{\'e} \textbf{15}, (2014) 345--367.
%%%%%%%%%%%%%%%%%%%%%%  I  %%%%%%%%%%%%%%%%%%%%%%




%%%%%%%%%%%%%%%%%%%%%%  J  %%%%%%%%%%%%%%%%%%%%%%



%%%%%%%%%%%%%%%%%%%%%%  K  %%%%%%%%%%%%%%%%%%%%%%
\bibitem[KM12]{KM12} M. K\"onenberg and O. Matte.
\emph{The mass-shell in the semi-relativistic Pauli-Fierz model}.  Ann. Henri Poincar{\'e} \textbf{15}, (2014) 863--915.%Preprint: arXiv:1204.5123.


%%%%%%%%%%%%%%%%%%%%%%  L  %%%%%%%%%%%%%%%%%%%%%%




%%%%%%%%%%%%%%%%%%%%%%  M  %%%%%%%%%%%%%%%%%%%%%%



%%%%%%%%%%%%%%%%%%%%%%  N  %%%%%%%%%%%%%%%%%%%%%%



%%%%%%%%%%%%%%%%%%%%%%  O  %%%%%%%%%%%%%%%%%%%%%%




%%%%%%%%%%%%%%%%%%%%%%  P  %%%%%%%%%%%%%%%%%%%%%%
\bibitem[Pi03]{Pi03} A. Pizzo. \emph{One-particle (improper) states in Nelson's massless model.} 
Ann. Henri Poincar\'e~\bf 4\rm, (2003) 439--486.

\bibitem[Pi05]{Pi05} A. Pizzo. \emph{Scattering of an infraparticle: the one particle sector
in Nelson's massless models}.
Ann. Henri Poincar\'e  \bf 6\rm, (2005) 553--606.

%%%%%%%%%%%%%%%%%%%%%%  Q  %%%%%%%%%%%%%%%%%%%%%%



%%%%%%%%%%%%%%%%%%%%%%  R  %%%%%%%%%%%%%%%%%%%%%%
%\bibitem[Ru62]{Ru62} D. Ruelle,  \emph{On the asymptotic condition in quantum 
%field theory}. Helv. Phys. Acta \bf 35\rm, (1962) 147--163.

\bibitem[RS2]{RS2} M. Reed and B. Simon. \emph{Methods of modern mathematical physics II. Fourier analysis, self-adjointness}.
Academic Press, 1980.



%%%%%%%%%%%%%%%%%%%%%%  S  %%%%%%%%%%%%%%%%%%%%%%

%\bibitem[Sp97]{Sp97} H.~Spohn, 
%\emph{Asymptotic completeness for Rayleigh scattering}. J.~Math. Phys. \textbf{38},  (1997) 2281--2296 .


%%%%%%%%%%%%%%%%%%%%%%  T  %%%%%%%%%%%%%%%%%%%%%%
%\bibitem[St]{St} A. Strominger, \emph{Lectures on the infrared structure of gravity and gauge theory}. arXiv:1703.05448
%%%%%%%%%%%%%%%%%%%%%%  U  %%%%%%%%%%%%%%%%%%%%%%



%%%%%%%%%%%%%%%%%%%%%%  V  %%%%%%%%%%%%%%%%%%%%%%




%%%%%%%%%%%%%%%%%%%%%%  W  %%%%%%%%%%%%%%%%%%%%%%



%%%%%%%%%%%%%%%%%%%%%%  X  %%%%%%%%%%%%%%%%%%%%%%



%%%%%%%%%%%%%%%%%%%%%%  Y  %%%%%%%%%%%%%%%%%%%%%%



%%%%%%%%%%%%%%%%%%%%%%  Z  %%%%%%%%%%%%%%%%%%%%%%




\end{thebibliography}
\end{document}